\documentclass[preprint,12pt,showpacs,aps,amsmath,amssymb,nofootinbib,superscriptaddress,hyperref,showkeys]{revtex4} 

\usepackage[utf8]{inputenc}

\usepackage{epsfig} 
\usepackage{wasysym} 
\usepackage{mathrsfs} 
\usepackage{graphicx} 
\usepackage{amsfonts} 
\usepackage{amsbsy} 
\usepackage{amscd} 
\usepackage{pstricks} 
\usepackage{multirow}
\usepackage{slashed} 
\usepackage{tikz}
\usetikzlibrary{decorations.pathmorphing,decorations.markings}
\newcommand{\beq}{\begin{equation}}
\newcommand{\eeq}{\end{equation}}
\newcommand{\bea}{\begin{eqnarray}}
\newcommand{\eea}{\end{eqnarray}}
\newcommand{\beas}{\begin{eqnarray*}}
\newcommand{\eeas}{\end{eqnarray*}}
\newcommand{\bi}{\begin{itemize}}
\newcommand{\ei}{\end{itemize}}

\def\tev{\,{\ifmmode\mathrm {TeV}\else TeV\fi}}
\def\gev{\,{\ifmmode\mathrm {GeV}\else GeV\fi}}
\def\to{\rightarrow}

\allowdisplaybreaks

\begin{document}

\title{Pair production of heavy neutrinos in next-to-leading order QCD at the hadron colliders in the inverse seesaw framework}

\author{Arindam Das\footnote{arindam@kias.re.kr}}
\affiliation{School of Physics, KIAS, Seoul 130-722, Korea}
\affiliation{Department of Physics \& Astronomy, Seoul National University 1 Gwanak-ro, Gwanak-gu, Seoul 08826, Korea}
\affiliation{Korea Neutrino Research Center, Bldg 23-312, Seoul National University, Sillim-dong, Gwanak-gu, Seoul 08826, Korea}

\preprint{\today}

\begin{abstract}
The explanation of the small neutrino mass can be depicted using some handsome models like type-I and inverse seesaw where the Standard Model gauge singlet heavy right handed neutrinos are deployed.
The common thing in these two models is a lepton number violating parameter, however, its order of magnitude creates a striking difference between them making the 
nature of the right handed heavy neutrinos a major play factor. In the type-I seesaw a large lepton number violating parameter involves the heavy right handed neutrinos in the form of Majorana fermions while
 a small lepton number violating parameter being involved in the inverse seesaw demands the pseudo-Dirac nature of the heavy right handed neutrinos. Such heavy neutrinos are accommodated in these models through the sizable mixings with the Standard Model light neutrinos. In this paper we consider 
the purely inverse seesaw scenario to study the pair production of the pseudo-Dirac heavy neutrinos followed by their various multilepton decay modes through the leading branching fraction at the Leading Order and 
Next-to-Leading Order QCD at the LHC with a center of mass energy of 13 TeV and a luminosity of 3000 fb$^{-1}$. We also consider a prospective 100 TeV hadron collider with luminosities of 3000 fb$^{-1}$ and 30000 fb$^{-1}$ respectively to study the process.
Using anomalous multilepton search performed by the CMS at the 8 TeV with 19.5 fb$^{-1}$ luminosity we show prospective search reaches of the mixing angles for the three lepton and four lepton events at the 13 TeV LHC 
and 100 TeV hadron collider. 
\end{abstract}


\maketitle
\tableofcontents
\clearpage

\section{Introduction}
\label{sec:intro}
The Large Hadron Collider(LHC) has successfully discovered the Higgs boson at the Run-I. Such observations play important roles to establish the properties of the Standard Model (SM), however, 
there are still some open questions yet to be answered. One of them is related to the neutrino mass. In the SM the neutrinos are considered to be massless. Experimental observations \cite{Davis:1964zz, Bahcall:1964ya, Davis:1968cp,Davis:1969zz, Davis:1973hw} being supported by well motivated theoretical works \cite{Bahcall:1963ha, Bahcall:1964gx, Sears:1964zz, Bahcall:1966gz,Bahcall:1967gz, Bahcall:1968jvj, Bahcall:1964zk, Bahcall:1976zz, Bahcall:1978fa, Bahcall:1980qx, Bahcall:1981zh} hinted that the fact of `masslessness' of the SM neutrinos is questionable. Such developments led to another important idea in the form of a dimension five operator \cite{Weinberg:1979sa} within the SM which invited a drastic change in the viewpoint of the neutrino physics through the birth of seesaw/ type-I seesaw mechanism \cite{seesaw0,seesaw1,seesaw2,seesaw3,seesaw4,seesaw5,seesaw6}. The recent days experiments verified the existence of the tiny SM light neutrino mass and the flavor mixings \cite{Hirata:1991ub, Davis:1994jw,Neut1,Neut2,Neut3,Neut4,Neut5,Neut6}.

In case of the seesaw mechanism the SM gauge singlet heavy right handed Majorana neutrinos induce a dimension five operator to produce a very small light Majorana neutrino masses. In this case the heavy Majorana mass term is used as a suppression factor to generate the light neutrino mass. Depending upon the variation of the seesaw scale from the intermediate to the electroweak scales, the Dirac Yukawa coupling $(Y_{D})$ varies from the electron Yukawa coupling up to the top quark Yukawa coupling, e.g., from $\mathcal{O}(10^{-6})$ to $\mathcal{O}(1)$ respectively.  At the high energy hadron colliders such as LHC in Run-II with a center of mass energy 13 TeV \cite{Kawamoto} and a hadron collider at the center of mass energy 100 TeV \cite{Arkani-Hamed:2015vfh,Mangano:2016jyj,Contino:2016spe,Han}, such heavy neutrinos could be produced if they have masses in the TeV scale or below. Being singlets under the SM gauge group, the heavy neutrinos can only be produced at the colliders from the charged current and neutral current interactions mediated by the weak gauge bosons only through the mixings via the Dirac Yukawa couplings. Having the neutrino masses in the TeV scale or lower in the type-I seesaw model, the Dirac Yukawa coupling becomes extremely small, e. g, $\mathcal{O}(10^{-6})$. Which is an unavoidable obstacle to observe a versatile repertoire of the heavy right handed Majorana neutrinos in this milieu.

There is a possible alternative to us which arrives in the form of inverse seesaw mechanism \cite{inverse-seesaw1, inverse-seesaw2, Malinsky:2009df,Hu:2011ac, Kang:2006sn} where a small lepton number violating parameter becomes responsible to generate the tiny neutrino mass. Due to the smallness of the lepton number violating parameter, the heavy neutrino is pseudo-Dirac in nature. The Dirac Yukawa coupling between the SM lepton doublets, Higgs doublet and the heavy right handed neutrino could be $\mathcal{O}(1)$ while satisfying the neutrino oscillation data. Such pseudo-Dirac neutrinos can be produced at the 13 TeV LHC and a hadron collider at the center of mass energy 100 TeV which can be the `The Bishop's Candlesticks' in the voyage of the Beyond the Standard Model (BSM) physics. 

Apart from the seesaw and the inverse seesaw mechanisms \cite{BhupalDev:2010he, An:2011uq, BhupalDev:2012ru, Banerjee:2015gca, Dev:2009aw, BhupalDev:2012zg,Dev:2013wba, Alva:2014gxa, Dev:2015kca, Dev:2016gvv, Antusch:2014woa,Antusch:2015mia,Antusch:2015rma, Antusch:2015gjw, Antusch:2016brq, Antusch:2016vyf, Fischer:2016rsh, Antusch:2016qby, Antusch:2016ejd, Mondal:2016kof, Mondal:2016czu, Banerjee:2013fga, Mondal:2012jv, Basso:2013jka, Hessler:2014ssa, Matsumoto:2010zg, ILC1, Gluza:2016qqv, Gluza:2015goa,  Arganda:2015ija, Abdallah:2015hma, Abdallah:2015uba, Khalil:2015naa, Elsayed:2011de, Abbas:2015zna, Khalil:2015wua, Huitu:2008gf, Keung:1983uu, Datta:1992qw, Datta:1993nm, Fong:2011xh, Dias:2011sq, Ibarra:2011xn, Batell:2015aha, Leonardi:2015qna, Dib:2015oka,Dib:2016wge,Dib:2014iga,Dib:2014pga, Cvetic:2012hd, Cvetic:2010rw, Cvetic:2016fbv, Agashe:2016ttz, Bambhaniya:2016rbb, Rink:2016knw, Ahn:2014gva, Ahn:2016hbn, Canetti:2012vf, Drewes:2012ma, Drewes:2013gca, Drewes:2015jna, Drewes:2015vma, Drewes:2016jae, Drewes:2016fjh,Rasmussen:2016njh, Lindner:2014oea, Lindner:2016lpp,Humbert:2015yva,Humbert:2015epa, Queiroz:2016qmc}, the SM can be extended by an SU(2) triplet scalar in the type-II seesaw \cite{Magg, Cheng:1980qt, Lazarides:1980nt, Mohapatra, Nesti, Chun:2012zu, Chun:2012jw, Dev:2013ff, Chun:2013vma, Gu, Haba:2016zbu,Han:2015hba, Han:2015sca, BhupalDev:2011gi} model. The triplet scalar has neutral and charged multiplets giving rise to not only the neutrino mass but also a rich phenomenology in terms of the multilepton modes \cite{Kumar:2015tna, Mitra:2016wpr, Babu:2016rcr, Hays:2017ekz}. There is another type of neutrino mass generation mechanism, commonly known as the  type-III seesaw where an SU(2) triplet fermion is introduced to extend the SM to generate the neutrino mass through lepton number violation \cite{Foot, Franceschini:2008pz, Abada:2008ea, Aguilar-Saavedra:2013twa, Eboli:2011ia, Ahn:2011pq, Biggio:2011ja, France1}. The production of such triplets at the NLO-QCD level has been studied in \cite{Ruiz:2015zca}. Apart from these possibilities there are studies on the higher-dimensional operators at the TeV scale which can be considered to generate the neutrino mass. Such models can also be tested at the LHC \cite{Babu:2009aq, Bambhaniya:2013yca, Gogoladze:2008wz} in near future.

In this paper we consider the purely inverse seesaw model to produce the heavy neutrinos in pair at the hadron colliders from the neutral current interaction. The draft is arranged as follows. In Sec.~2 we discuss the neutrino mass generation mechanisms and the interaction Lagrangians between the right handed heavy neutrinos and the SM gauge bosons. In Sec.~3 we give the scale dependent Leading Order (LO) and Next-to-Leading Order QCD (NLO-QCD) cross sections of the pseudo-Dirac heavy neutrino pair production at the hadron colliders. We discuss the various decay modes of the heavy neutrinos at the scale dependent LO and NLO-QCD processes in Sec.~4. We give prospective scale dependent upper bound on the mixing angles between the heavy neutrinos and the SM light neutrinos at the 13 TeV LHC with 3000 fb$^{-1}$ luminosity and at a 100 TeV hadron collider with 3000 fb$^{-1}$ and 30000 fb$^{-1}$ luminosities respectively. To do this we use the anomalous multilepton search performed by the CMS at the 8 TeV LHC with 19.5 fb$^{-1}$ luminosity\cite{Chatrchyan:2014aea}. The conclusions are written in Sec.~6.

\section{Neutrino mass generation mechanism}
\label{sec:numass}

An SM-singlet right handed heavy Majorana neutrino $N_R^{\beta}$ , with     
 $\beta$ as the flavor index , is introduced in the type-I seesaw model. 
 The SM lepton doublets $(\ell_{L}^{\alpha})$, SM Higgs doublet $\Phi$ and $N_R^{\beta}$ are
coupled through the Dirac Yukawa coupling $Y_{D}$.
The relevant part of the Lagrangian can be written as
\bea
\mathcal{L}^{\rm{seesaw}} \supset -Y_D^{\alpha\beta} \overline{\ell_L^{\alpha}}\Phi N_R^{\beta} 
                   -\frac{1}{2} m_N^{\alpha\beta} \overline{N_R^{\alpha C}} N_R^{\beta}  + \rm{H. c.} .
\label{typeI}
\eea
After the spontaneous Electroweak symmetry breaking (EWSB)
   by the vacuum expectation value (VEV), 
   $ \Phi =\begin{pmatrix} \frac{v}{\sqrt{2}} \\  0 \end{pmatrix}$, 
    we get the Dirac mass matrix as $m_{D}= \frac{Y_D v}{\sqrt{2}}$.
Using the Dirac and Majorana mass matrices 
  the neutrino mass matrix is written as 
\bea
m_{\nu}^{\rm{seesaw}}=\begin{pmatrix}
0&&m_{D}\\
m_{D}^{T}&&m_{N}
\end{pmatrix}.
\label{typeInu}
\eea
Diagonalizing Eq.~\ref{typeInu} we obtain the seesaw formula
 for the light Majorana neutrinos as 
\bea
m_{\nu}^{\rm{seesaw}} \simeq - m_{D} m_{N}^{-1} m_{D}^{T}.
\label{seesawI}
\eea
For the heavy neutrino mass $\mathcal{O}(100~\rm{GeV)}$, we may find Dirac Yukawa coupling $\mathcal{O}(\sim 10^{-6})$  with $m_{\nu}^{\rm{seesaw}}\sim 0.1$ eV.

On the other hand in the inverse seesaw(ISS) mechanism, the light Majorana neutrino mass is generated through a tiny lepton number violating mass term.
The relevant part of the Lagrangian is written as
\bea
\mathcal{L}^{\rm{ISS}} \supset - Y_D^{ab} \overline{\ell_L^{a}} \Phi N_R^{b}- m_N^{ab} \overline{S_L^{a}} N_R^{b} -\frac{1}{2} \mu_{ab} \overline{S_L^{a}}S_L^{b^{C}} + \rm{H. c.} ,
\label{InvYuk}
\eea 
where  $N_R^{a}$ and $S_L^{b}$ are two SM-singlet heavy neutrinos
   with the same lepton numbers, $m_N$ is the Dirac mass matrix, and
   $\mu$ is a small lepton number violating Majorana mass matrix.
After the EWSB we get the neutrino mass matrix
\bea
m_{\nu}^{\rm{ISS}}=\begin{pmatrix}
0&&m_{D}&&0\\
m_{D}^{T}&&0&&m_{N}^{T}\\
0&&m_{N}&&\mu
\end{pmatrix}.
\label{InvMat}
\eea
Diagonalizing Eq.~\ref{InvMat} we obtain the light neutrino mass eigenvalue as
\bea
m_{\nu}^{\rm{ISS}} \simeq m_{D} m_{N}^{-1}\mu m_{N}^{-1^{T}} m_{D}^{T}.
\label{numass}
\eea
It is clear from the Eq.~\ref{numass} that the light neutrino mass is proportional
to the small lepton number violating mass term, $\mu$.
Which allows the $m_{D}m_{N}^{-1}$ parameter
  to be $\mathcal{O}(1)$ even for the heavy right handed neutrino at the electroweak scale.
  Due to the smallness of the scale of the lepton number violating mass term $\mu$
in comparison to the scale of $m_{N}$,
  the right handed heavy neutrinos become the pseudo-Dirac fermions in nature.
Which is a striking difference between the type-I and inverse seesaw mechanisms \footnote{The order of smallness of the lepton number violating parameter in the inverse seesaw case has been studied in Ref.~\cite{Das:2012ze} for different flavor structures and general parameterizations.}.
See, for example,  \cite{Chen:2011hc,Das:2012ze,Das:2014jxa, Das:2015toa, Das:2016akd,Dev:2015pga,Degrande:2016aje, Das:2016hof} for the 
  interaction Lagrangians and the partial decay widths 
  of the heavy neutrinos. 
  
   In this context it must be mentioned that such scenarios can also be found with equal importance for another neutrino mass generation model called linear seesaw(LS)\cite{Gu:2010xc,Zhang:2009ac,Hirsch:2009mx,Khan:2012zw}. To study such scenario, we generalize the ISS mechanism which results into a neutrino mass matrix in the $(\nu_{L}, N_{L}^{c}, S_{L})$ basis such as 
 \bea
 m_{\nu}^{\rm{gen}}=\begin{pmatrix}
0&&m_{D}&&\delta^{T}\\
m_{D}^{T}&&m&&m_{N}^{T}\\
\delta&&m_{N}&&\mu
\end{pmatrix}.
\label{genInvMat} 
\eea
In the general mass matrix given in Eq.~\ref{genInvMat} if $\delta\to 0$, $\delta^{T}\to 0$ and $m\to 0$ (or extremely small compared to the other elements so that we can neglect them due to simplicity)\cite{Gogoladze:2008wz}, then a purely inverse seesaw scenario is arrived. A pure linear seesaw can be obtained while $m \to 0$ and $\mu \to 0$. Hence diagonalizing Eq.~\ref{genInvMat} we can write the light neutrino mass as
\bea
m_{\nu}^{\rm{LS}} = m_{D}^{T}m_{N}^{-1}\delta+\delta m_{N}^{-1} m_{D}.
\label{LS}
\eea  
 Depending upon the choice of the generations we can consider a minimal linear seesaw scenario, as carefully cultivated in \cite{Khan:2012zw}. In Eq.~\ref{genInvMat} there is another scenario which is obtained using
 $m \to 0$. Such scenario is a mixed situation where the contributions from the ISS and the LS are both possible. In such case the light neutrino mass term is a combination of Eqs.~\ref{numass} and \ref{LS}, however, the ranks of the matrices can be fixed with the choice and the simplicity of the models. There is another possibility in the inverse seesaw mechanism when $\mu \to 0$ and the mass term $m$ can be generated radiatively \cite{Pilaftsis:1991ug, Dev:2012sg, Dev:2012bd}, satisfying the neutrino oscillation data. A two loop analysis of the neutrino mass generation mechanism in the B$-$L framework has been studied in 
 \cite{Bazzocchi:2010dt}. In our paper we specially consider the purely inverse seesaw model for the pair production of the heavy neutrinos at the high energy colliders.
 
 For the above neutrino mass models, the flavor eigenstate $(\nu)$, of the SM neutrino can be expressed as a 
  combination of the light Majorana neutrino mass eigenstate $(\nu_m)$ and 
  that of the heavy Majorana neutrino $(N_m)$. Such dependence can be expresses as
\bea 
  \nu \simeq  \nu_m  + V_{\ell N} N_m,  
\eea 
where $V_{\ell N}$  is the mixing between the SM neutrino and the SM gauge singlet right handed heavy neutrino assuming
a small mixing, $|V_{\ell N}| \ll 1$.  Thus, the charged current interaction for the heavy neutrino in terms of the mass eigenstates
can be expressed as
\bea 
\mathcal{L}^{CC} \supset 
 -\frac{g}{\sqrt{2}} W_{\mu}
  \bar{e} \gamma^{\mu} P_L   V_{\ell N} N_m  + \rm{H.c}., 
\label{CC}
\eea
where $e$ denotes the three generations of the charged leptons in the vector form, and 
  $P_L =\frac{1}{2} (1- \gamma_5)$ is the projection operator. 
Similarly, in terms of the mass eigenstates the neutral current interaction is written as
\bea 
\mathcal{L}^{NC} \supset 
 -\frac{g}{2 c_w}  Z_{\mu} 
\left[ 
  \overline{N_m} \gamma^{\mu} P_L  |V_{\ell N}|^2 N_m 
+ \left\{ 
  \overline{\nu_m} \gamma^{\mu} P_L V_{\ell N}  N_m 
  + \rm{H.c.} \right\} 
\right] , 
\label{NC}
\eea
 where $c_w=\cos \theta_w$ with $\theta_w$ being the weak mixing angle. 
In this paper we concentrate on the neutral current interaction which is responsible for the heavy neutrino pair production 
\bea
\mathcal{L}^{NC}_{\rm{pair}}\supset  -\frac{g}{2 c_w}  Z_{\mu} 
 \overline{N_m} \gamma^{\mu} P_L  |V_{\ell N}|^2 N_m .
 \label{pair}
  \eea
  The production cross section is proportional to the fourth power of the mixing angle $(|V_{\ell N}|^{4})$. This process is phenomenologically very rich 
  due to the pair production of the heavy neutrinos and the further decays of the heavy states into various modes, .

\begin{figure}
\begin{center}
\includegraphics[scale=0.35]{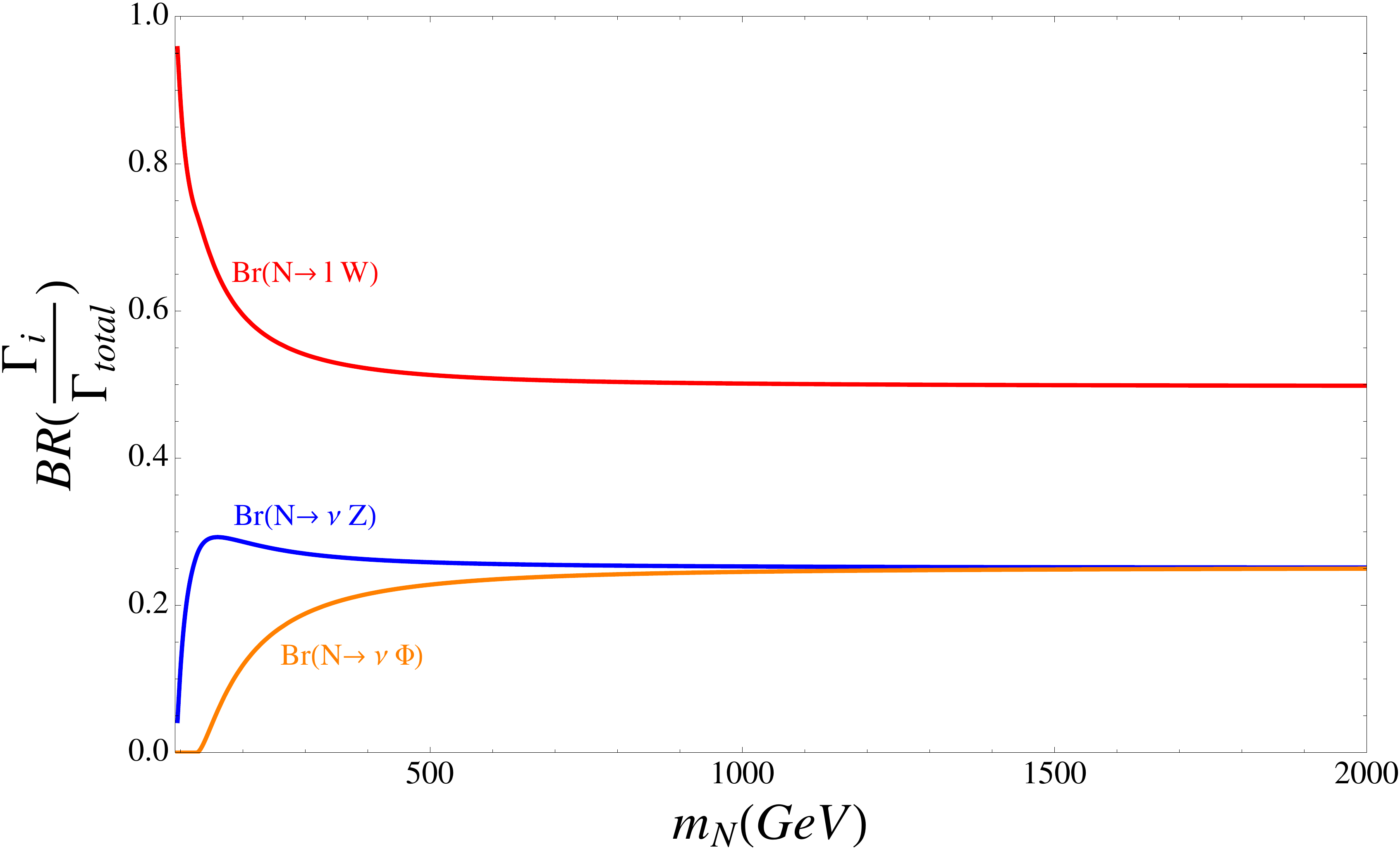}
\caption{Branching rations of different decay modes of the right handed heavy neutrino as a function of the its mass $(m_{N})$.}
\label{fig:BR}
\end{center}
\end{figure}

 The partial decay widths of the heavy neutrino into the weak gauge bosons $(W$ and $Z)$ and SM Higgs boson $(\Phi)$ with associated leptons $(W\ell)$ and missing transverse energy $(Z\nu, \Phi \nu)$  are given as 
\bea
\Gamma(N \rightarrow \ell W) 
 &=& \frac{g^2 |V_{\ell N}|^{2}}{64 \pi} 
 \frac{ (m_{N}^2 - m_W^2)^2 (m_{N}^2+2 m_W^2)}{m_{N}^3 m_W^2} ,
\nonumber \\
\Gamma(N \rightarrow \nu_\ell Z) 
 &=& \frac{g^2 |V_{\ell N}|^{2}}{128 \pi c_w^2} 
 \frac{ (m_{N}^2 - m_Z^2)^2 (m_{N}^2+2 m_Z^2)}{m_{N}^3 m_Z^2} ,
\nonumber \\
\Gamma(N \rightarrow \nu_\ell \Phi) 
 &=& \frac{ |V_{\ell N}|^2 (m_{N}^2-m_h^2)^2}{32 \pi m_{N}} 
 \left( \frac{1}{v }\right)^2.
\label{widths}
\eea 
 with $m_W$, $m_Z$ and $m_\Phi$ as the $W$, $Z$ and $\Phi$ masses respectively. The partial decay width of heavy neutrino into the charged gauge bosons being twice as large as that of the neutral one due to the two degrees of freedom $(W^{\pm})$. The branching ratios $BR_i \left(= {\Gamma_{i}}/{\Gamma_{\rm total}}\right)$ of each decay modes of the right handed heavy neutrino as a function of its mass $(m_{N})$ are plotted in Fig.~\ref{fig:BR}.
It should be interestingly noted that for larger values of $m_{N}$, the branching ratios can be obtained as 
\bea
BR\left(N\rightarrow  W \ell \right) : BR\left(N\rightarrow  Z \nu \right) : BR\left(N\rightarrow \Phi \nu \right) \simeq 2: 1: 1.
\eea  
and such ratios has to be maintained for correct and flawless analysis\footnote{ In this context, we should mention that in Ref.\cite{Kang:2015uoc} the pair production of the heavy neutrino has been studied in the $Z^{\prime}$ inspired model  which comes from the $U(1)^{\prime}$ extensions of the SM where the mass generation of the SM neutrinos will be possible only through the seesaw mechanism. The heavy neutrinos will also have three generations to get rid of the $U(1)^{\prime}$ anomalies. Due to the Seesaw mechanism, the flavors will be carried out through the Dirac Yukawa coupling and as a result the Flavor non-Democratic (FND) case will be evolved. The striking difference with this paper is, we are working on the inverse seesaw framework where the 
lepton flavor violation mass term is too small so that a Flavor Democratic (FD) case can be observed keeping the Dirac Yukawa coupling as a diagonal matrix while $\mu$ is non-diagonal. The effect of $\mu$ is too small, so that it will have no effect in the signal. In case of Ref.\cite{Kang:2015uoc}, they have worked with the Majorana right handed neutrinos and seesaw mechanism. In case of seesaw mechanism FD case can never be observed. Inverse seesaw is the only possible extension to study the neutrino mass generation technique where both of the FND and FD cases could be observed, however, in the FND case lepton flavor violation constraints will be applied as discussed in \cite{Das:2012ze}.}.
\section{Scale dependent pair production of the heavy neutrinos at the hadron colliders at the LO and NLO-QCD levels}
\label{sec:formulas}
\begin{figure}
\begin{center}
\includegraphics[scale=1.0]{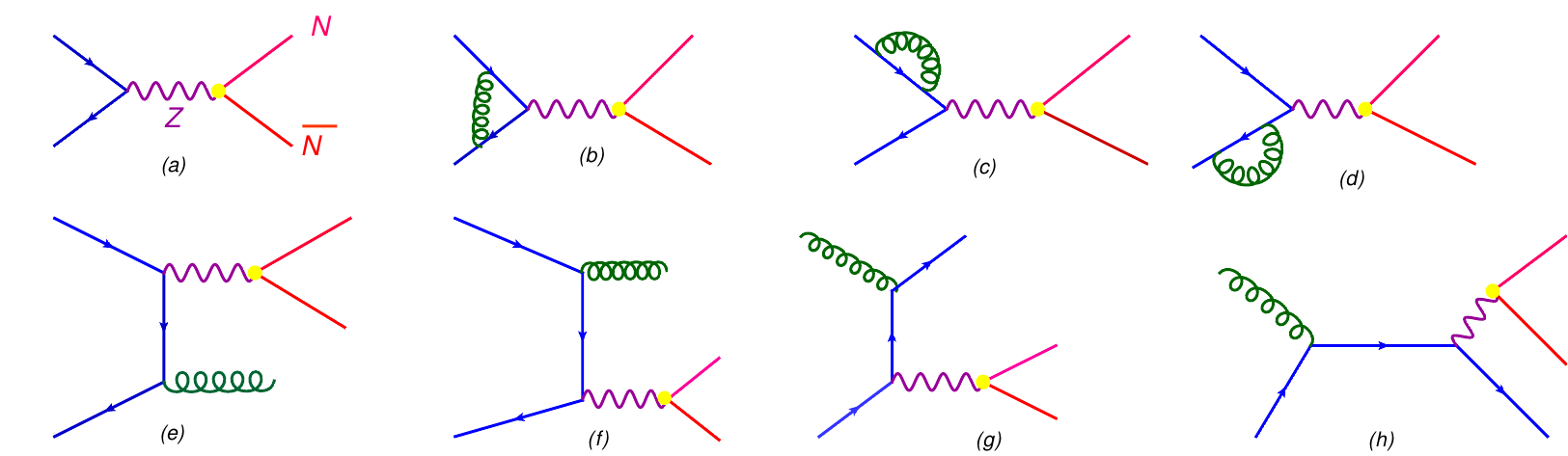}
\end{center}
\caption{Pair production of the heavy neutrinos in the inverse seesaw framework at the LO and NLO-QCD orders. The Born process/ LO is given in $(a)$. NLO diagrams including the virtual corrections $(b-d)$ and real emissions $(e-h)$ are also shown.}
\label{LO-NLO}
\end{figure}
Implementing the inverse seesaw model in the event generator {\tt MadGraph5-aMC@NLO} \cite{MG, MG5, aMC}, we calculate the production cross section of the heavy neutrino pair
at the LO and NLO-QCD levels. The relevant Feynman diagrams of the production 
processes are given Fig.~\ref{LO-NLO} where the LO production (Born level) channel is shown in $(a)$. In Fig.~\ref{LO-NLO}$(b-d)$ and in Fig.~\ref{LO-NLO}$(e-h)$
the virtual corrections and the real emission processes are shown respectively. The event generator uses {\tt MADLOOP}\cite{madloop} to calculate the one loop processes
using Ossola-Papadopoulos-Pittau {\tt OPP}\cite{opp} integrand-reduction procedure implemented in {\tt CutTools}\cite{CutTools}. 
The Born process and the real emission processes are calculated using {\tt MadFKS}\cite{madfks} through the integration and matching scheme of the {\tt MC@NLO}.
The subtraction method has been taken care of by {\tt FKS}\cite{FKS} formalism in {\tt MadFKS}. {\tt PYTHIA6.4} for the LO and {\tt PYTHIA6Q} for the NLO processes\cite{Pyth}
bundled with {\tt MadGraph} perform the showering and the hadronization of the events using the {\tt anti-$k_{T}$} algorithm which clusters the jets using {\tt FastJet}\cite{FJ}.
In calculating the parton level cross sections we used {\tt CTEQ6L1 (CTEQ6M)} parton distribution functions (PDF) for the LO (NLO) processes through the {\tt MadGraph5-aMC@NLO}
choosing $\alpha_s(m_Z) = 0.130$ in {\tt CTEQ6L1} for LO and  $\alpha_s(M_Z) = 0.1180$ in {\tt CTEQ6M} for NLO as given in \cite{Madevent}. We have considered that $m_Z = 91.188$ GeV, $m_W = 80.423$  GeV and $G_F = 1.166 \times 10^{-5}$  GeV$^{-2}$ as the electroweak input parameters. The $\alpha_{QED} = 1/132.54$ and $\sin^2{\theta_W} = 0.22217$ are computed via LO electroweak relations.
 \begin{figure}
\begin{center}
\includegraphics[scale=0.18]{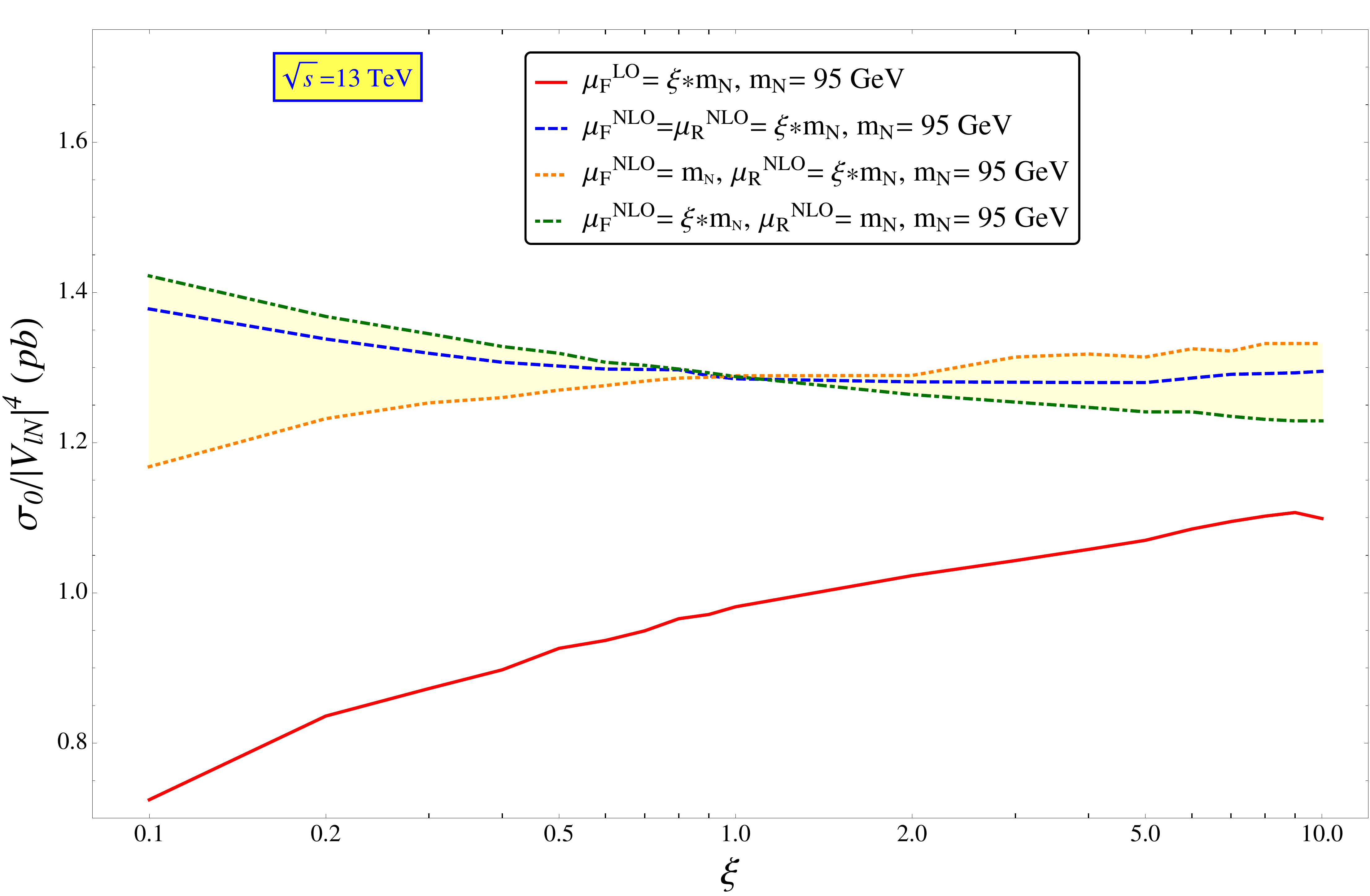}
\includegraphics[scale=0.18]{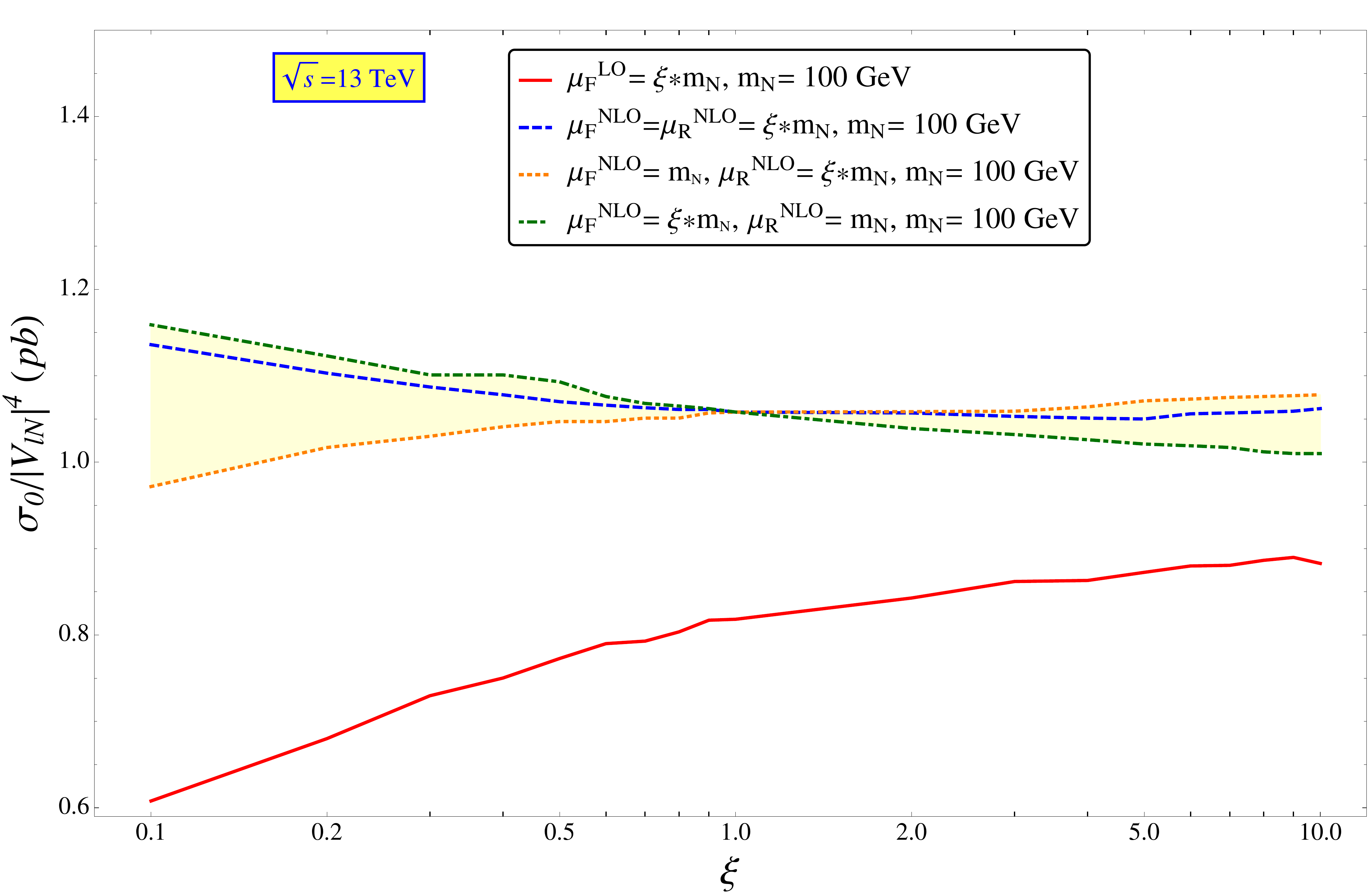}\\
\includegraphics[scale=0.18]{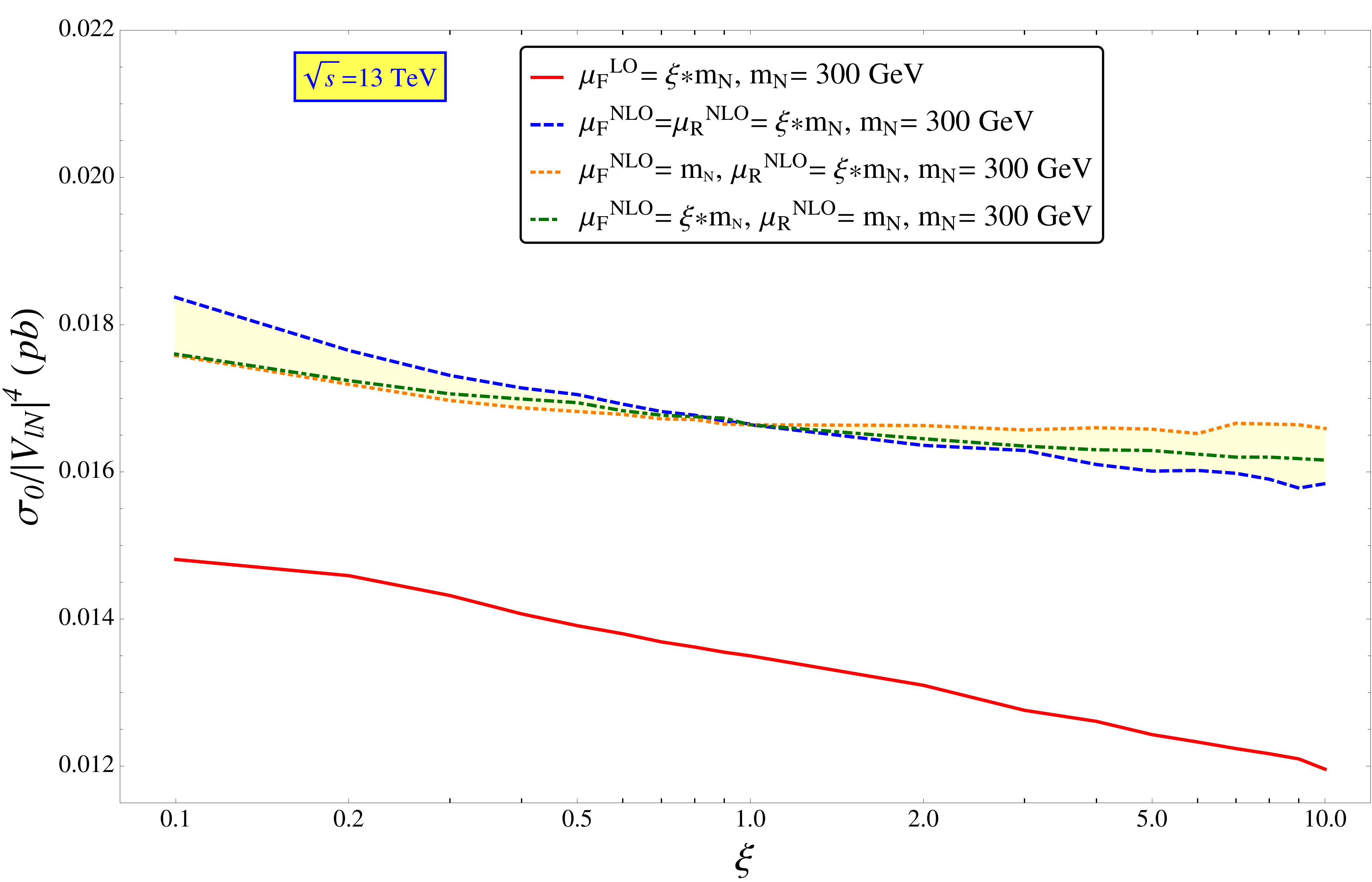}
\includegraphics[scale=0.18]{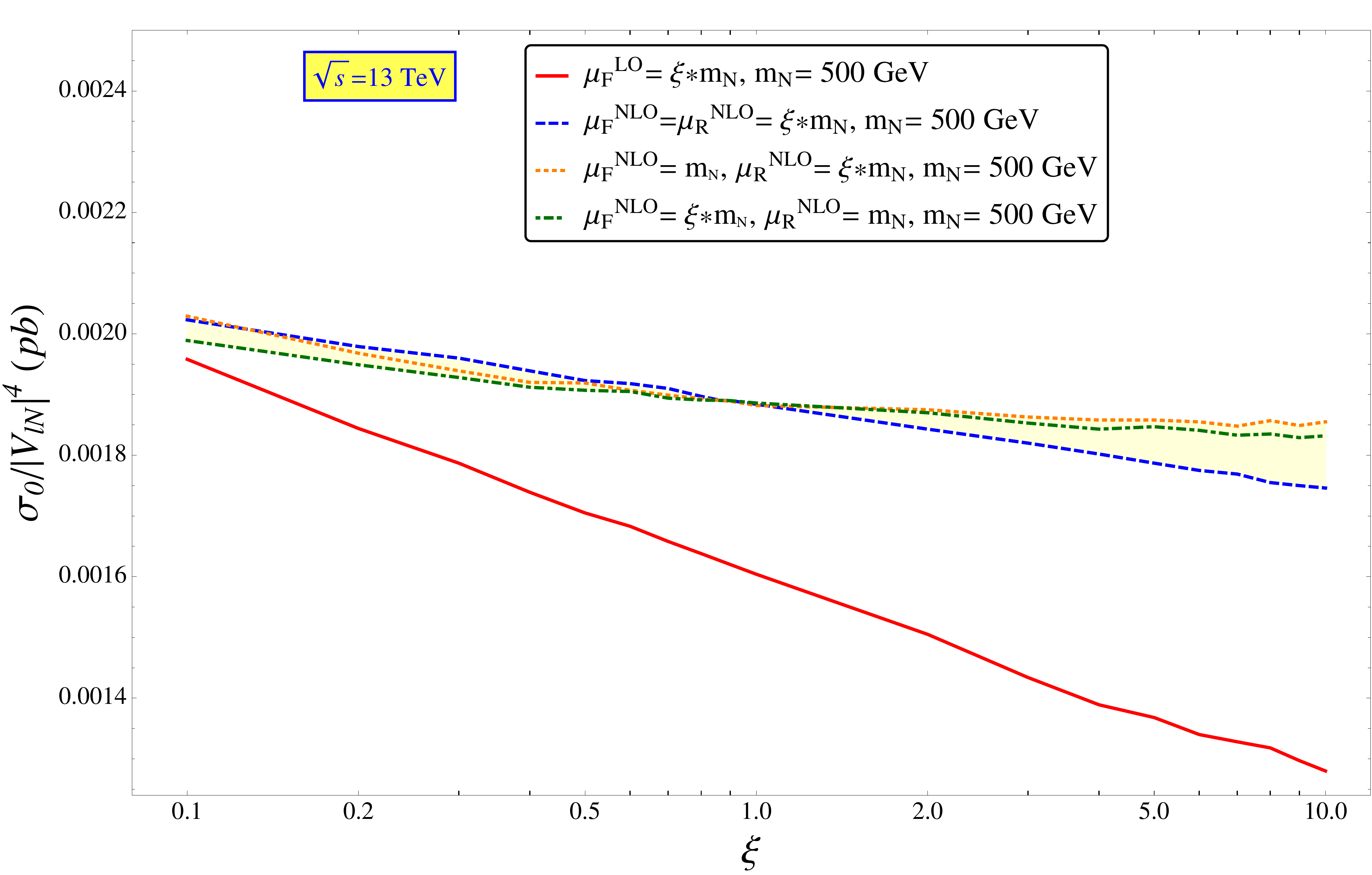}
\end{center}
\caption{Scale variation of the heavy neutrino pair production at the LO and NLO-QCD levels at the 13 TeV LHC different masses of the heavy neutrino. The upper left panel stands for $m_{N}=95$ GeV and the upper right panel stands for $m_{N}=100$ GeV. Whereas the lower left panel stands for $m_{N}=300$ GeV and the lower right panel stands for $m_{N}=500$ GeV.}
\label{HC13}
\end{figure}

We have calculated the cross-sections fixing the mass of the heavy neutrino $(m_{N})$ at the LO level varying the factorization scale ($\mu_F$). The LO cross section varies due to the variation in the factorization scale $(\mu_{F})$ because of the PDFs according to
\bea
\mu_{F}^{\rm{LO}} =\xi \ast m_{N}    \, \, \, \,  \, \, \, \, \text{with} \, \,\, \,  0.1\le \xi \le 10.
\label{muF}
\eea
where  $\xi$ is the scale factor. On the other hand the NLO cross section depends not only on the $\mu_F$ but also on the renormalization scale $(\mu_R)$. 
The effect of $\mu_F$ comes in the NLO cross section through the PDFs whereas that of the $\mu_R$ is involved in the NLO cross section due to the strong coupling, $\alpha_{s}(\mu_R)$.

The scale dependent cross sections have been produced being normalized by the fourth power of the mixing angle $(|V_{\ell N}|^{4})$ which is indicated from Eq.~\ref{pair}.
In Fig.~\ref{HC13} we systematically produce the heavy neutrino pair at the 13 TeV LHC at the LO level for the different masses such as $m_{N}=95$ GeV, $100$ GeV,
$300$ GeV and $500$ GeV with $0.1\leq \xi \leq 10$ according to Eq.~\ref{muF} at the LO.
We have also displayed the theoretical scale uncertainties of the NLO-QCD processes for the variations in $\mu_{F}$ and  $\mu_{R}$ in Fig.~\ref{HC13}.
For the NLO-QCD analysis we have made three choices between $\mu_{F}$ and $\mu_{R}$, which are
\bea
\mu_{F}^{\rm{NLO}}&=& \mu_{R}^{\rm{NLO}}=\xi\ast m_{N} \nonumber\\
\mu_{F}^{\rm{NLO}}&=&m_{N}, \mu_{R}^{\rm{NLO}}=\xi\ast m_{N} \nonumber\\
\mu_{F}^{\rm{NLO}}&=&\xi \ast m_{N}, \mu_{R}^{\rm{NLO}}= m_{N}. 
\label{muR}
\eea
$\mu_{F}^{\rm{NLO}}$ and $\mu_{R}^{\rm{NLO}}$ are the factorization and renormalization scales at the NLO-QCD level.
\begin{figure}
\begin{center}
\includegraphics[scale=0.18]{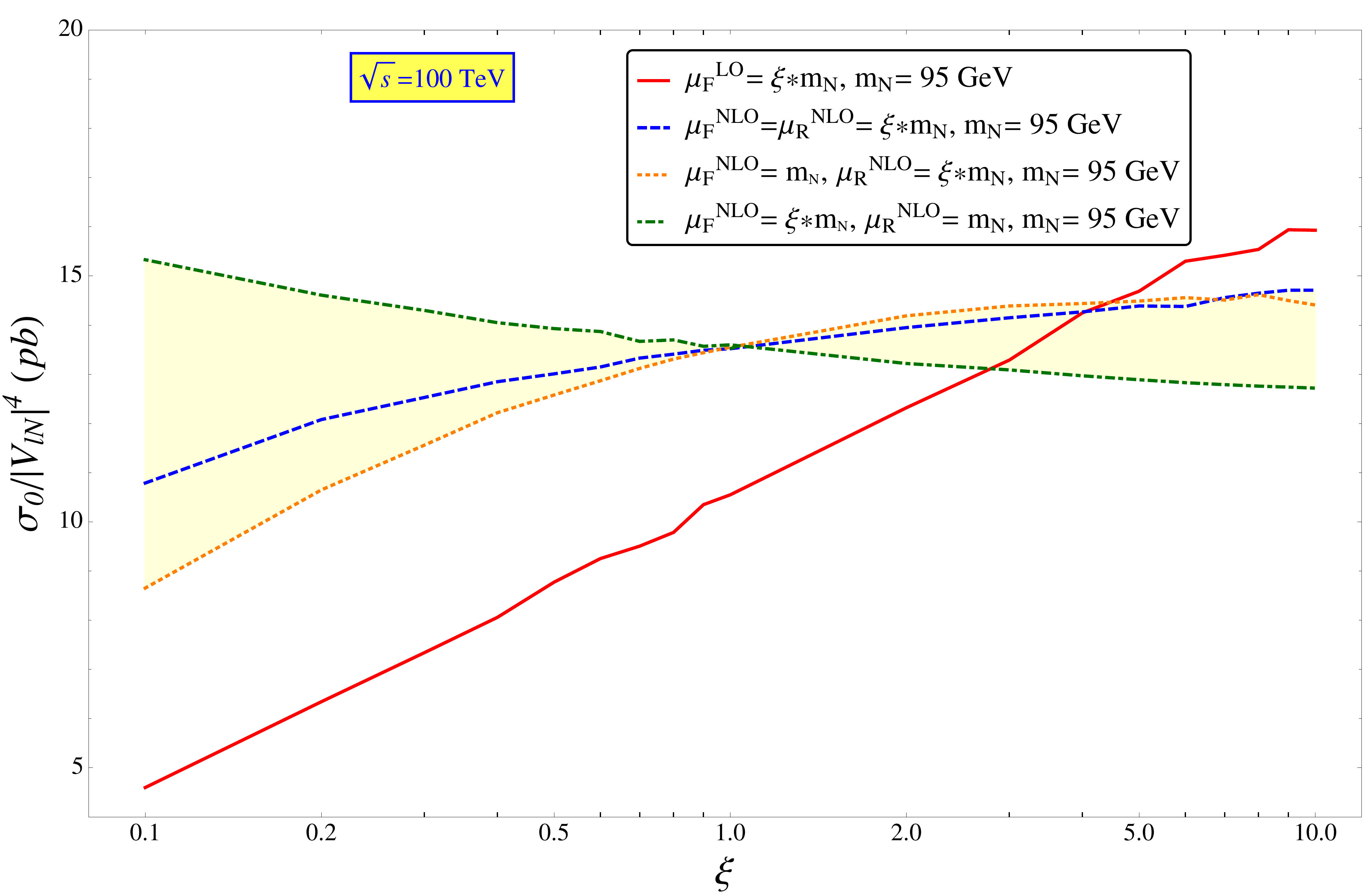}
\includegraphics[scale=0.18]{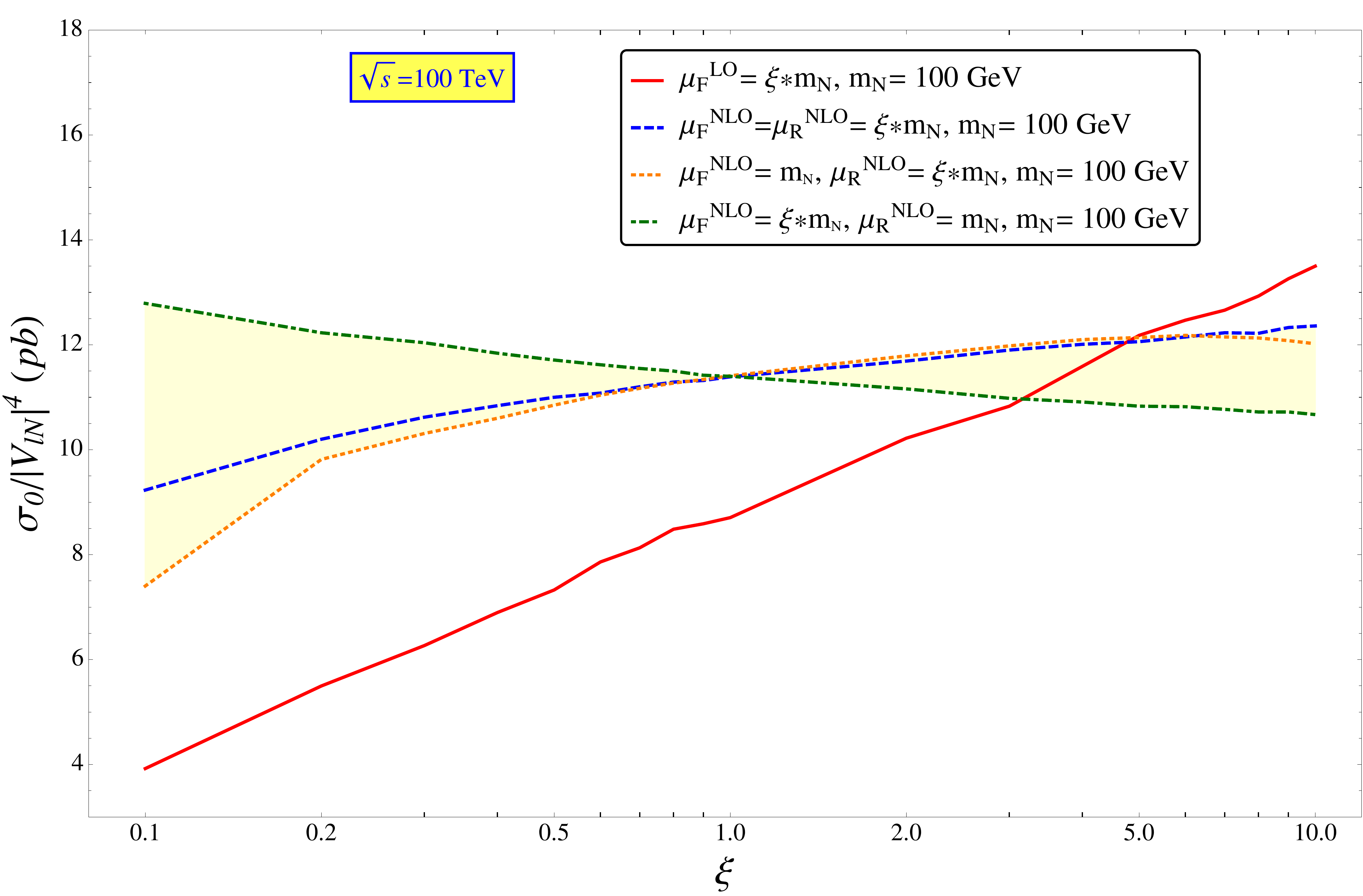}\\
\includegraphics[scale=0.18]{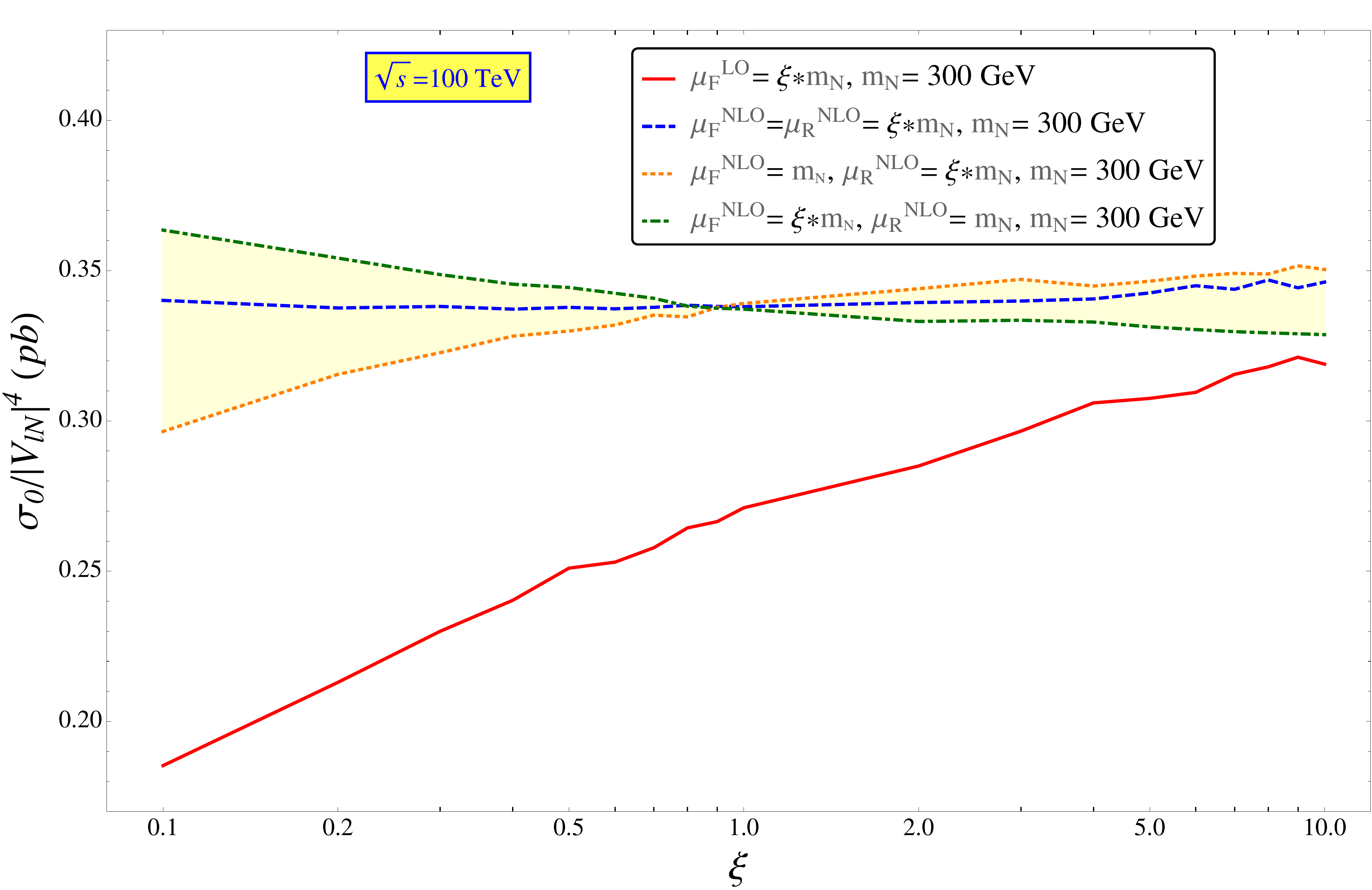}
\includegraphics[scale=0.18]{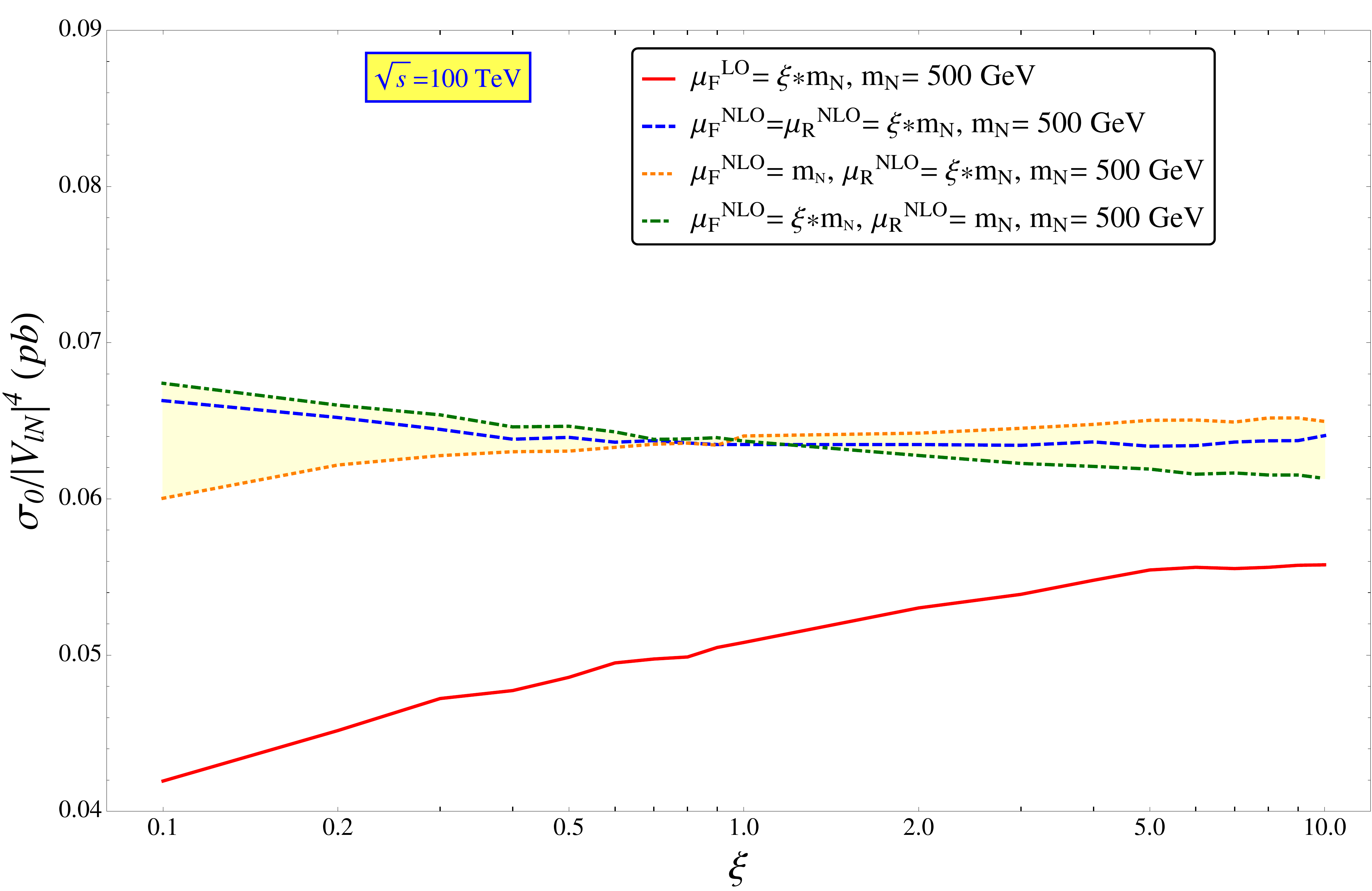}
\end{center}
\caption{Scale variation of the heavy neutrino pair production at the LO and NLO-QCD levels at the 100 TeV LHC for the different heavy neutrino masses. The upper left panel stands for $m_{N}=95$ GeV and the upper right panel stands for $m_{N}=100$ GeV. Whereas the lower left panel stands for $m_{N}=300$ GeV and the lower right panel stands for $m_{N}=500$ GeV.}
\label{HC100}
\end{figure} 

For $m_{N}= 95$ GeV, the LO cross-section$(\sigma^{LO})$ varies up to a factor of $1.43$ while the scale varies within a range $0.1\leq \xi \leq10$.
A considerable amount of theoretical uncertainty has been noticed from the LO results because of the quark-antiquark $(q\overline{q})$ PDFs are involved at the 
partonic level. In this case each PDF is dependent upon $\mu_{F}^{\rm{LO}}$ and $\xi$ according to Eq.\ref{muF}.
 On the other hand in the NLO-QCD processes, the $\xi$ dependence comes from the PDFs through $\mu_{F}^{\rm{NLO}}$ and also from the strong coupling, $\alpha_{s}(\mu_{R})$
according to Eq.~\ref{muR} at the parton level. 
The NLO PDFs are involved in the quark-antiquark $(q\overline{q})$, quark-gluon $(qg)$ and antiquark-gluon $(\overline{q}g)$ interactions through the strong coupling 
constant and hence the strong scale dependent part cancels among themselves. 
As a result a soft scale dependence has been observed for the NLO-QCD processes compared to the LO process.
At the 13 TeV in Fig.~\ref{HC13} with $m_N=95$ GeV the cross section at the NLO-QCD process is increasing when the factorization scale is fixed at $m_N$
but the renormalization scale is varying as $\xi\ast m_N$. In the other two choices the cross section is decreasing depending up on the choices. Where as for $m_N=100$ GeV the cross sections are more stable
and the same situation is noticed for $m_N=300$ GeV and $m_N=500$ GeV. However, for higher masses the NLO cross sections are stable. 

Following the Eqs.~\ref{muF} and \ref{muR} we have also performed the scale dependent pair production of the right handed heavy neutrinos at the 100 TeV hadron collider. 
In Fig.~\ref{HC100} we have shown the production cross sections for the LO and NLO-QCD processes normalized by $|V_{\ell N}|^{4}$ for four different choices of the right handed 
heavy neutrino masses like the 13 TeV LHC case. Compared to the 13 TeV case, a very sharp scale dependence of a factor of 3 has been noticed at the LO level for the masses 
$95$ GeV, $100$ GeV and $300$ GeV whereas for the $500$ GeV the scale dependence at the LO level is weaker compared to others. In the NLO-QCD level, we have noticed weaker scale dependence compared to the 
LO case. However, for $m_N=500$ GeV at the NLO-QCD level the cross sections are more stable compared to the cases $m_N < 500$ GeV. This is due to the effect of $\alpha_s(\xi\ast m_N)$.
\begin{figure}
\begin{center}
\includegraphics[scale=0.33]{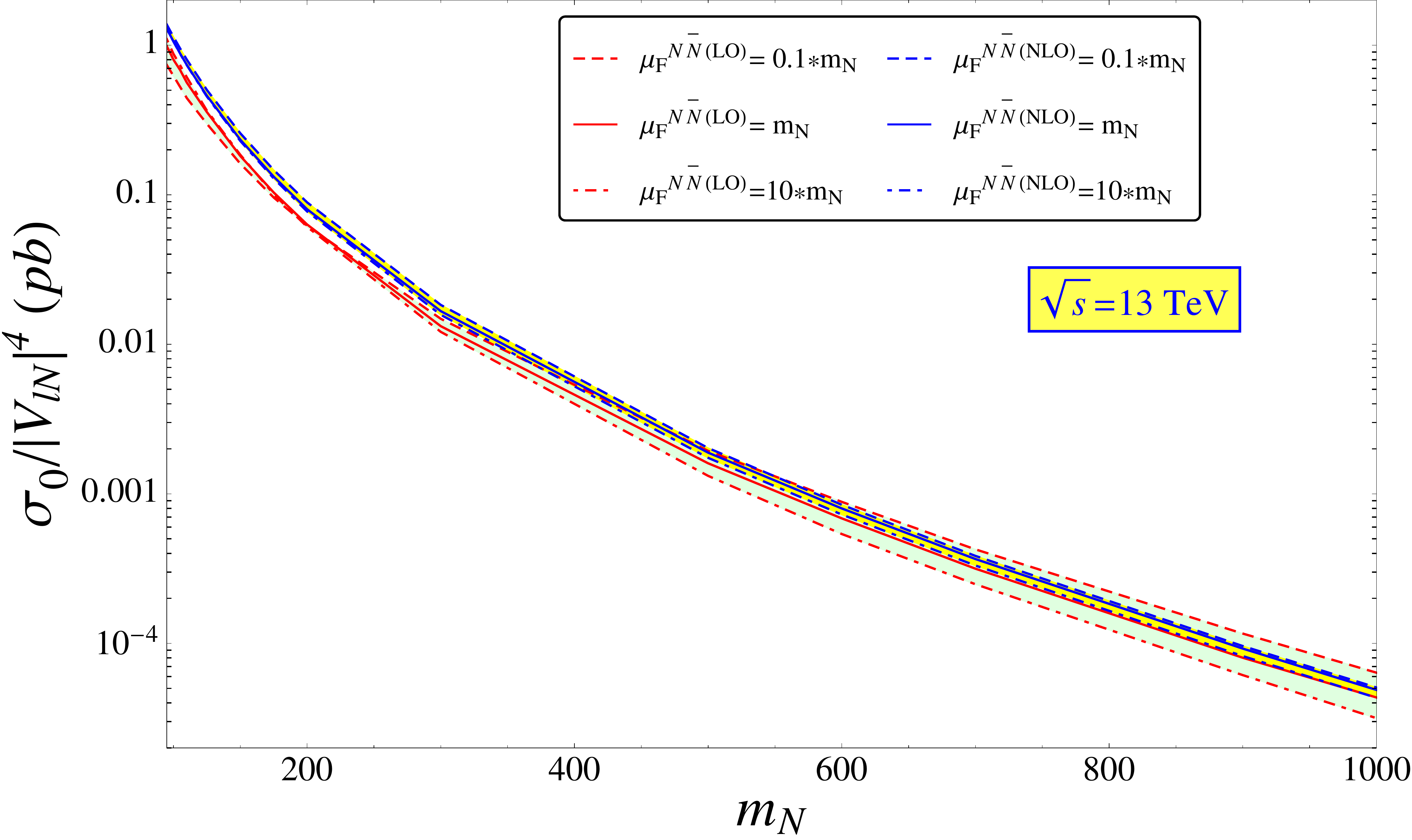}
\end{center}
\caption{Scale dependent heavy neutrino pair production cross section as a function of the heavy neutrino mass at the 13 TeV LHC at the LO and NLO-QCD levels. The effect of the scale variations at the LO and NLO-QCD levels are shown as bands. The production cross section is normalized by $|V_{\ell N}|^{4}$.}
\label{HC13all}
\end{figure}  
\begin{figure}
\begin{center}
\includegraphics[scale=0.33]{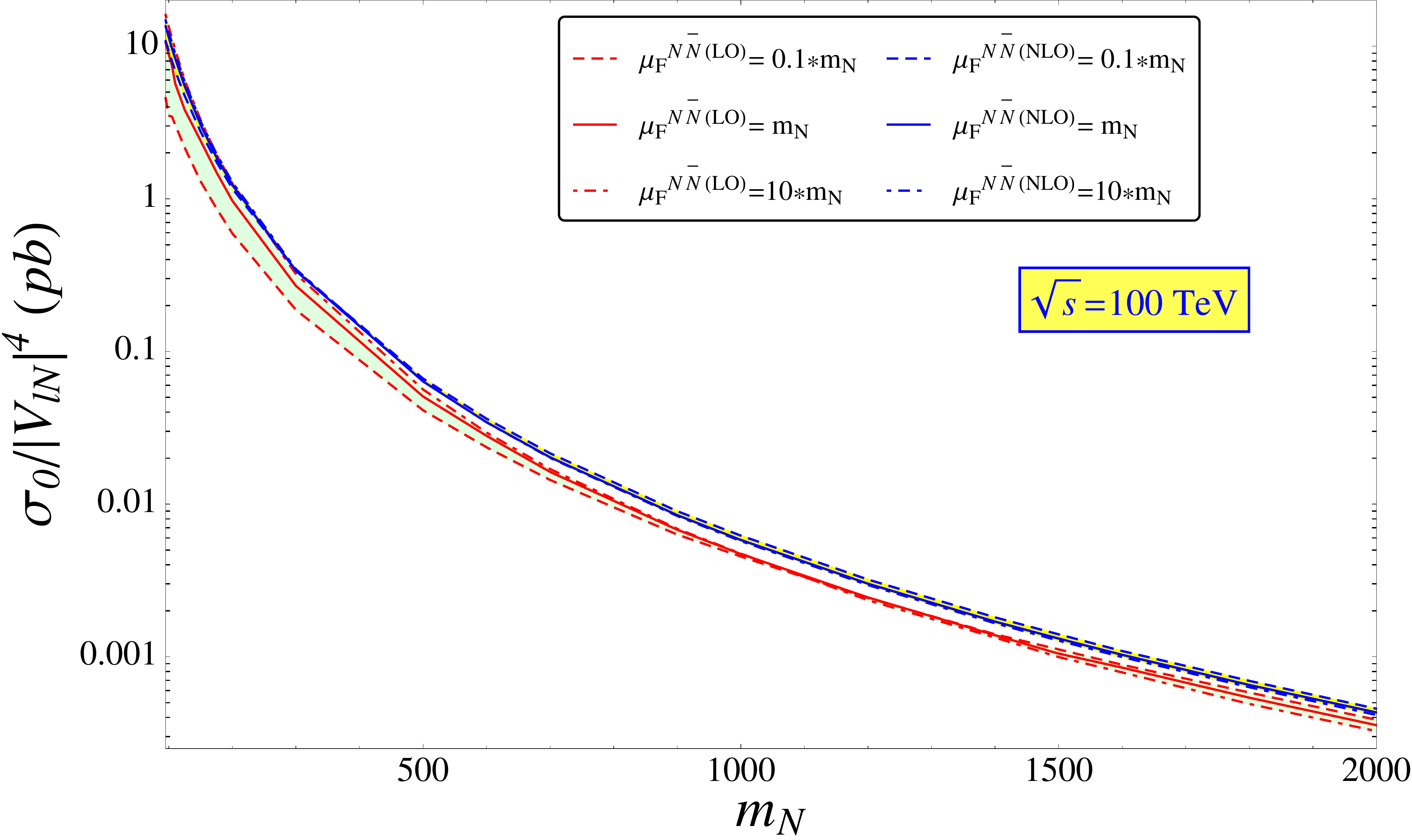}
\end{center}
\caption{Scale dependent heavy neutrino pair production cross section as a function of the heavy neutrino mass at the 100 TeV hadron collider at the LO and NLO-QCD levels. The effect of the scale variations at the LO and NLO-QCD levels are shown as bands. The production cross section is normalized by $|V_{\ell N}|^{4}$.}
\label{HC100all}
\end{figure}  
The scale variation in the pair production cross section as a function $m_N$ has been shown in Fig.~\ref{HC13all} for the 13 TeV LHC and that for the 100 TeV hadron collider is shown in Fig.~\ref{HC100all}. In the further analysis we will be considering the $13$ TeV LHC at the 3000 fb$^{-1}$ luminosity which is also called the High Luminosity LHC (HL-LHC).  The 100 TeV hadron collider will be a new age proposed high energy collider. We will take 3000 fb$^{-1}$ and 30000 fb$^{-1}$ luminosities into our account for further analyses at 100 TeV.
\section{Various decay modes of the pair produced heavy neutrinos}
\label{sec:decays}
In this section we will discuss various decay modes of the heavy neutrinos in the inverse seesaw framework. The possible decay modes are
\bea
p p &\to& N \overline{N}, \nonumber \\
        \, \, \, \,  \, \, \, \,&&   N \to \ell_{1}^{-} W^{+}, W^{+} \to j j \nonumber \\
        \, \, \, \,  \, \, \, \,&&   \overline{N} \to \ell_{2}^{+} W^{-}, W^{-} \to j j   \, \, \, \,  \, \, \, (a) \nonumber \\              
 p p &\to& N \overline{N}, \nonumber \\
        \, \, \, \,  \, \, \, \,&&   N \to \ell_{1}^{-} W^{+}, W^{+} \to \ell_{2}^{+} \nu \nonumber \\
        \, \, \, \,  \, \, \, \,&&   \overline{N} \to \ell_{3}^{+} W^{-}, W^{-} \to j j   \, \, \, \,  \, \, \,(b) \nonumber  \\     
 p p &\to& N \overline{N}, \nonumber \\
        \, \, \, \,  \, \, \, \,&&   N \to \ell_{1}^{-} W^{+}, W^{+} \to \ell_{2}^{+}  \nu \nonumber \\
        \, \, \, \,  \, \, \, \,&&   \overline{N} \to \ell_{3}^{+} W^{-}, W^{-} \to \ell_{4}^{-} \overline{\nu}
         \, \, \, \,  \, \, \,(c) \nonumber  \\     
 \label{decay1}
 \eea
The leading decay mode of the heavy neutrino irrespective of mass is $N\to W \ell$ as shown in Fig.~\ref{fig:BR} is taken into our account. From this mode we have allowed $W$ boson to decay
hadronically and leptonically respectively. Depending upon the decay modes of the $W$ boson we have various final states like $2\ell+ 2j$ from Eq.~\ref{decay1}(a), $3\ell+2j+\rm{MET}$ from Eq.~\ref{decay1}(b) and $4\ell+\rm{MET}$
from Eq.~\ref{decay1}(c). We have written these modes in the descending order of the cross sections after the complete decay. However, in Eq.~\ref{decay1}(b), the cross section can be doubled taking other $W$ decaying leptonically while the remaining one will show a hadronic decay.
\begin{figure}
\begin{center}
\includegraphics[scale=0.4]{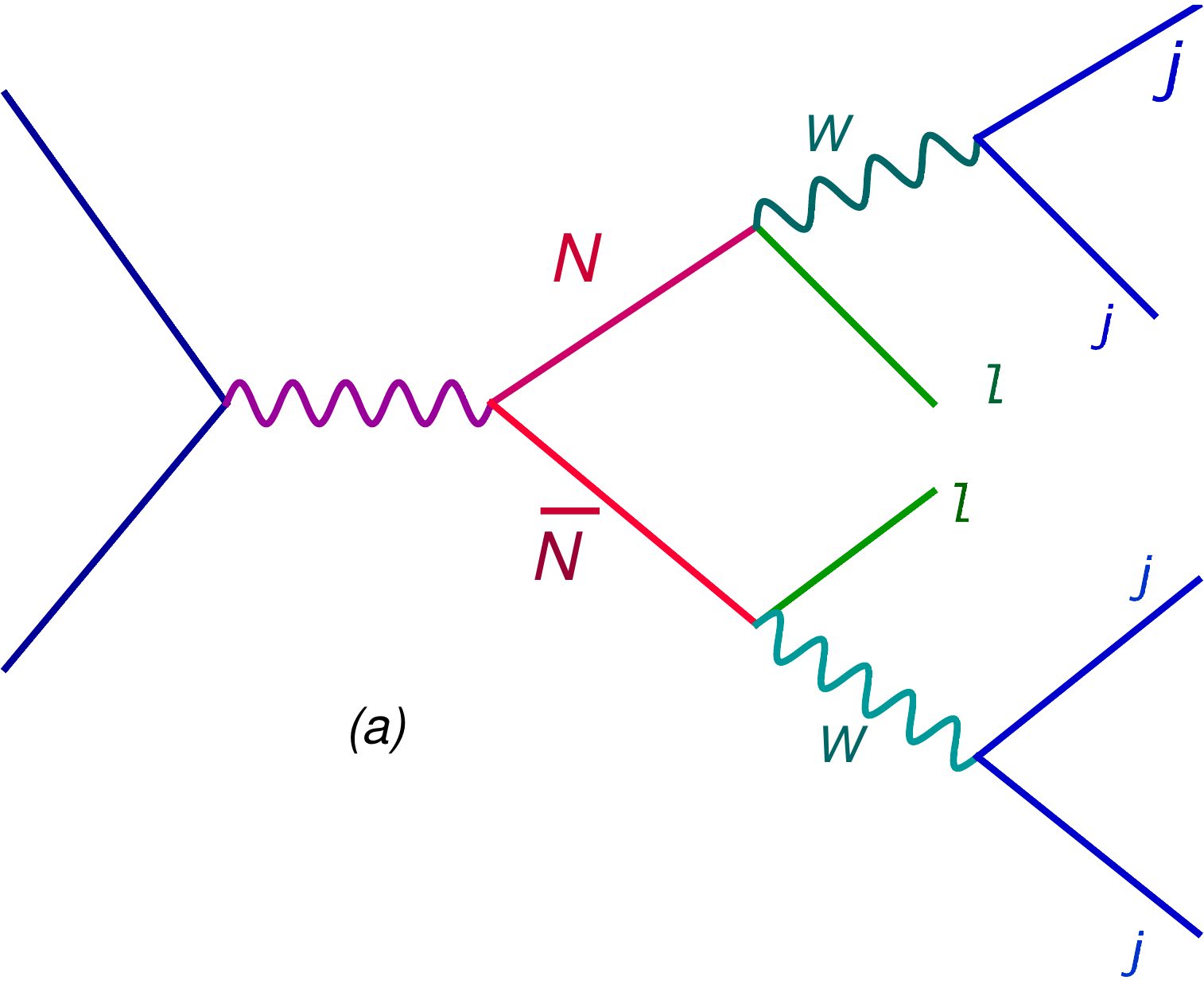}
\includegraphics[scale=0.4]{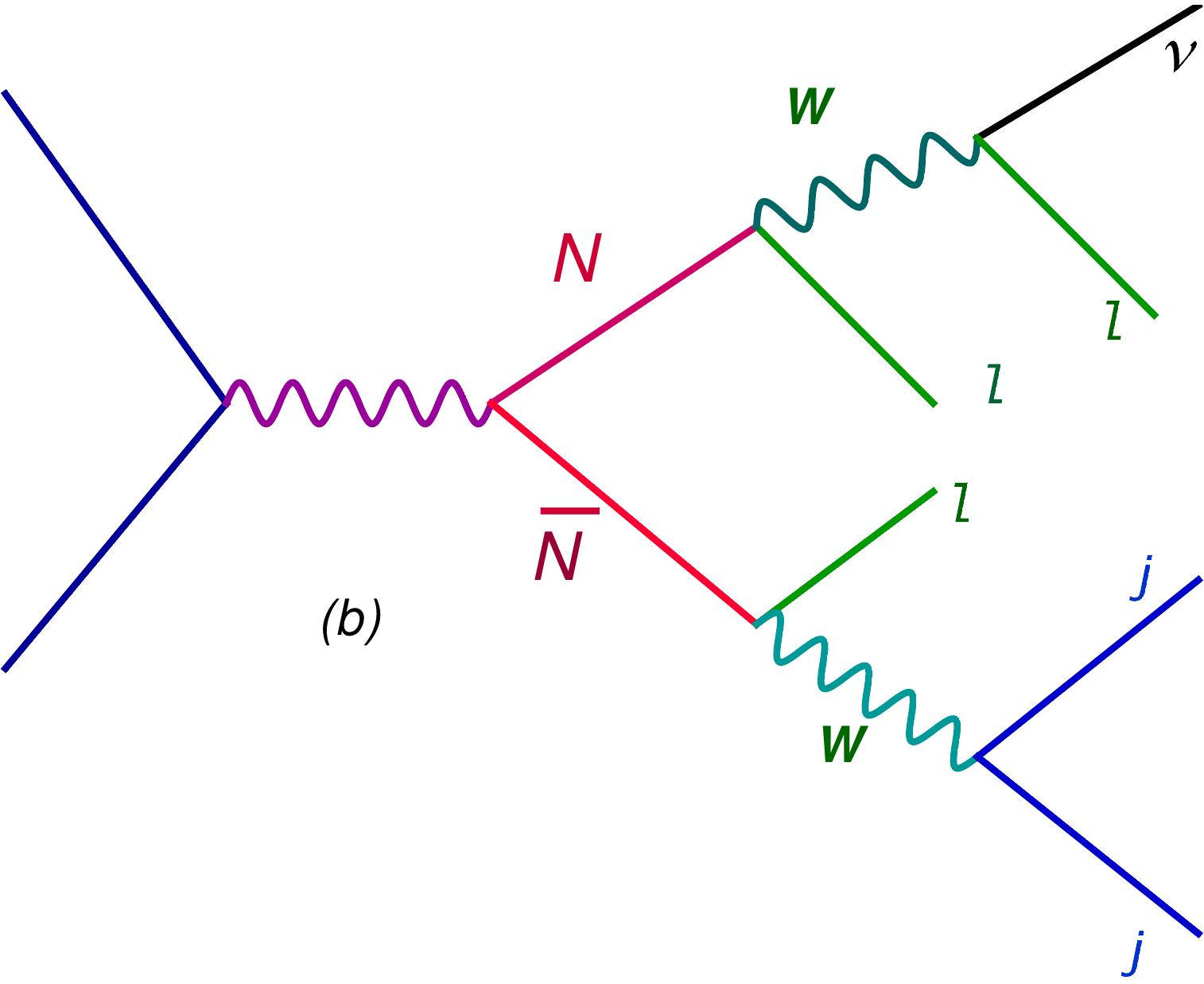}\\
\includegraphics[scale=0.4]{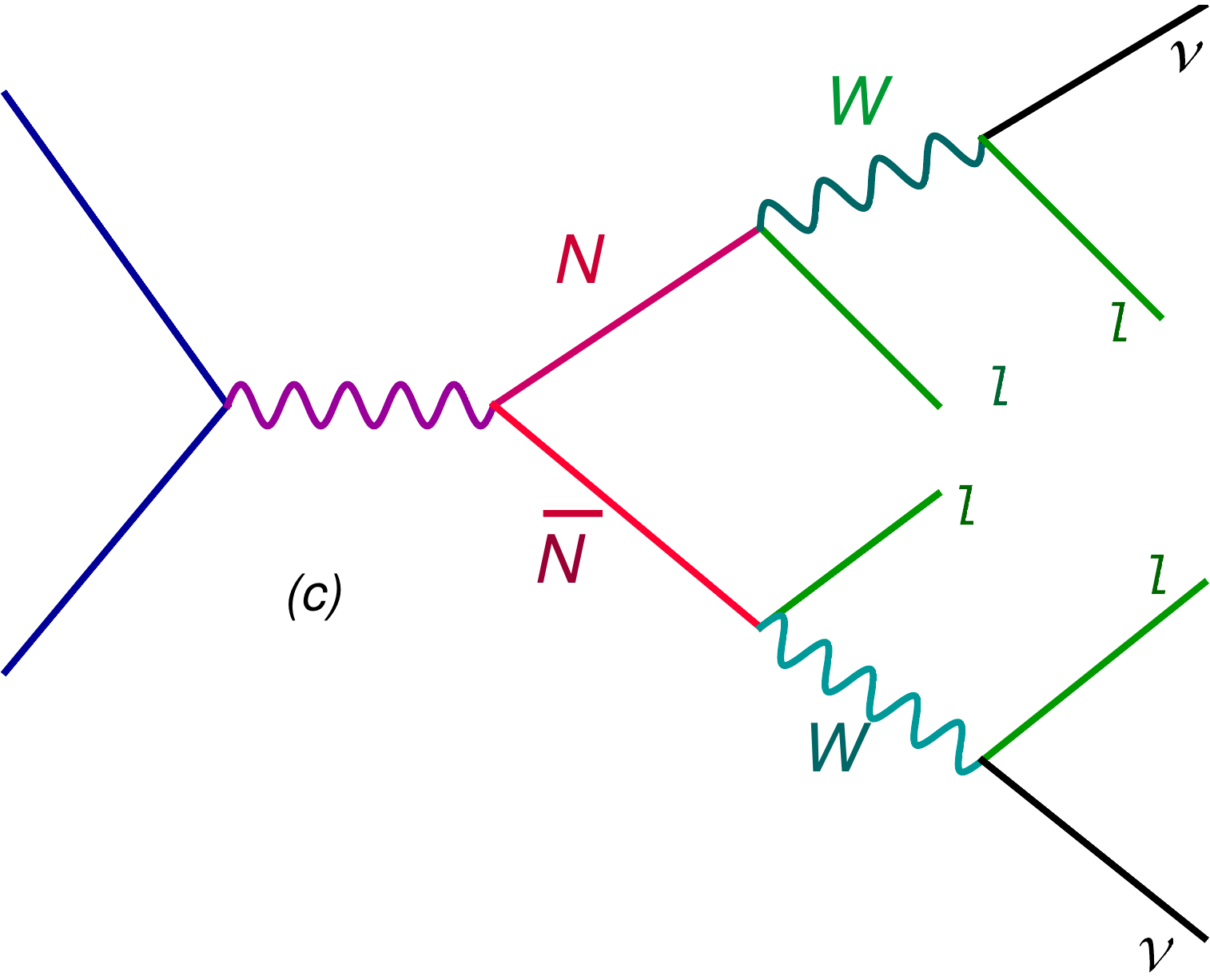}
\end{center}
\caption{Different decay modes of the heavy neutrinos after the pair production. We consider only the $N\to W\ell$ mode because it is always maximum branching ratio according to Fig.~\ref{fig:BR}}
\label{HC131}
\end{figure} 

There are several other decay modes of the heavy neutrinos through the $Z$ boson such as $N \to Z \nu$ which can be expressed as
\bea
p p &\to& N \overline{N}, \nonumber \\
        \, \, \, \,  \, \, \, \,&&   N \to \nu Z, Z \to j j  \nonumber \\
        \, \, \, \,  \, \, \, \,&&   \overline{N} \to \nu Z, Z \to j j   \, \, \, \,  \, \, \, (a) \nonumber \\              
 p p &\to& N \overline{N}, \nonumber \\
        \, \, \, \,  \, \, \, \,&&   N \to \nu Z, Z \to j j \nonumber \\
        \, \, \, \,  \, \, \, \,&&   \overline{N} \to \nu Z, Z \to \ell \ell   \, \, \, \,  \, \, \,(b) \nonumber  \\     
 p p &\to& N \overline{N}, \nonumber \\
        \, \, \, \,  \, \, \, \,&&   N \to \nu Z, Z \to \ell \ell \nonumber \\
        \, \, \, \,  \, \, \, \,&&   \overline{N} \to \nu Z, Z \to \ell \ell   \, \, \, \,  \, \, \,(c). \nonumber  \\     
\label{decay2}        
\eea
The heavy neutrino can display an on-shell decay into the SM Higgs with MET only if $m_{N} > m_{\Phi}$. In such cases the decay mode of the SM Higgs could be taken into $b\overline{b}$ as its leading mode with $m_{\Phi}=125~\rm{GeV}$ while $4b+\rm{MET}$ final state.  We did not consider these modes in the present analysis. For the Higgs mediated cases a complete study has been made in Ref.~\cite{Kang:2015uoc}.

In this context, we can mention that the pair production of the heavy neutrinos is also possible from the $U(1)^{\prime}$ extensions of the SM where the new $U(1)^{\prime}$ gauge coupling squared $(g^{\prime^{2}})$ will be involved in the cross section. The pair production is studied from the BSM gauge boson, commonly known as $Z^{\prime}$. Pair production of the heavy neutrino followed by the leading decay mode of the heavy neutrino $(N \to W \ell)$, one can study the several mutilepton channels according to Eq.~\ref{decay1}. In such models, the heavy neutrino is naturally Majorana while generating the neutrino mass using the canonical seesaw mechanism from the $U(1)^{\prime}$ symmetry breaking \cite{Kang:2015uoc}\footnote{In Ref.\cite{Kang:2015uoc} the authors have studied $pp\to Z^{\prime}\to NN$ using $N\to W\ell$ and $N\to h\nu$ followed $h\to b\overline{b}$. In our analysis we always have the both $N$ decaying into $W\ell$ only, however, the $W$ decays either leptonically or hadronically because whatever be the situation. $N\to W\ell$ has the highest branching ratio. Where as $N\to h \nu$ will be much less than that, see Fig.~\ref{fig:BR}. For $M_N ~\mathcal{O}(100~\rm{GeV})$, the $N\to h\nu$ mode has extremely small branching ratio. In our case we have $Z$ mediated heavy neutrino pair-production channel where the mixing angle is directly involved in the production cross section. Using such pair-production channel in the inverse seesaw framework, we are probing the mixing angle for the heavy neutrinos at the colliders and these are not discussed in \cite{Kang:2015uoc} as they were studying the $pp\to Z^{\prime} \to NN$ mode in the $U(1)^{\prime}$ inspired model. They have also studied the $3\ell$ mode from the $Z^{\prime}$. However, the scale dependent variations in the LO and NLO studies are also not discussed in Ref~\cite{Kang:2015uoc}.}.
\subsection{$2\ell+4j$ final state}
\begin{figure}
\begin{center}
\includegraphics[scale=0.3]{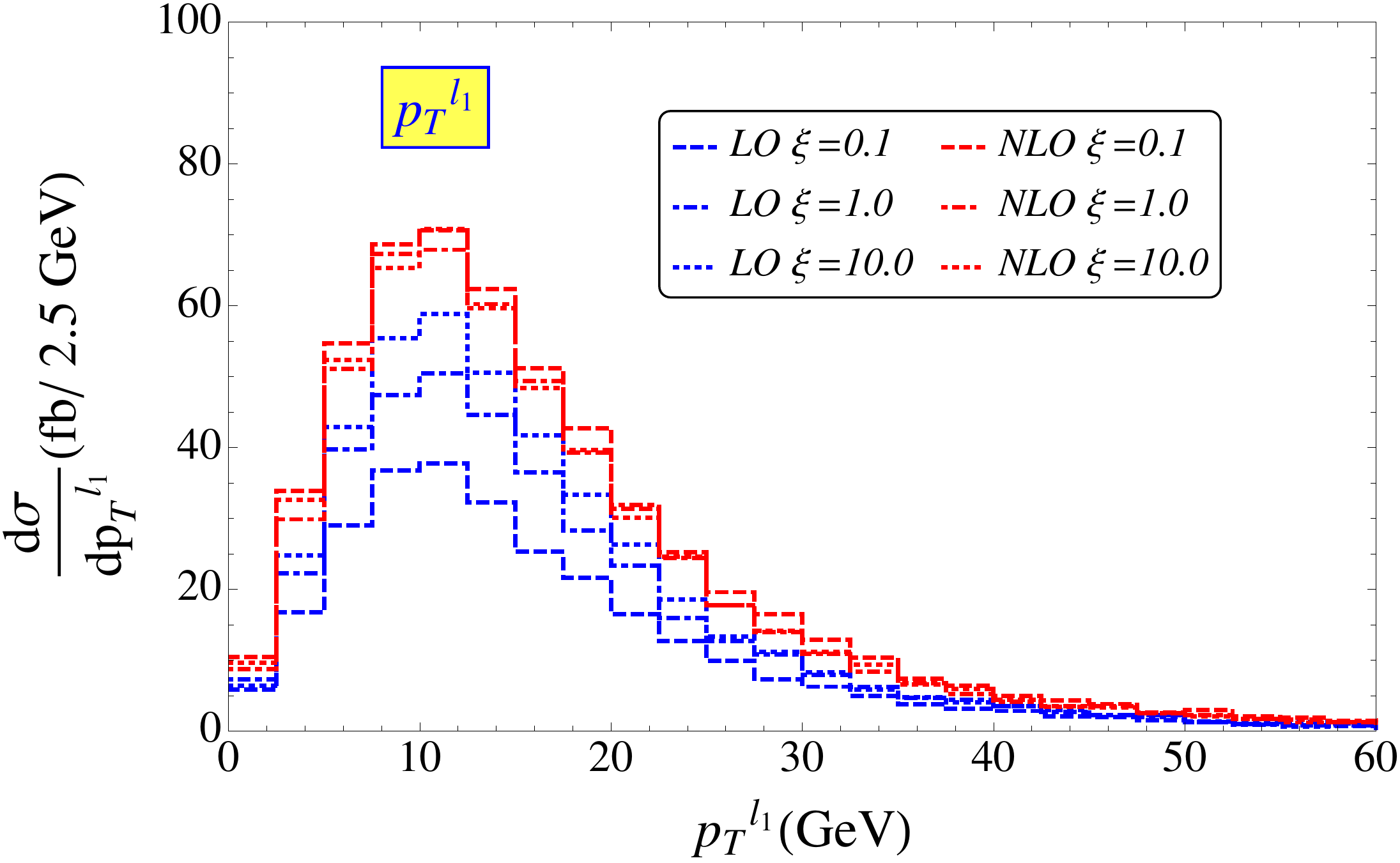}
\includegraphics[scale=0.3]{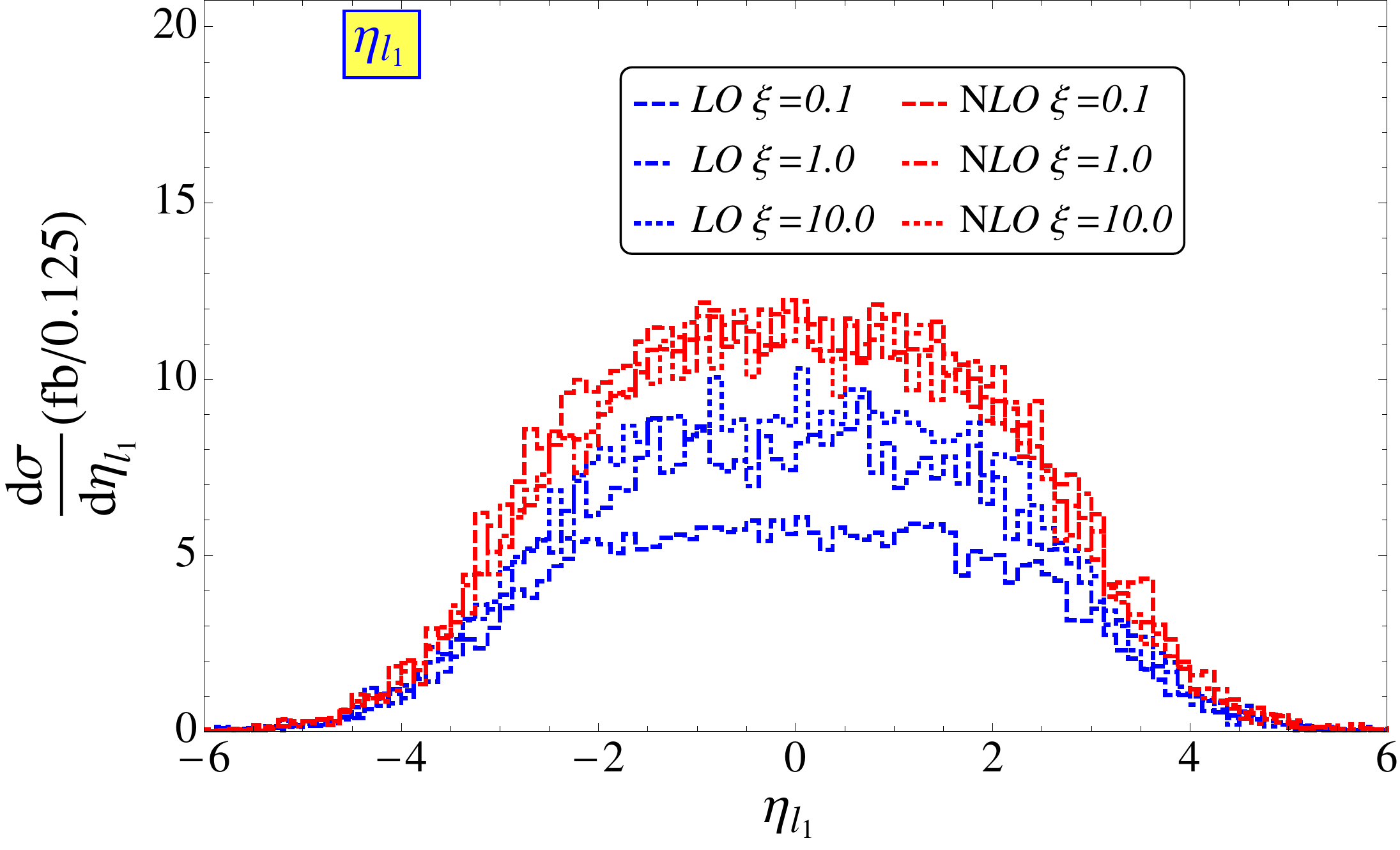}\\
\includegraphics[scale=0.3]{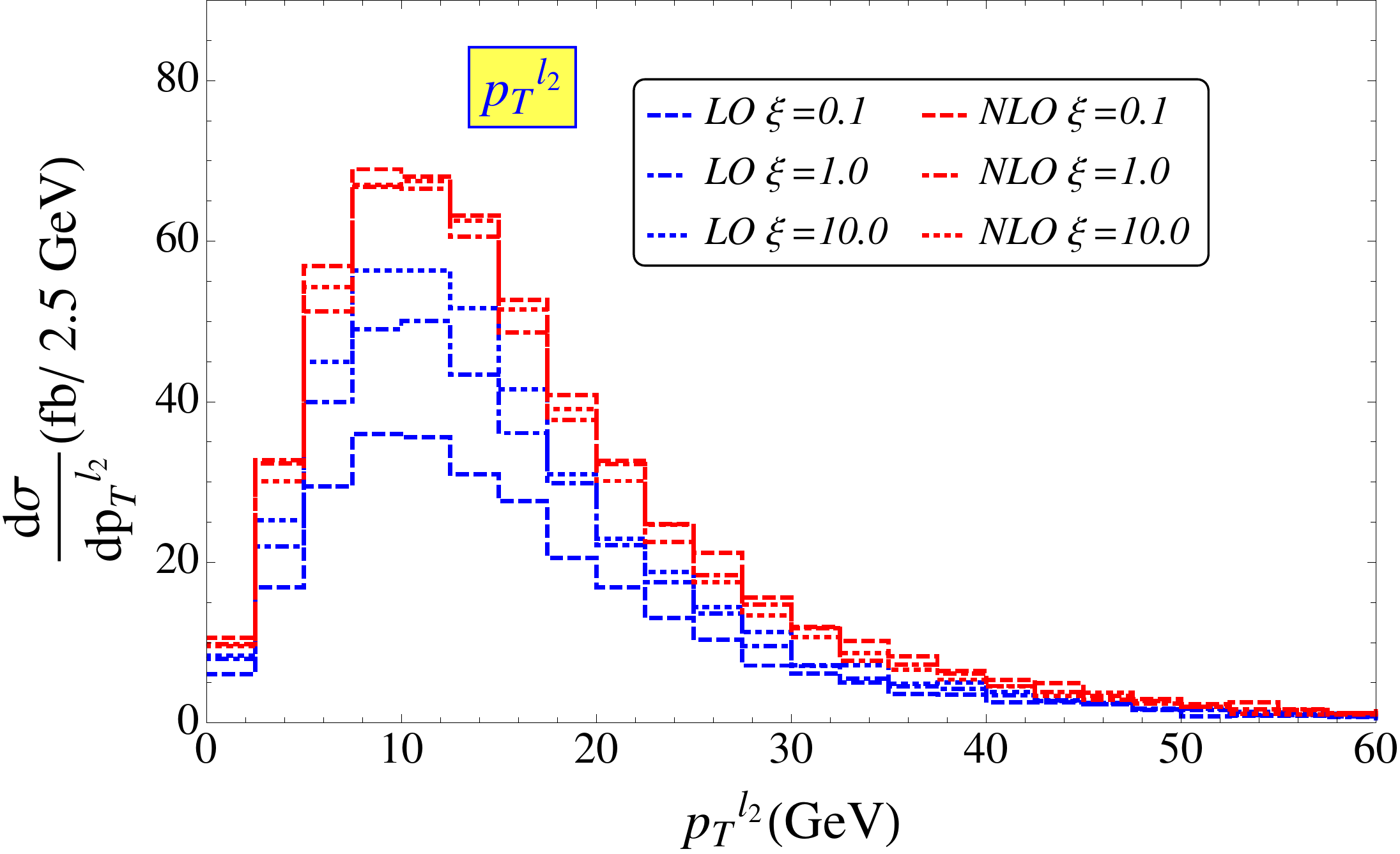}
\includegraphics[scale=0.3]{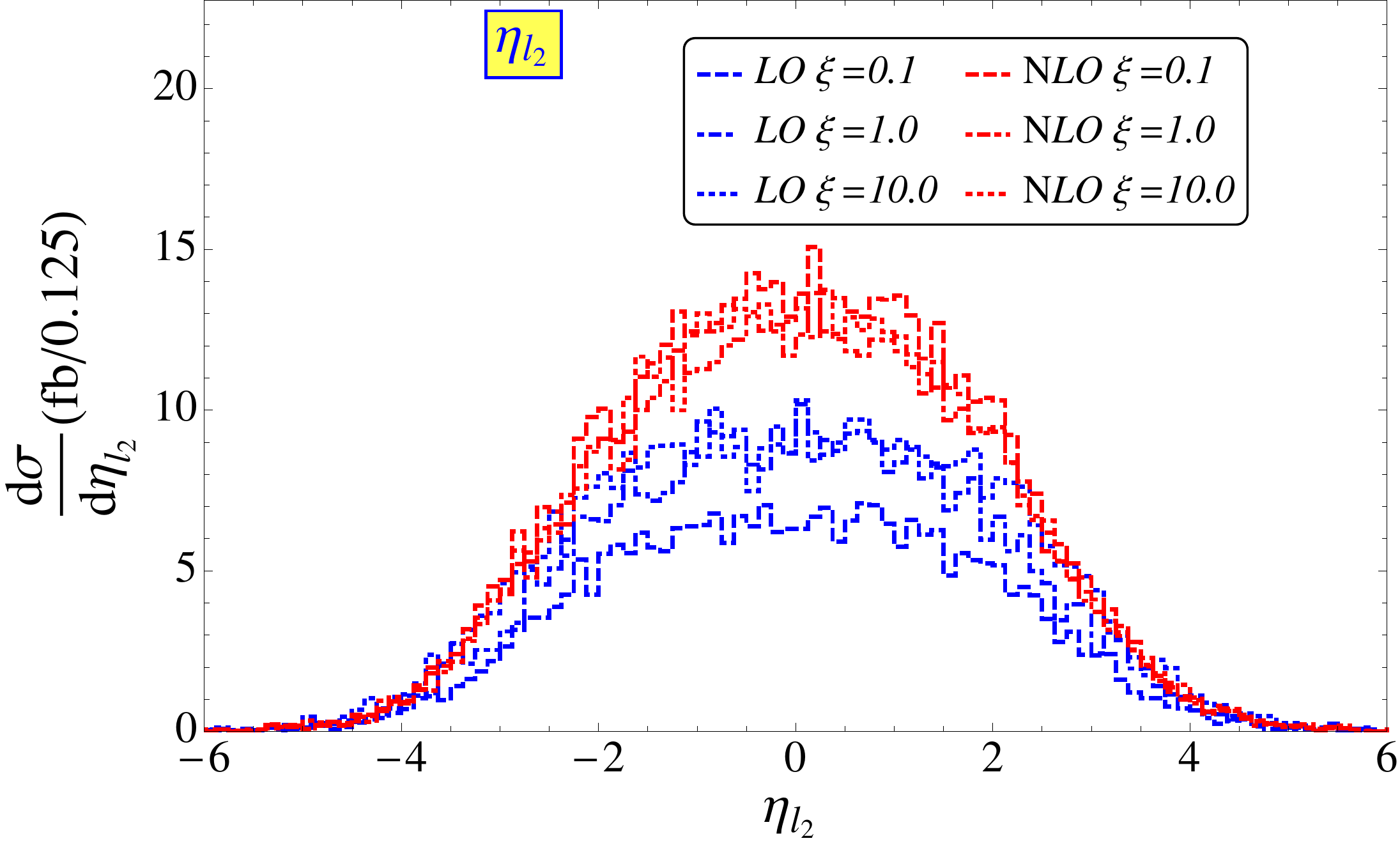}
\end{center}
\caption{Scale dependent LO and NLO-QCD $p_T^{\ell}$ (left column) and $\eta^{\ell}$ (right column) distributions of the heavy neutrino pair production followed by the decays of the heavy neutrinos into 
$2\ell+4j$ channel at the 13 TeV LHC for $m_N=95$ GeV.} 
\label{2l_95_0}
\end{figure} 
\begin{figure}
\begin{center}
\includegraphics[scale=0.3]{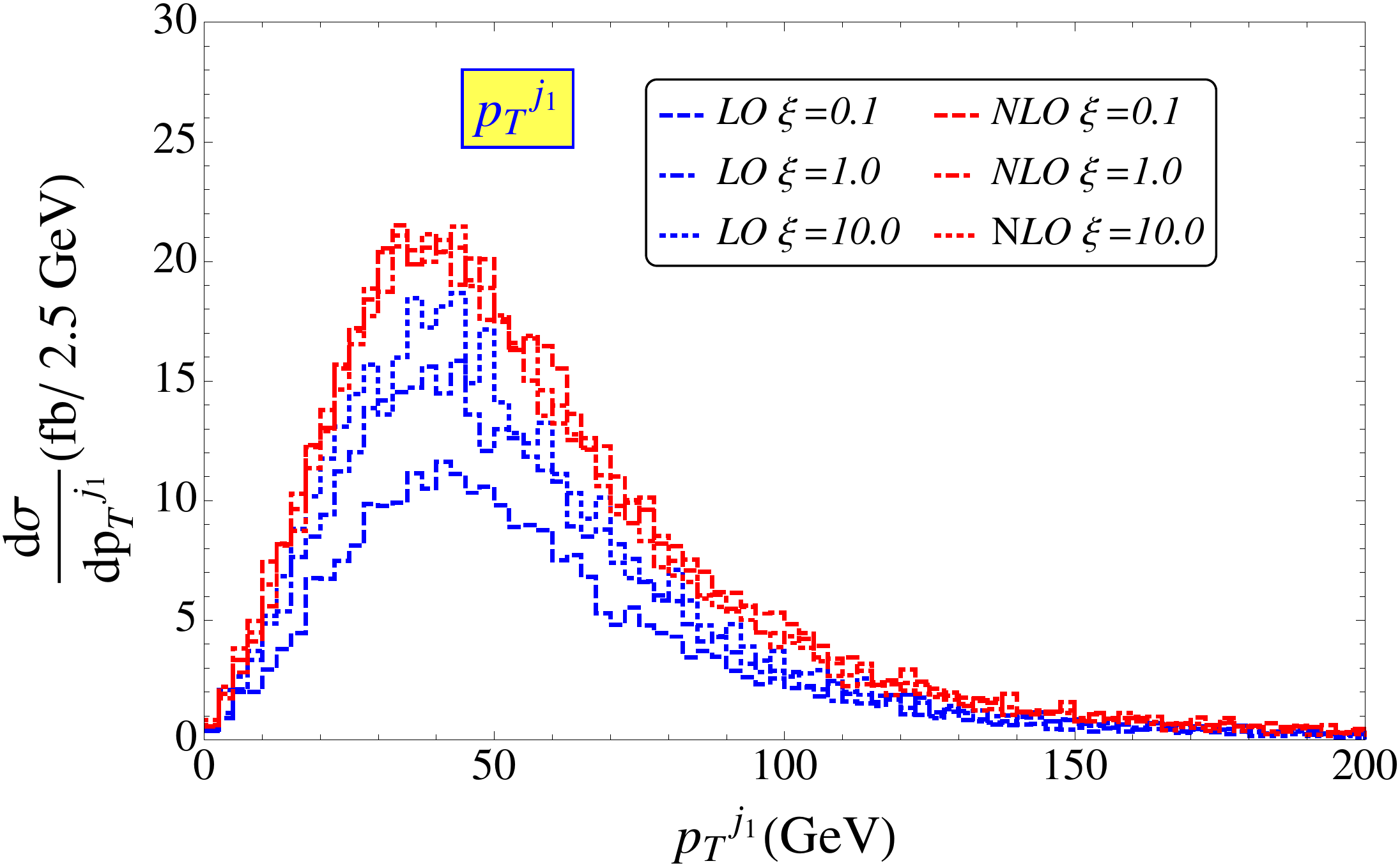}
\includegraphics[scale=0.3]{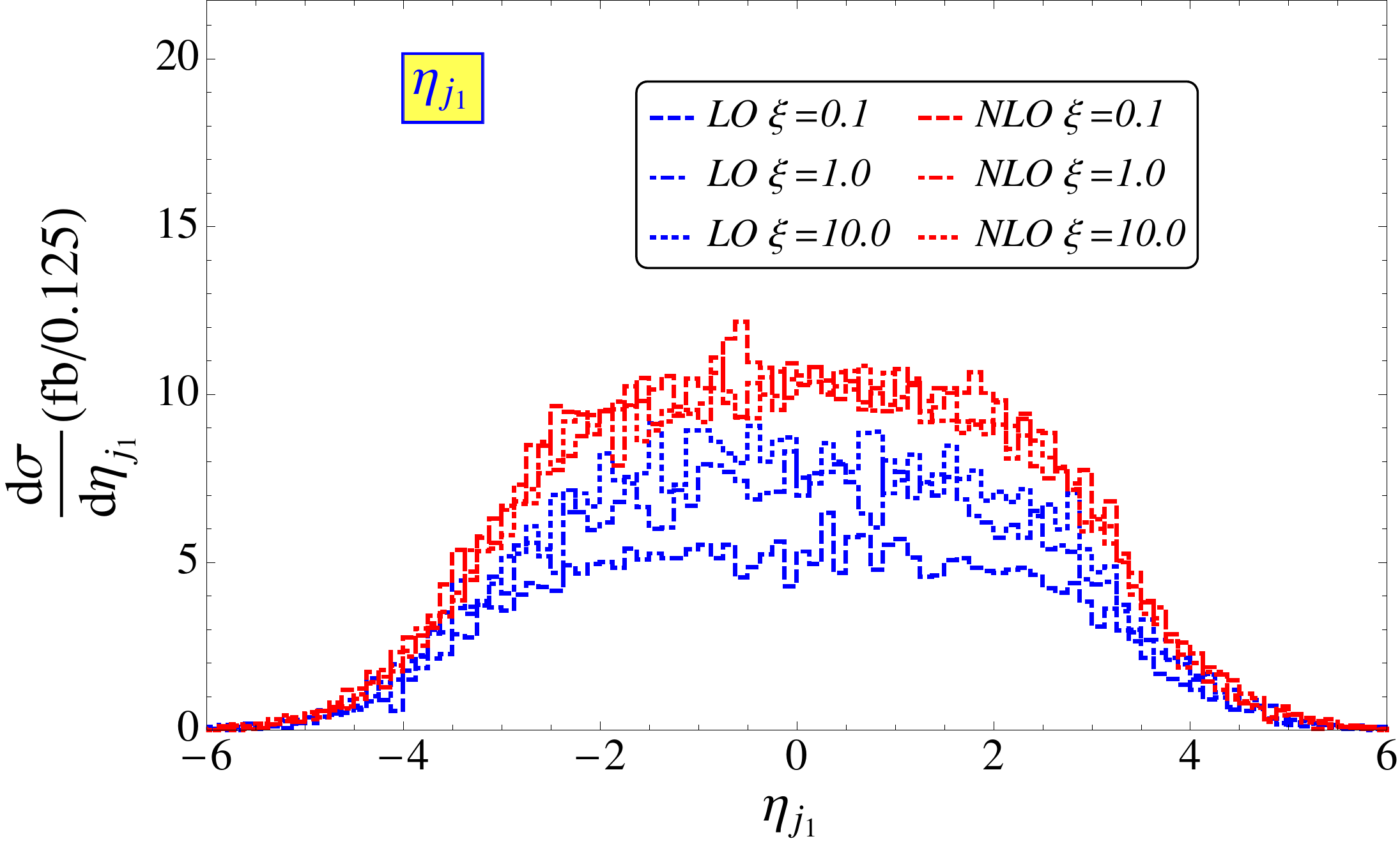}\\
\includegraphics[scale=0.32]{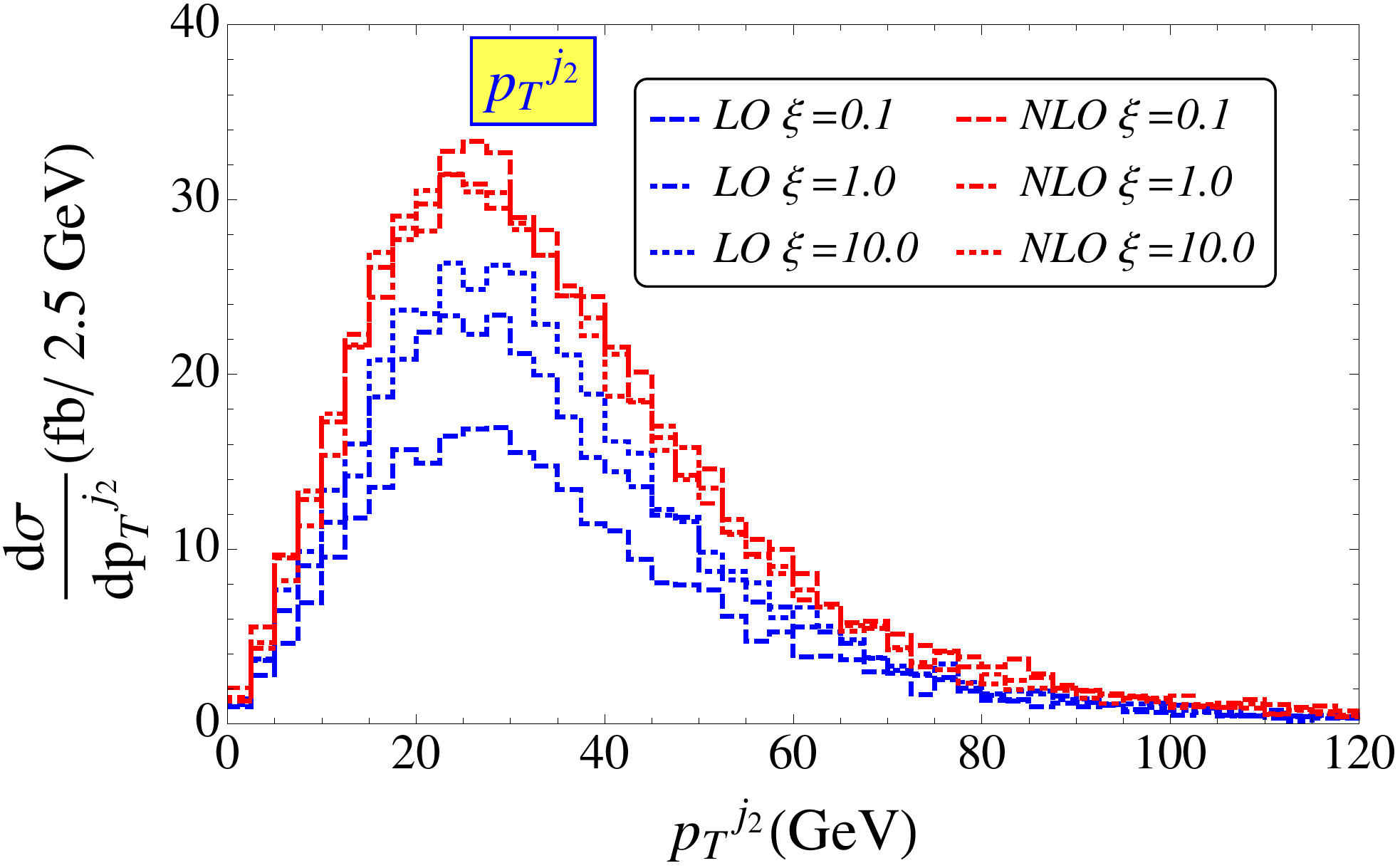}
\includegraphics[scale=0.3]{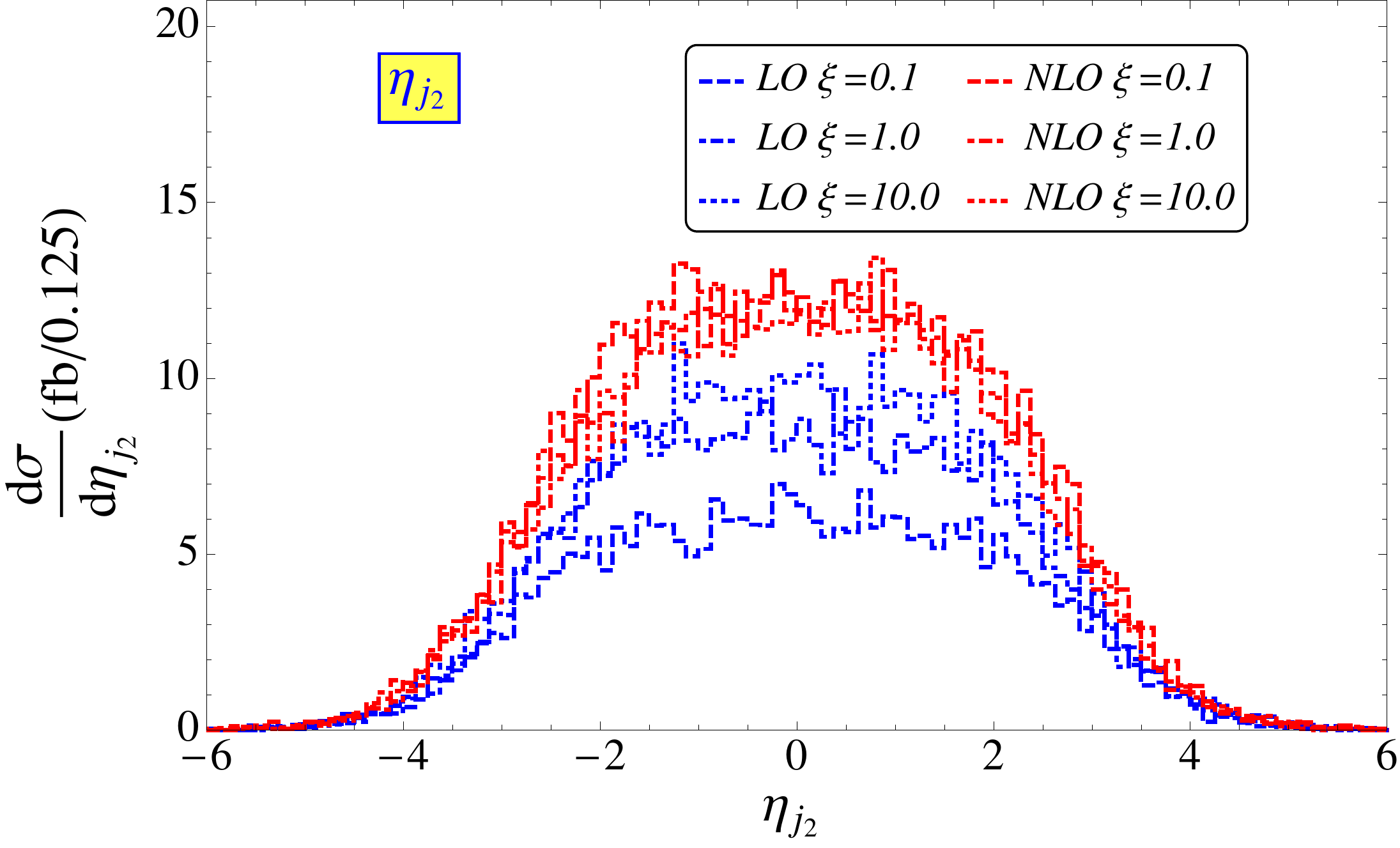}\\
\includegraphics[scale=0.3]{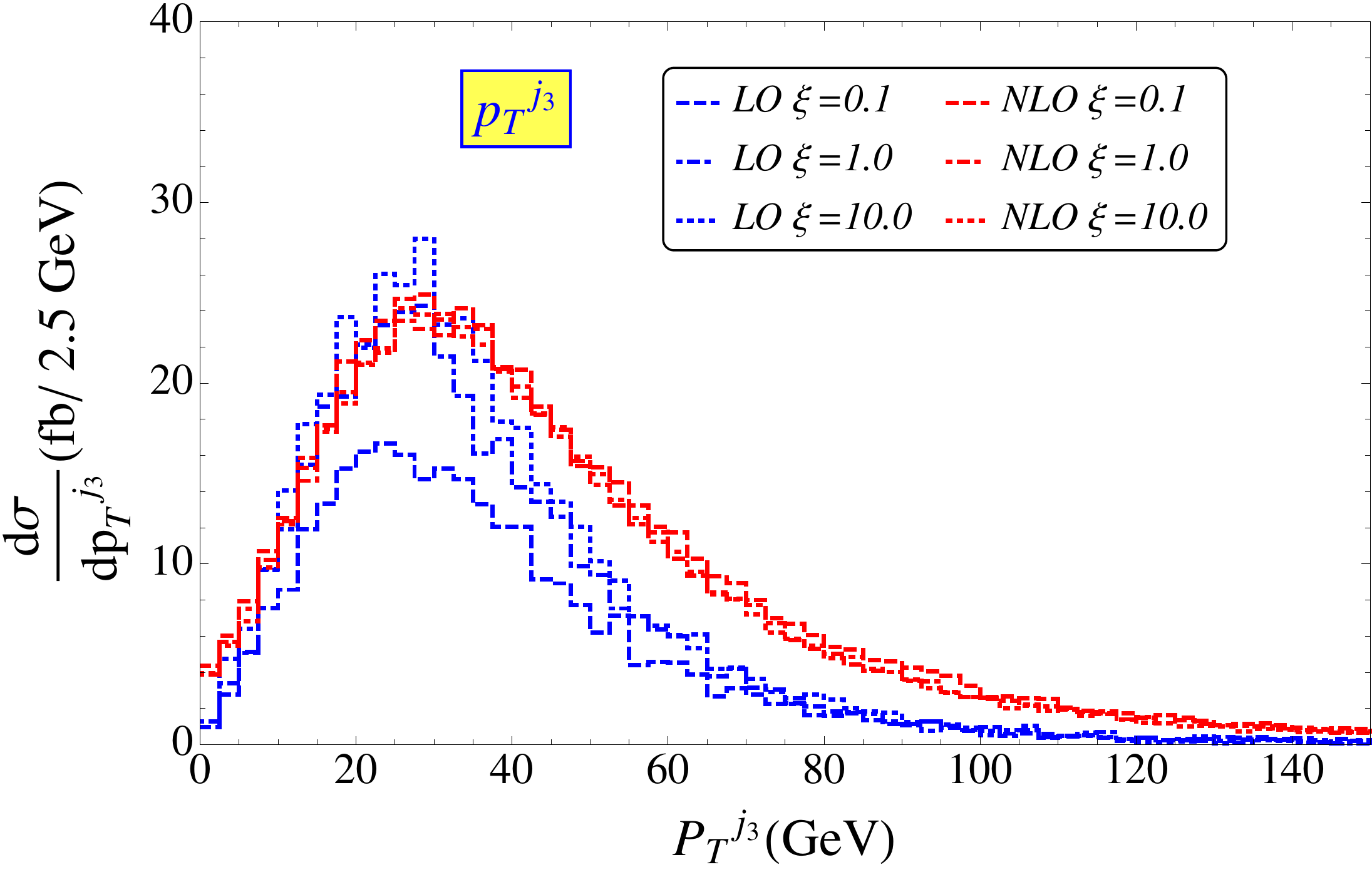}
\includegraphics[scale=0.3]{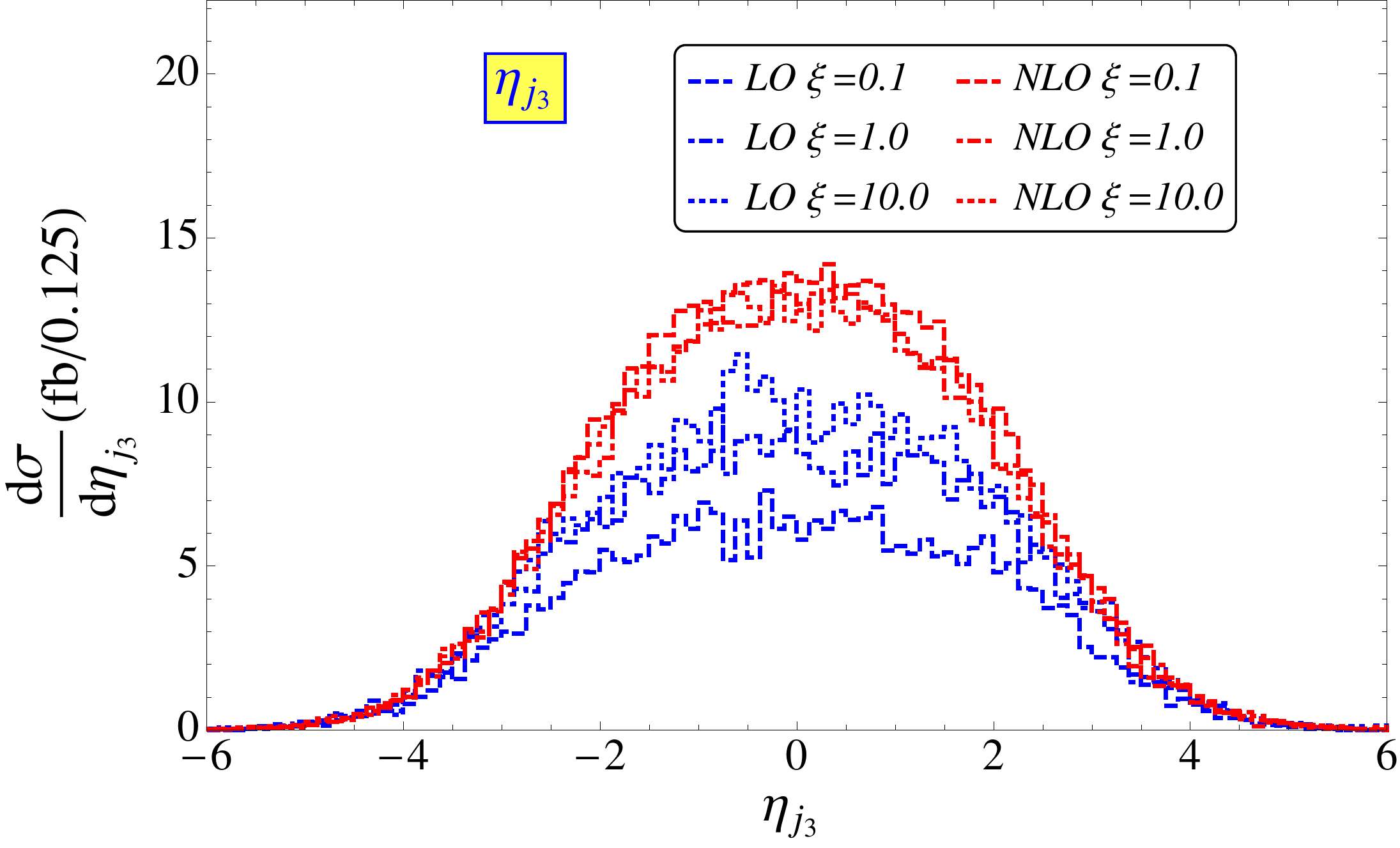}\\
\includegraphics[scale=0.3]{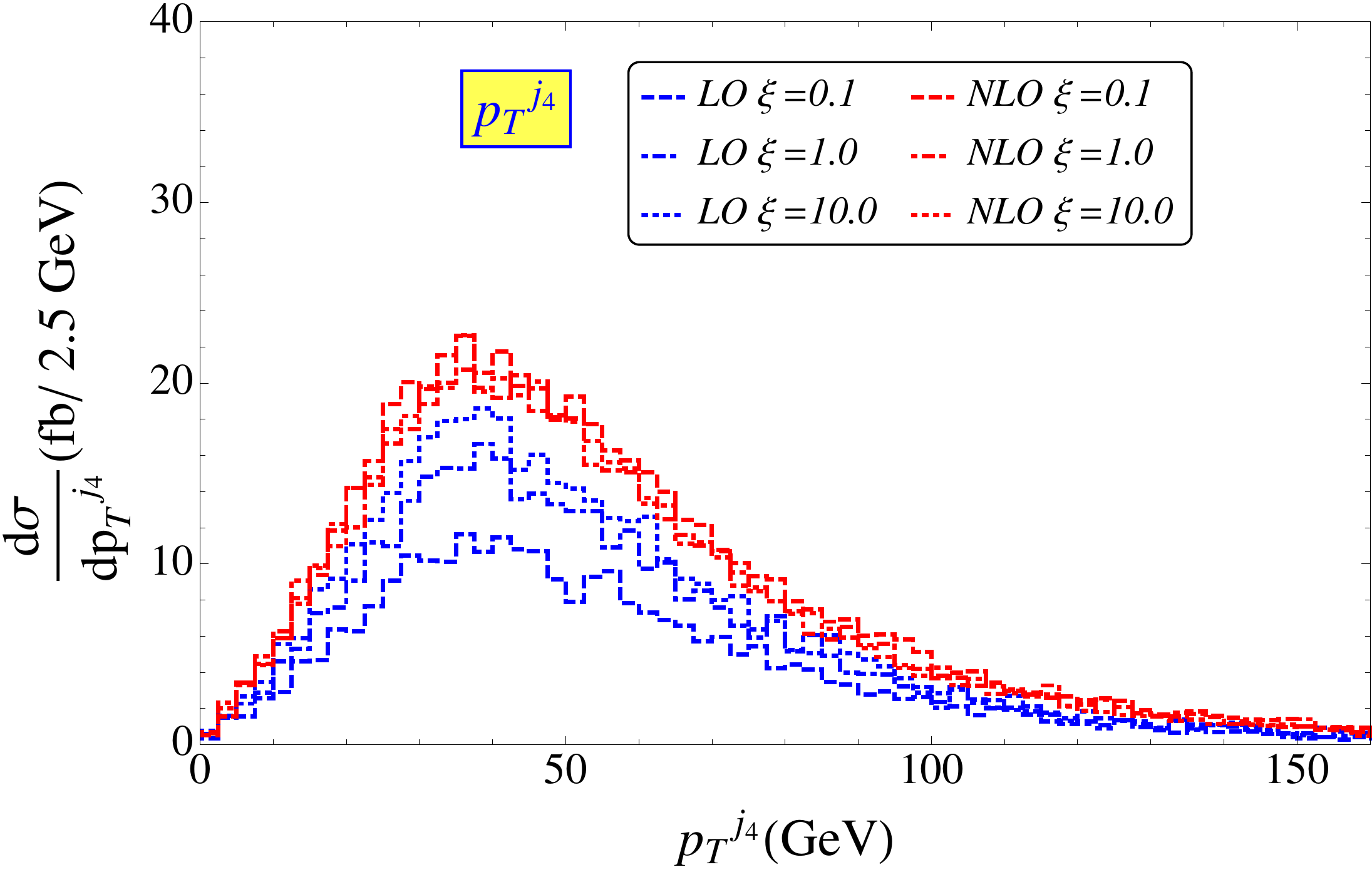}
\includegraphics[scale=0.3]{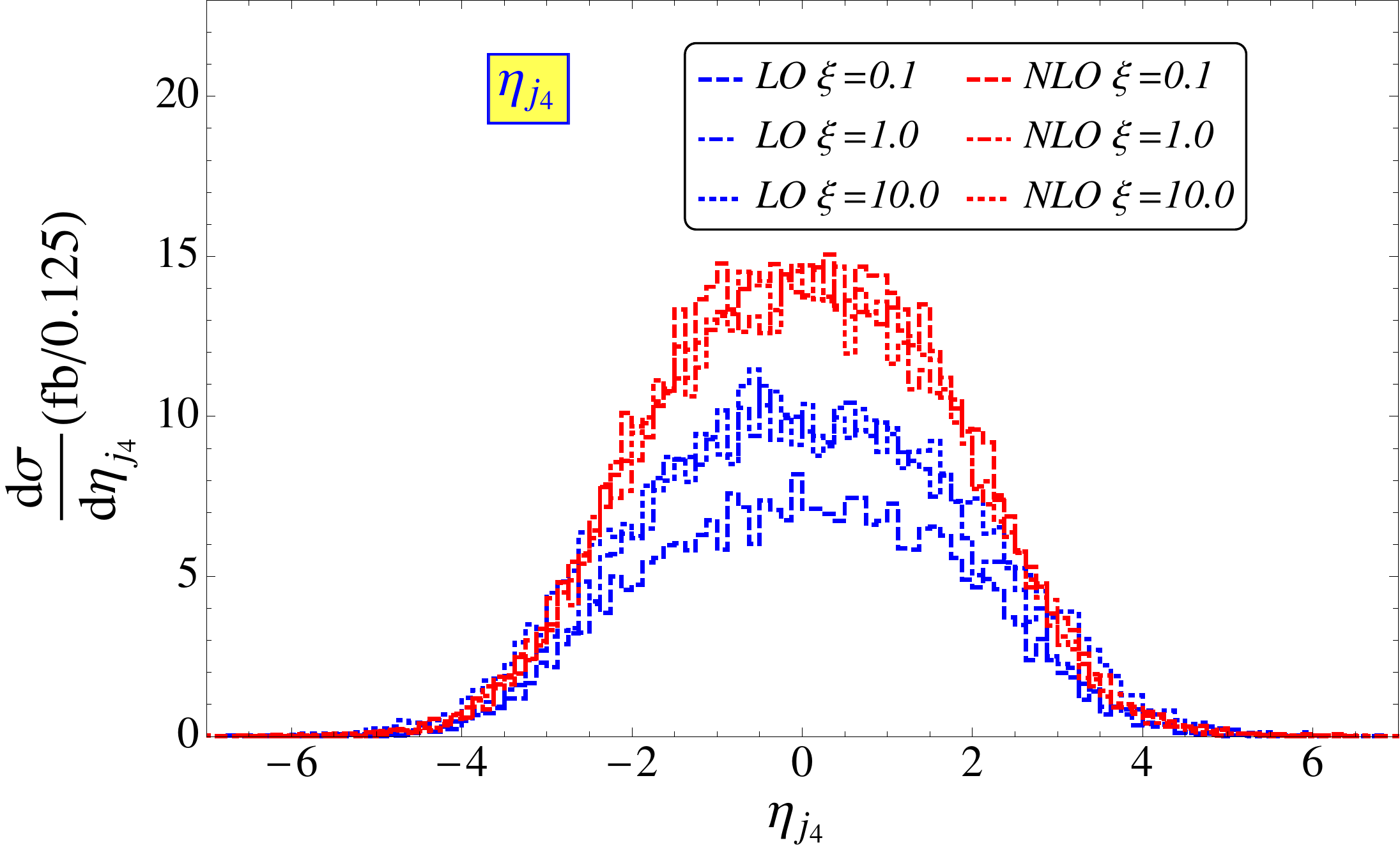}
\end{center}
\caption{Scale dependent LO and NLO-QCD $p_T^{j}$ (left column) and $\eta^{j}$ (right column) distributions of the heavy neutrino pair production followed by the decays of the heavy neutrinos into 
$2\ell+4j$ channel at the 13 TeV LHC for $m_N=95$ GeV.} 
\label{2l_95_1}
\end{figure} 
\begin{figure}
\begin{center}
\includegraphics[scale=0.4]{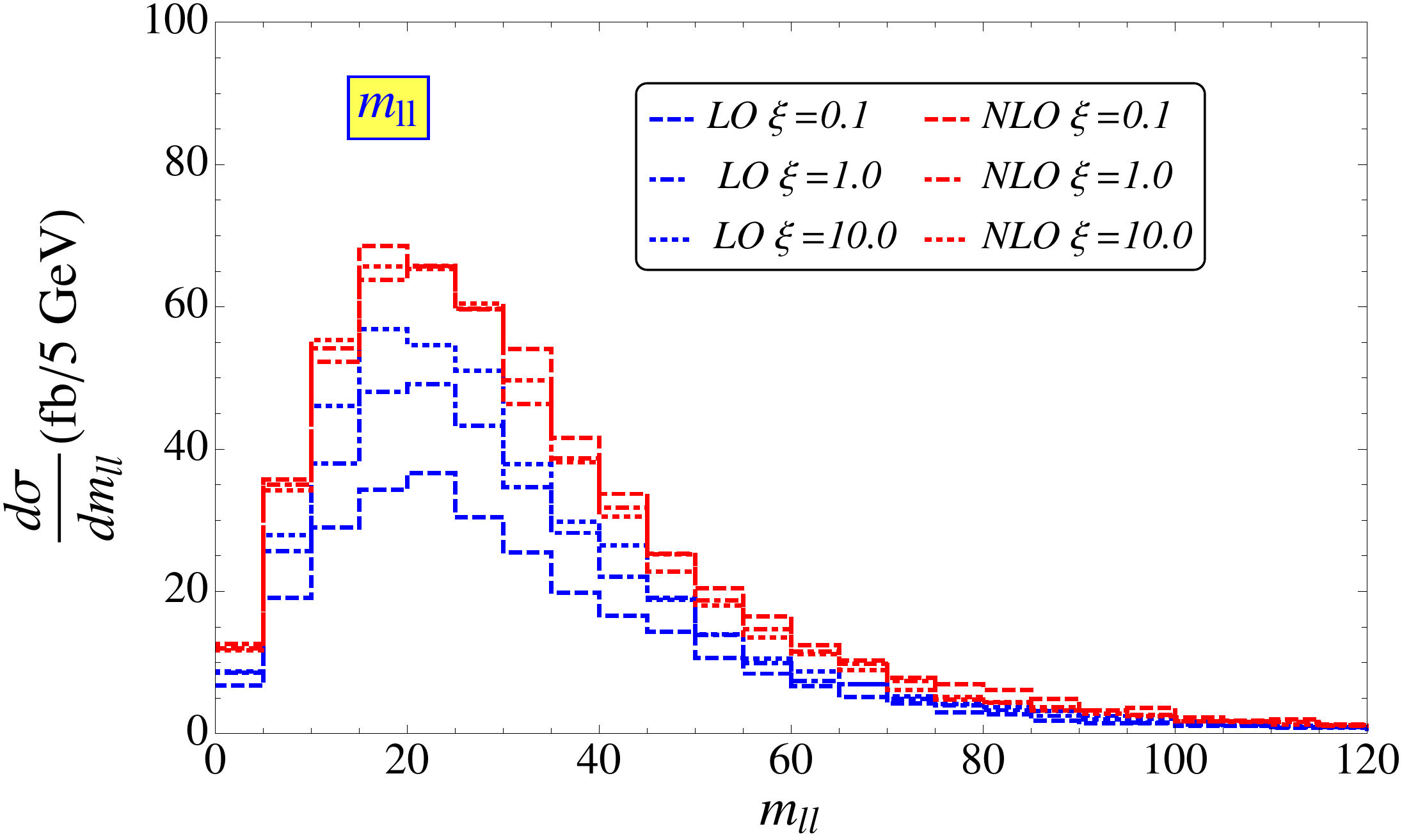}
\end{center}
\caption{Scale dependent LO and NLO-QCD $m_{\ell\ell}$ distributions of the heavy neutrino pair production followed by the decays of the heavy neutrinos into 
$2\ell+4j$ channel at the 13 TeV LHC for $m_N=95$ GeV.}
\label{2l_95_2}
\end{figure} 
\begin{figure}
\begin{center}
\includegraphics[scale=0.4]{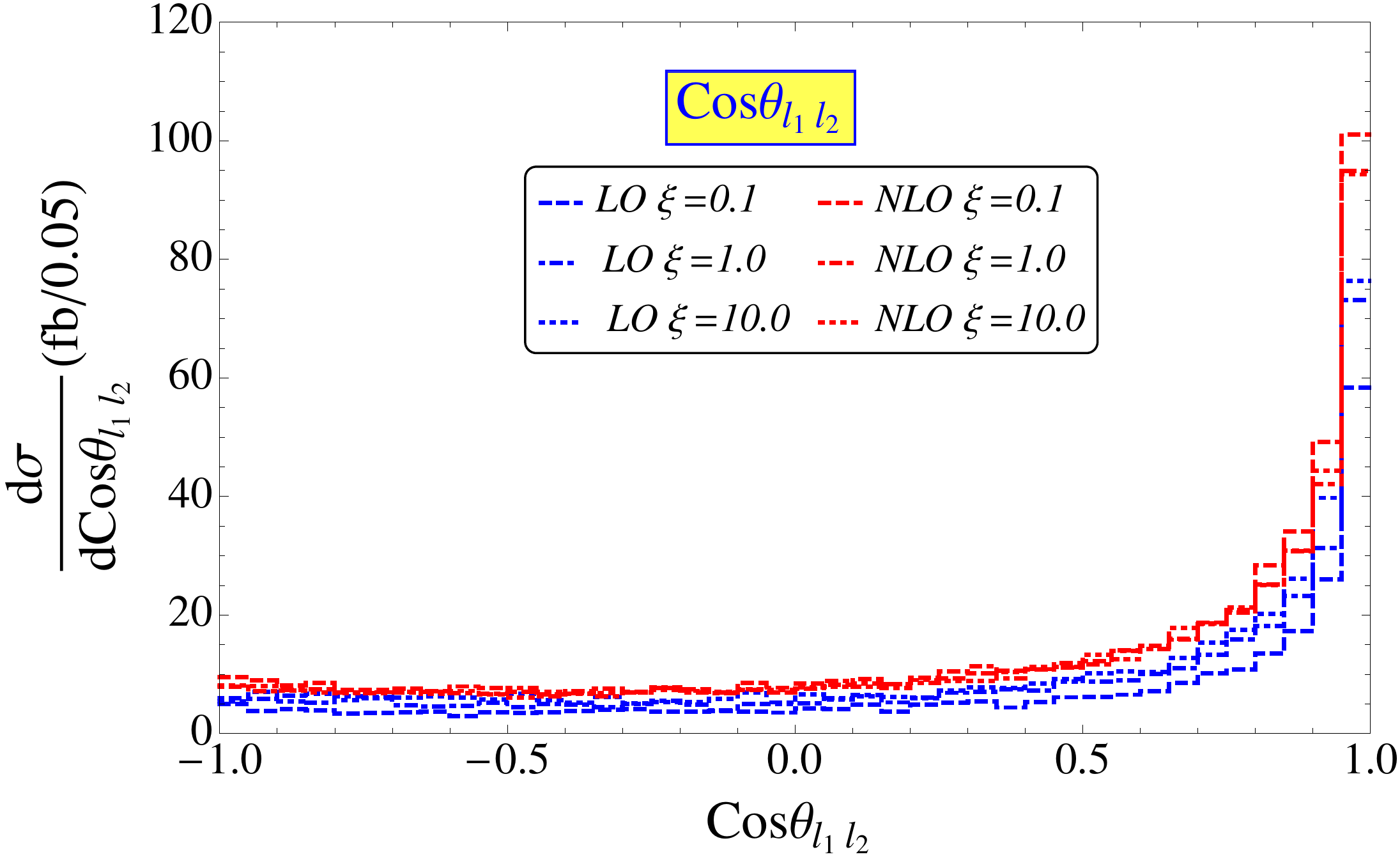}
\end{center}
\caption{Scale dependent LO and NLO-QCD $\cos\theta_{\ell\ell}$ distributions of the heavy neutrino pair production followed by the decays of the heavy neutrinos into 
$2\ell+4j$ channel at the 13 TeV LHC for $m_N=95$ GeV.} 
\label{2l_95_3}
\end{figure} 
\begin{figure}
\begin{center}
\includegraphics[scale=0.23]{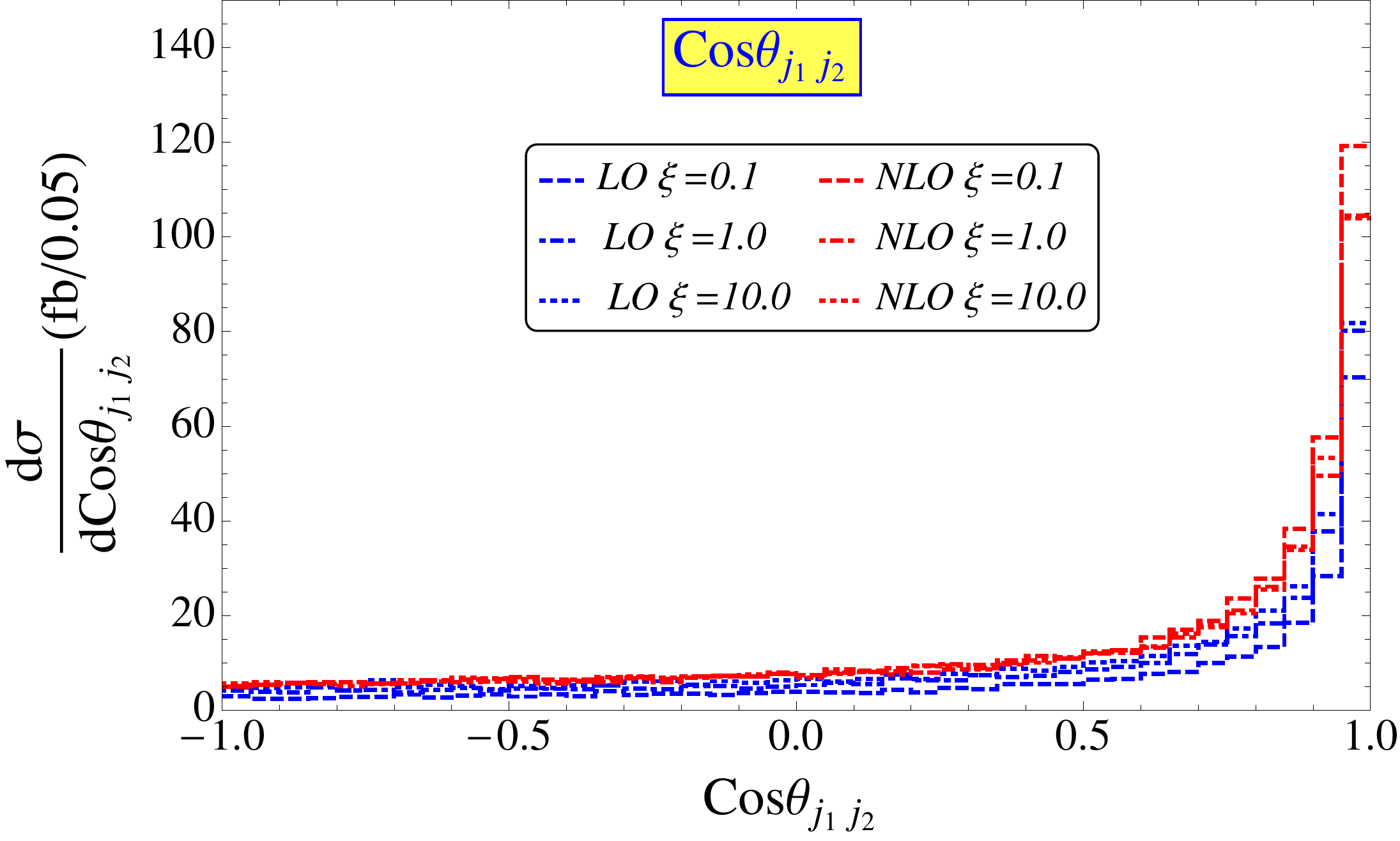}
\includegraphics[scale=0.23]{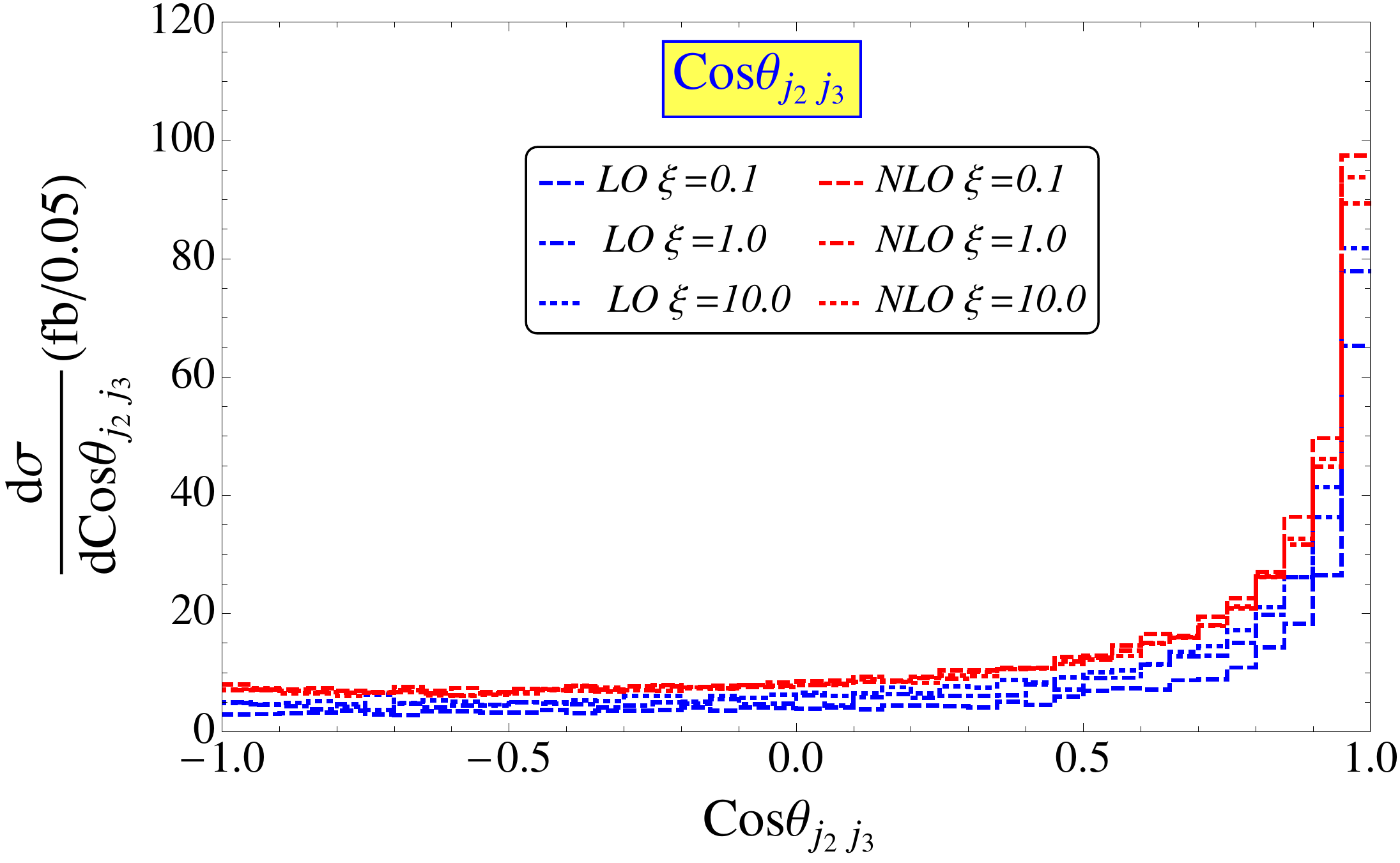}
\includegraphics[scale=0.23]{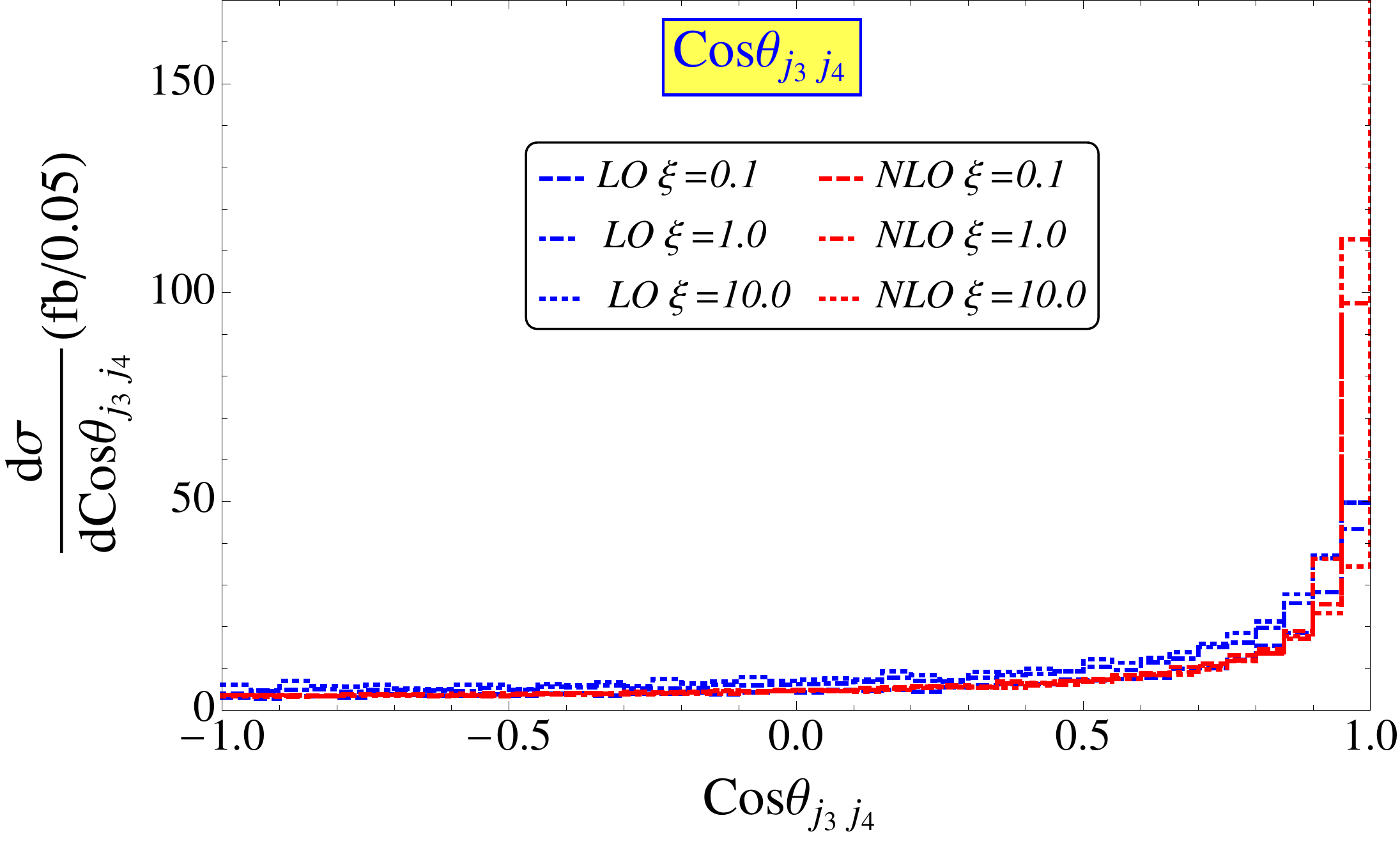}\\
\includegraphics[scale=0.23]{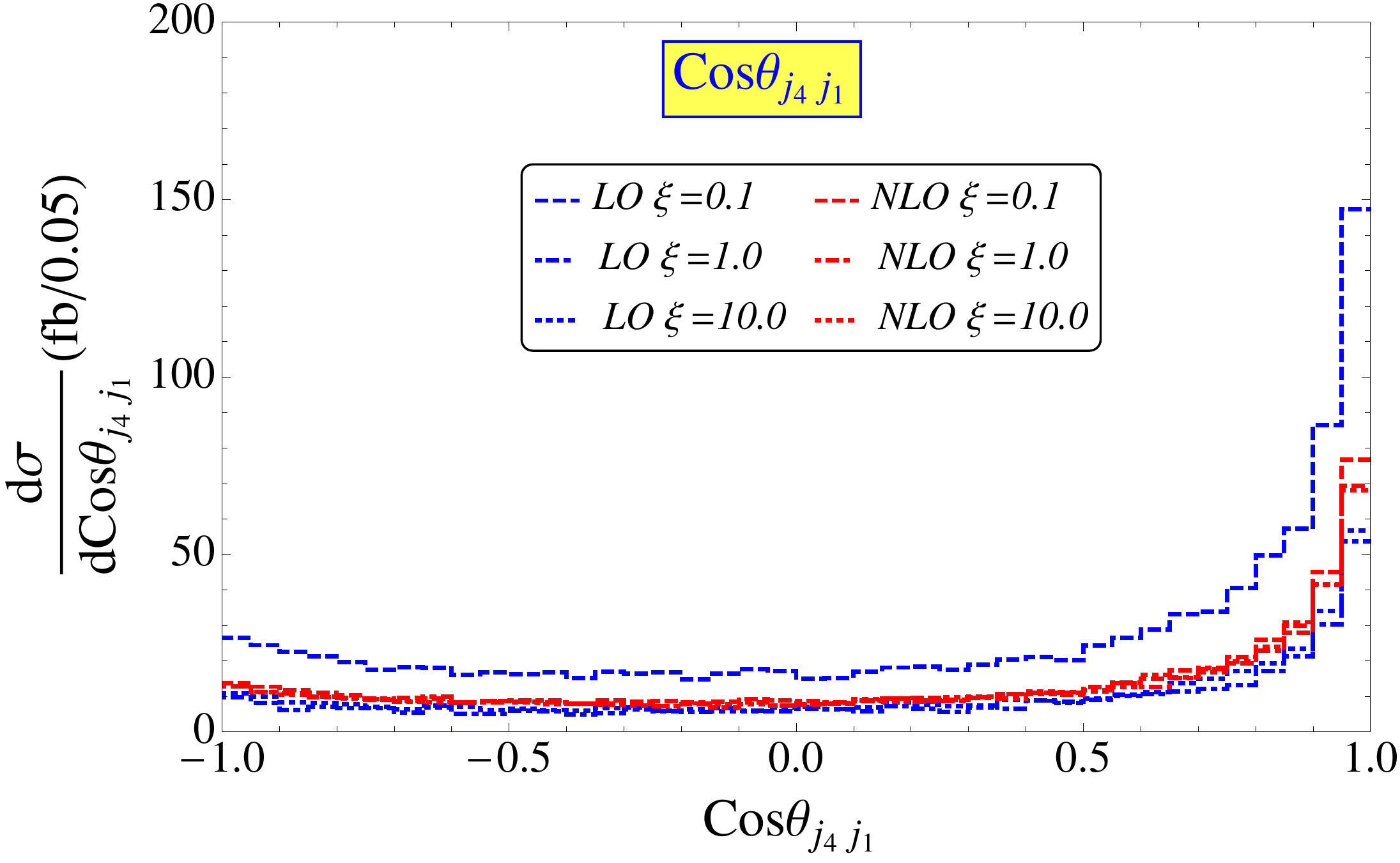}
\includegraphics[scale=0.23]{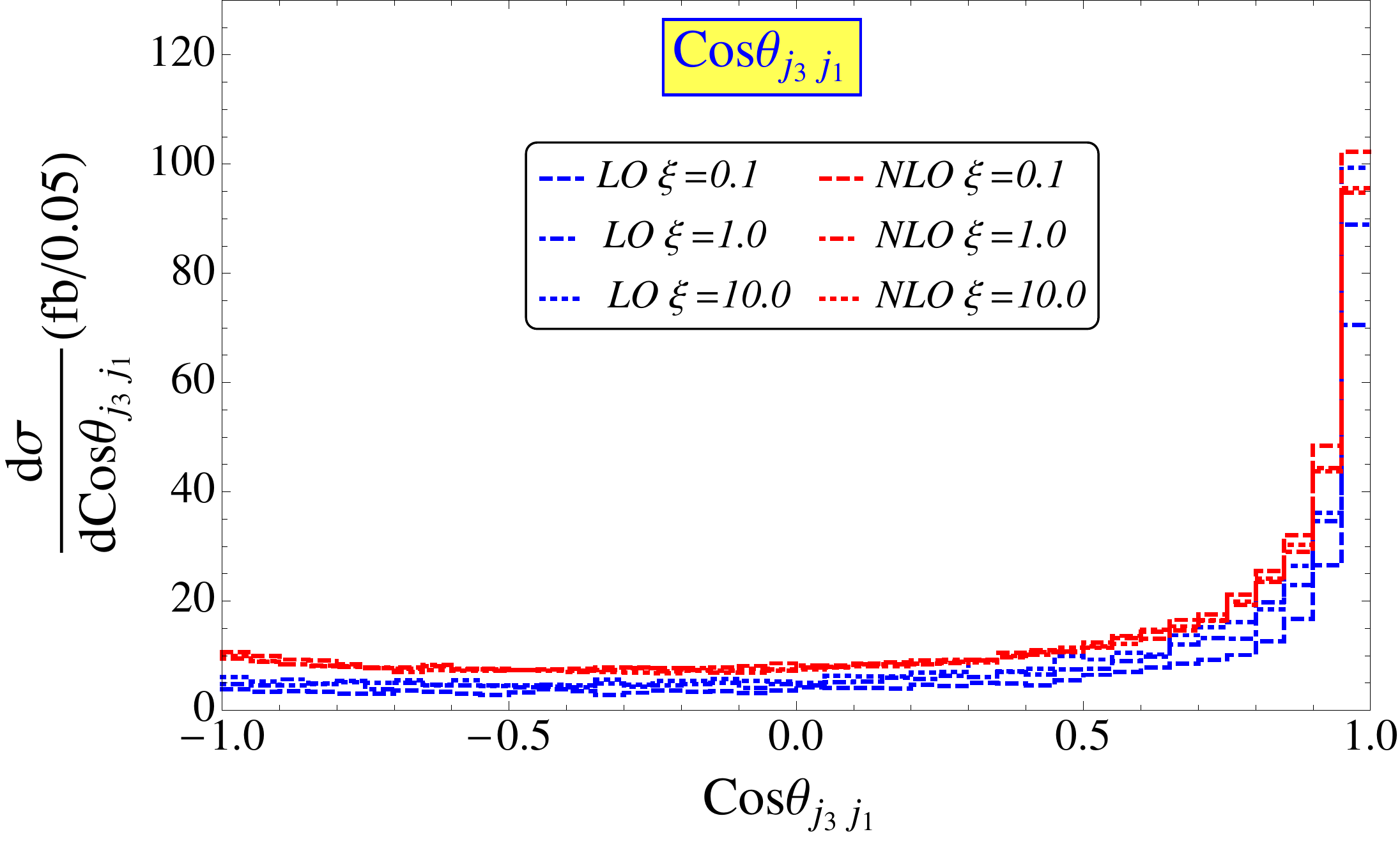}
\includegraphics[scale=0.23]{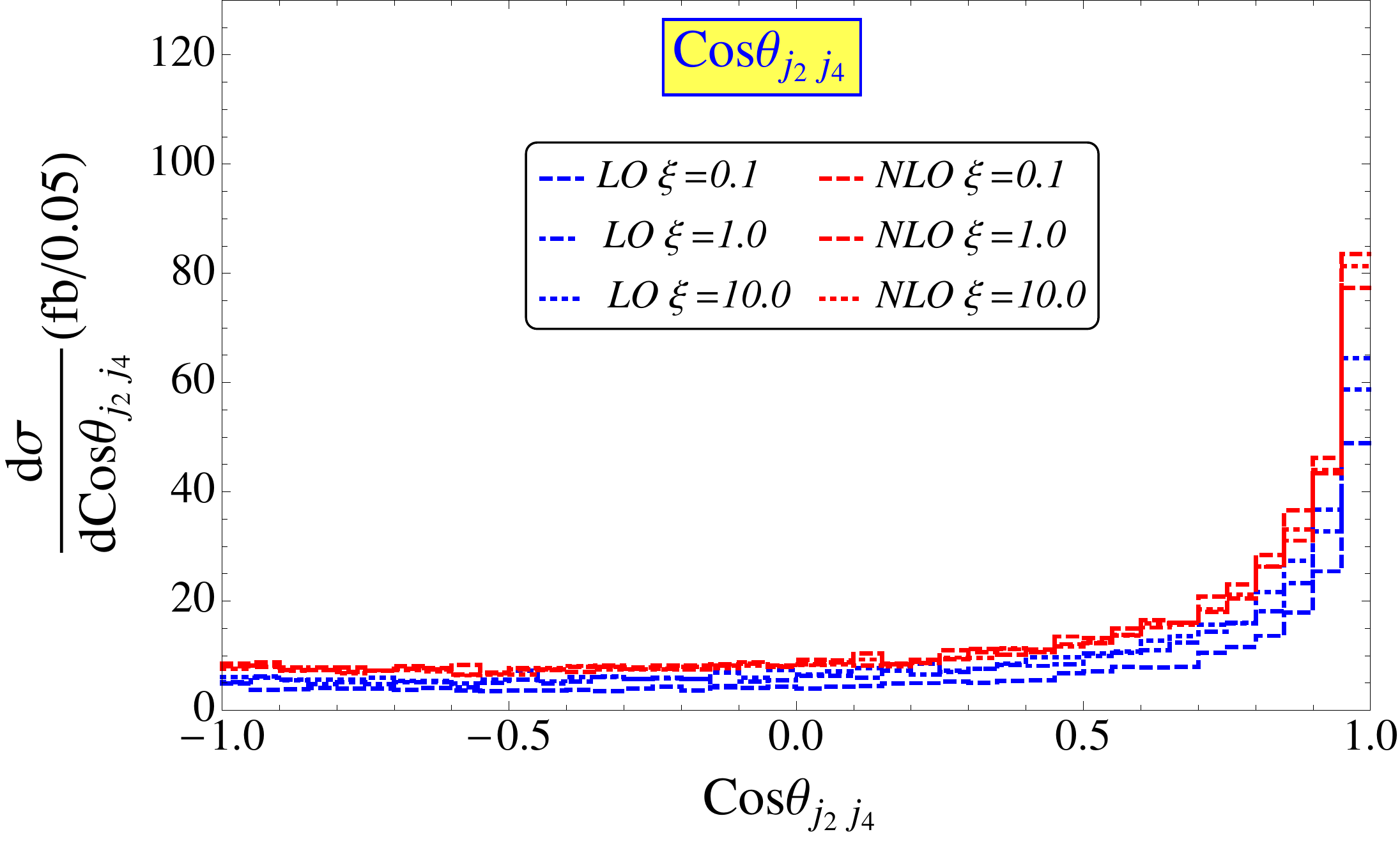}
\end{center}
\caption{Scale dependent LO and NLO-QCD $\cos\theta_{jj}$ distributions of the heavy neutrino pair production followed by the decays of the heavy neutrinos into 
$2\ell+4j$ channel at the 13 TeV LHC for $m_N=95$ GeV.} 
\label{2l_95_4}
\end{figure} 
\begin{figure}
\begin{center}
\includegraphics[scale=0.4]{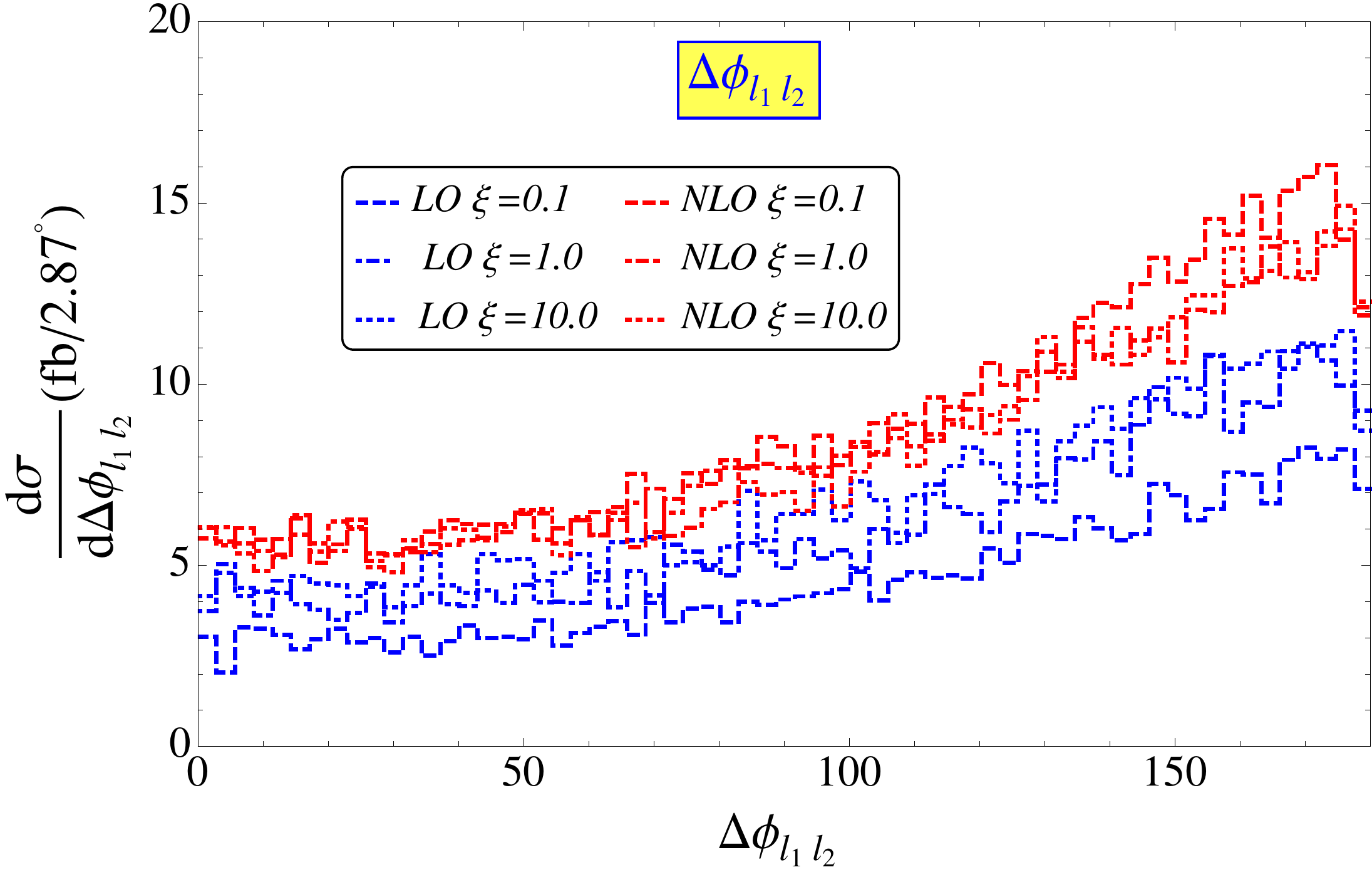}
\end{center}
\caption{Scale dependent LO and NLO-QCD $\Delta\phi_{\ell\ell}$ distributions of the heavy neutrino pair production followed by the decays of the heavy neutrinos into 
$2\ell+4j$ channel at the 13 TeV LHC for $m_N=95$ GeV.} 
\label{2l_95_5}
\end{figure} 
\begin{figure}
\begin{center}
\includegraphics[scale=0.23]{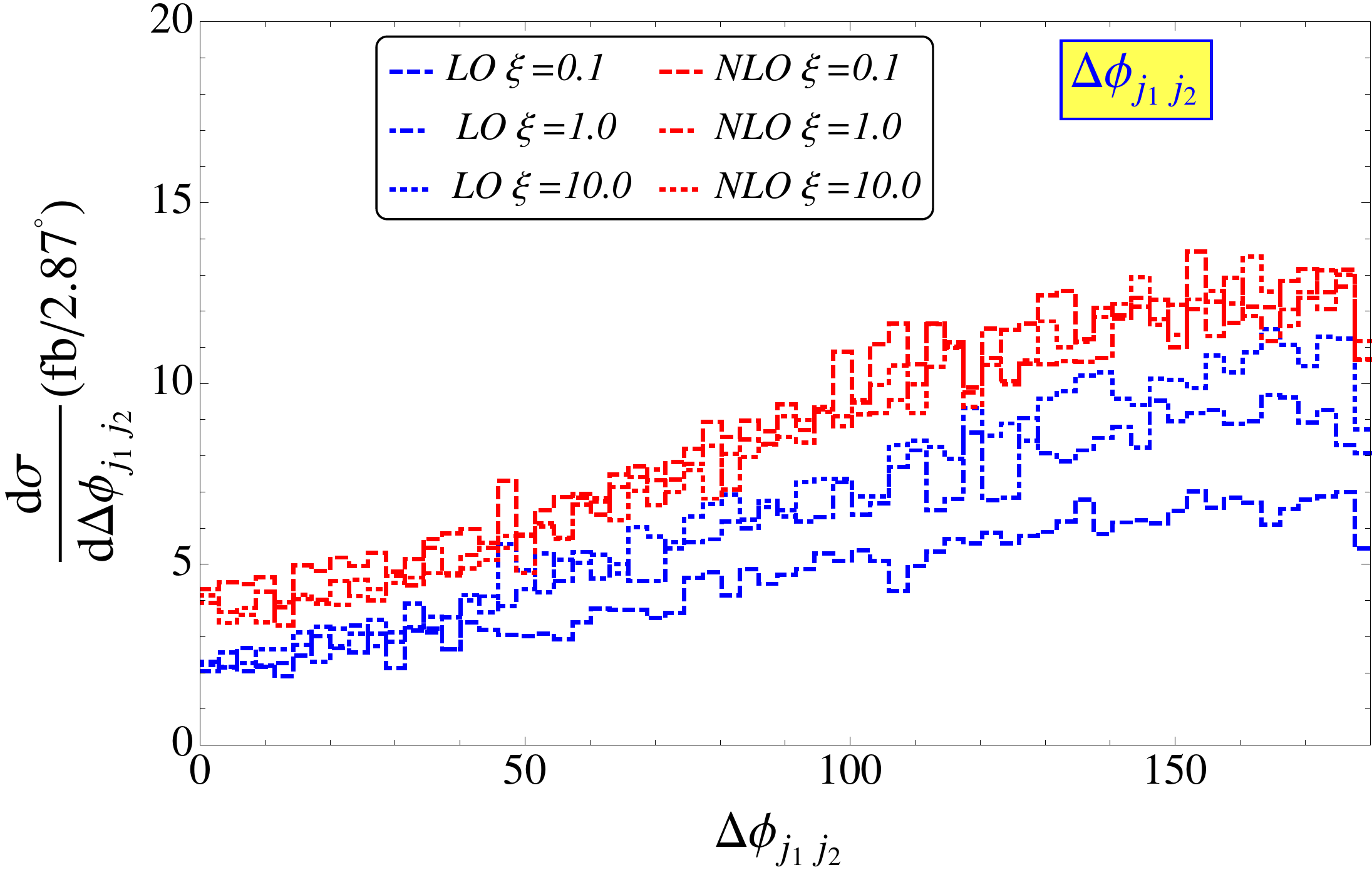}
\includegraphics[scale=0.23]{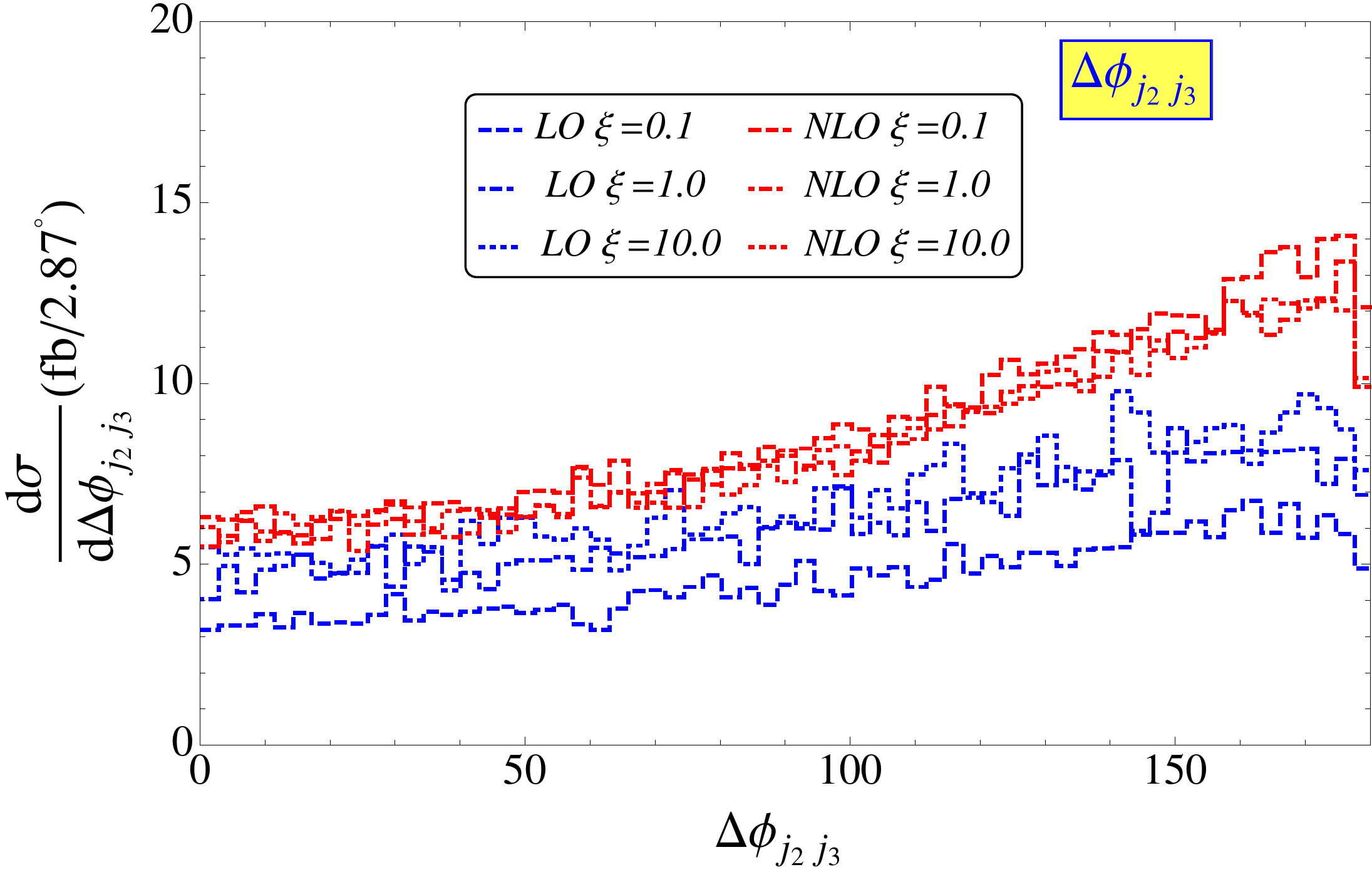}
\includegraphics[scale=0.23]{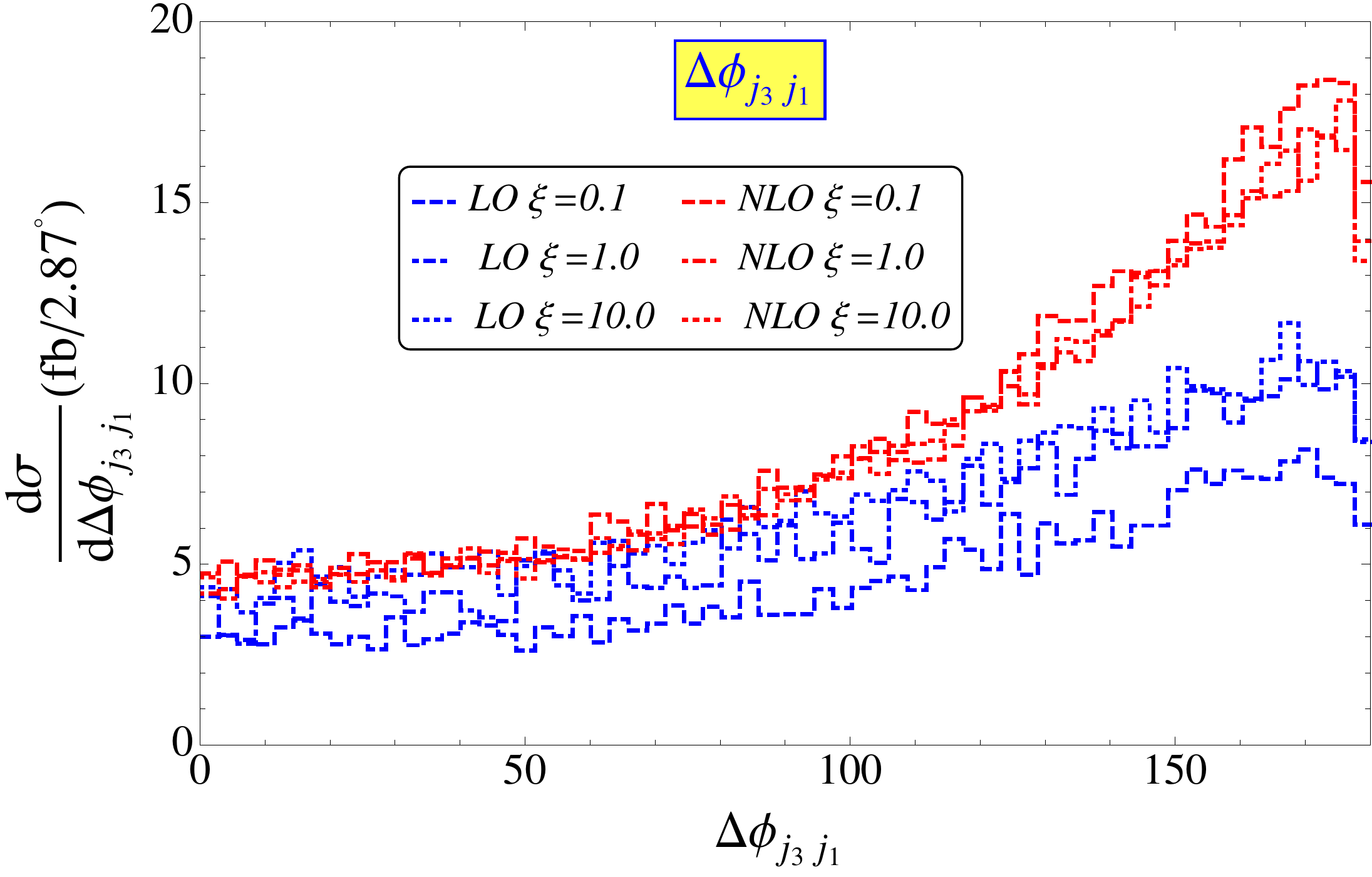}\\
\includegraphics[scale=0.23]{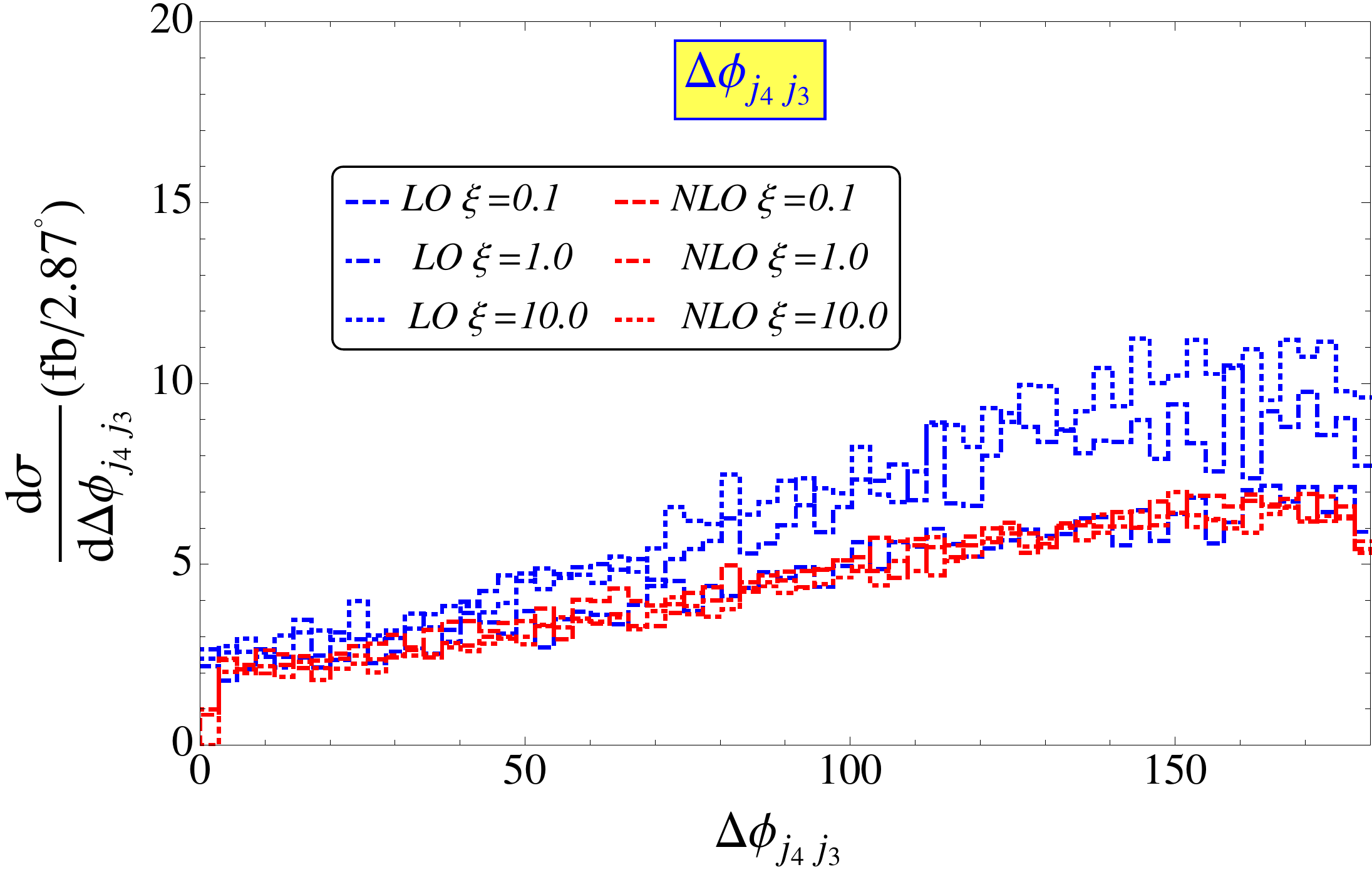}
\includegraphics[scale=0.23]{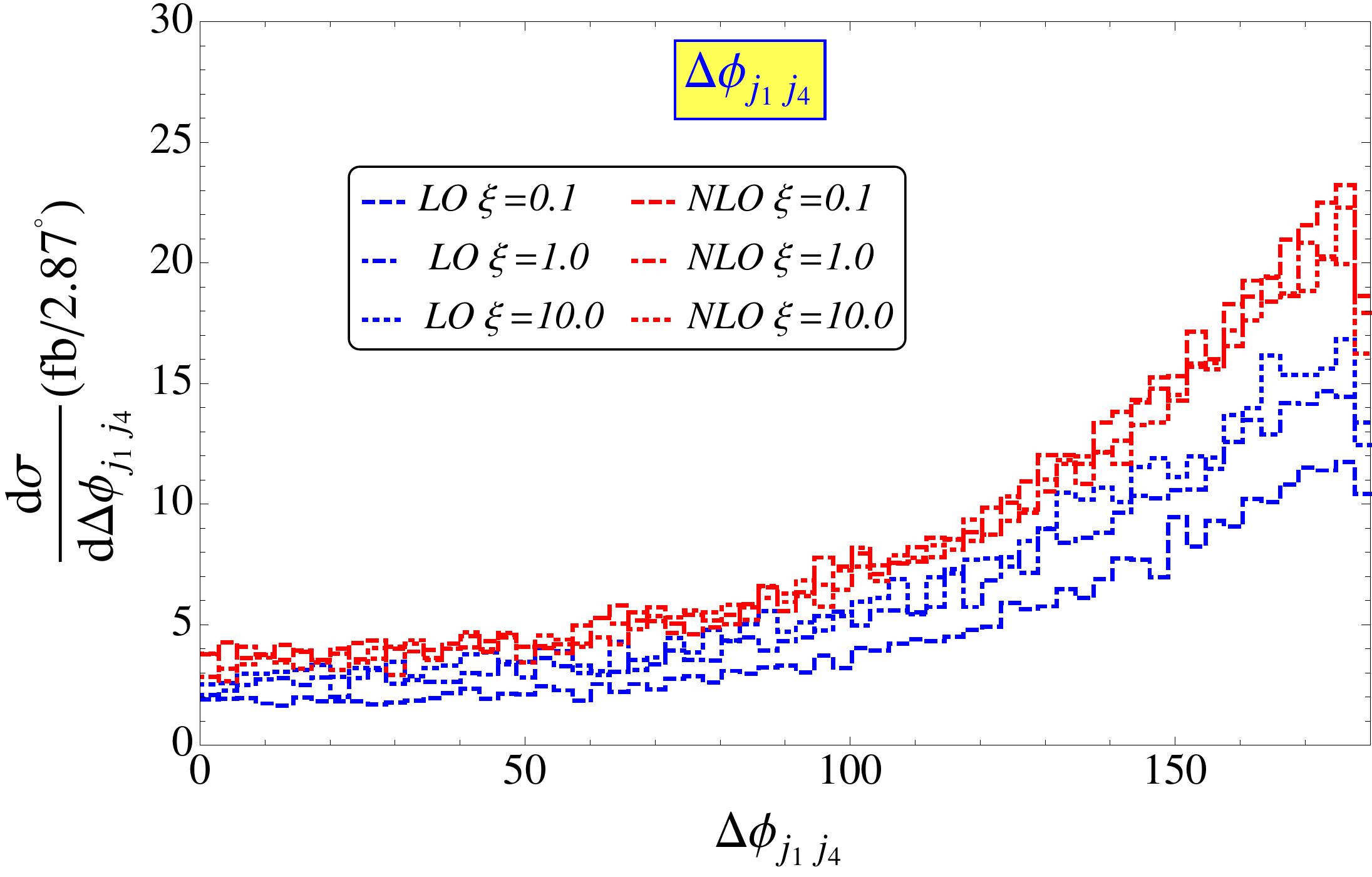}
\includegraphics[scale=0.23]{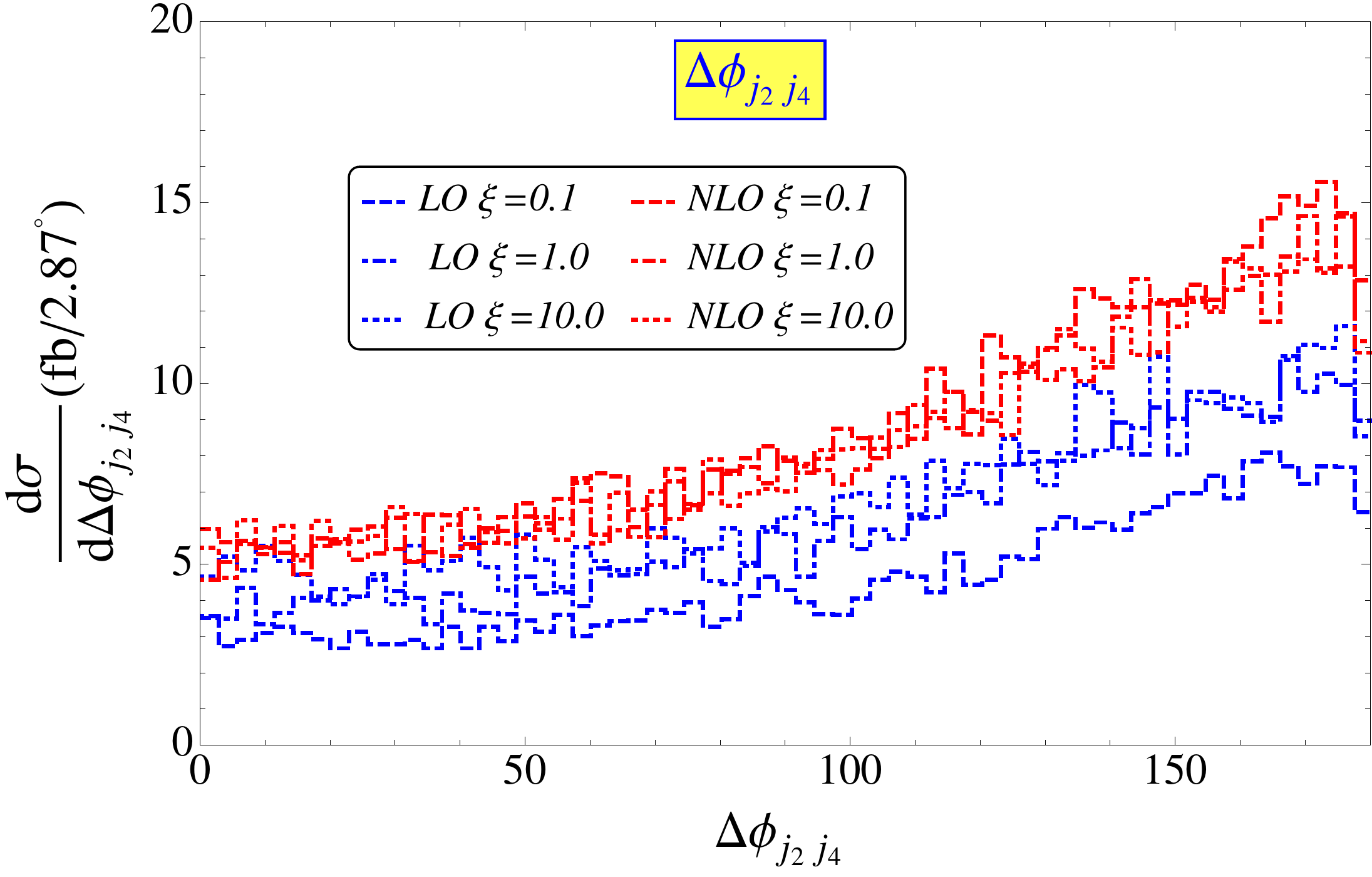}
\end{center}
\caption{Scale dependent LO and NLO-QCD $\Delta\phi_{jj}$ distributions of the heavy neutrino pair production followed by the decays of the heavy neutrinos into 
$2\ell+4j$ channel at the 13 TeV LHC for $m_N=95$ GeV.} 
\label{2l_95_6}
\end{figure} 
\begin{figure}
\begin{center}
\includegraphics[scale=0.23]{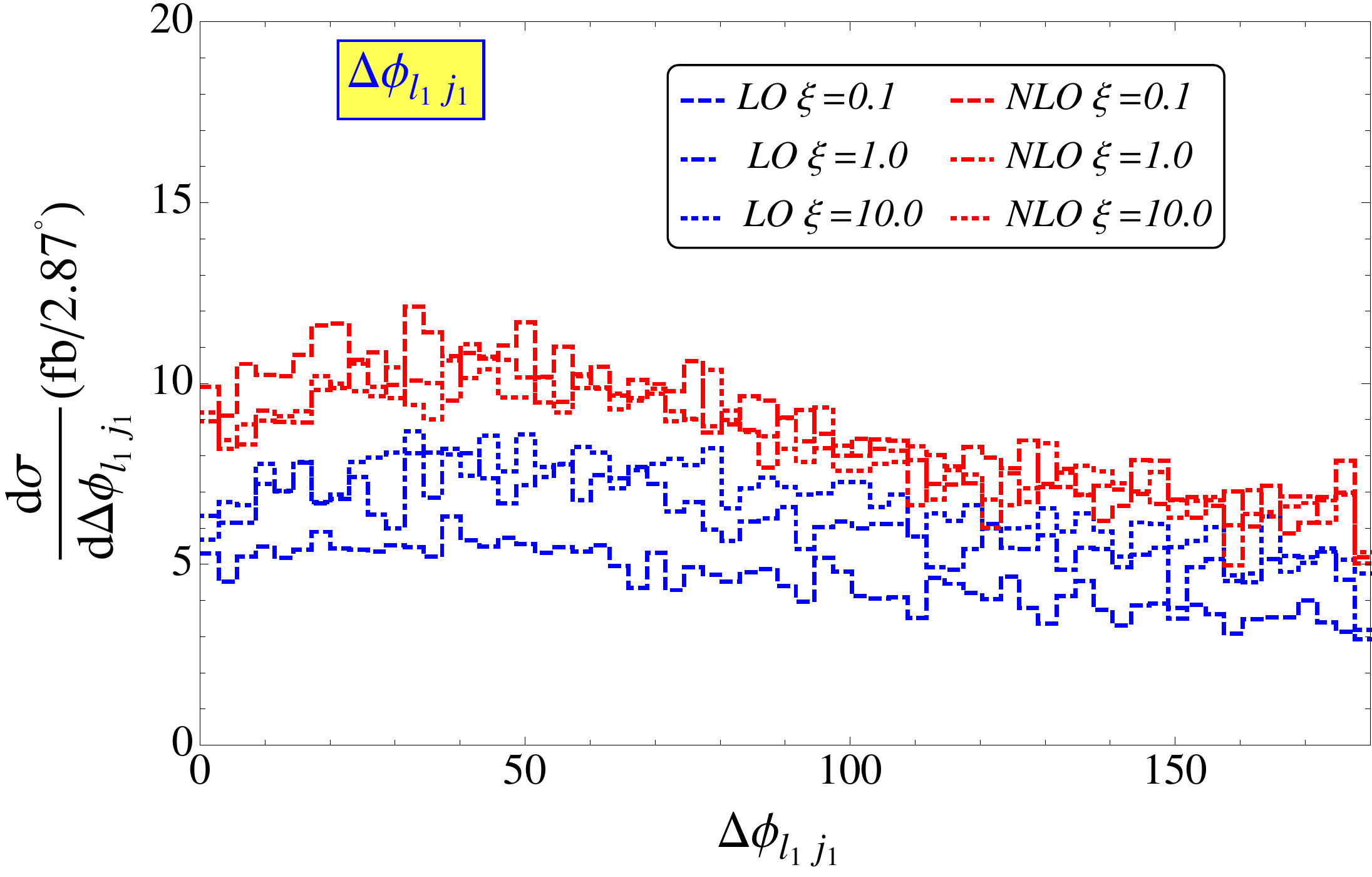}
\includegraphics[scale=0.23]{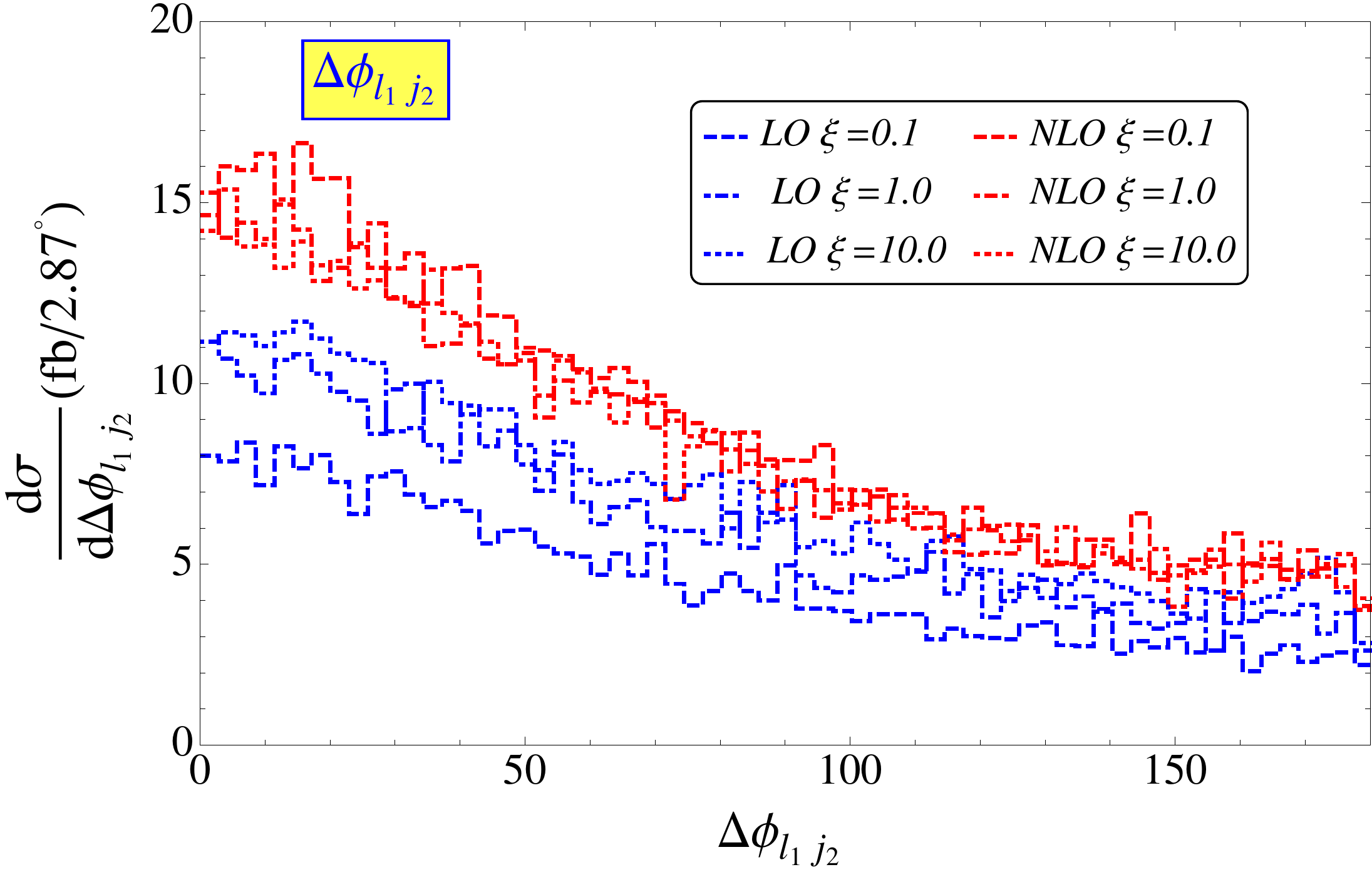}
\includegraphics[scale=0.23]{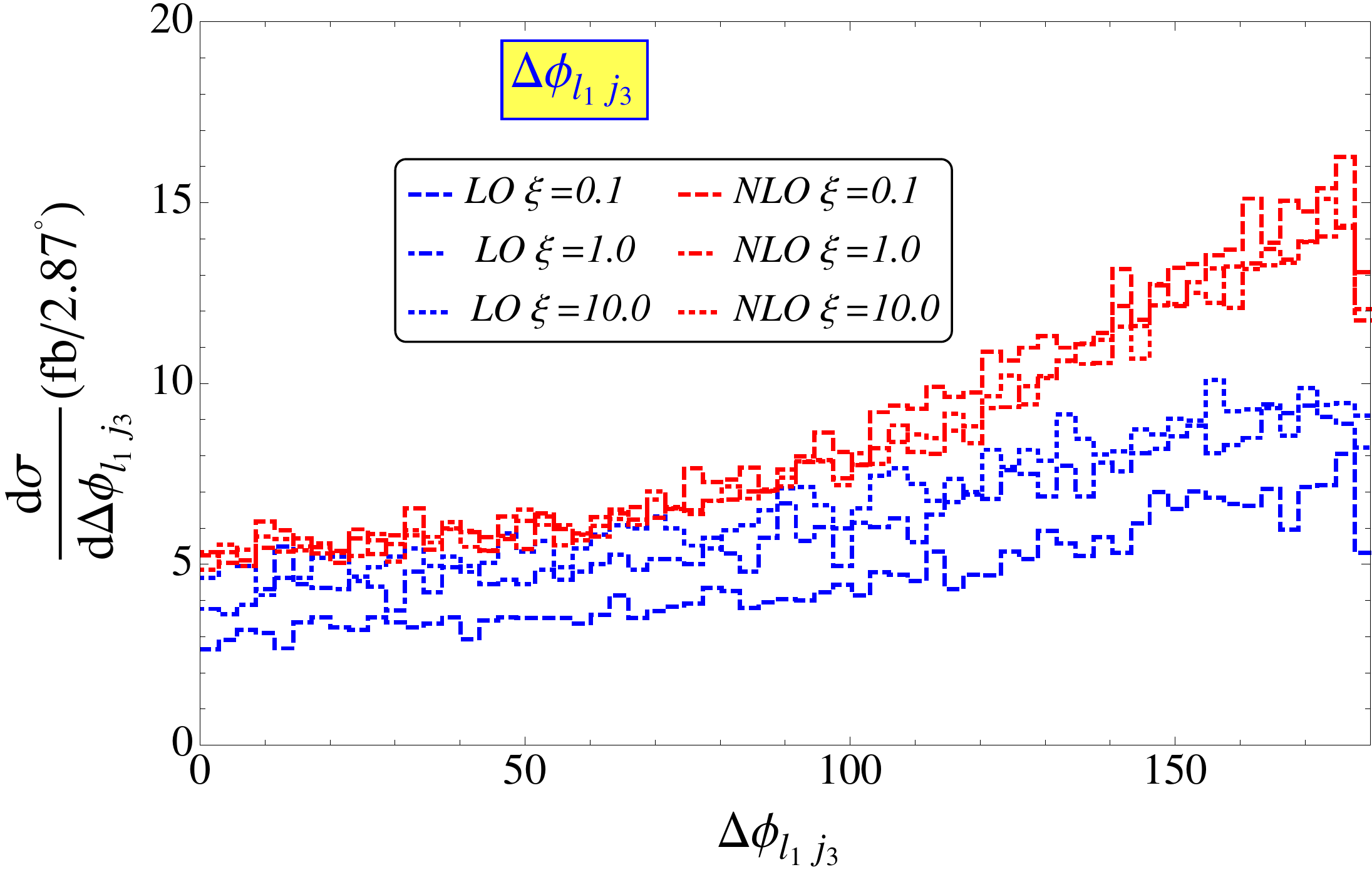}\\
\includegraphics[scale=0.23]{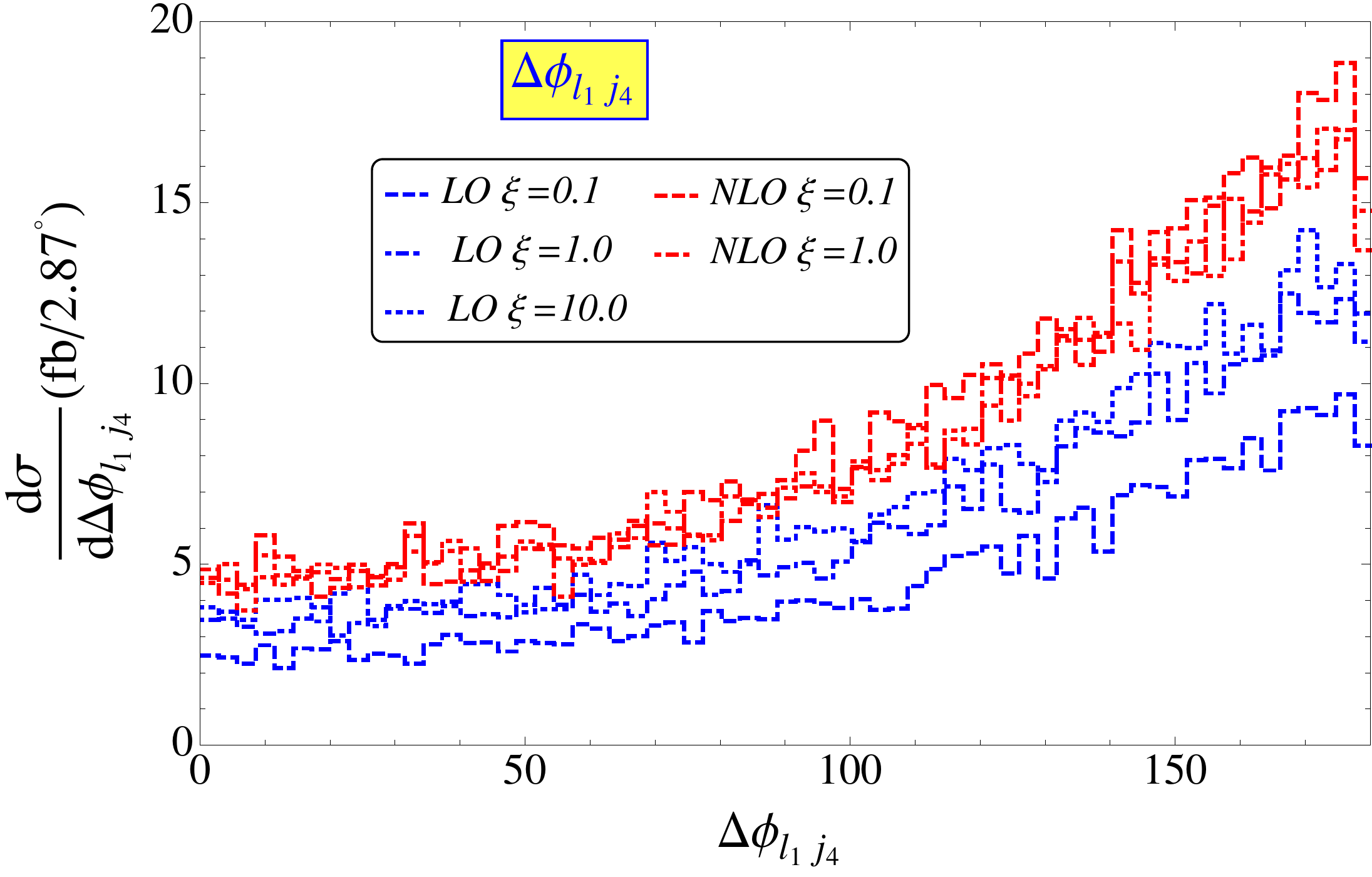}
\includegraphics[scale=0.23]{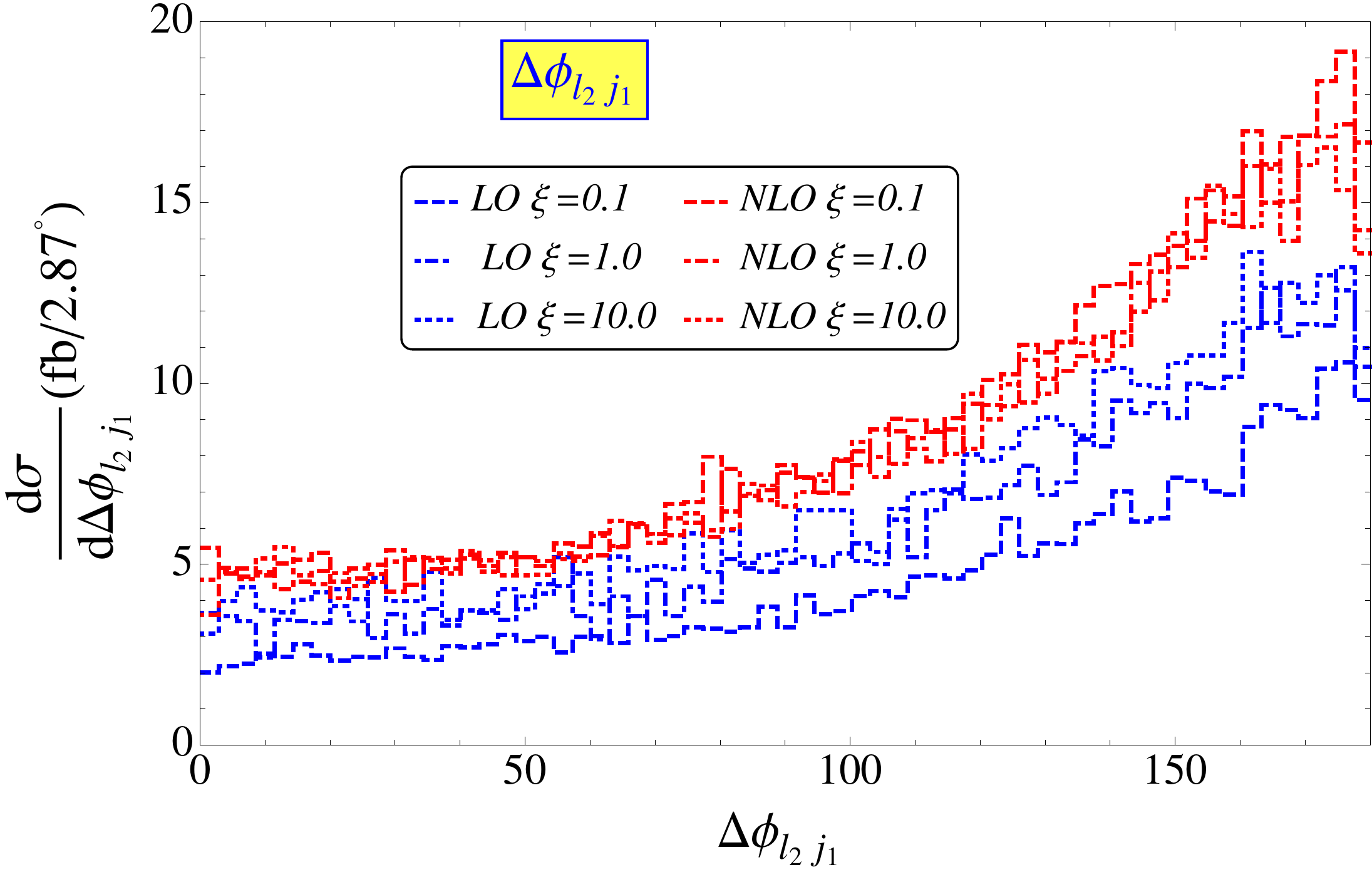}
\includegraphics[scale=0.23]{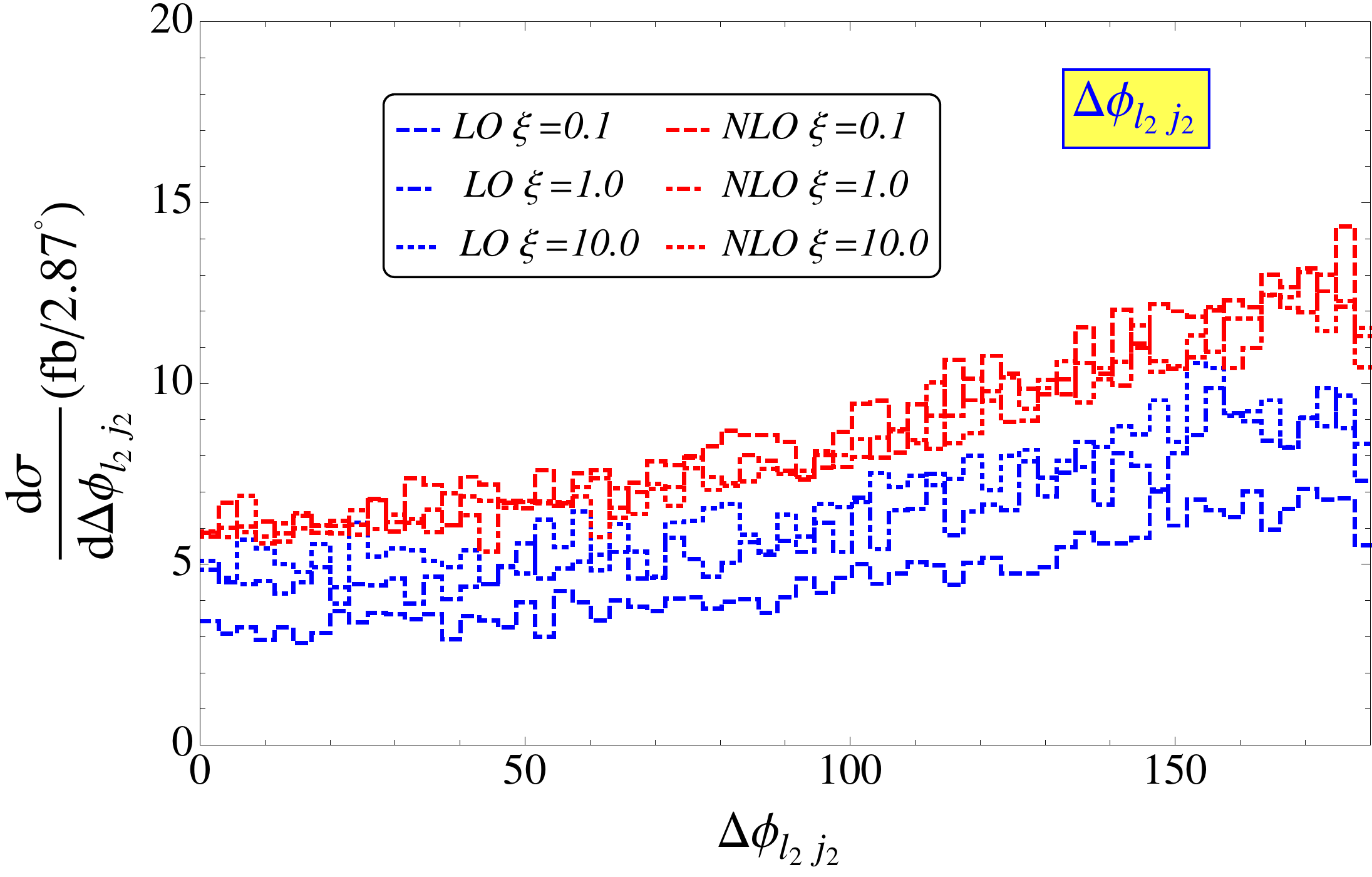}\\
\includegraphics[scale=0.23]{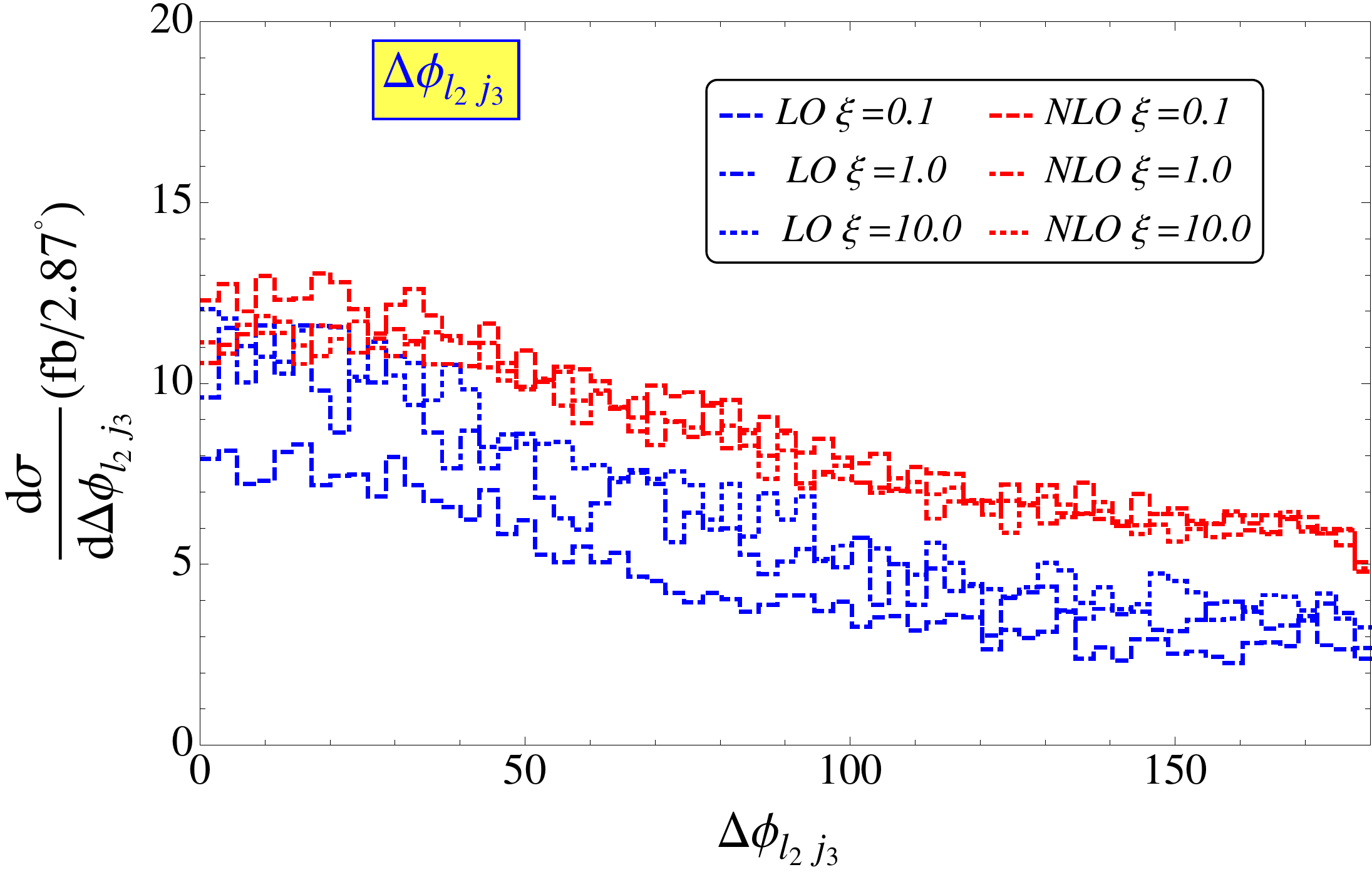}
\includegraphics[scale=0.23]{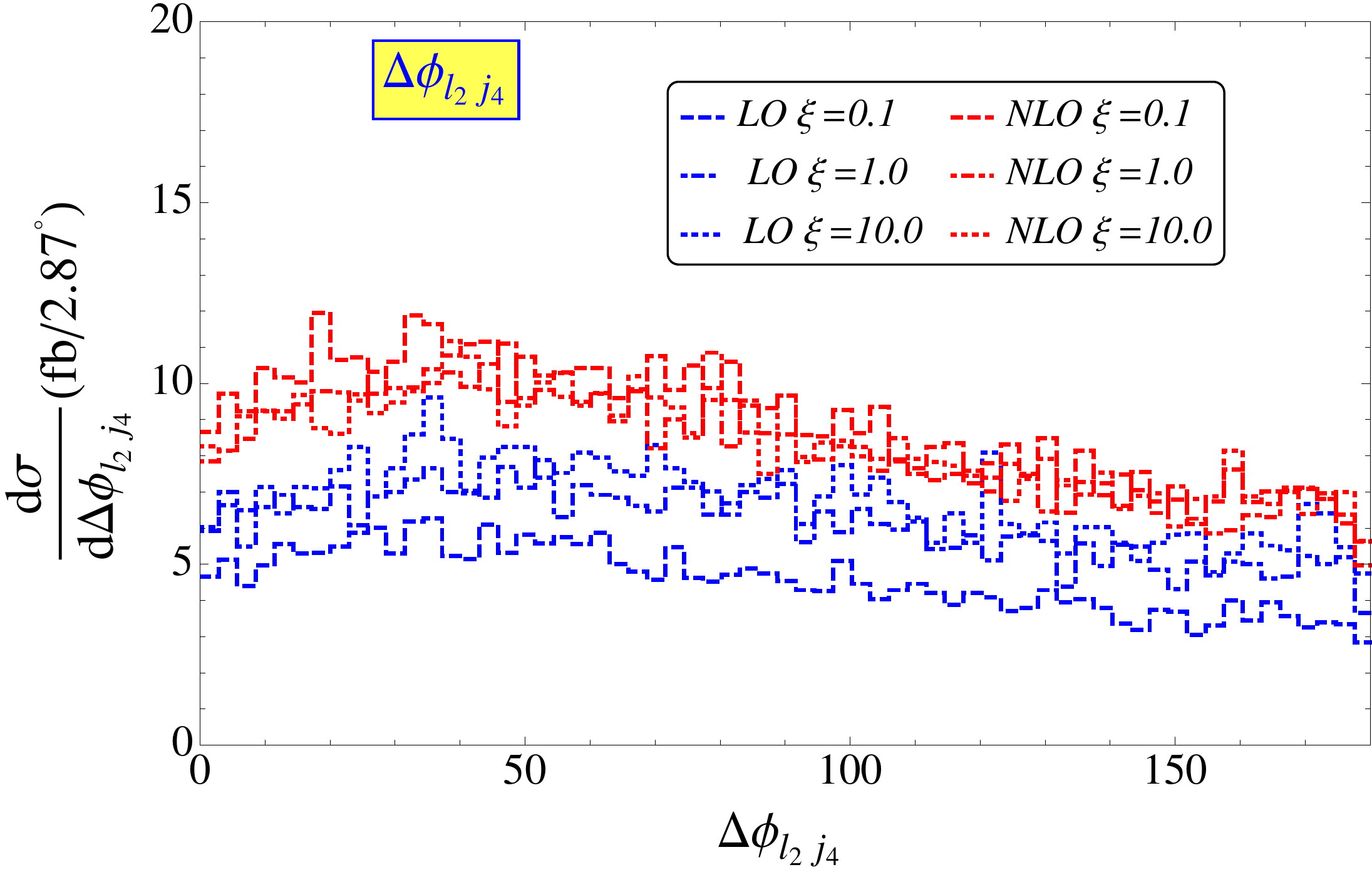}
\end{center}
\caption{Scale dependent LO and NLO-QCD $\Delta\phi_{\ell j}$ distributions of the heavy neutrino pair production followed by the decays of the heavy neutrinos into 
$2\ell+4j$ channel at the 13 TeV LHC for $m_N=95$ GeV.} 
\label{2l_95_7}
\end{figure} 

We consider the $2\ell+4j$ final state from Eq.~\ref{decay1}(a) as it produces the maximum cross section. 
In Figs.~\ref{2l_95_1}-\ref{2l_95_7} we have the scale dependent distributions of different kinematic variables for the LO and NLO-QCD processes using Eqs.\ref{muF} and \ref{muR} respectively while the pair produced heavy neutrinos show $2\ell+4j$ final state at the 13 TeV LHC with $m_N=95$ GeV. To study this final state it is important to mention from Fig.~\ref{2l_95_0} that one can use $p_T^{\ell} > 15$ GeV and $|\eta^{\ell}| < 2.5$ as the leptons will be obtained at the central region. 
 All the four jets are coming from the $W$. It is expected that each of the jets will have peak around $40$ GeV where as a $p_{T}^{j}$ cut about $p_{T}^{j} > 25$ GeV could be acceptable. The jets will also populate the central region. Therefore a pseudo-rapidity cut $|\eta^{j}| < 2.5$ will be good to accept most of the signal events. The scale dependent $p_T^{j}$ and $\eta^{j}$ distributions at the LO and NLO-QCD levels are shown in Fig.~\ref{2l_95_1} .The signal event will carry a pair of opposite sign same flavor (OSSF) dilepton which is has been plotted in Fig.~\ref{2l_95_2}. The maximum number of events are distributed below the  $Z$-pole, the invariant mass $(m_{\ell\ell})$ could be selected using $m_{\ell\ell} < (m_{Z}-15)~\rm{GeV}$  to avoid the OSSF dilepton from the $Z$ boson in the SM background. However, a realistic search for such channel from this model has not been performed yet.

 The LO and NLO-QCD scale dependent distributions of the cosine of the angle between the two leptons $(\cos\theta_{\ell_1\ell_2})$ are plotted in Fig.~\ref{2l_95_3} and the populated area can be obtained when $\cos\theta_{\ell_1\ell_2}>0.5$. The scale dependent distributions of the LO and NLO-QCD process for the angles between all the combinations of the jets $\cos\theta_{jj}$ are shown in Fig.~\ref{2l_95_4} giving rise to a populated area for the events when $\cos\theta_{jj} > 0.5$.

 The azimuthal angular difference between the two leptons $\Delta\phi_{\ell_1\ell_2}$ are shown in Fig.~\ref{2l_95_5} where as that among the jets $(\Delta\phi_{jj})$are shown in Fig.~\ref{2l_95_6}. This distributions show that the maximum number of events are obtained when $\Delta\phi$ is close to $180^{0}$. Scale dependent distributions of $\Delta\phi_{\ell j}$ are shown in Fig.~\ref{2l_95_7} for the LO and the NLO-QCD levels. The distributions for $\Delta\phi_{\ell_1 j_2}$ and $\Delta\phi_{\ell_2 j_3}$ are mostly probable when $\Delta\phi$ close to zero. The distributions for $\Delta\phi_{\ell_1 j_1}$ and $\Delta\phi_{\ell_2 j_4}$ are almost flat however the number of events with $\Delta\phi_{\ell j} < 50^{0}$ are little greater than those when $\Delta\phi_{\ell j}> 50^{0}$. The other distributions for $\Delta\phi_{\ell j}$ show the highest probability of getting the events when $\Delta\phi_{\ell j} > 75^{0}$. Such distributions are utilized to fix the separation cuts between the leptons $(\Delta R_{\ell \ell})$, between the jets $(\Delta R_{jj})$ and between the lepton-jets $(\Delta R_{\ell j})$ because $\Delta R=\sqrt{(\Delta{\eta})^2+ (\Delta{\phi})^2}$ which is generally used between 0.3-0.4 for the hadron colliders.

\begin{figure}
\begin{center}
\includegraphics[scale=0.3]{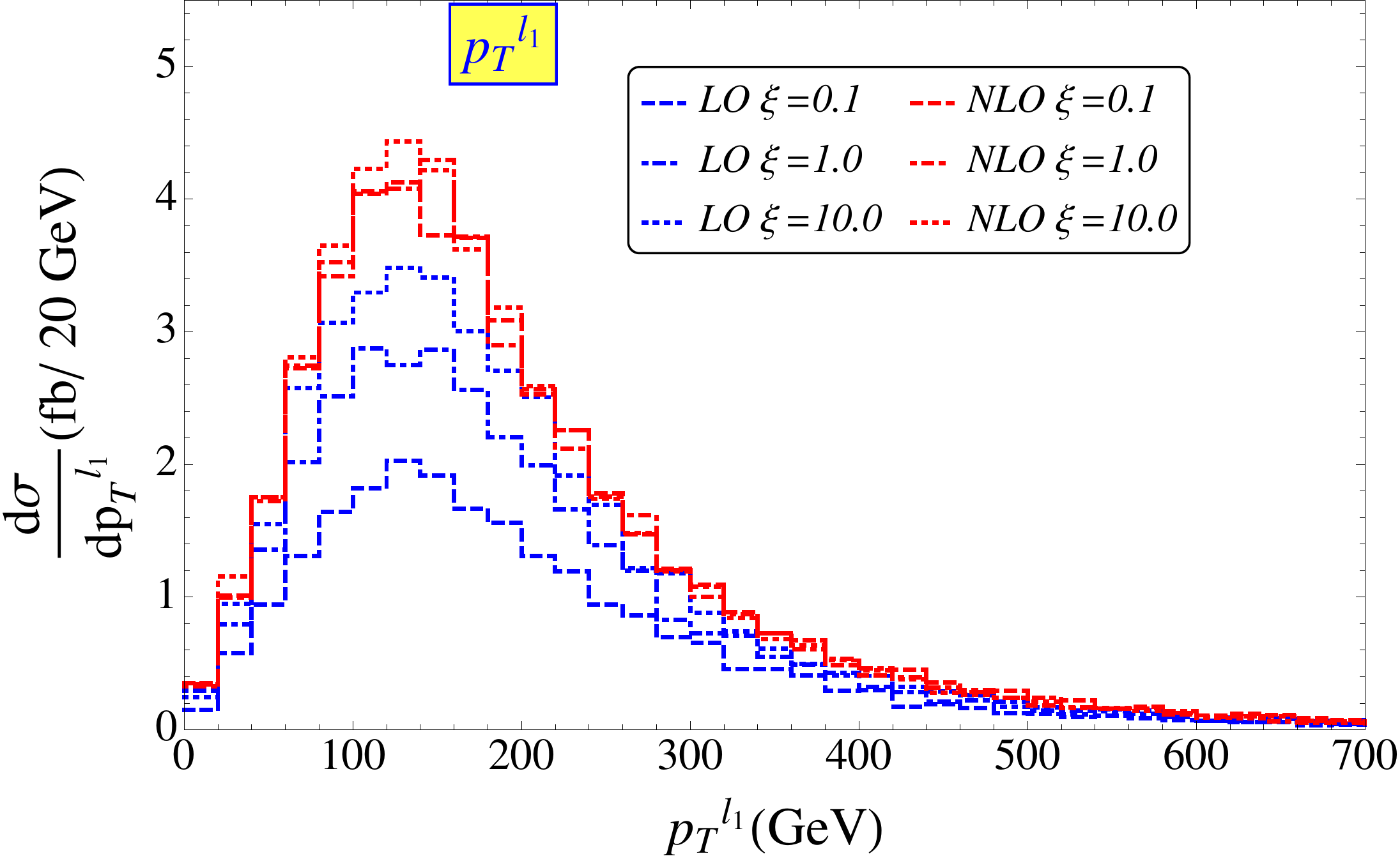}
\includegraphics[scale=0.3]{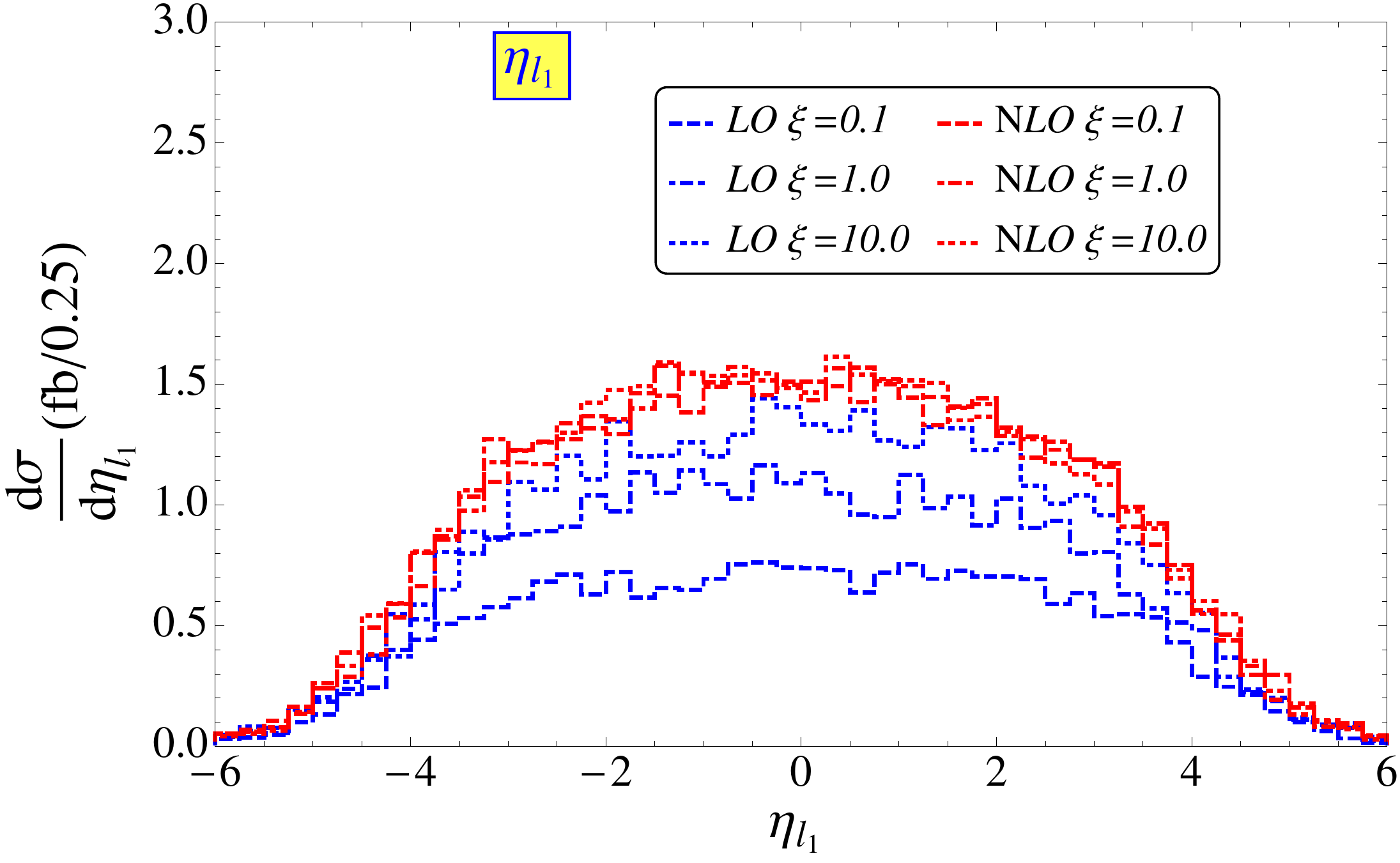}\\
\includegraphics[scale=0.3]{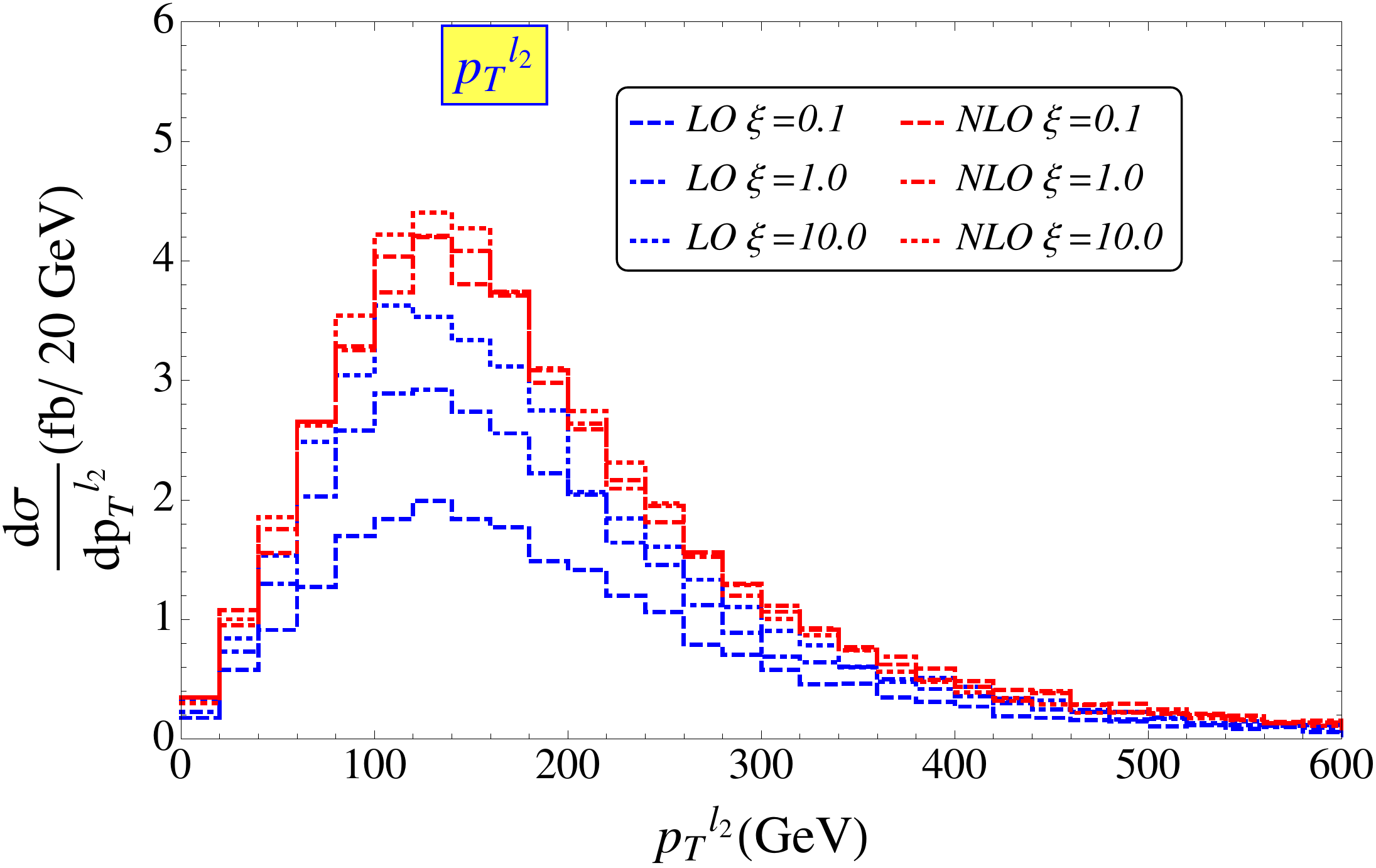}
\includegraphics[scale=0.3]{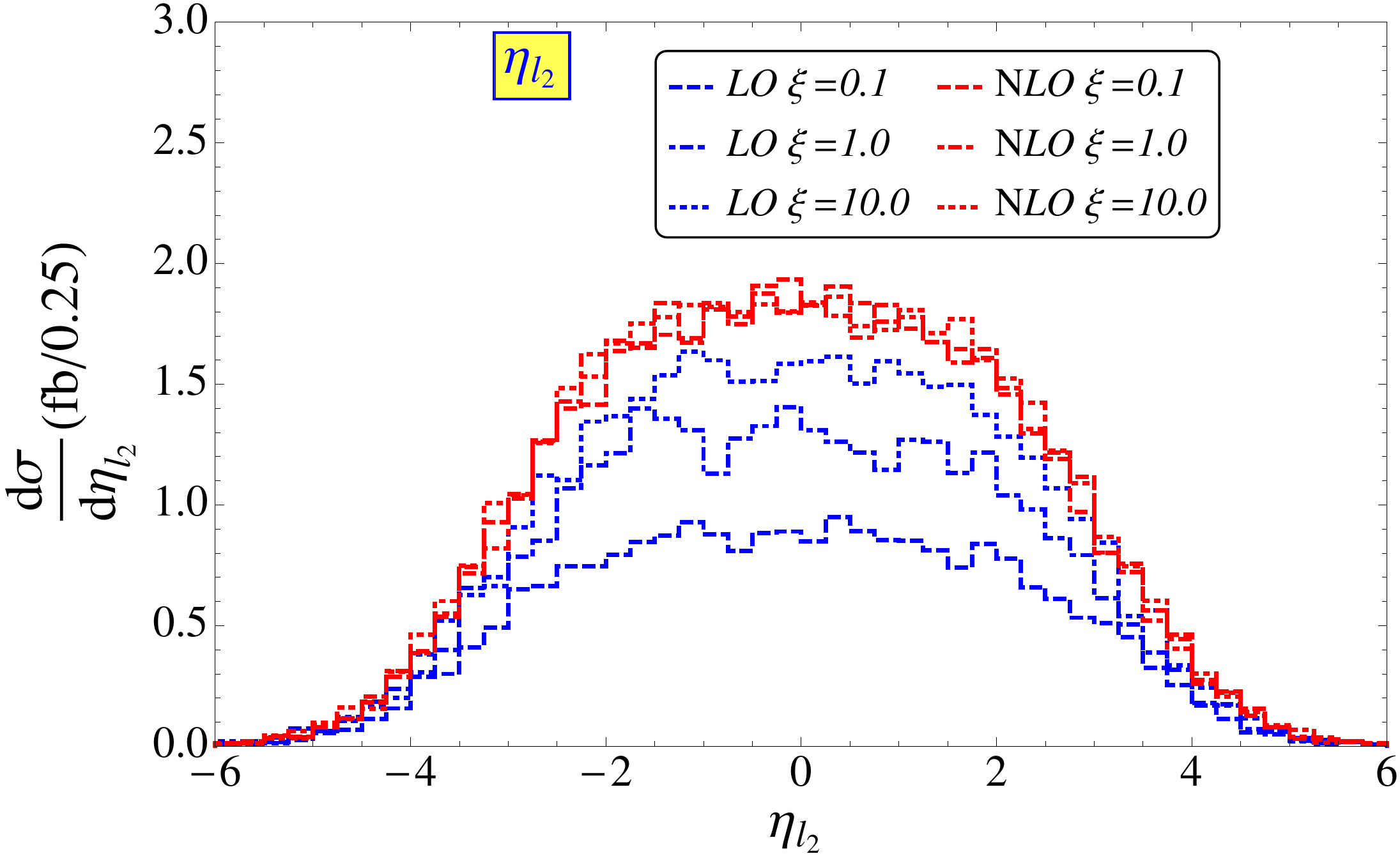}
\end{center}
\caption{Scale dependent LO and NLO-QCD $p_T^{\ell}$ (left column) and $\eta^{\ell}$ (right column) distributions of the heavy neutrino pair production followed by the decays of the heavy neutrinos into 
$2\ell+4j$ channel at the 100 TeV hadron collider for $m_N=300$ GeV.}
\label{2l_300_0}
\end{figure}

\begin{figure}
\begin{center}
\includegraphics[scale=0.3]{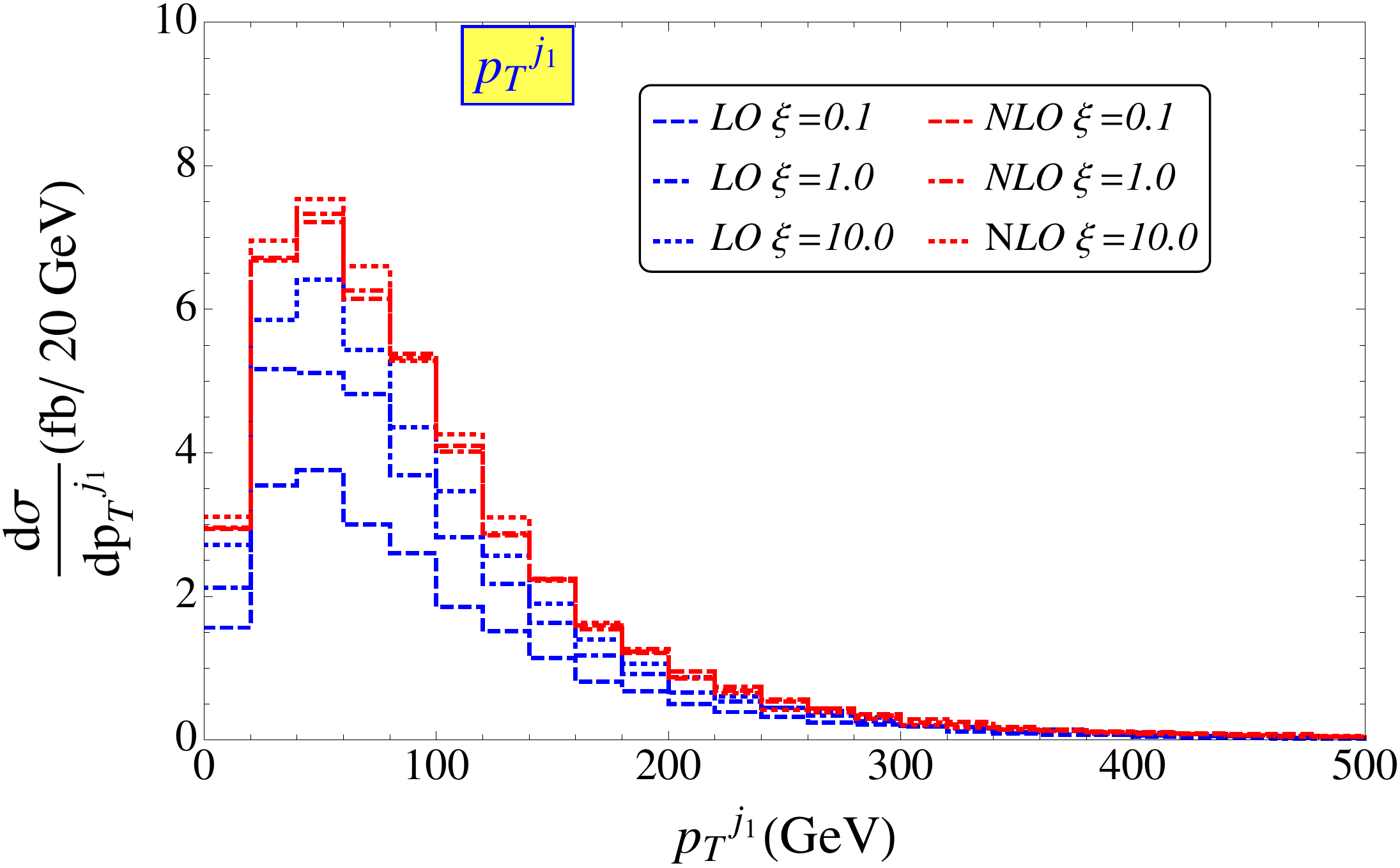}
\includegraphics[scale=0.3]{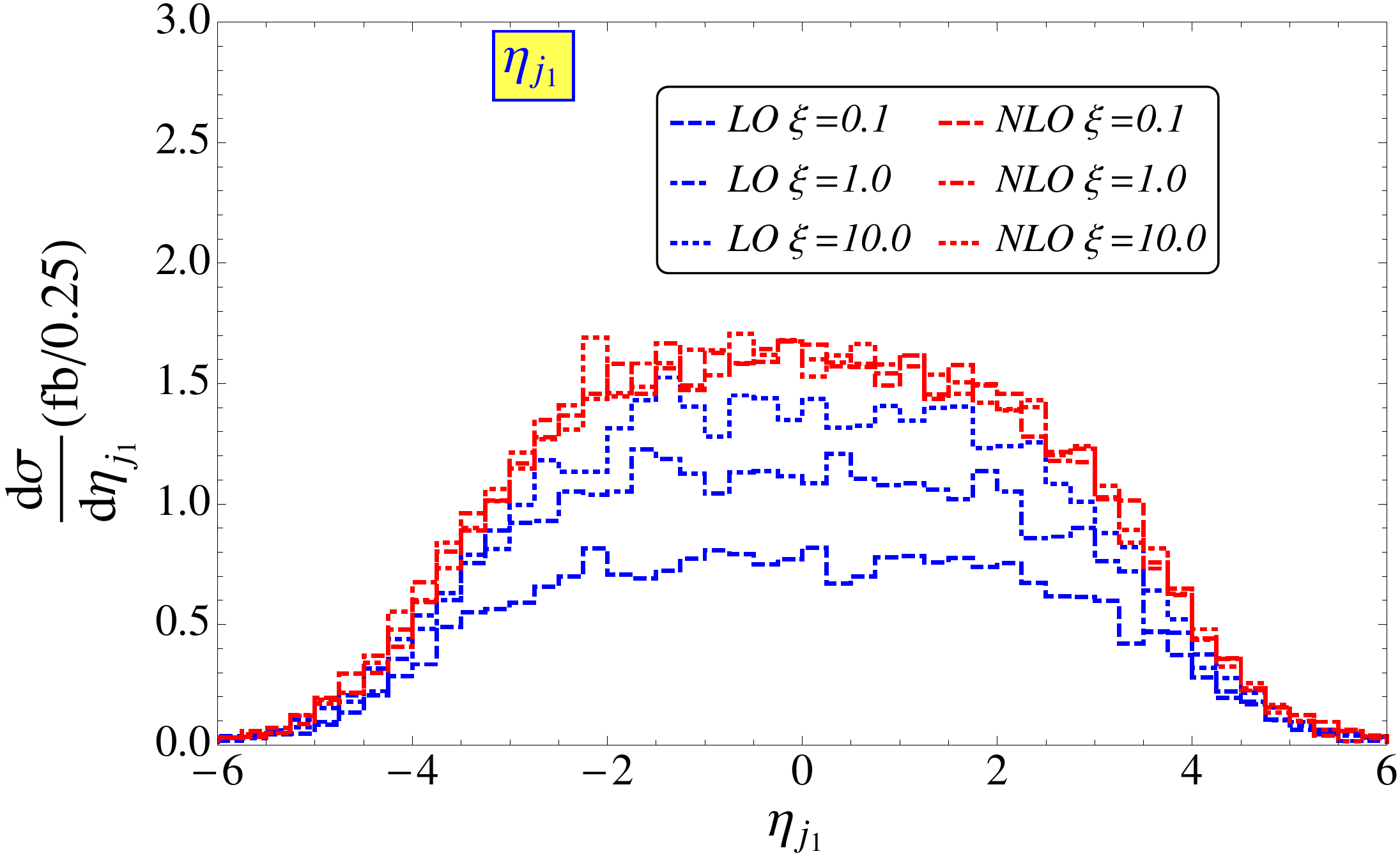}\\
\includegraphics[scale=0.3]{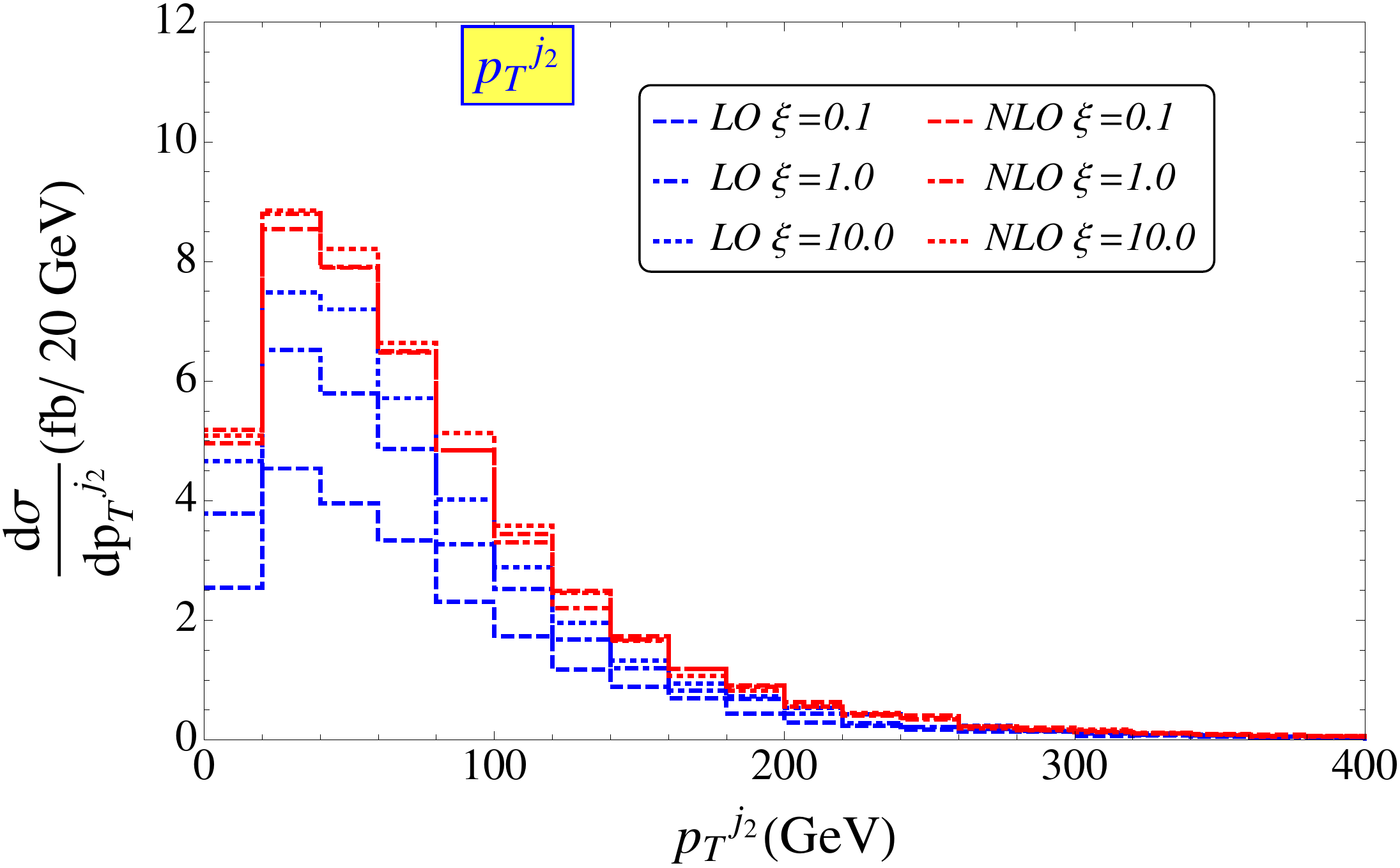}
\includegraphics[scale=0.3]{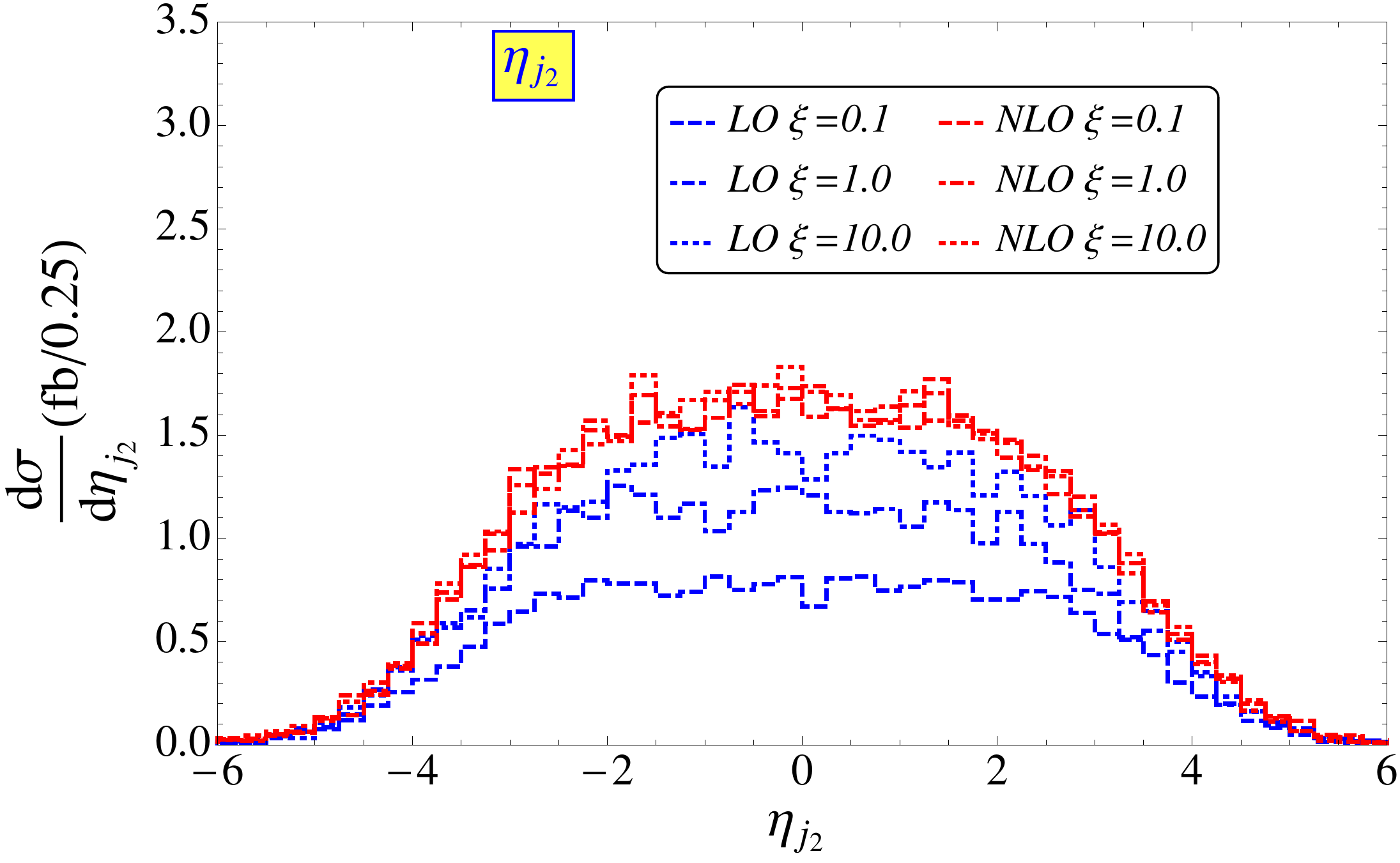}\\
\includegraphics[scale=0.3]{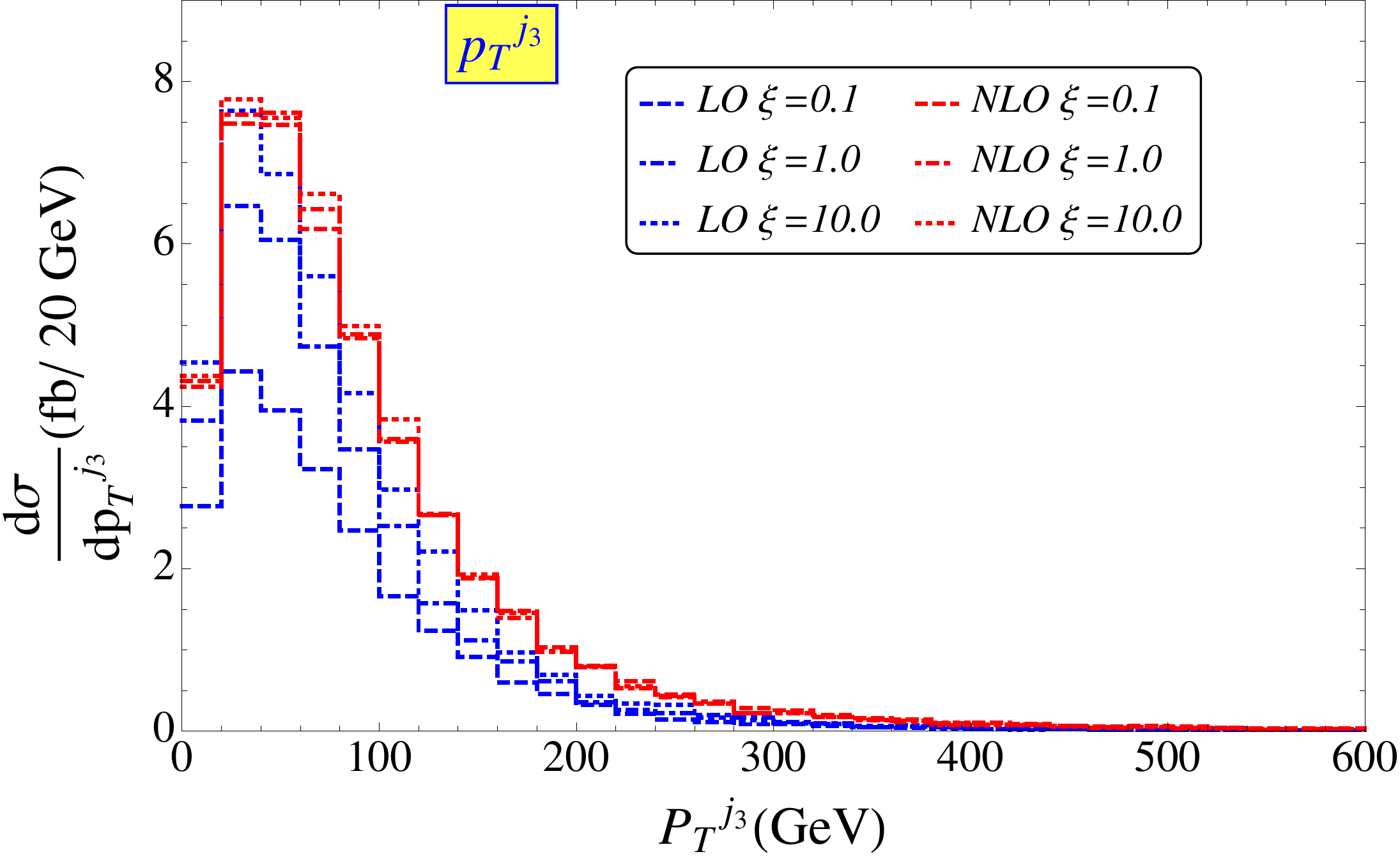}
\includegraphics[scale=0.3]{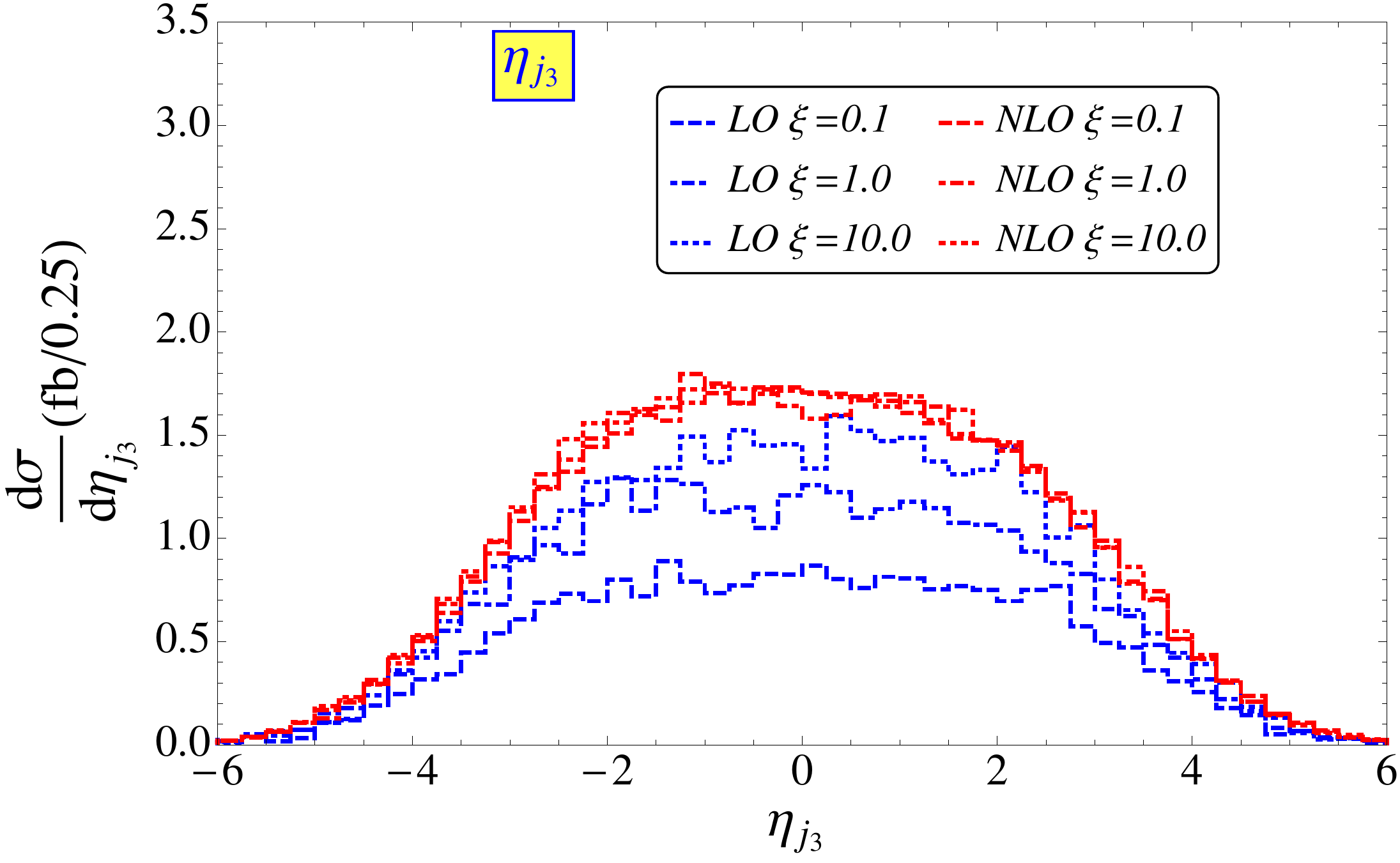}\\
\includegraphics[scale=0.3]{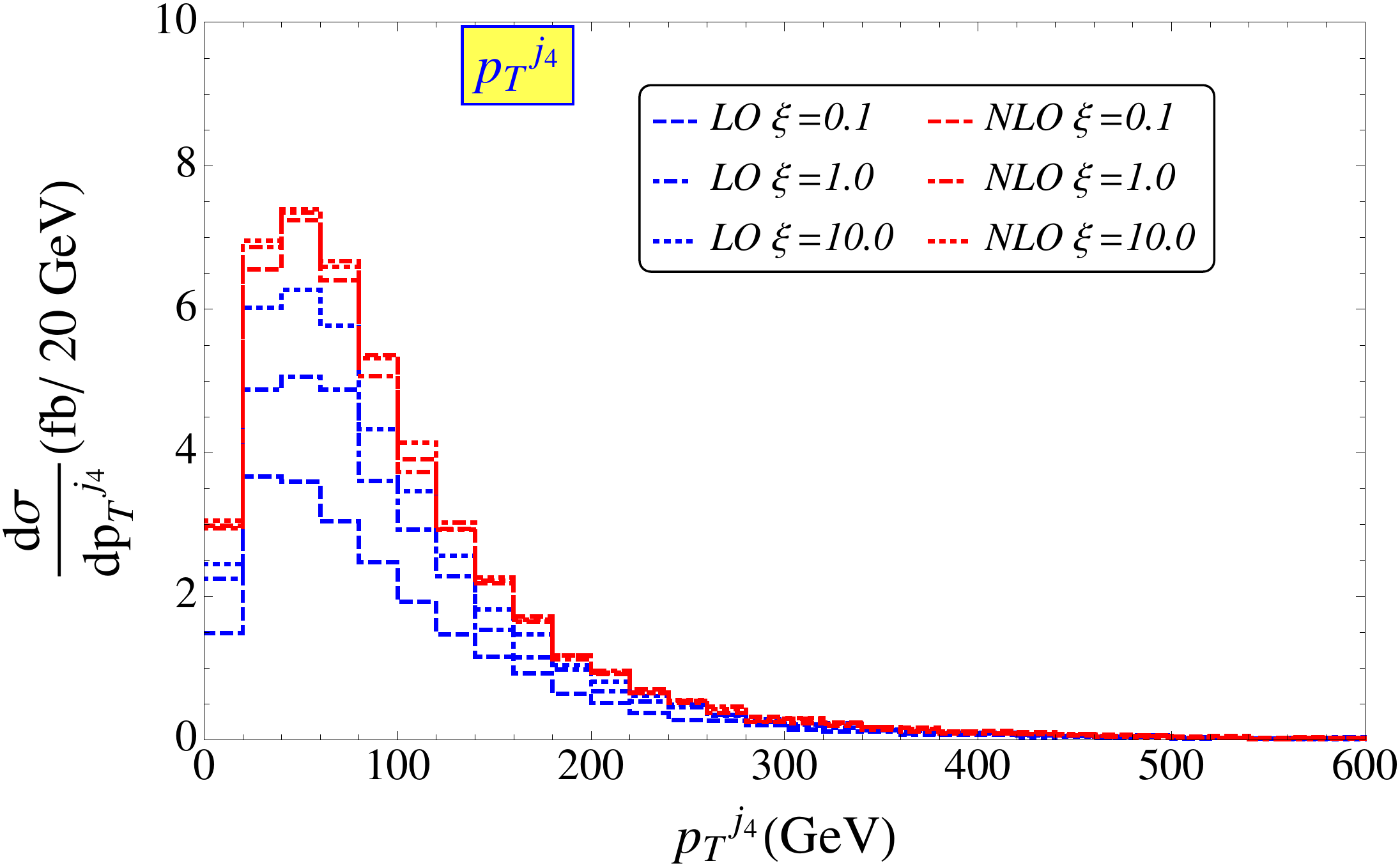}
\includegraphics[scale=0.3]{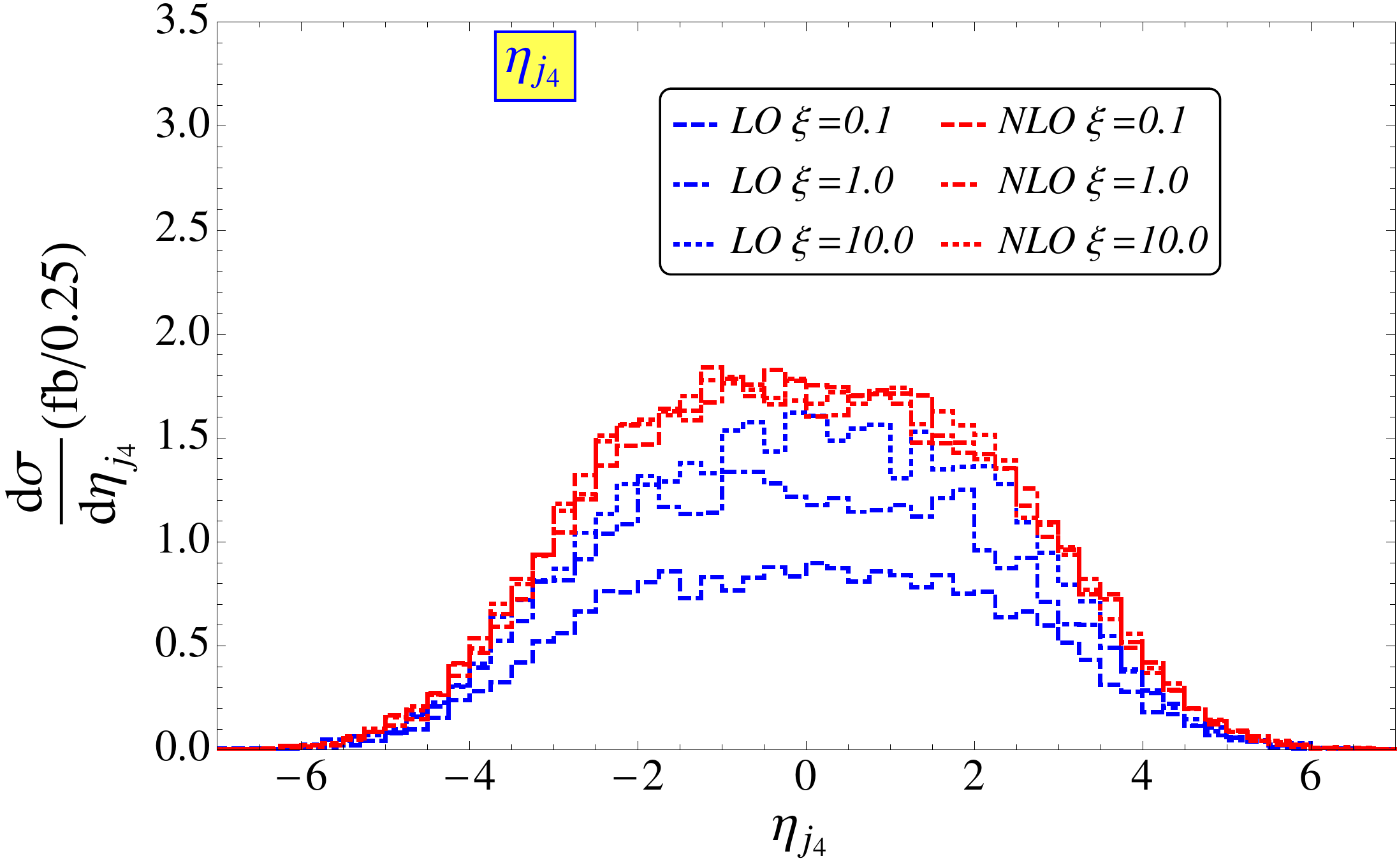}
\end{center}
\caption{Scale dependent LO and NLO-QCD $p_T^{j}$ (left column) and $\eta^{j}$ (right column) distributions of the heavy neutrino pair production followed by the decays of the heavy neutrinos into 
$2\ell+4j$ channel at the 100 TeV hadron collider for $m_N=300$ GeV.}
\label{2l_300_1}
\end{figure}

We have studied the dilepton plus four jet signal events for the proposed 100 TeV hadron collider with $m_{N}=300$ GeV. The $p_T^{\ell}$ and $\eta^{\ell}$ distributions are shown in Fig.~\ref{2l_300_0}. $p_{T}^{\ell} > 90$ GeV and $|\eta^{\ell}|< 2.5$ could be good to select the leptons in the signal events repeating the same $p_T^j$ and $\eta^j$ cuts used for the 13 TeV LHC because the jets in the signal events are coming from the $W$ boson. The corresponding scale dependent distributions for the jets at the 100 TeV hadron collider at $m_N=300$ GeV are given in Fig.~\ref{2l_300_1}.
\begin{figure}
\begin{center}
\includegraphics[scale=0.40]{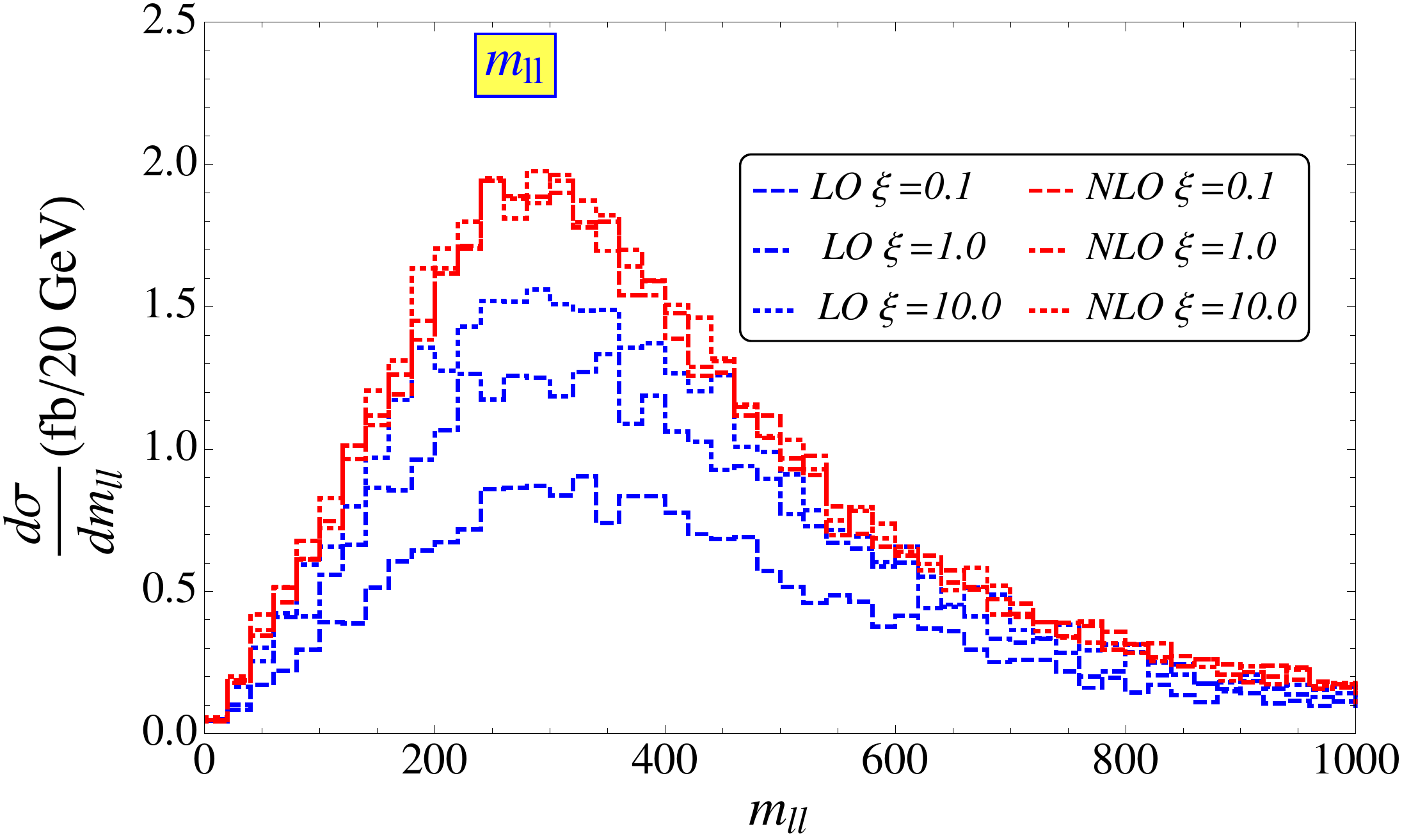}
\end{center}
\caption{Scale dependent LO and NLO-QCD $E_T^{\rm{miss}}$ distributions of the heavy neutrino pair production followed by the decays of the heavy neutrinos into 
$2\ell+4j$ channel at the 100 TeV hadron collider for $m_N=300$ GeV.}
\label{2l_300_2}
\end{figure} 
For the invariant mass $(m_{\ell \ell})$ cuts using the OSSF pairs, $m_{\ell\ell} > (m_{Z} +15)$~\rm{GeV} could be used to accept the signal events to avoid the SM backgrounds from the $Z$-pole.

Due to the dependence of $\mu_{F}$ on the scale factor $\xi$ from the PDFs, a strong scale variation in the LO is observed at both of the colliders whereas a comparatively soft scale variation has been noticed in the NLO-QCD level as the strong coupling depends upon the $\mu_R$ as $\alpha_s(\mu_R)$.
\subsection{$3\ell+2j+\rm{MET}$ final state}
From Eq.~\ref{decay1}(b) we study the $3\ell+2j+\rm{MET}$ where one of the $W$s coming from the $N$ will show leptonic decay mode and the remaining one will display hadronic decay mode respectively.

\begin{figure}
\begin{center}
\includegraphics[scale=0.33]{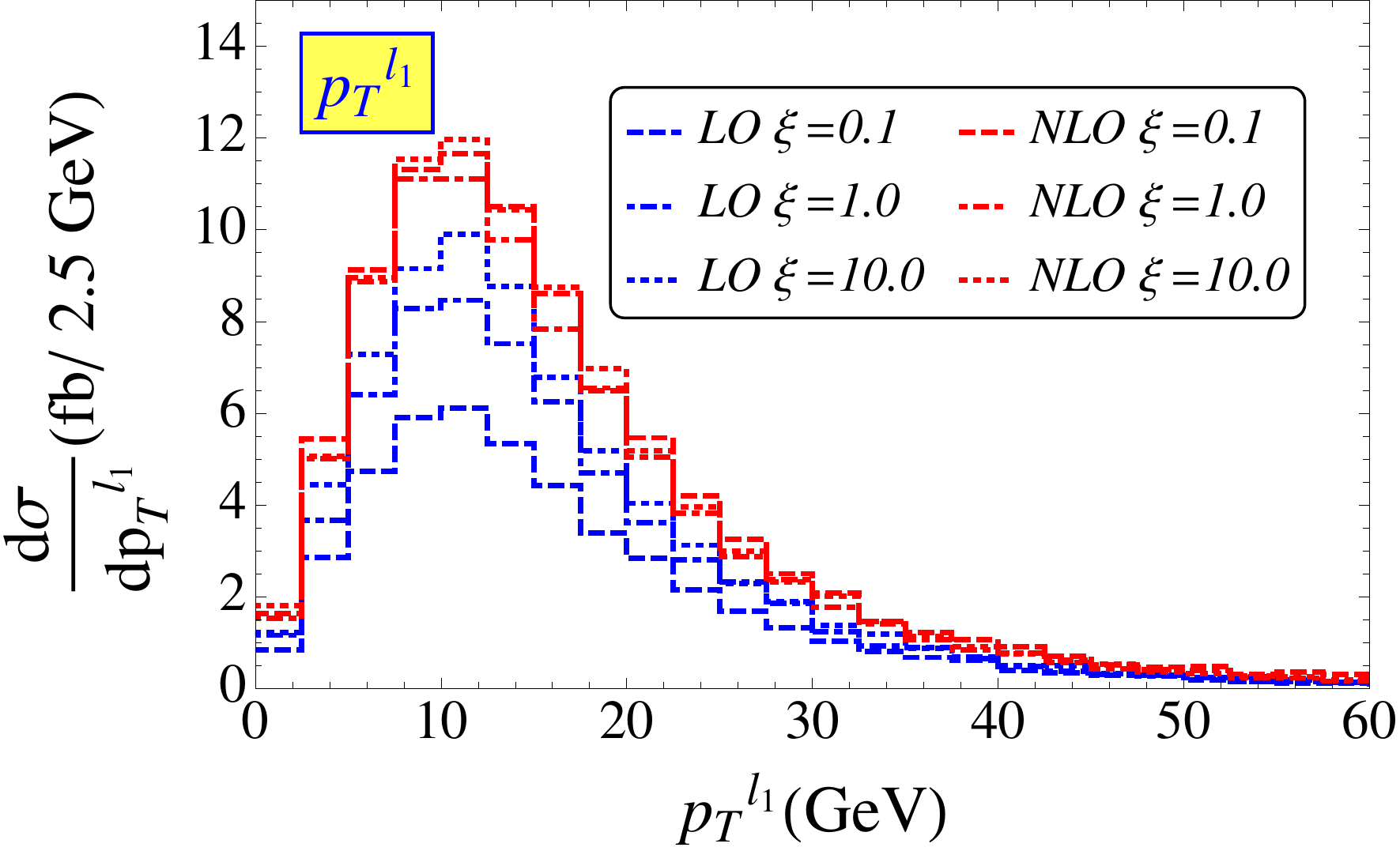}
\includegraphics[scale=0.34]{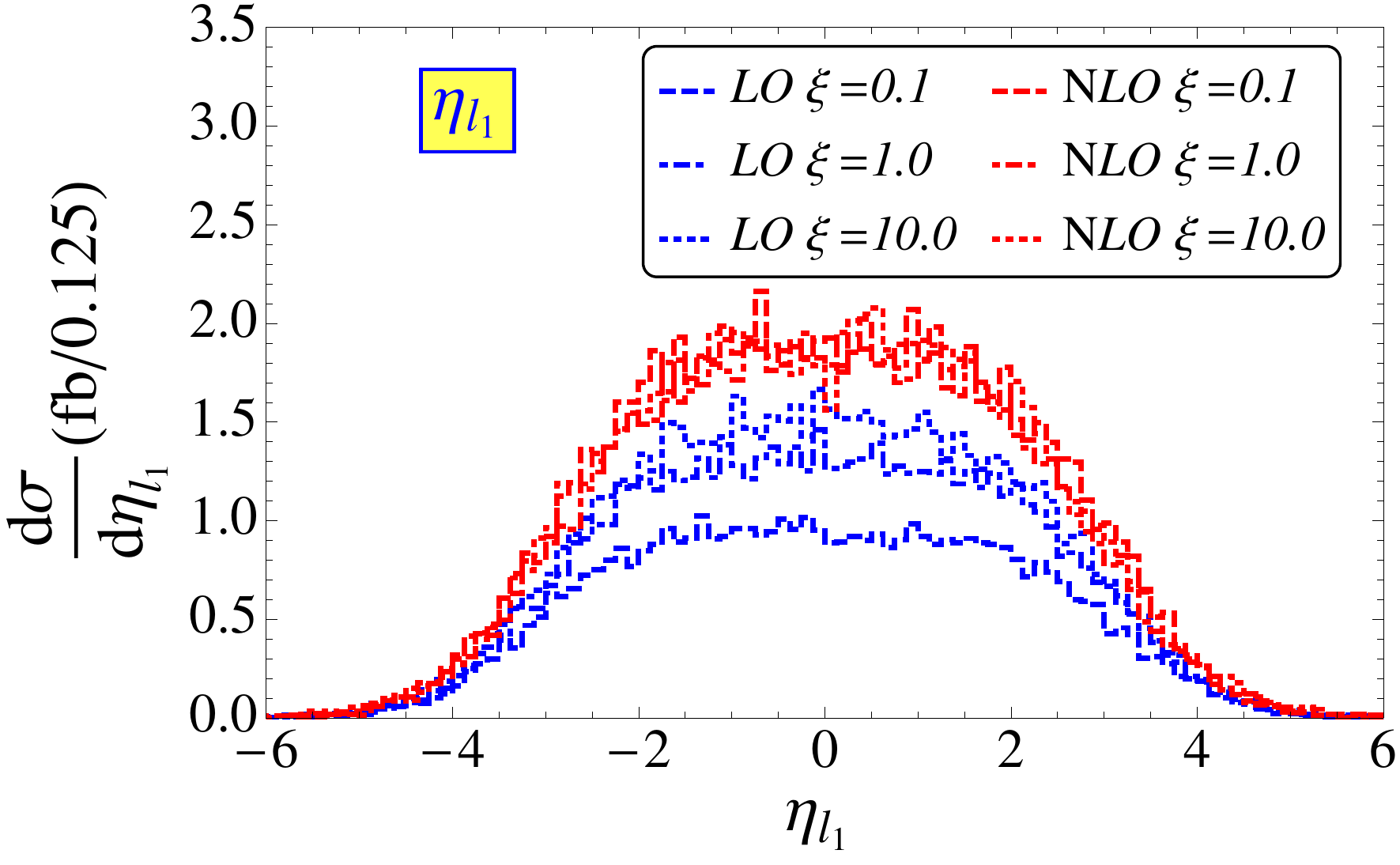}\\
\includegraphics[scale=0.3]{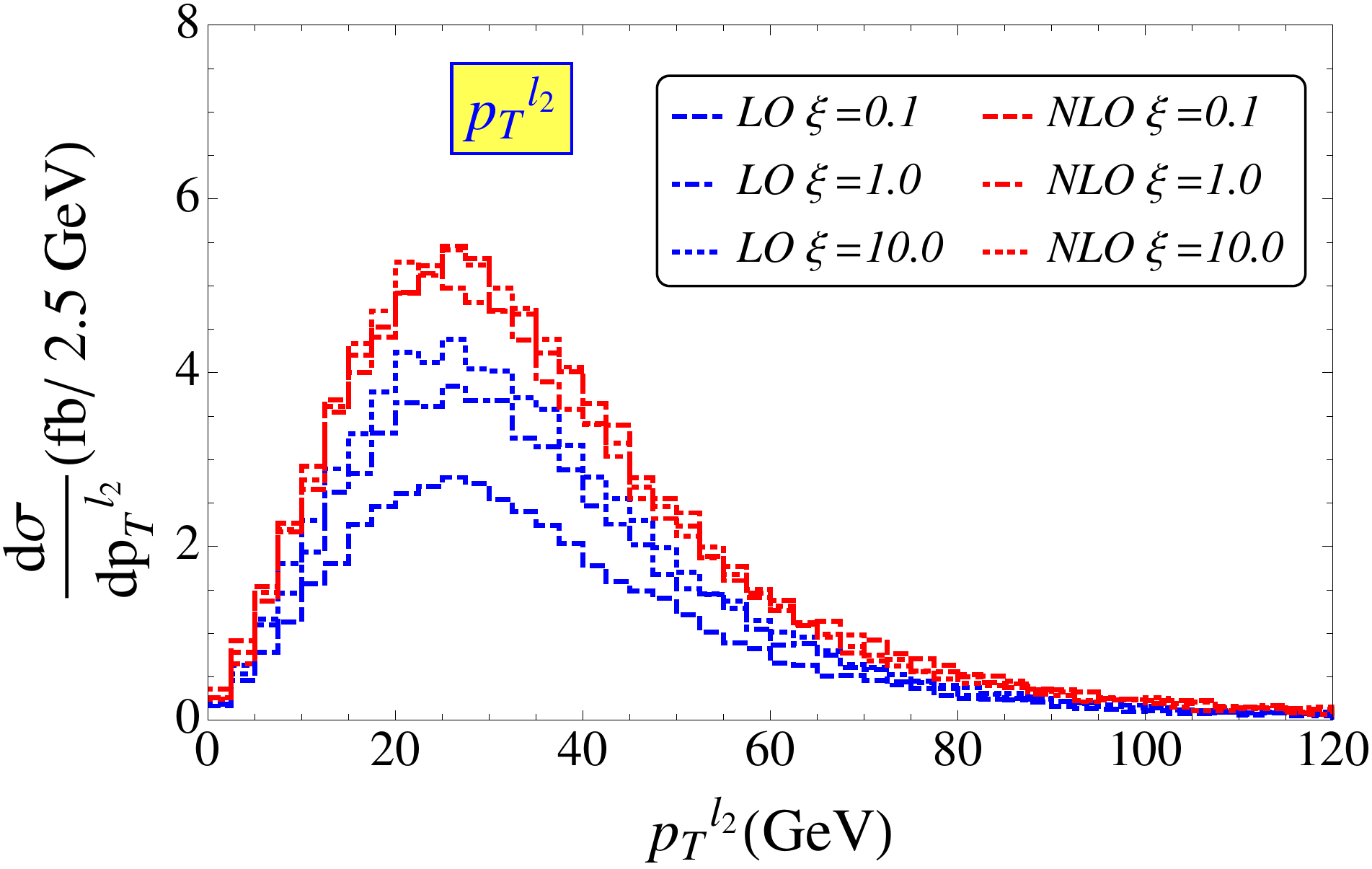}
\includegraphics[scale=0.29]{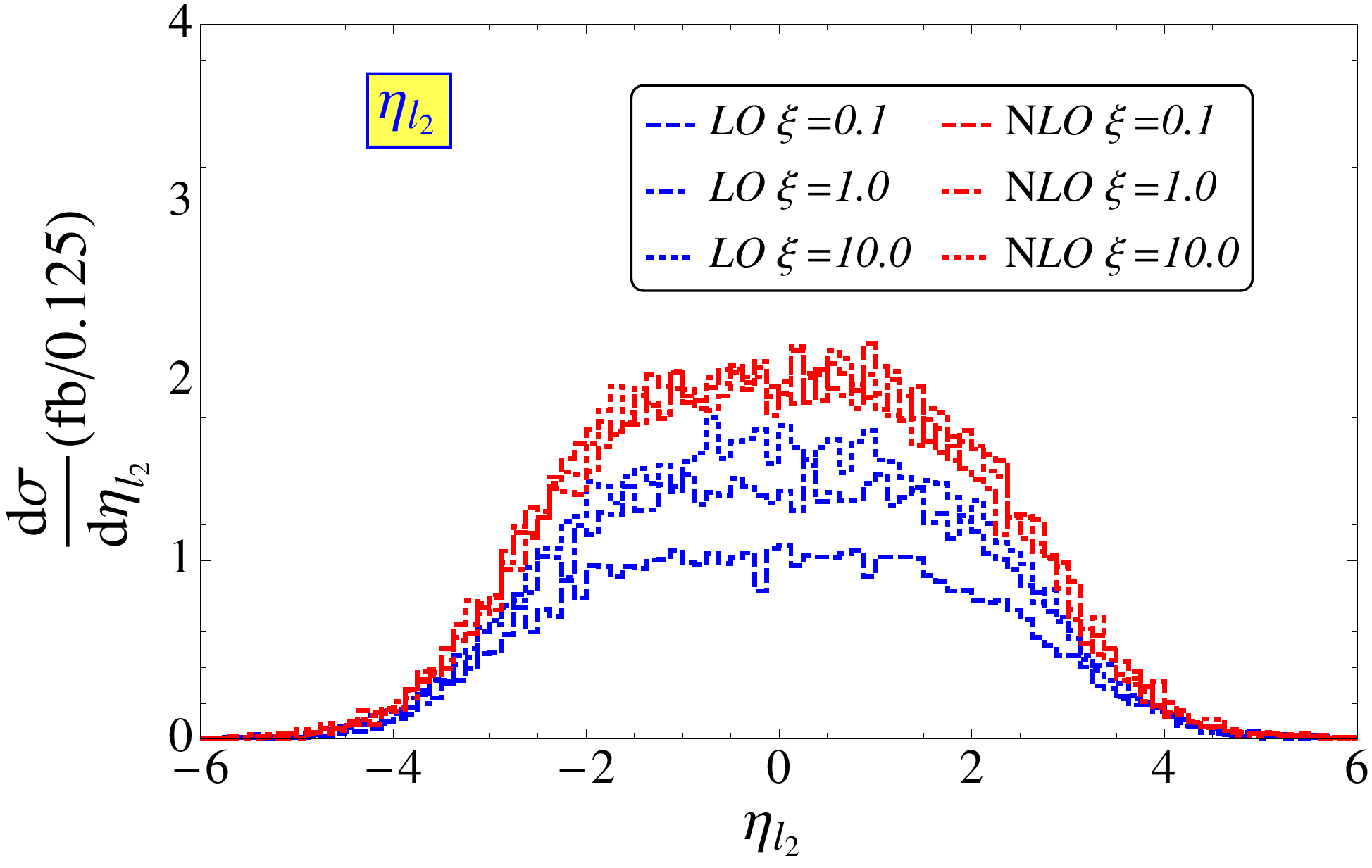}\\
\includegraphics[scale=0.3]{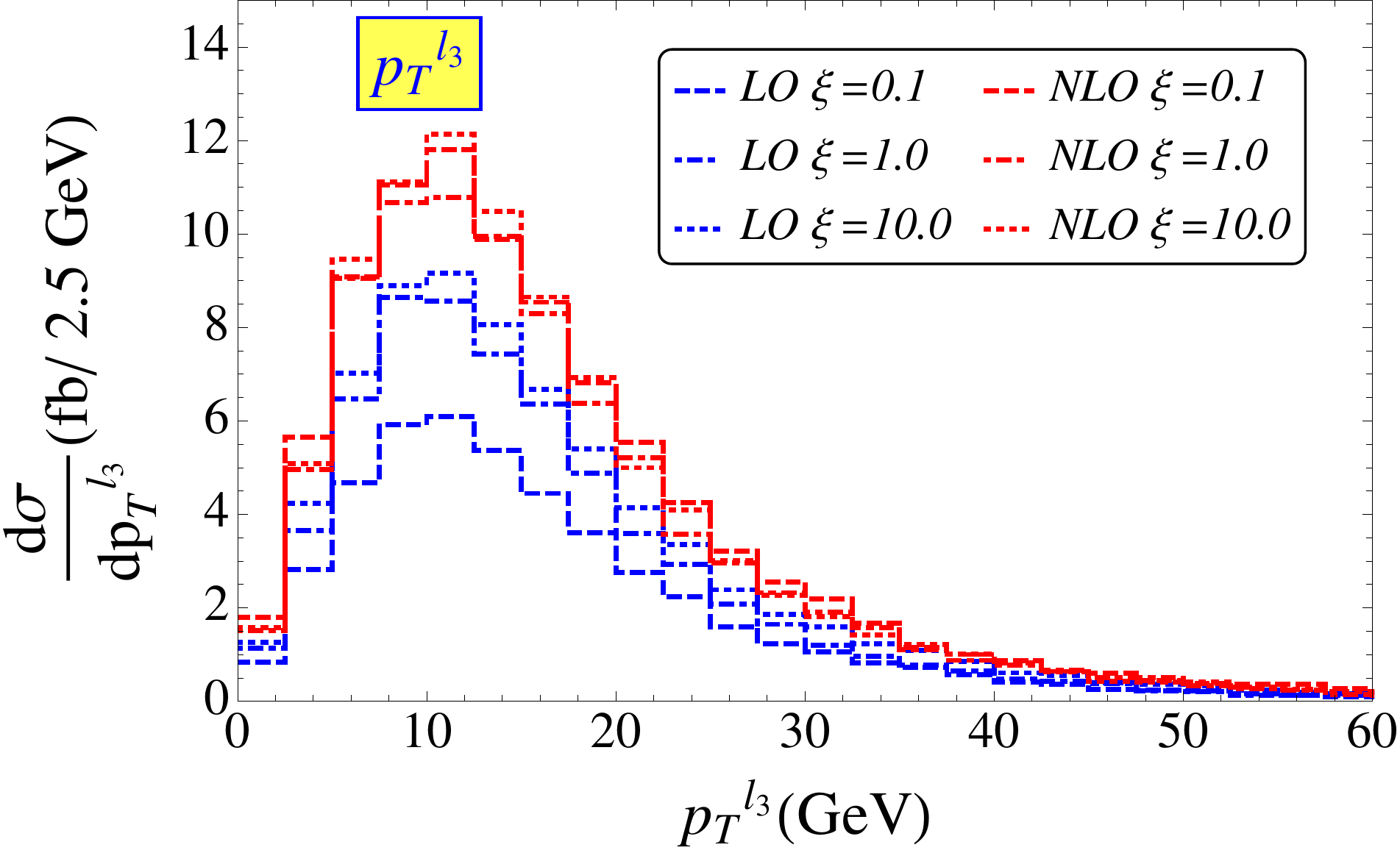}
\includegraphics[scale=0.3]{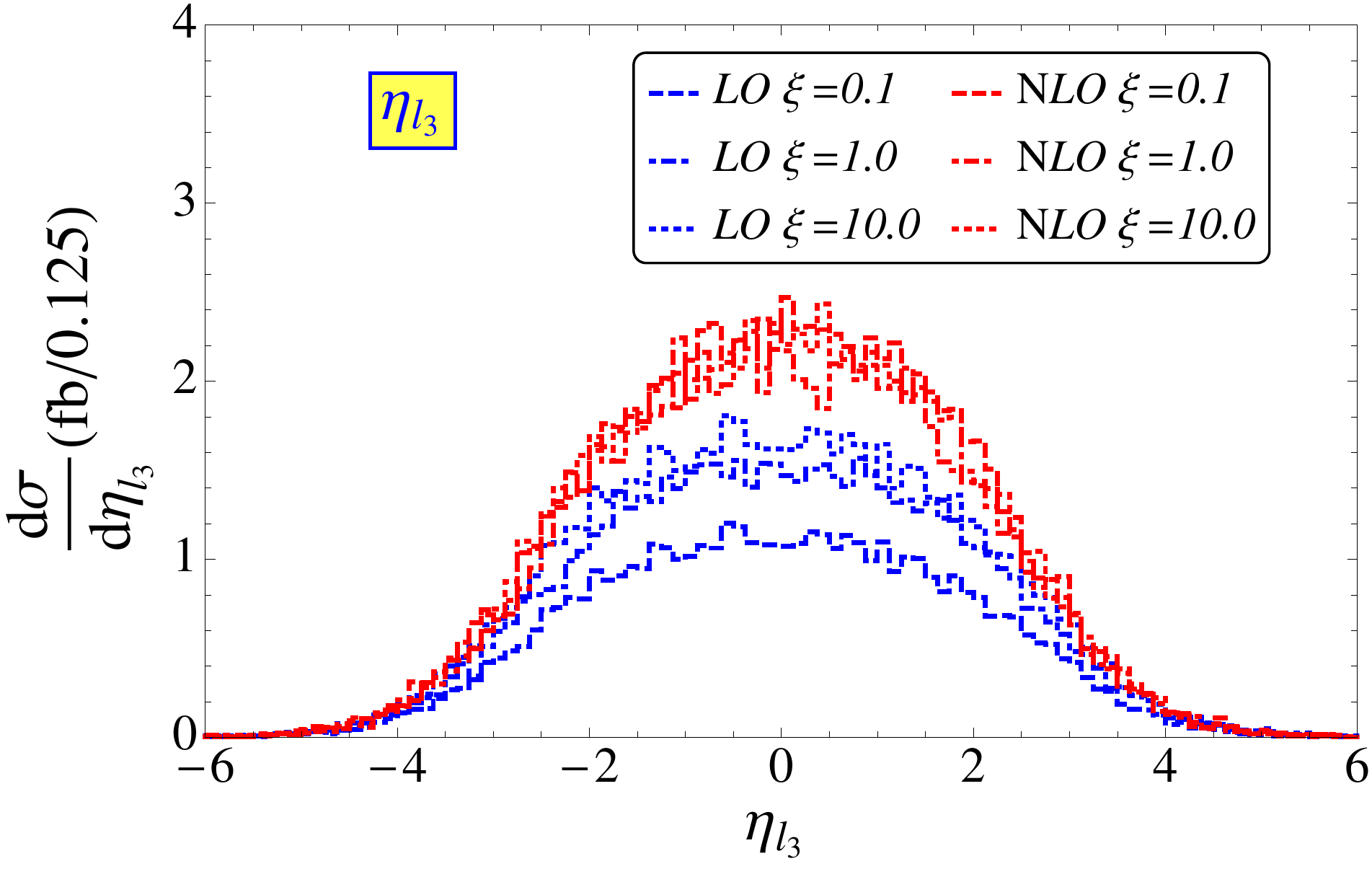}
\end{center}
\caption{Scale dependent LO and NLO-QCD $p_T^{\ell}$(left column) and $\eta^{\ell}$(right column) distributions of the heavy neutrino pair production followed by the decays of the heavy neutrinos into 
$3\ell+\rm{MET}+2j$ channel at the 13 TeV LHC for $m_N=95$ GeV.}
\label{HC 95-3l1}
\end{figure} 
\begin{figure}
\begin{center}
\includegraphics[scale=0.3]{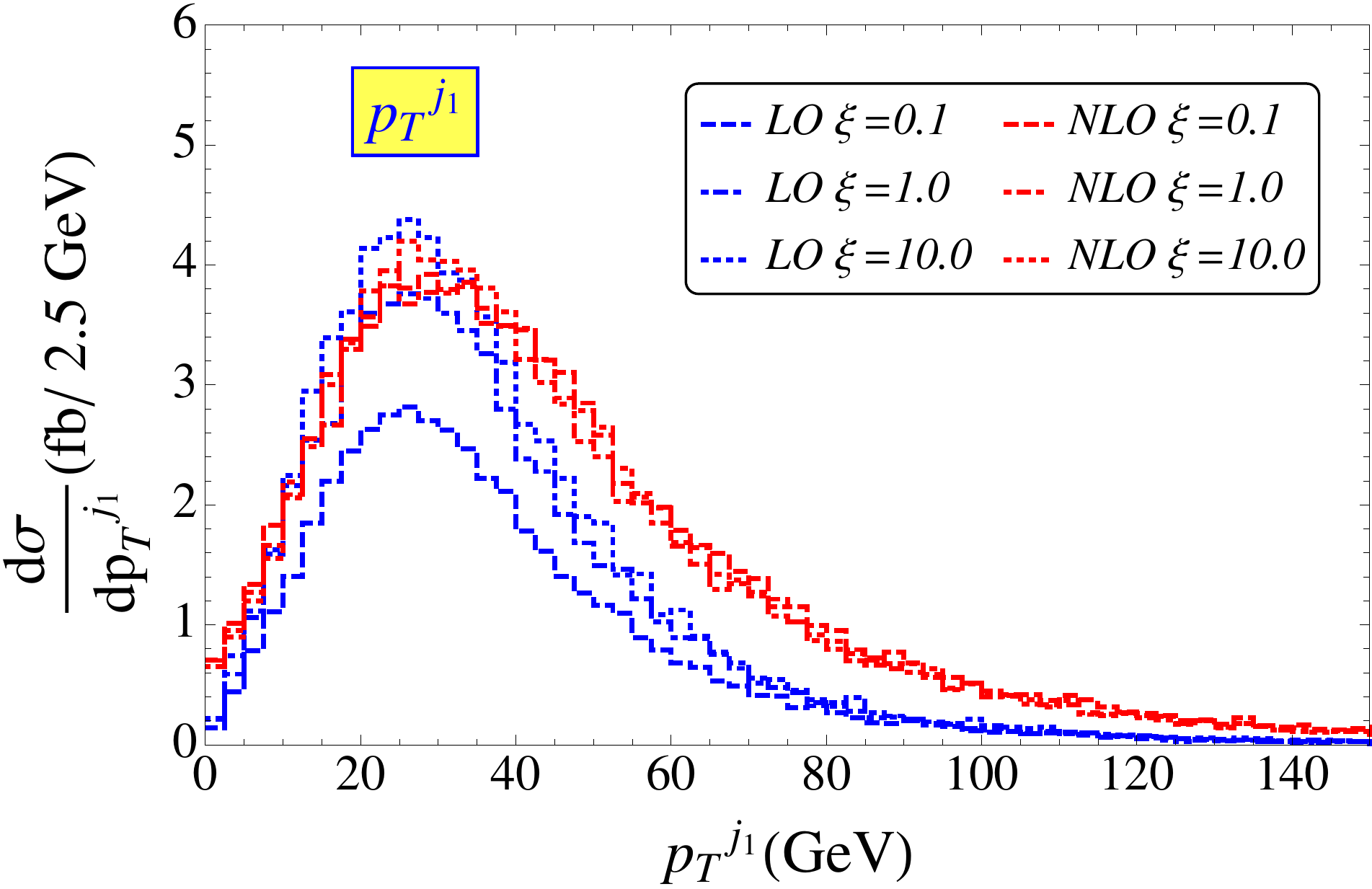}
\includegraphics[scale=0.3]{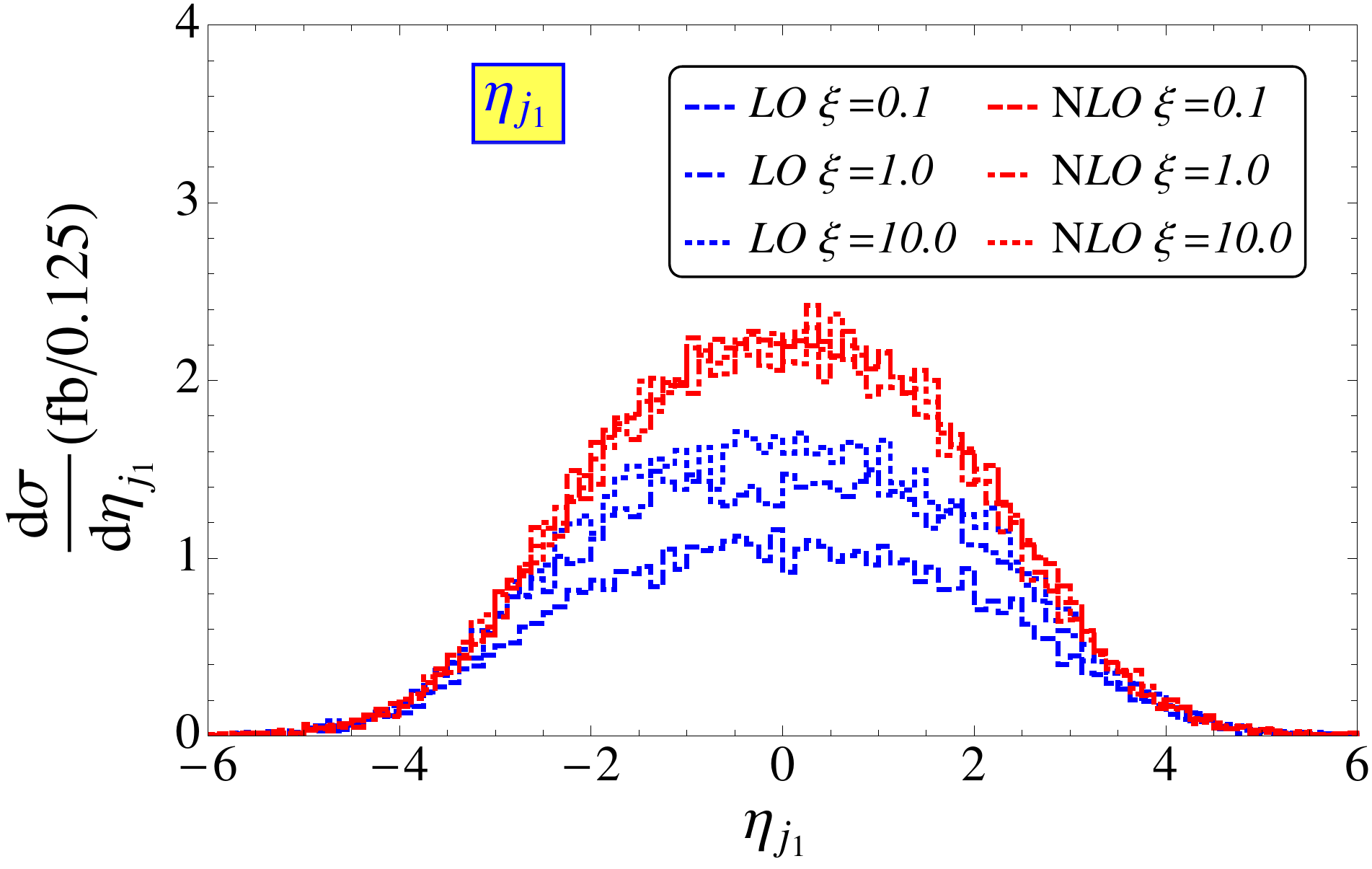}\\
\includegraphics[scale=0.3]{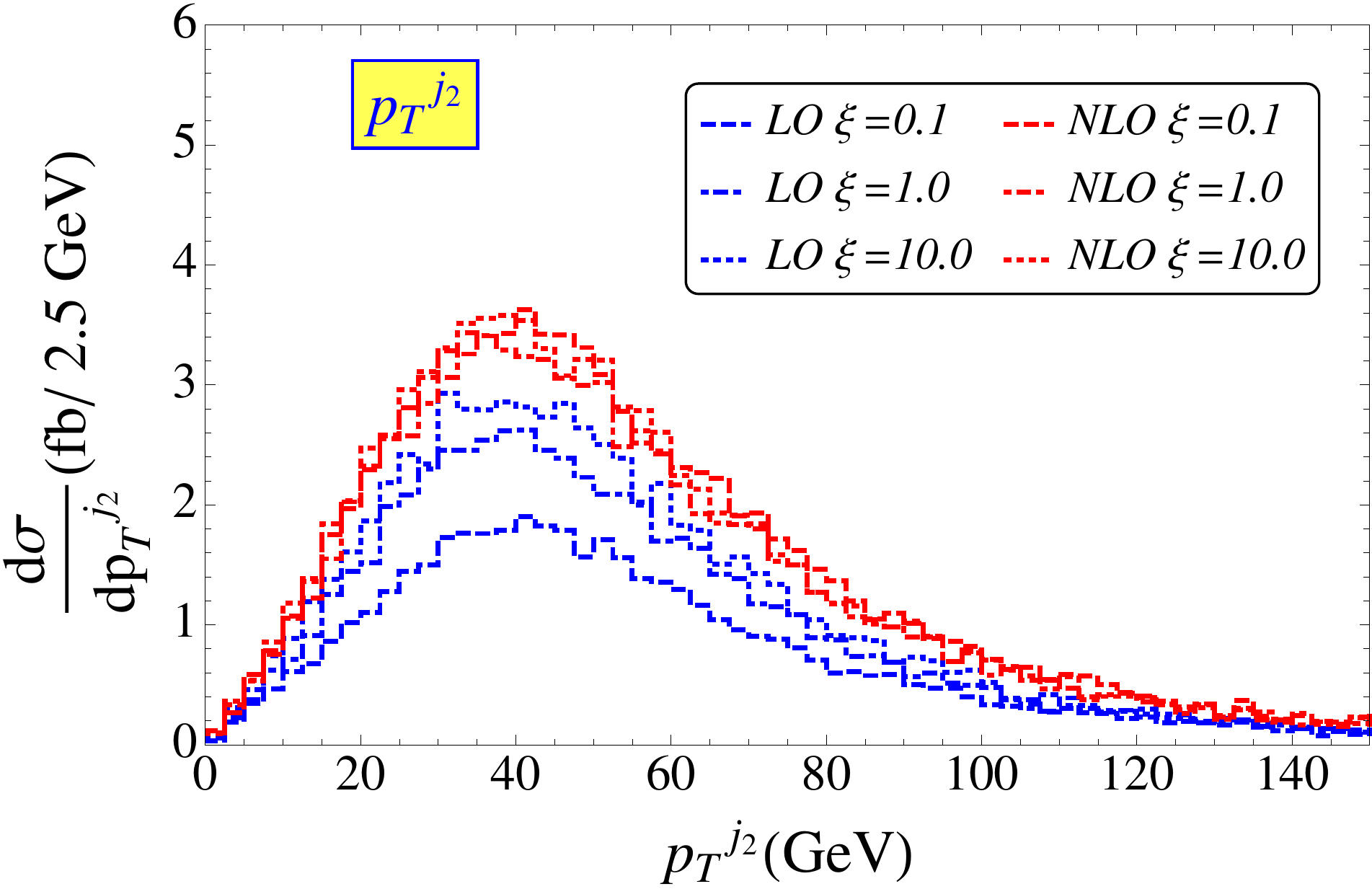}
\includegraphics[scale=0.3]{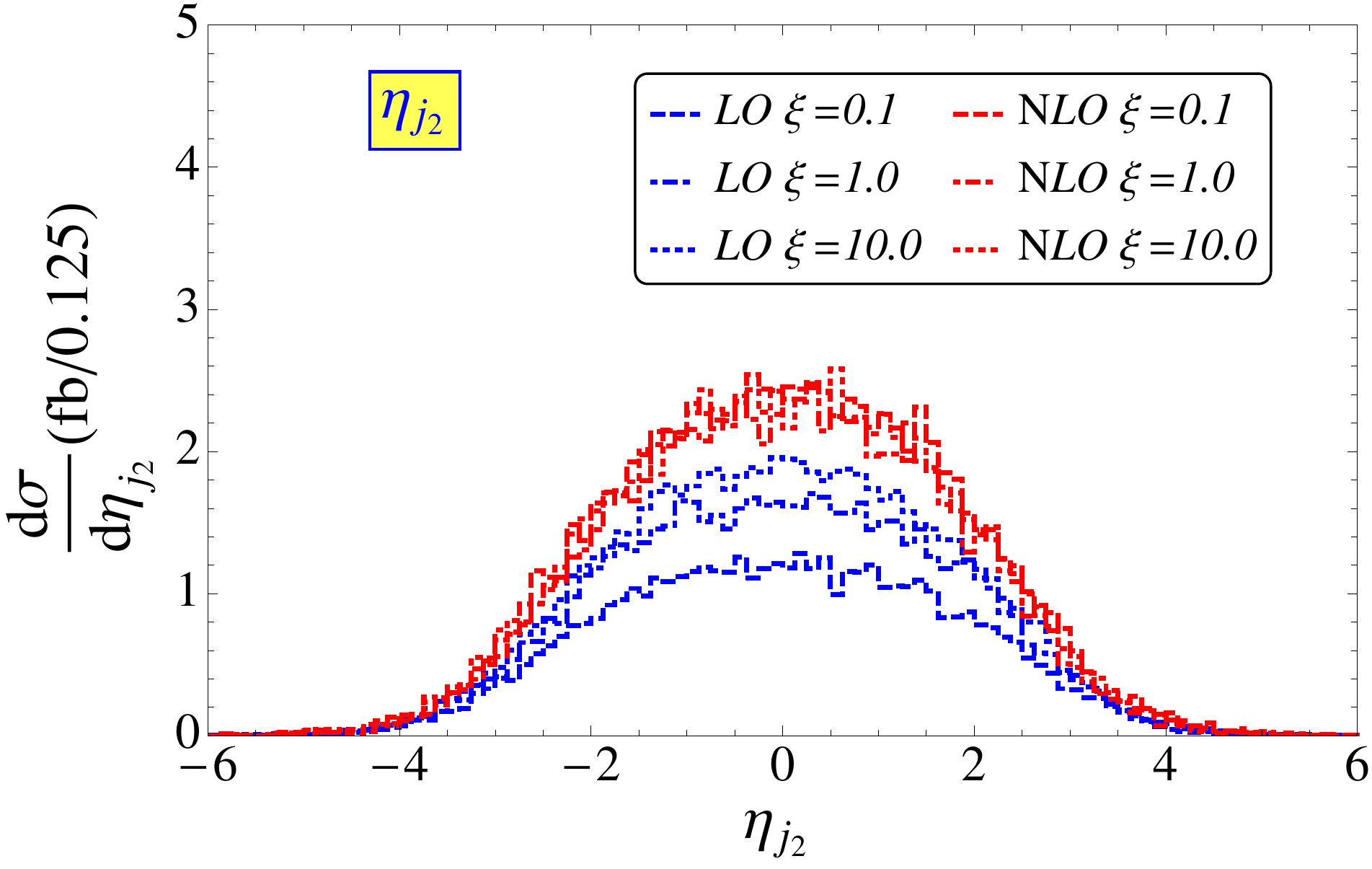}
\end{center}
\caption{Scale dependent LO and NLO-QCD $p_T^{j}$(left column) and $\eta^{j}$(right column) distributions of the heavy neutrino pair production followed by the decays of the heavy neutrinos into 
$3\ell+\rm{MET}+2j$ channel at the 13 TeV LHC for $m_N=95$ GeV.}
\label{HC 95-3l2}
\end{figure} 

The $p_T^{\ell}$ and $\eta^{\ell}$  distributions for the 13 TeV LHC case with $m_{N}=95$ GeV are shown in Fig.~\ref{HC 95-3l1} for the scale dependent LO and NLO-QCD processes. $\ell_2$ is produced after the decay of the $W$ boson which can be considered as the leading lepton. For the leading lepton, a selection of $p_{T}^{\ell, \rm{leading}} > 20~\rm{GeV}$ could be applicable to accept most of the signal events with respect to the SM backgrounds. For the trailing leptons coming from the $N$ decays could have the transverse momentum cut $p_{T}^{\ell,\rm{trailing}}> 15$ GeV. As the jets are coming from the $W$, the transverse momenta of the jets are likely to have peaks around $40$ GeV and the pseudo-rapidity cuts for the jets can be considered as $|\eta^{j}| < 2.5$. The scale dependent LO and NLO-QCD histograms are shown in Fig.~\ref{HC 95-3l2}.

\begin{figure}
\begin{center}
\includegraphics[scale=0.40]{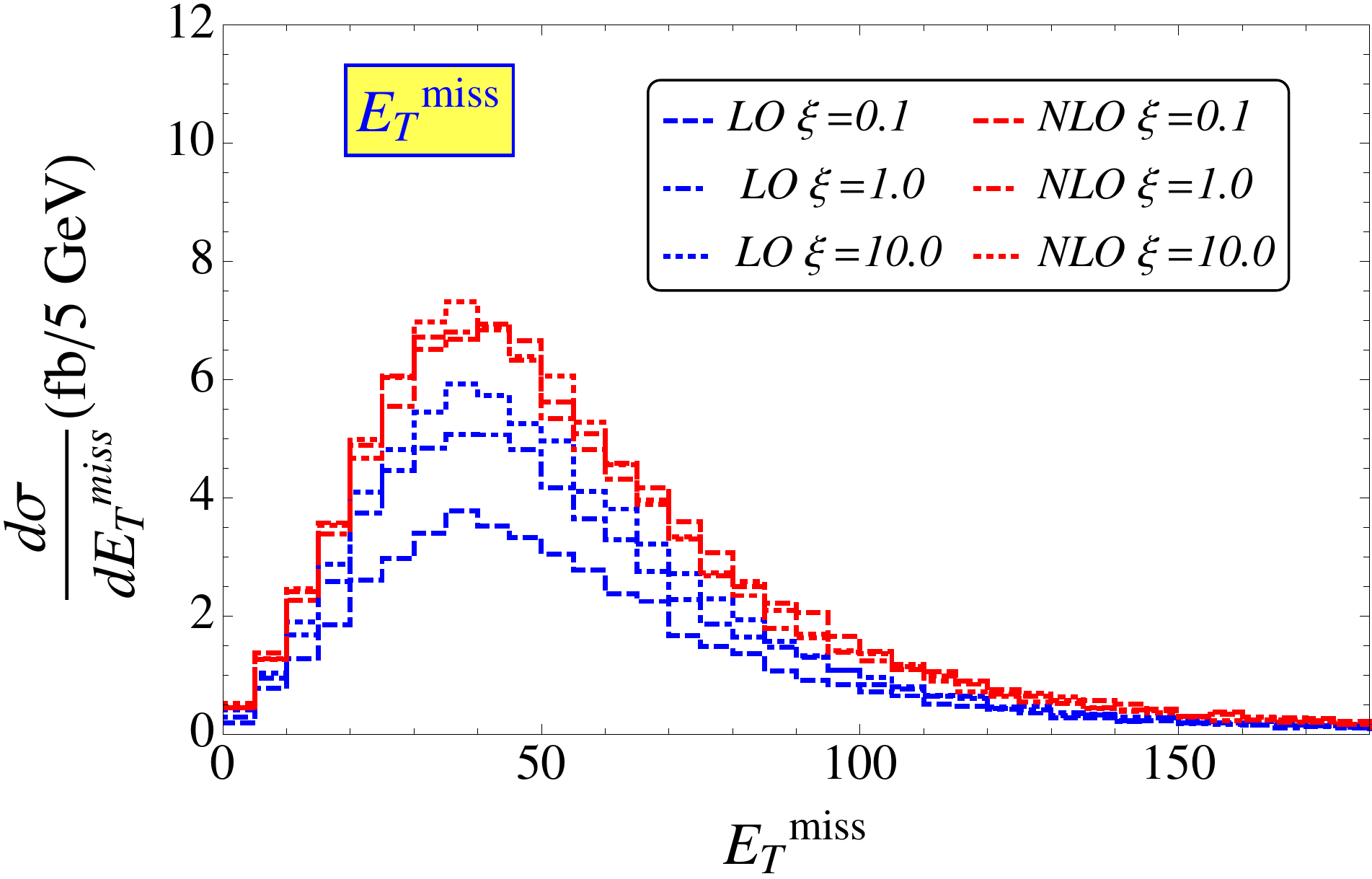}
\end{center}
\caption{Scale dependent LO and NLO-QCD $E_T^{\rm{miss}}$ distributions of the heavy neutrino pair production followed by the decays of the heavy neutrinos into 
$3\ell+\rm{MET}+2j$ channel at the 13 TeV LHC for $m_N=95$ GeV.}
\label{HC 95-3l3}
\end{figure}

Depending upon the scale dependent $E_{T}^{\rm{miss}}$ distributions at the LO and NLO-QCD levels from Fig.~\ref{HC 95-3l3} two regions can be selected for the study of the signal and the backgrounds.  The regions are shown in Tab.~\ref{R}.
\begin{table}[ht]
\begin{center}
\begin{tabular}{ccc}
     &&Selections  \\
\hline
Region-I&&$E_{T}^{\rm{miss}} < 50~\rm{GeV}$\\
Region-II&&$50~\rm{GeV} < E_{T}^{\rm{miss}} < 100~\rm{GeV}$\\
\hline
\end{tabular}
\end{center}
\caption{Selection regions based on $E_{T}^{\rm{miss}} $}
\label{R}
\end{table} 

\begin{figure}
\begin{center}
\includegraphics[scale=0.25]{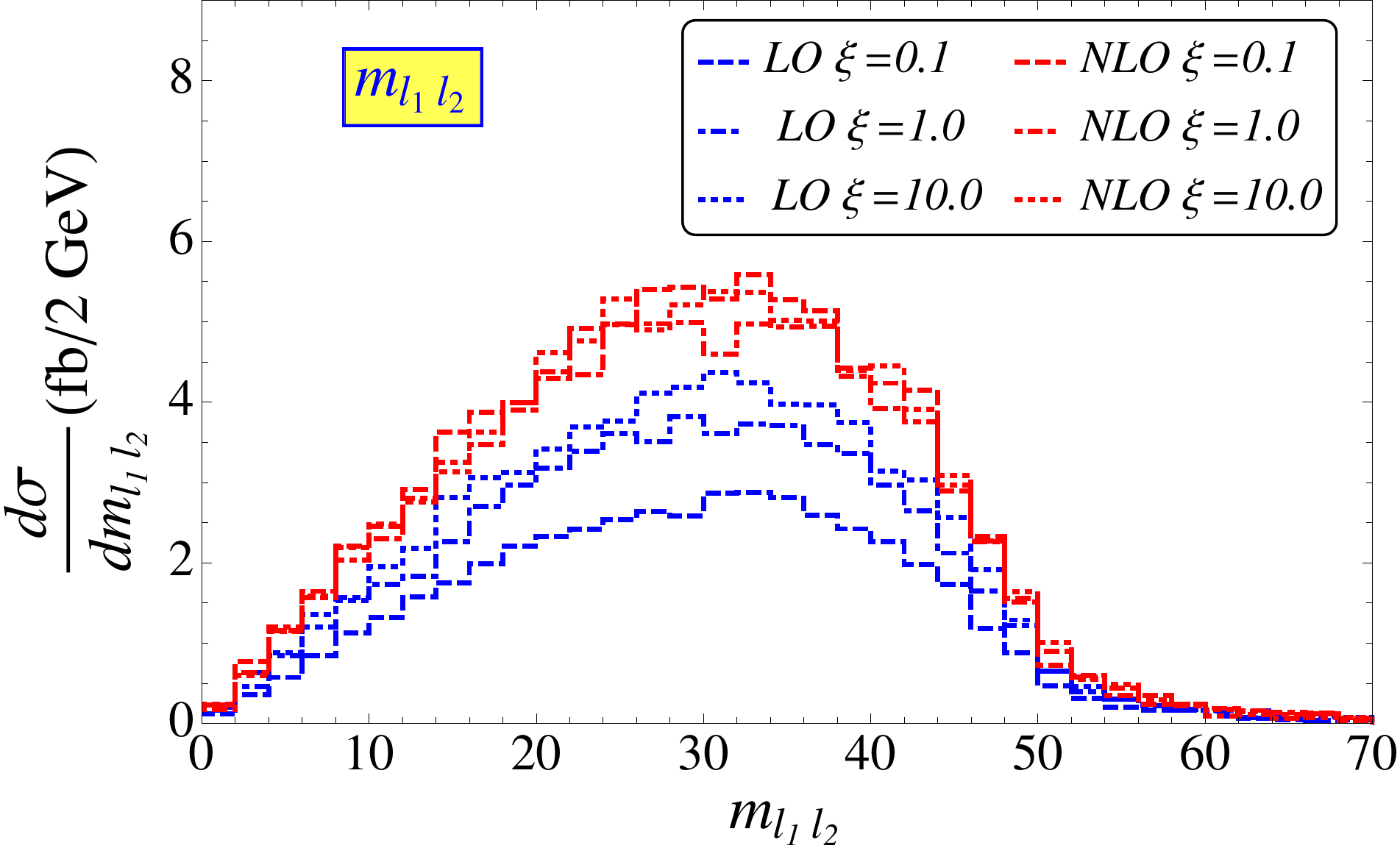}
\includegraphics[scale=0.25]{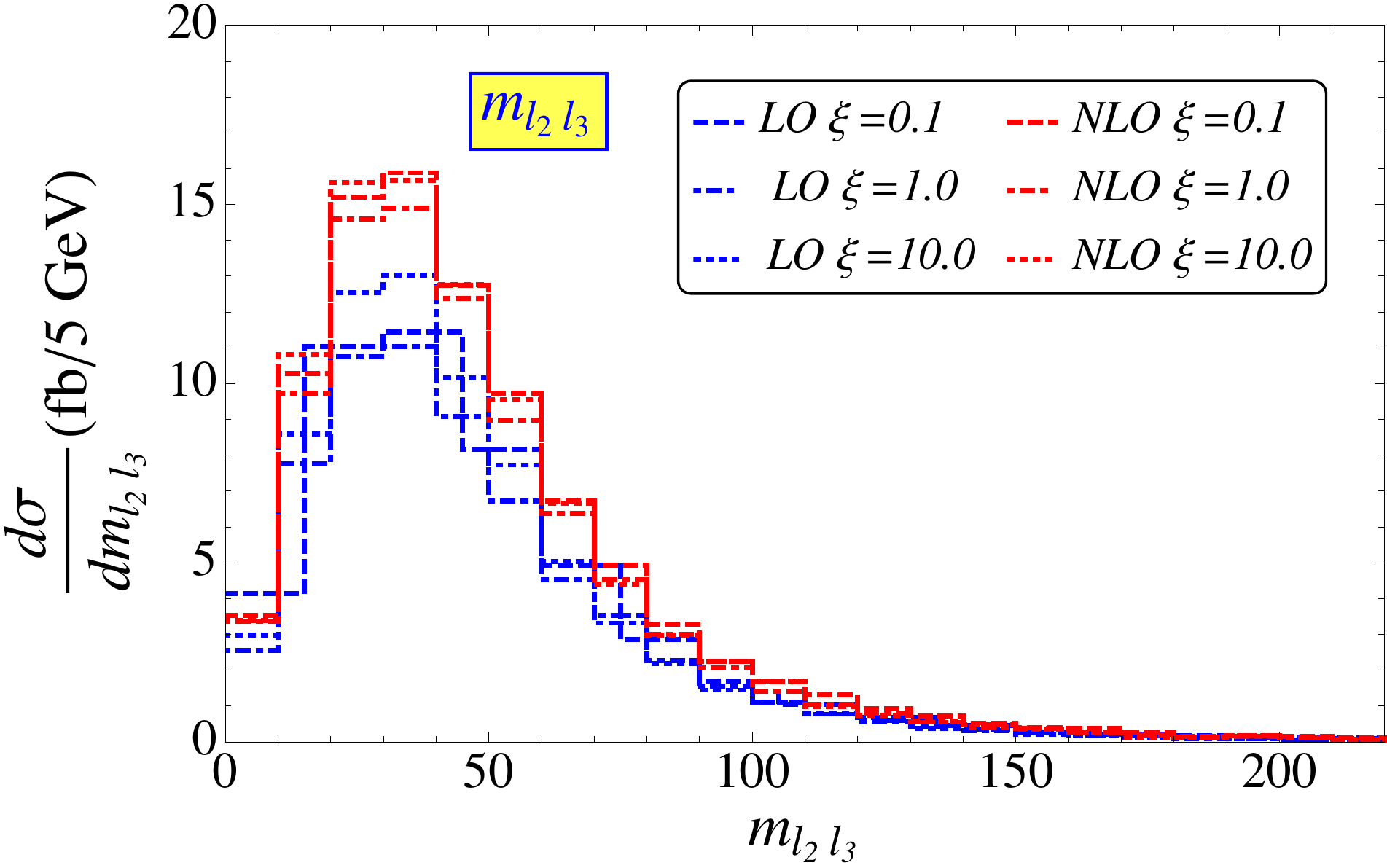}
\includegraphics[scale=0.25]{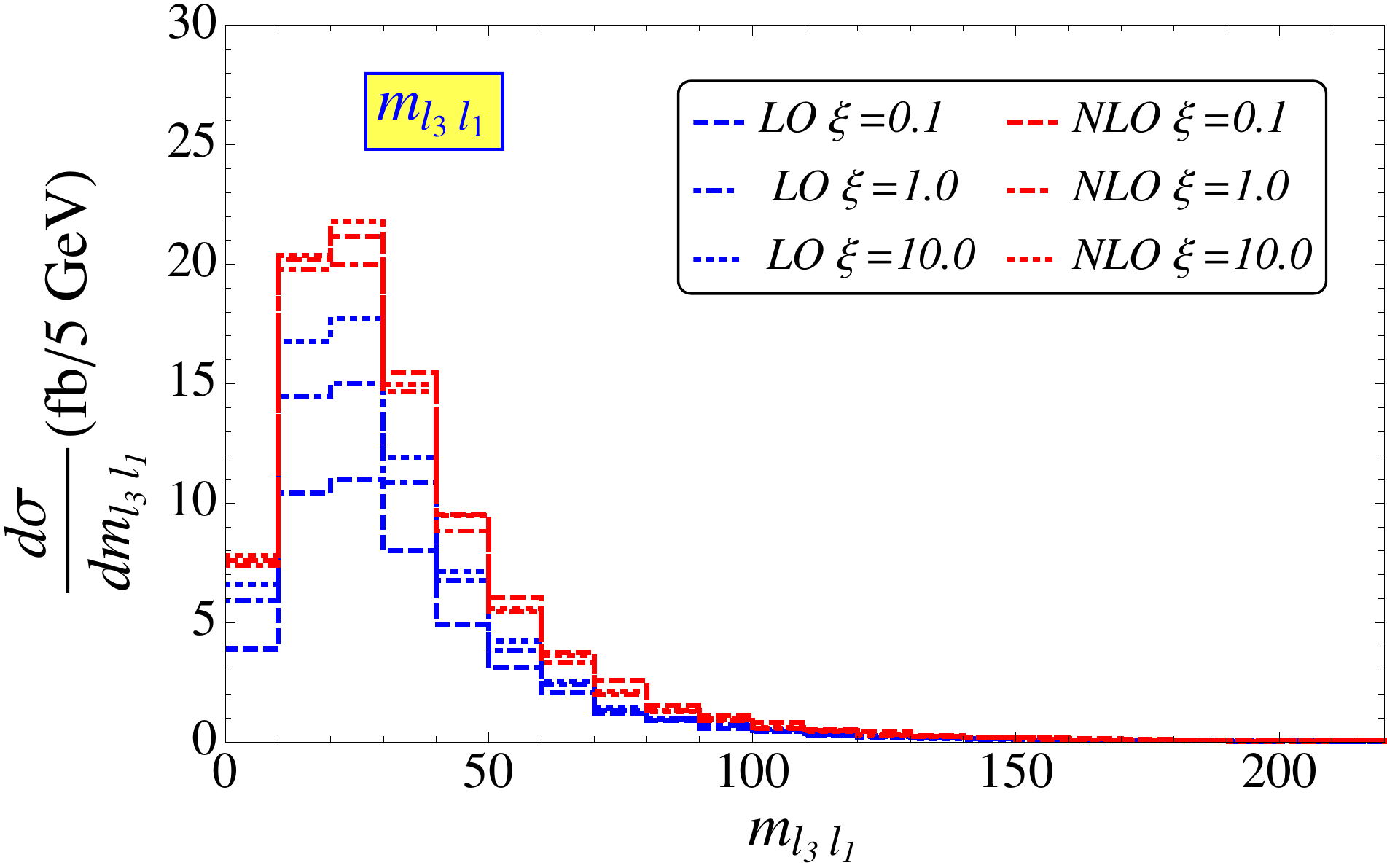}
\end{center}
\caption{Scale dependent LO and NLO-QCD $m_{\ell\ell}$ distributions of the heavy neutrino pair production followed by the decays of the heavy neutrinos into 
$3\ell+\rm{MET}+2j$ channel at the 13 TeV LHC for $m_N=95$ GeV.}
\label{HC 95-3l4}
\end{figure}

The invariant mass distributions for the three combinations of the leptons are given in Fig.~\ref{HC 95-3l4}. In the trilepton event $\ell_{1}^{-}\ell_{2}^{+}\ell_{3}^{+}$ there are two same sign leptons which make 
two OSSF pairs $\ell_{1}^{-}\ell_{2}^{+}$ and $\ell_{1}^{-}\ell_{3}^{+}$, however, there could be one OSSF pair due to 
\bea
W^{+} \to \ell^{\prime^{+}}_{2} \nu
\eea
where $\ell_{2}^{+}\neq \ell^{\prime^{+}}_{2}$. Therefore the trilepton mode will look like
\bea         
 p p &\to& N \overline{N}, \nonumber \\
        \, \, \, \,  \, \, \, \,&&   N \to \ell_{1}^{-} W^{+}, W^{+} \to \ell_{2}^{+} \nu / \ell^{\prime^{+}}_{2} \nu \nonumber \\
        \, \, \, \,  \, \, \, \,&&   \overline{N} \to \ell_{3}^{+} W^{-}, W^{-} \to j j .
\label{decay121}        
\eea
These OSSF pairs could be utilized for giving the invariant mass cut to allow the signal region below the $Z$ pole such as $m_{\ell\ell} < (m_{Z}- 15)~\rm{GeV}$ in the trilepton mode \footnote{There is another possible trilepton mode which could be written as 
\bea        
p p &\to& N \overline{N}, N \to \ell_{1}^{-} W^{+}, W^{+} \to j j,  \overline{N} \to \ell_{2}^{+} W^{-}, W^{-} \to  \ell_{3}^{-} \overline{\nu} / \ell^{\prime^{-}}_{3} \overline{\nu}
         \label{decay122}
 \eea
 Where $\ell_{3}$ will be leading lepton}.

\begin{figure}
\begin{center}
\includegraphics[scale=0.25]{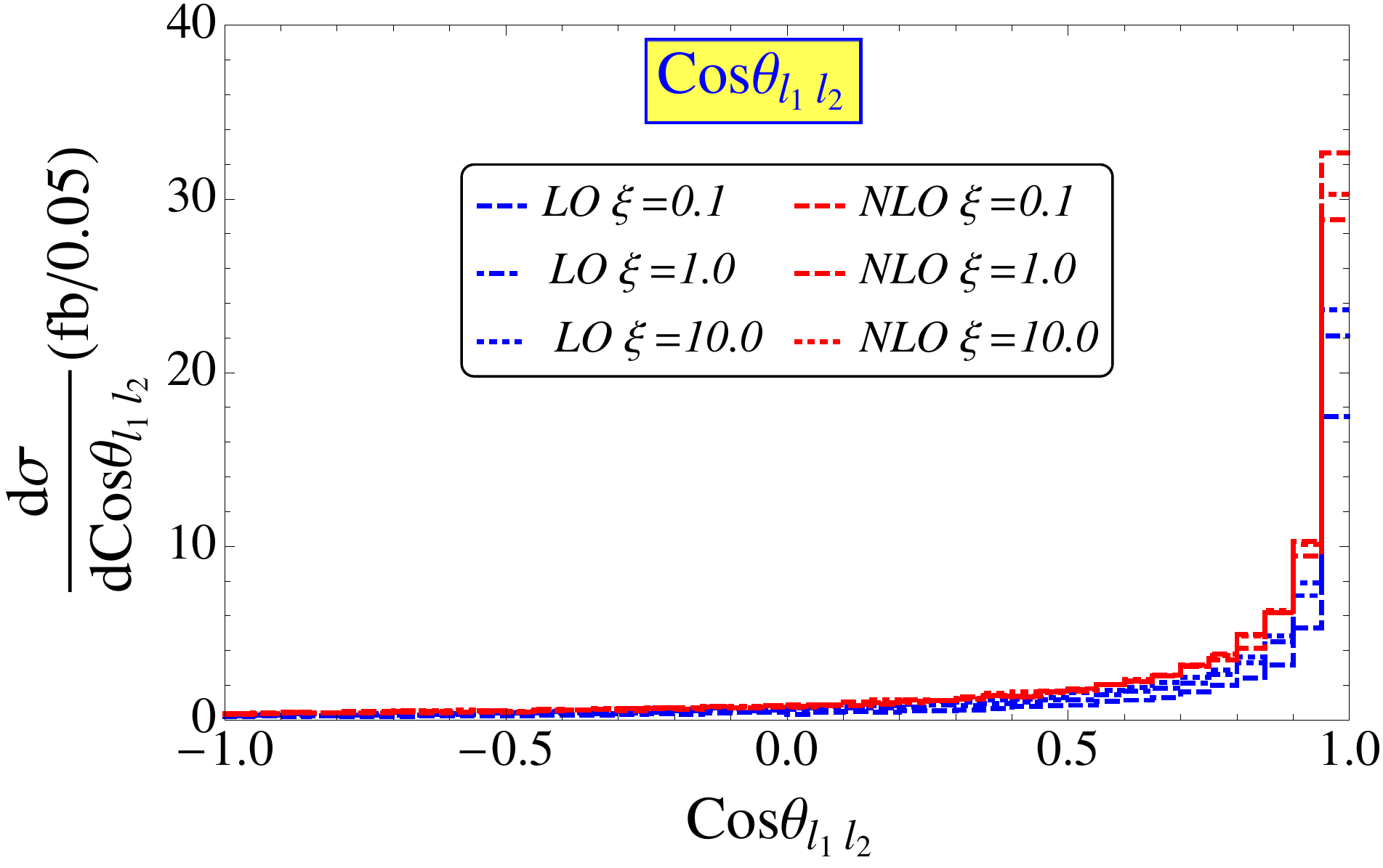}
\includegraphics[scale=0.25]{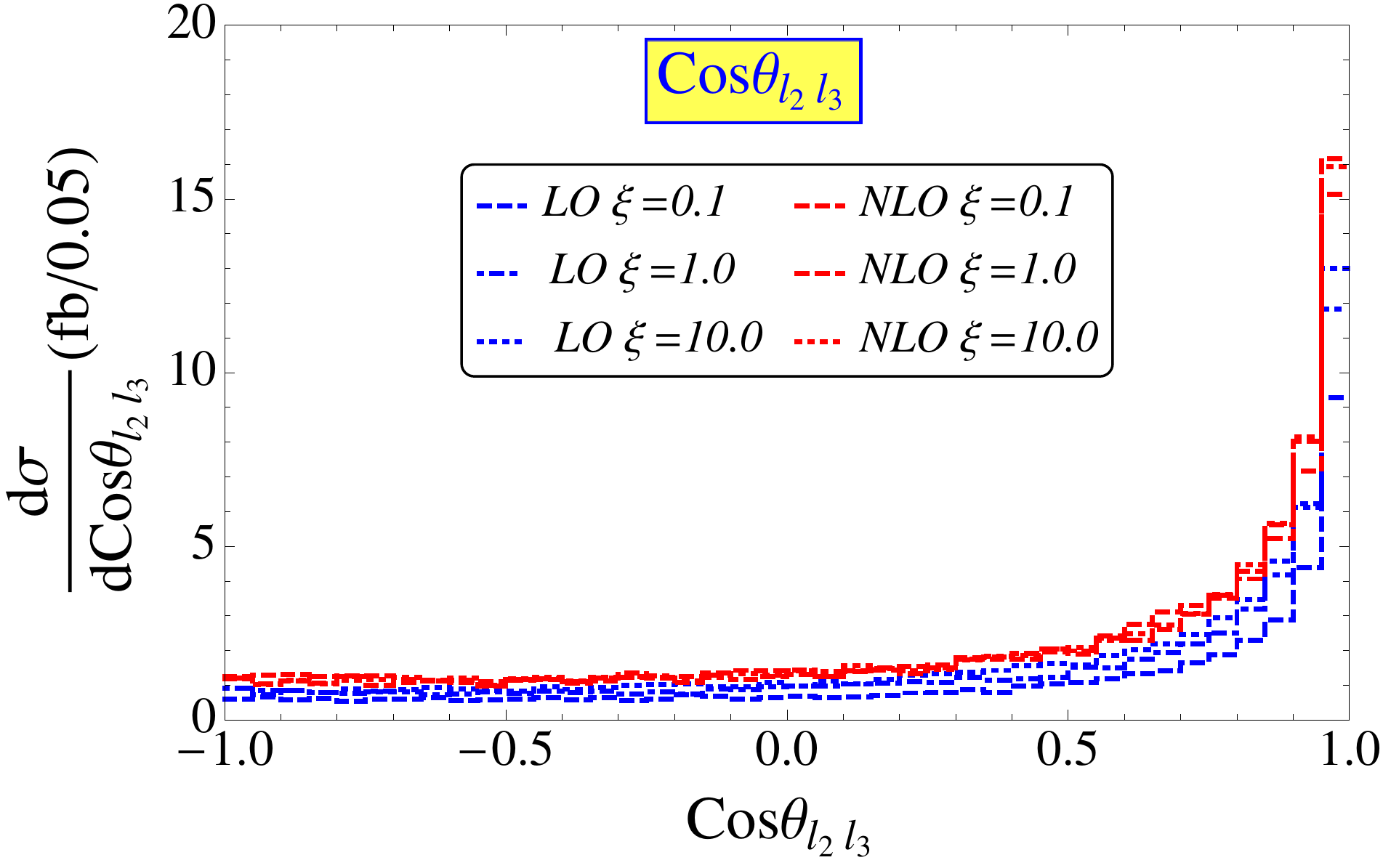}
\includegraphics[scale=0.25]{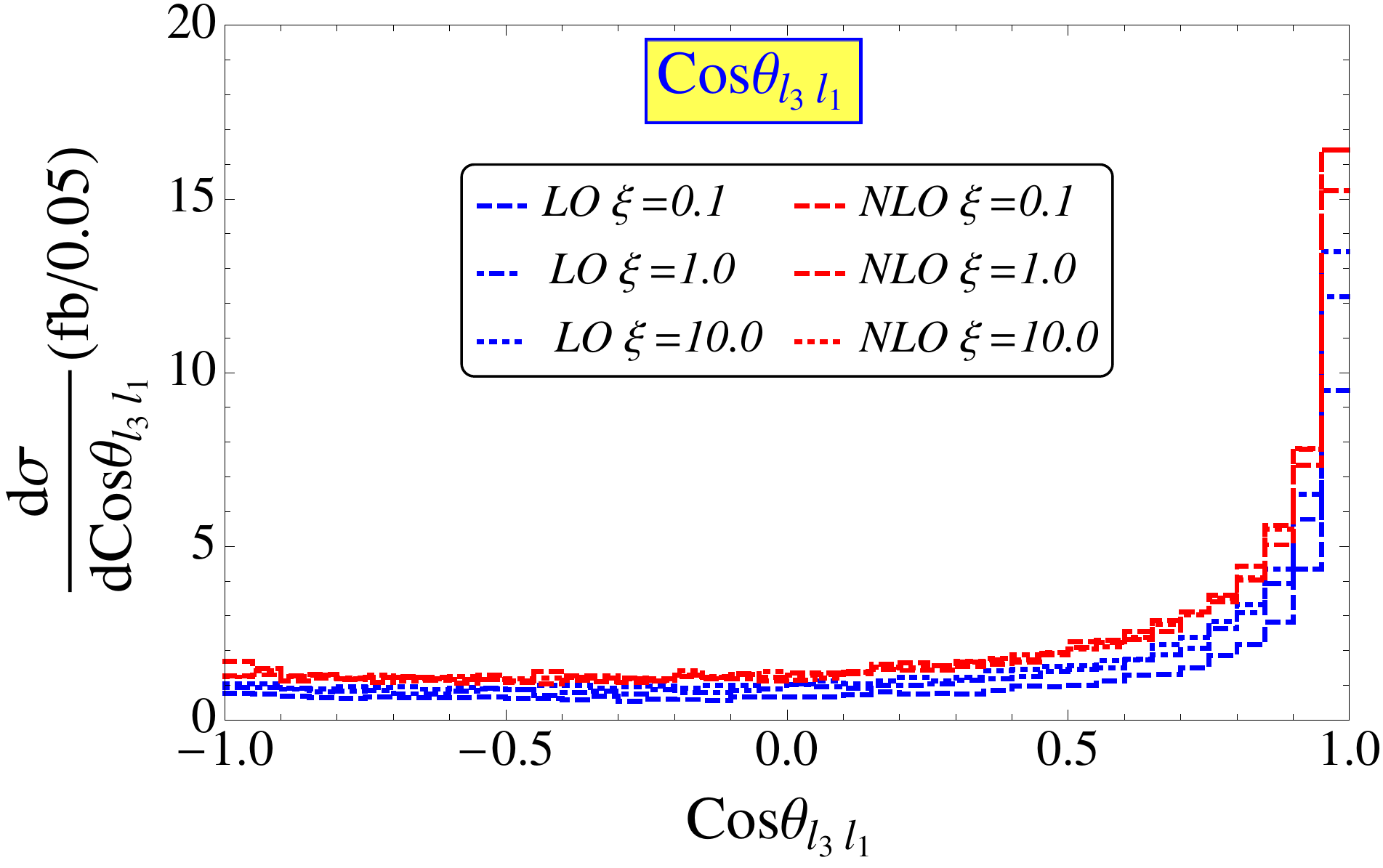}
\end{center}
\caption{Scale dependent LO and NLO-QCD $\cos\theta_{\ell\ell}$ distributions of the heavy neutrino pair production followed by the decays of the heavy neutrinos into 
$3\ell+\rm{MET}+2j$ channel at the 13 TeV LHC for $m_N=95$ GeV.}
\label{HC 95-3l5}
\end{figure}

\begin{figure}
\begin{center}
\includegraphics[scale=0.4]{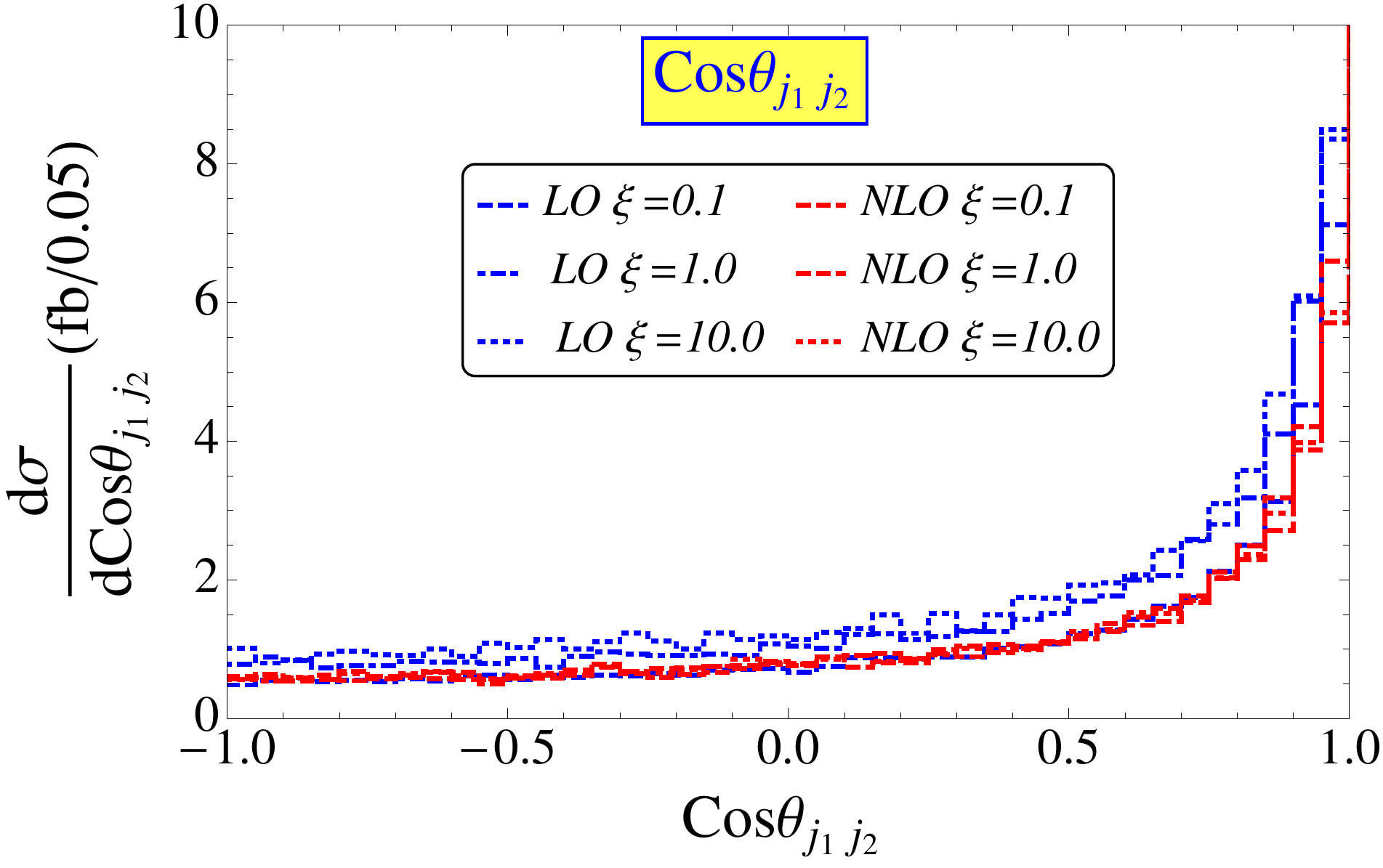}
\end{center}
\caption{Scale dependent LO and NLO-QCD $\cos\theta_{jj}$ distributions of the heavy neutrino pair production followed by the decays of the heavy neutrinos into 
$3\ell+\rm{MET}+2j$ channel at the 13 TeV LHC for $m_N=95$ GeV.}
\label{HC 95-3l6}
\end{figure}

\begin{figure}
\begin{center}
\includegraphics[scale=0.26]{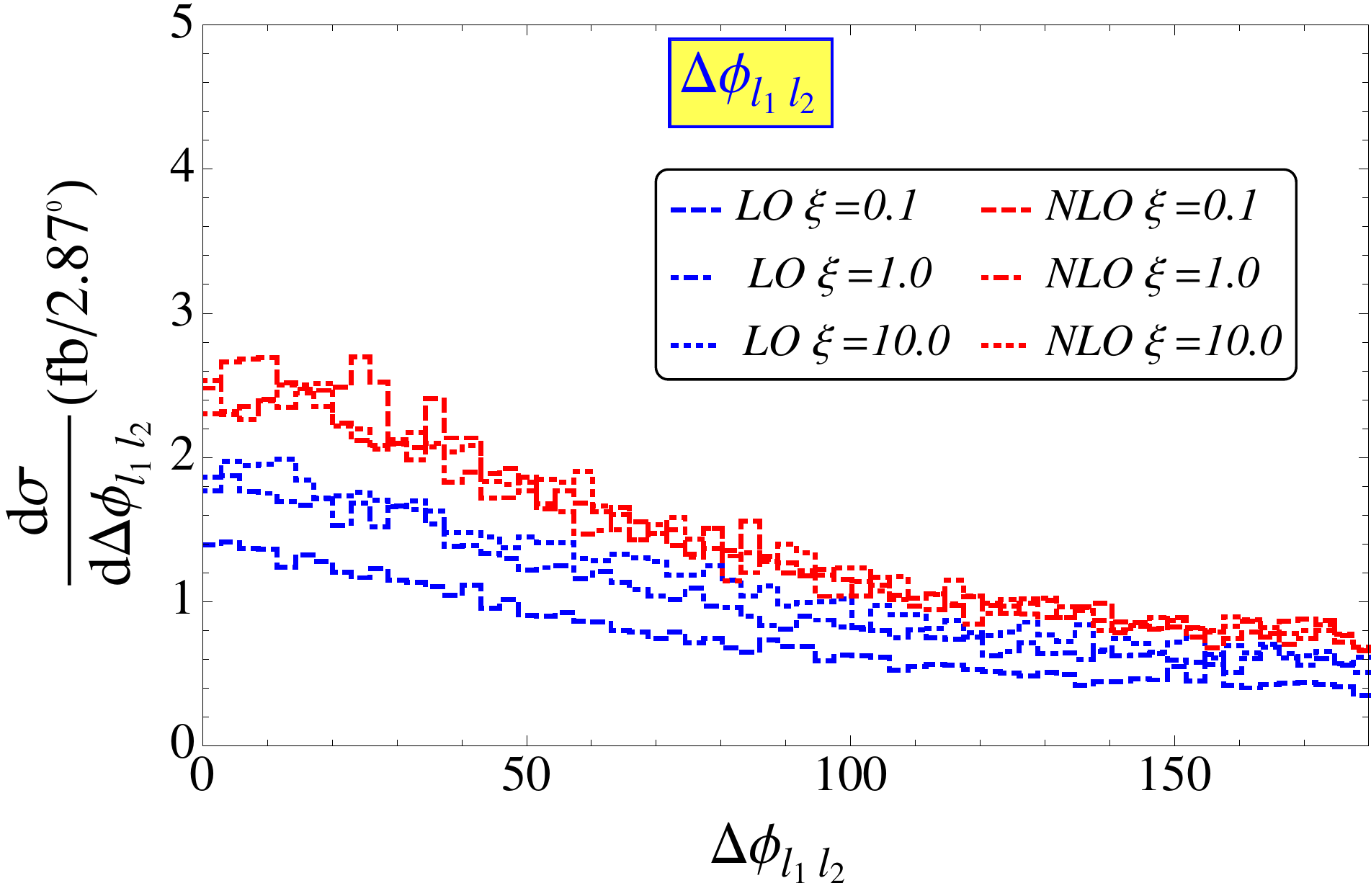}
\includegraphics[scale=0.26]{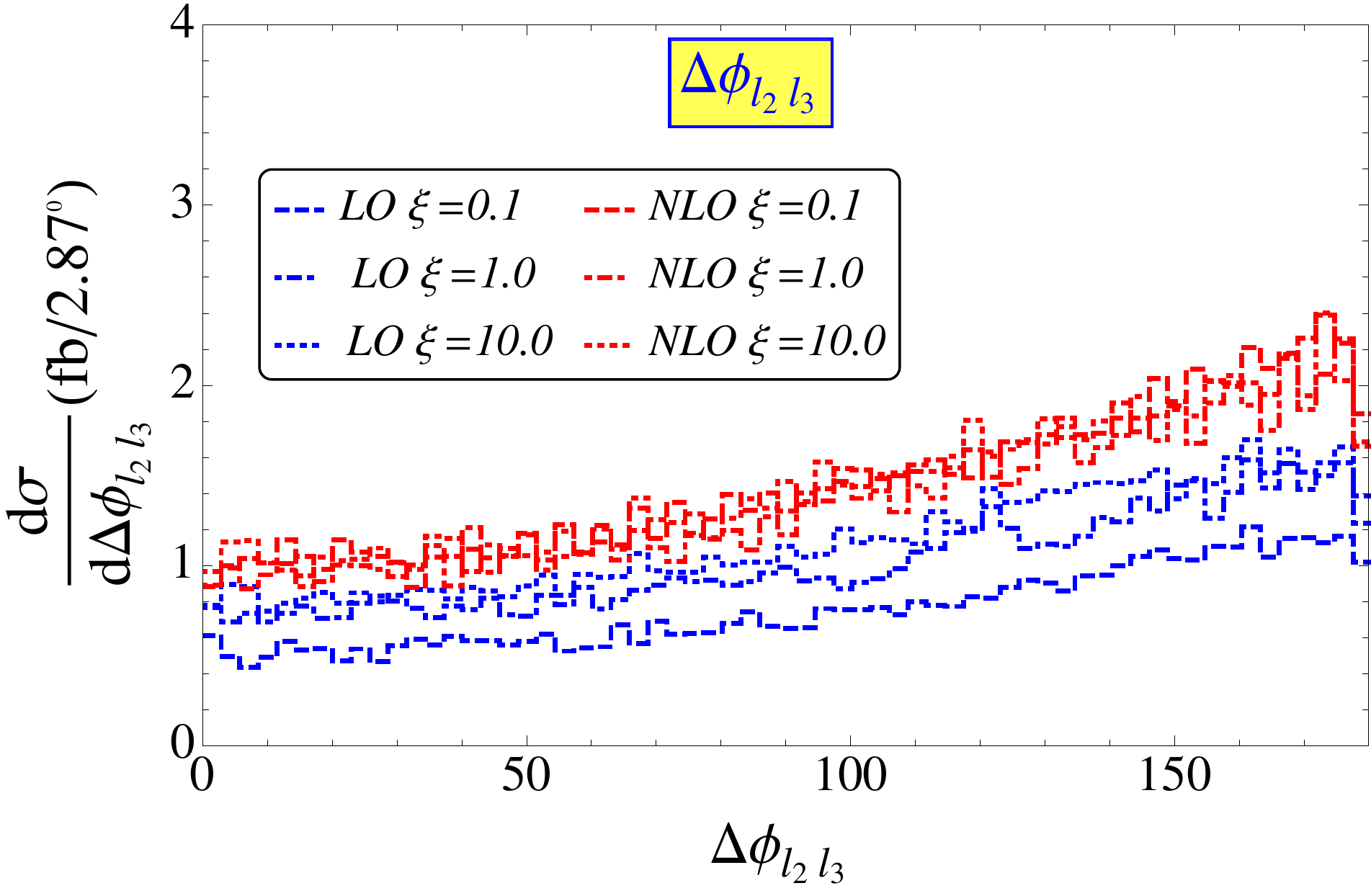}
\includegraphics[scale=0.26]{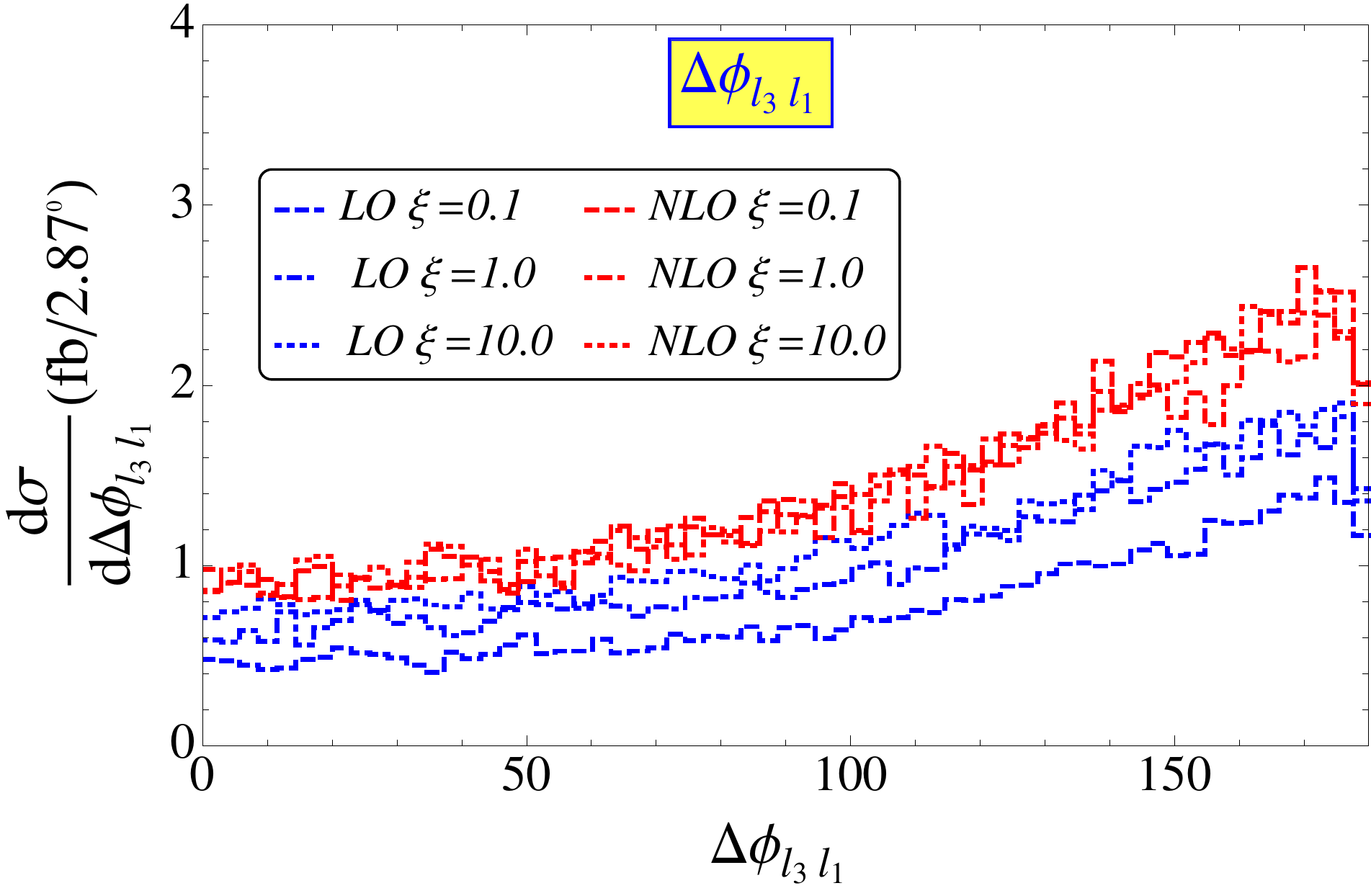}
\end{center}
\caption{Scale dependent LO and NLO-QCD $\Delta\phi_{\ell\ell}$ distributions of the heavy neutrino pair production followed by the decays of the heavy neutrinos into 
$3\ell+\rm{MET}+2j$ channel at the 13 TeV LHC for $m_N=95$ GeV.}
\label{HC 95-3l7}
\end{figure}

\begin{figure}
\begin{center}
\includegraphics[scale=0.40]{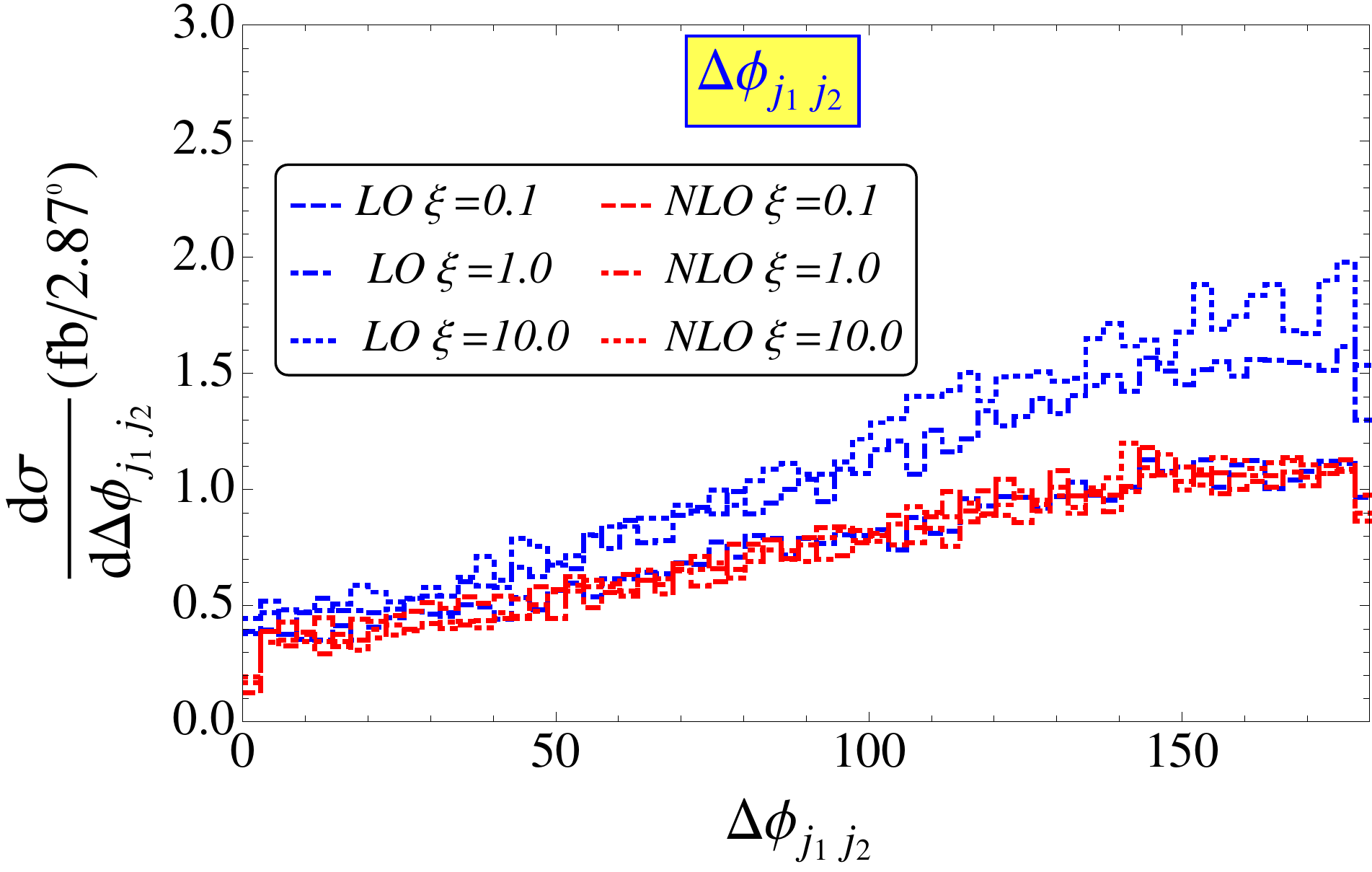}
\end{center}
\caption{Scale dependent LO and NLO-QCD $\Delta\phi_{jj}$ distributions of the heavy neutrino pair production followed by the decays of the heavy neutrinos into 
$3\ell+\rm{MET}+2j$ channel at the 13 TeV LHC for $m_N=95$ GeV.}
\label{HC 95-3l8}
\end{figure}

The scale dependent LO and NLO-QCD distributions of the angles between the different combinations of the leptons are shown in Fig.~\ref{HC 95-3l5} and those between the jets are shown in Fig.~\ref{HC 95-3l6}. 
It shows that most of the events have been stored at the region $\cos\theta_{\ell\ell / \ell j} > 0.5$. The difference between the azimuthal angles of the leptons are plotted in Fig.~\ref{HC 95-3l7} and those between the 
jets are plotted in Fig.~\ref{HC 95-3l8}. In $\Delta\phi_{\ell_{1}\ell_{2}}$ most of the events are obtained $\Delta\phi_{\ell \ell} < 100^{0}$ and the opposite thing happens for $\Delta\phi_{\ell_{2}\ell_{3}}$, $\Delta\phi_{\ell_{3}\ell_{1}}$ 
and $\Delta\phi_{jj}$ for the jets from Fig.~\ref{HC 95-3l8}.


\begin{figure}
\begin{center}
\includegraphics[scale=0.30]{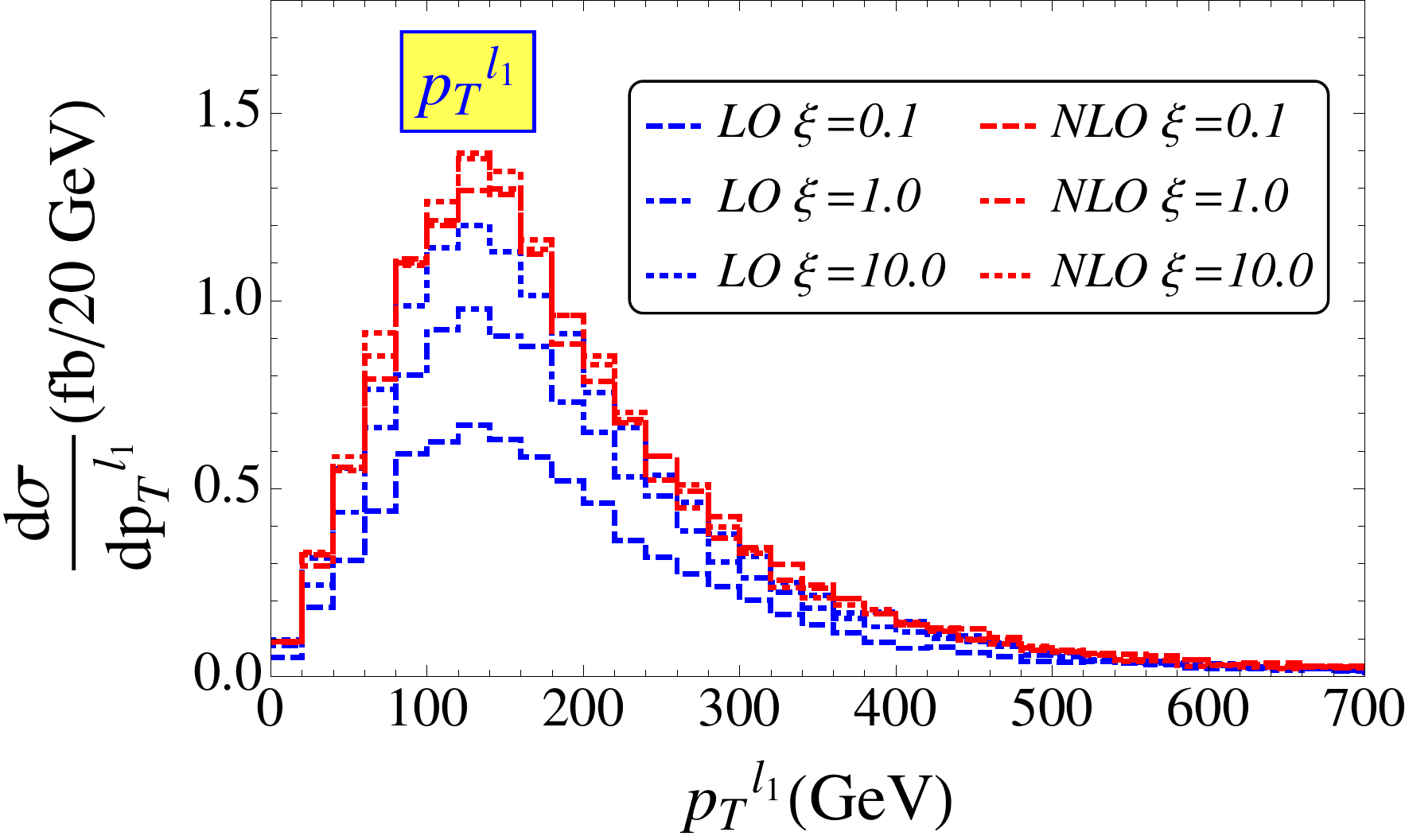}
\includegraphics[scale=0.30]{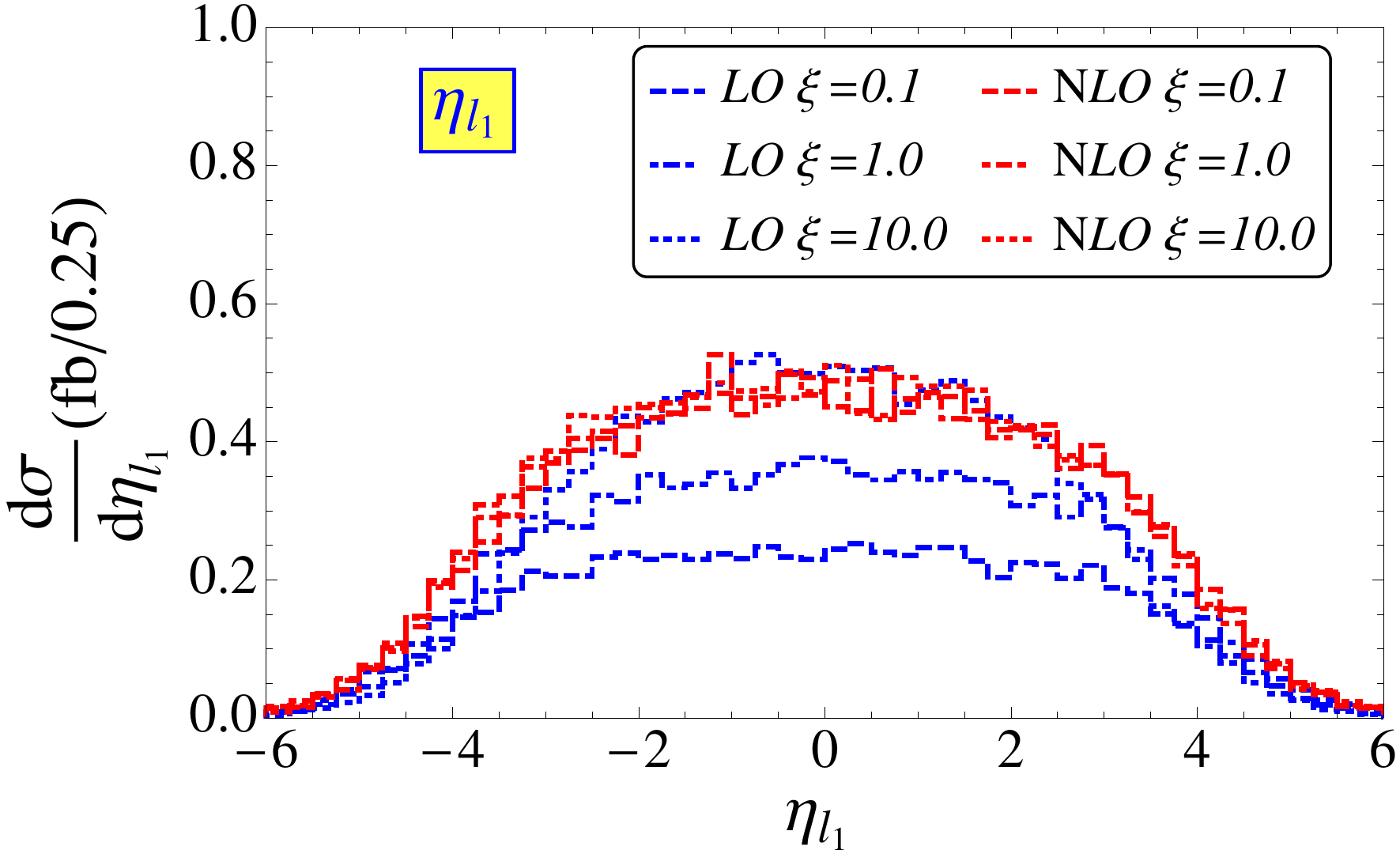}\\
\includegraphics[scale=0.30]{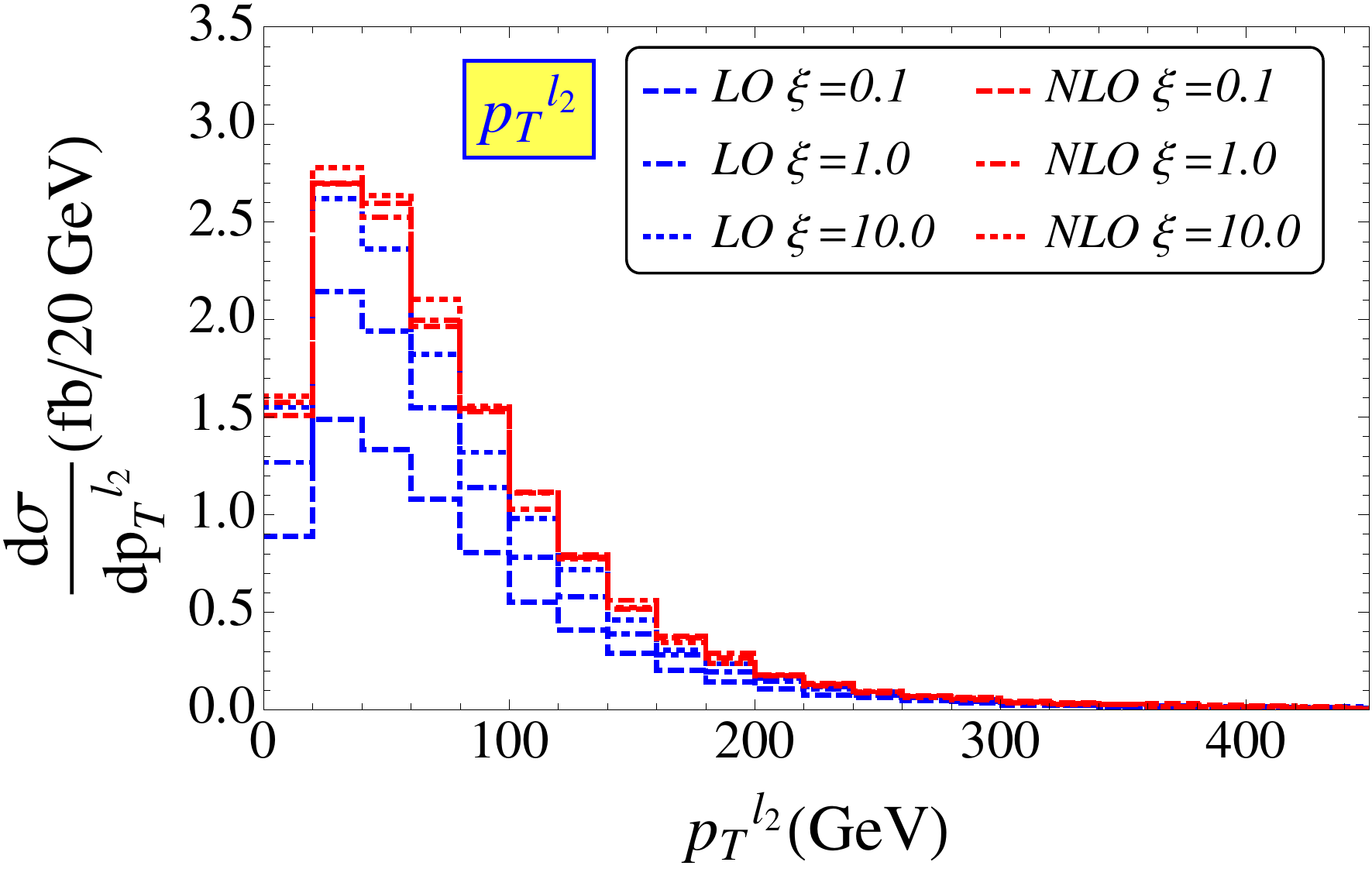}
\includegraphics[scale=0.30]{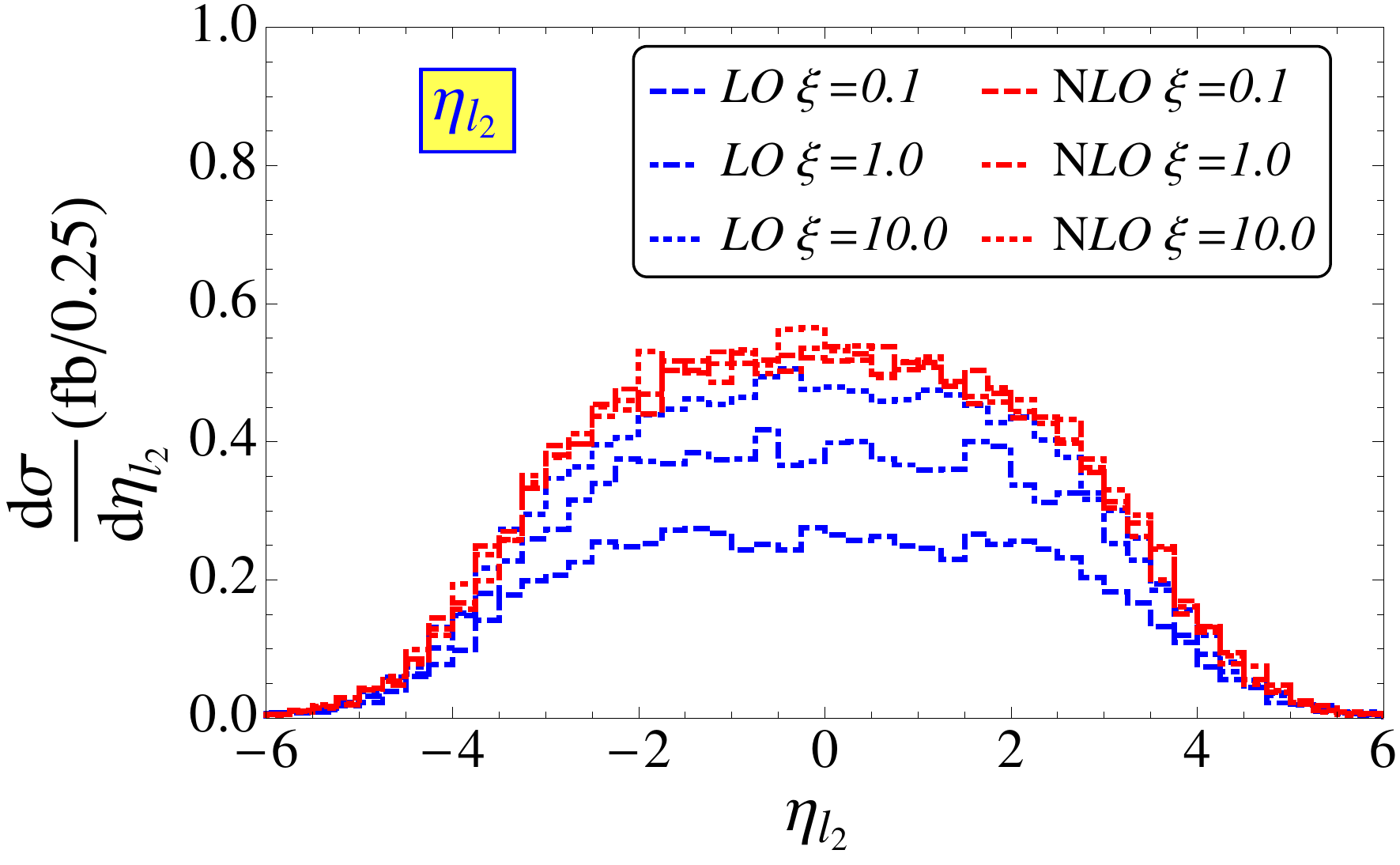}\\
\includegraphics[scale=0.30]{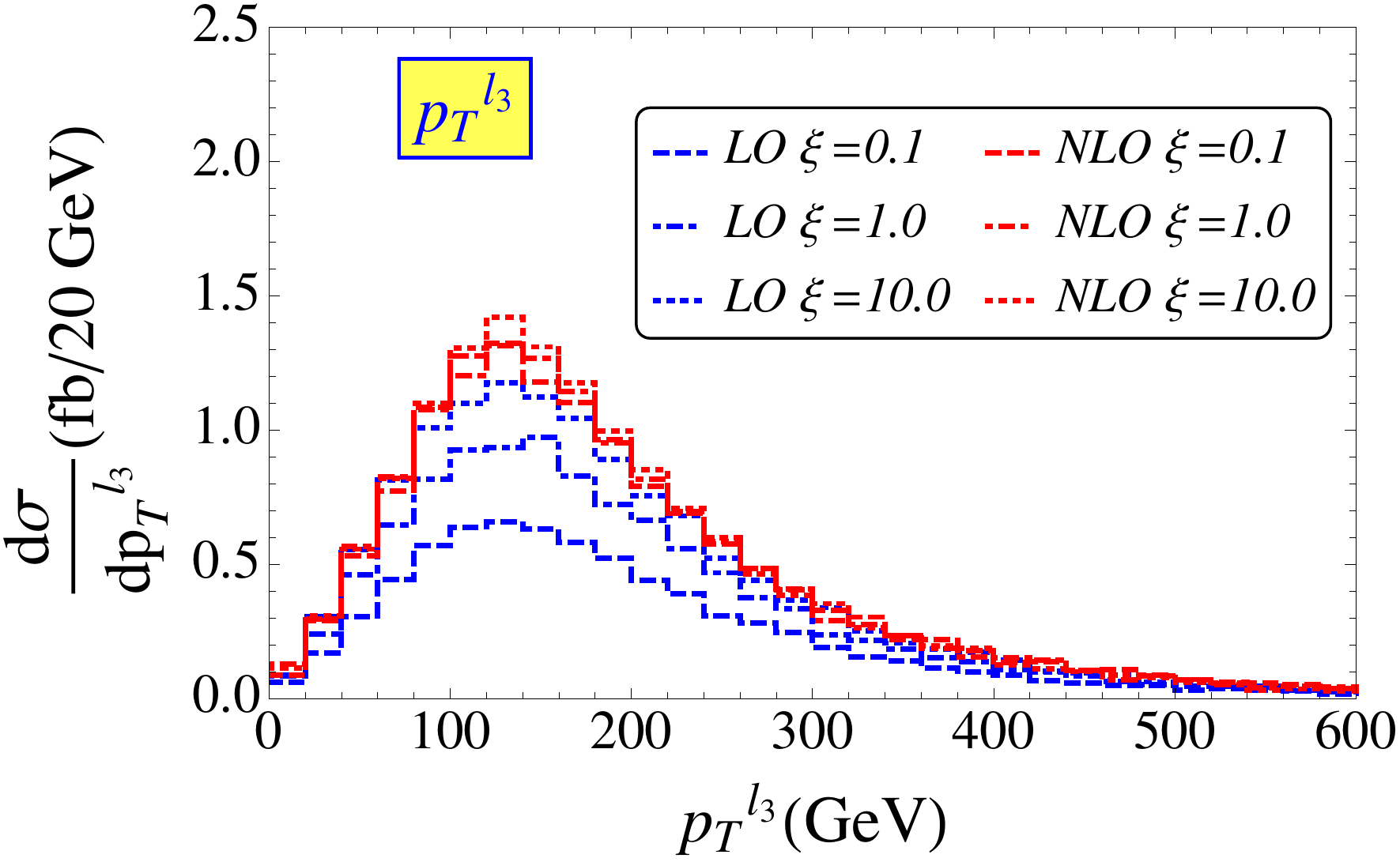}
\includegraphics[scale=0.30]{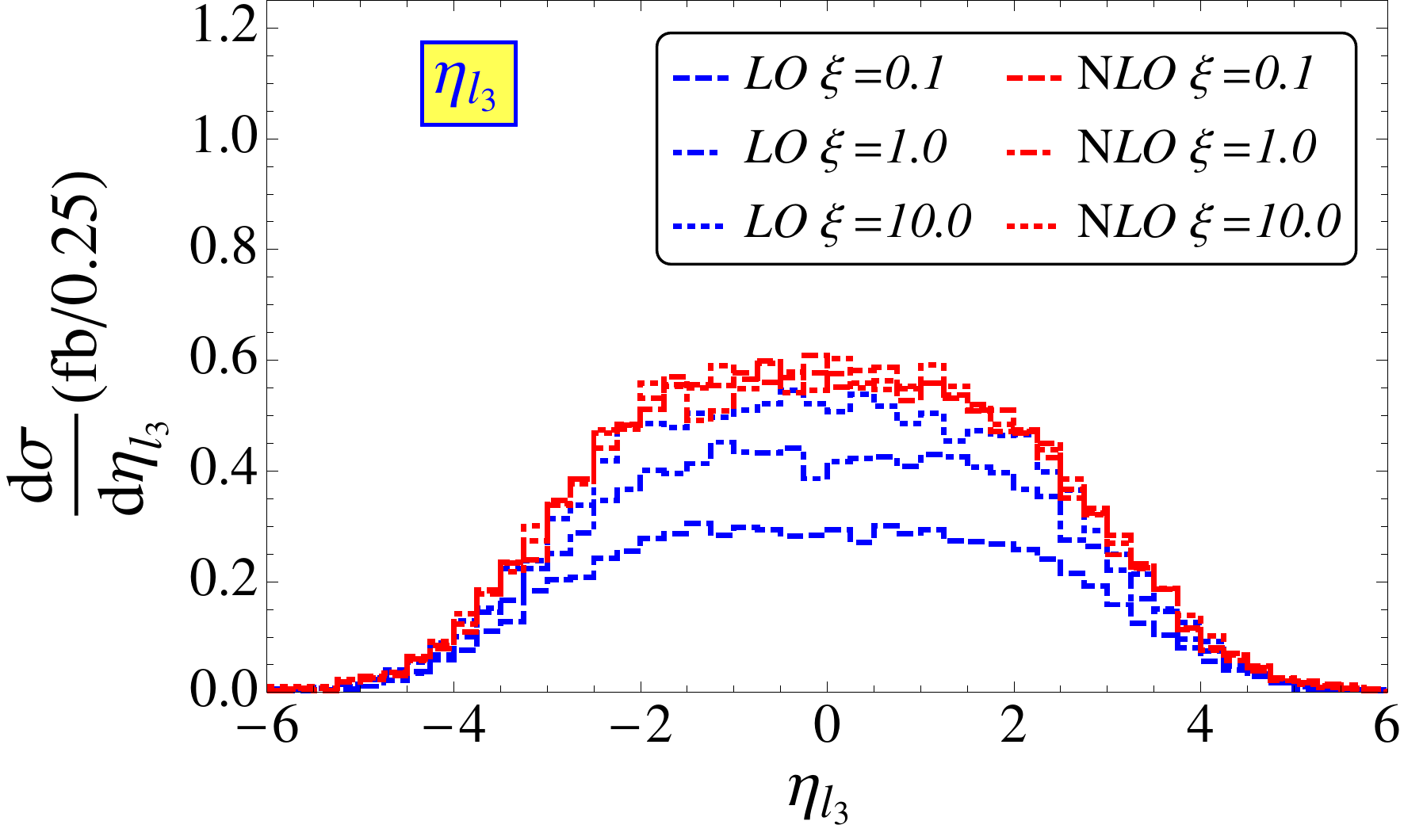}
\end{center}
\caption{Scale dependent LO and NLO-QCD $p_T^{\ell}$(left column) and $\eta^{\ell}$ (right column) distributions of the heavy neutrino pair production followed by the decays of the heavy neutrinos into 
$3\ell+\rm{MET}+2j$ channel at the 100 TeV hadron collider for $m_N=300$ GeV.}
\label{HC 300-3l1}
\end{figure}

To study the behavior of the trilepton decay mode plus MET and two jets at the 100 TeV, we consider $m_{N}=300$ GeV. The 
$p_T^\ell$ and $\eta^{\ell}$ distributions for the leptons are plotted in Fig.~\ref{HC 300-3l1}. For the leading leptons a transverse momentum cut $p_T^{\ell,\rm{leading}} > 90$ GeV and for the trailing lepton a transverse momentum cut $p_{T}^{\ell, \rm{trailing}} > 30$ GeV could be applicable. A pseudo-rapidity cut for the leptons, $|\eta^{\ell} |< 2.5$ could be considered to accept the signal events.
\begin{figure}
\begin{center}

\includegraphics[scale=0.30]{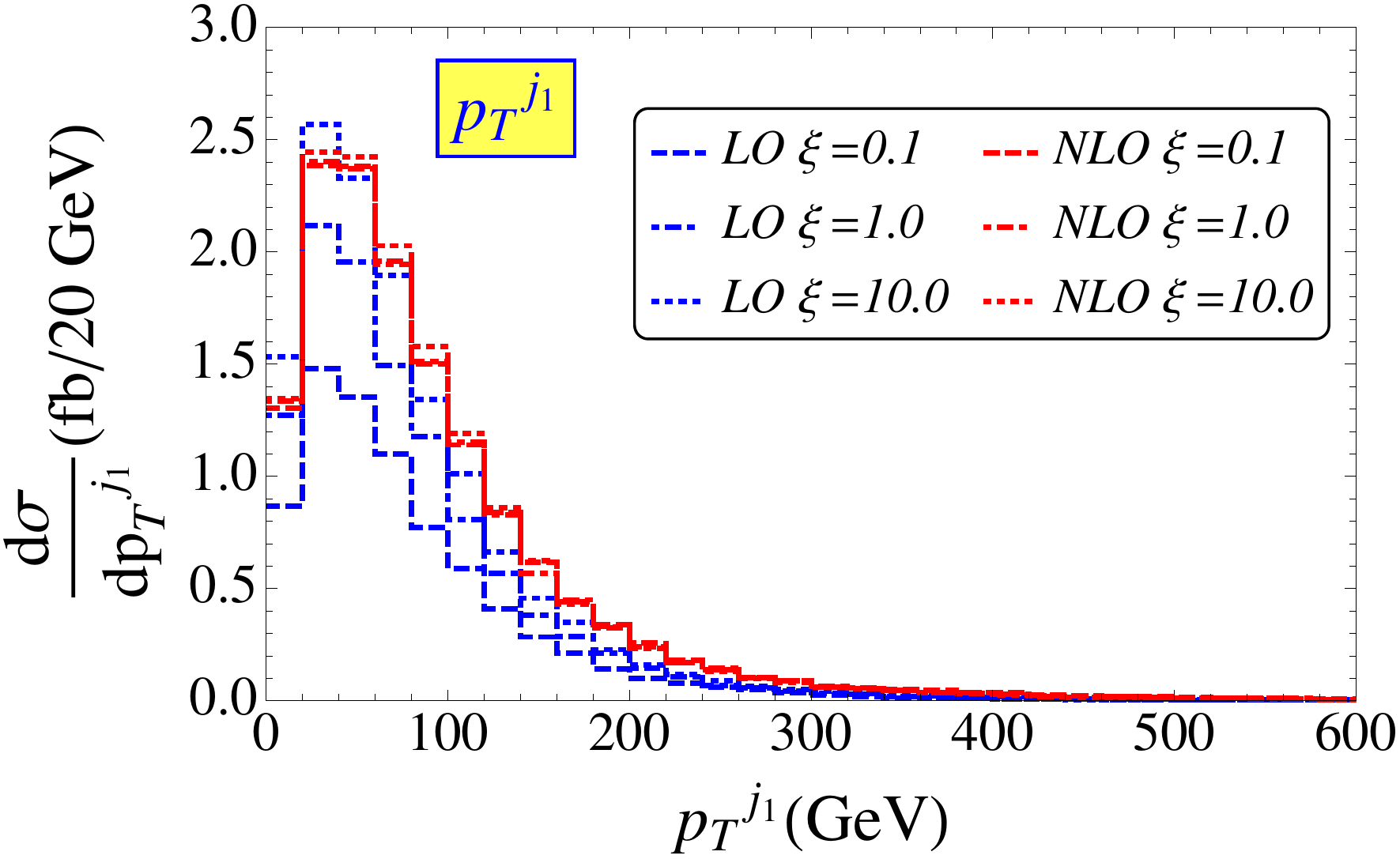}
\includegraphics[scale=0.30]{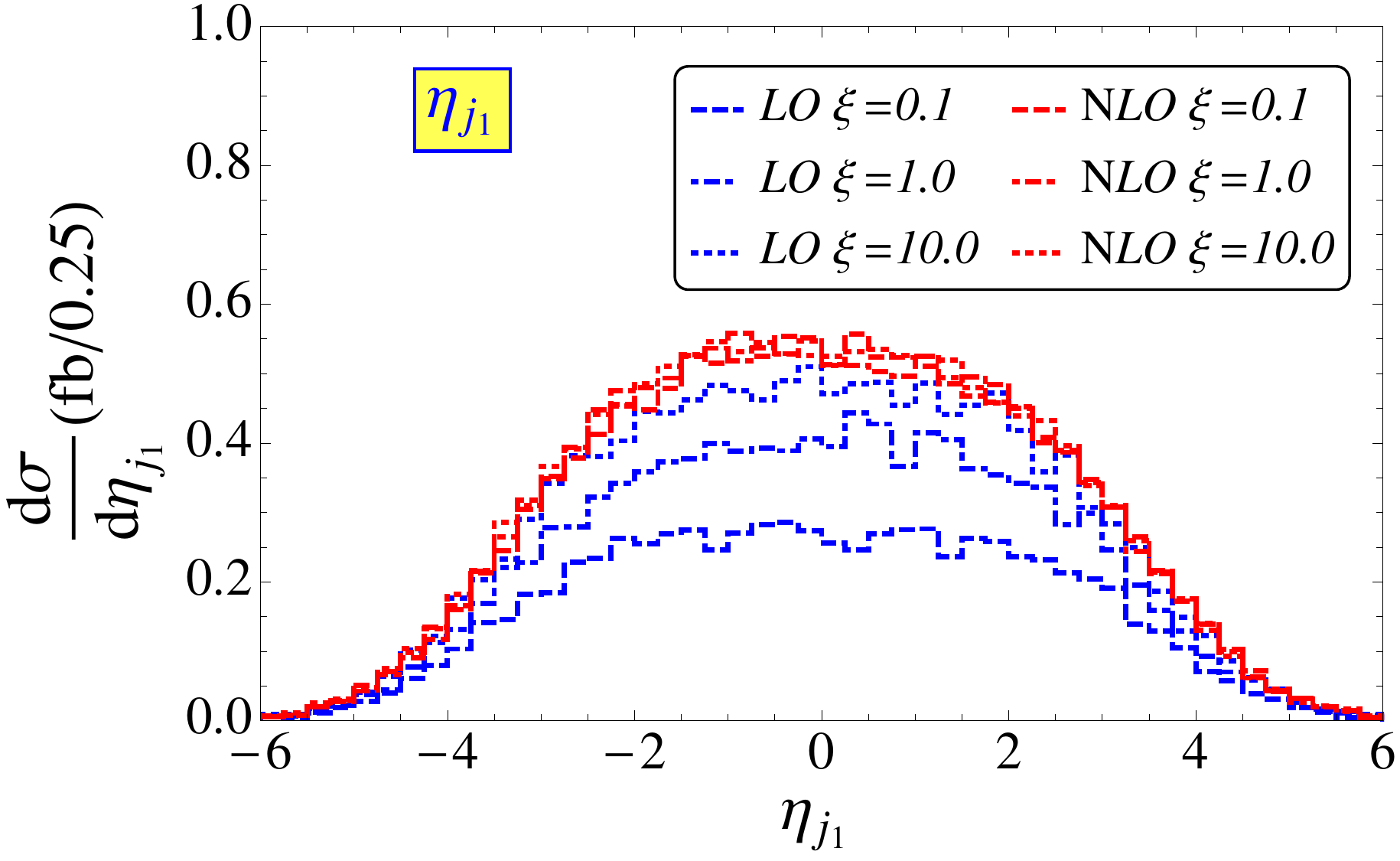}\\
\includegraphics[scale=0.30]{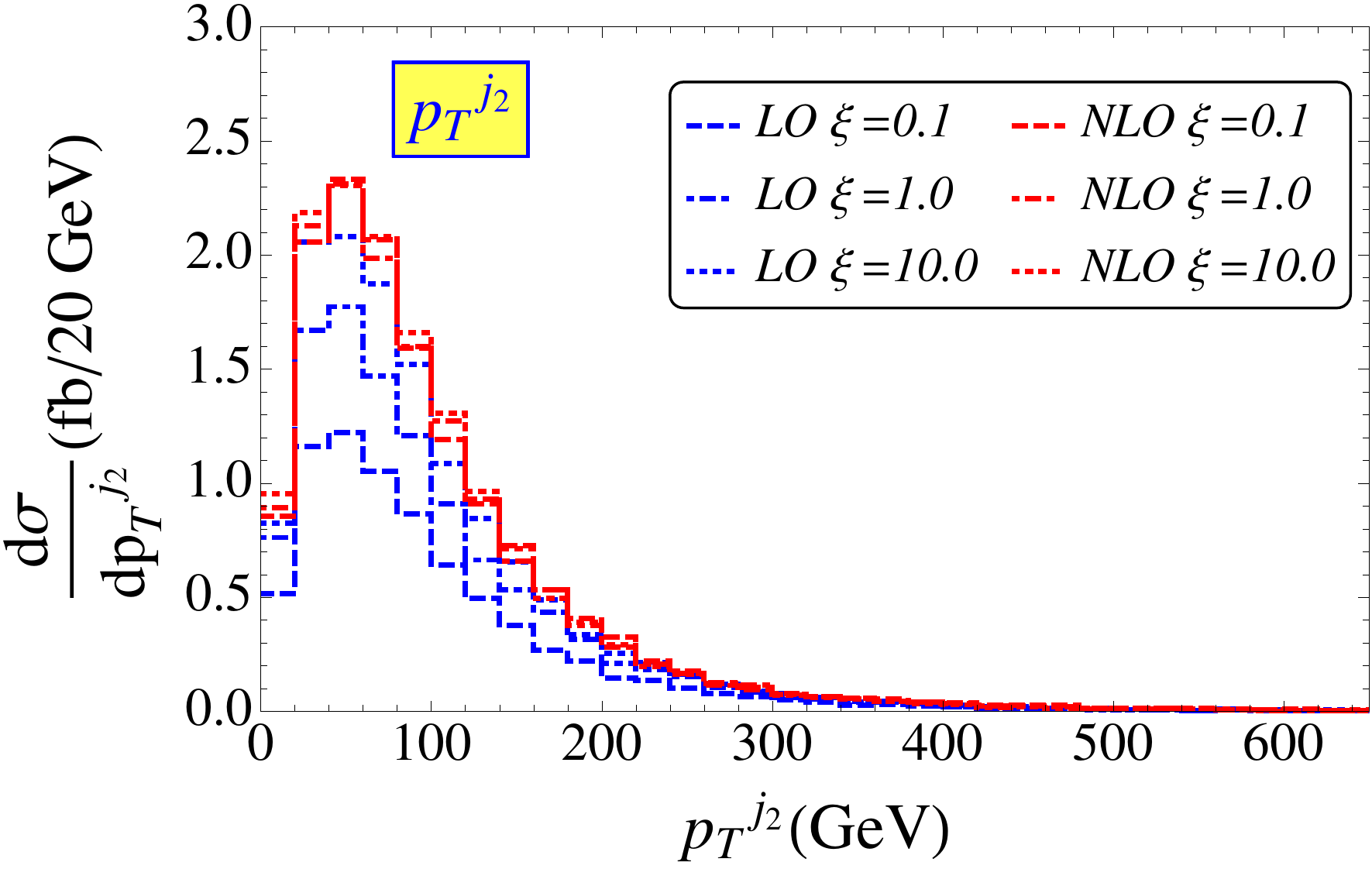}
\includegraphics[scale=0.30]{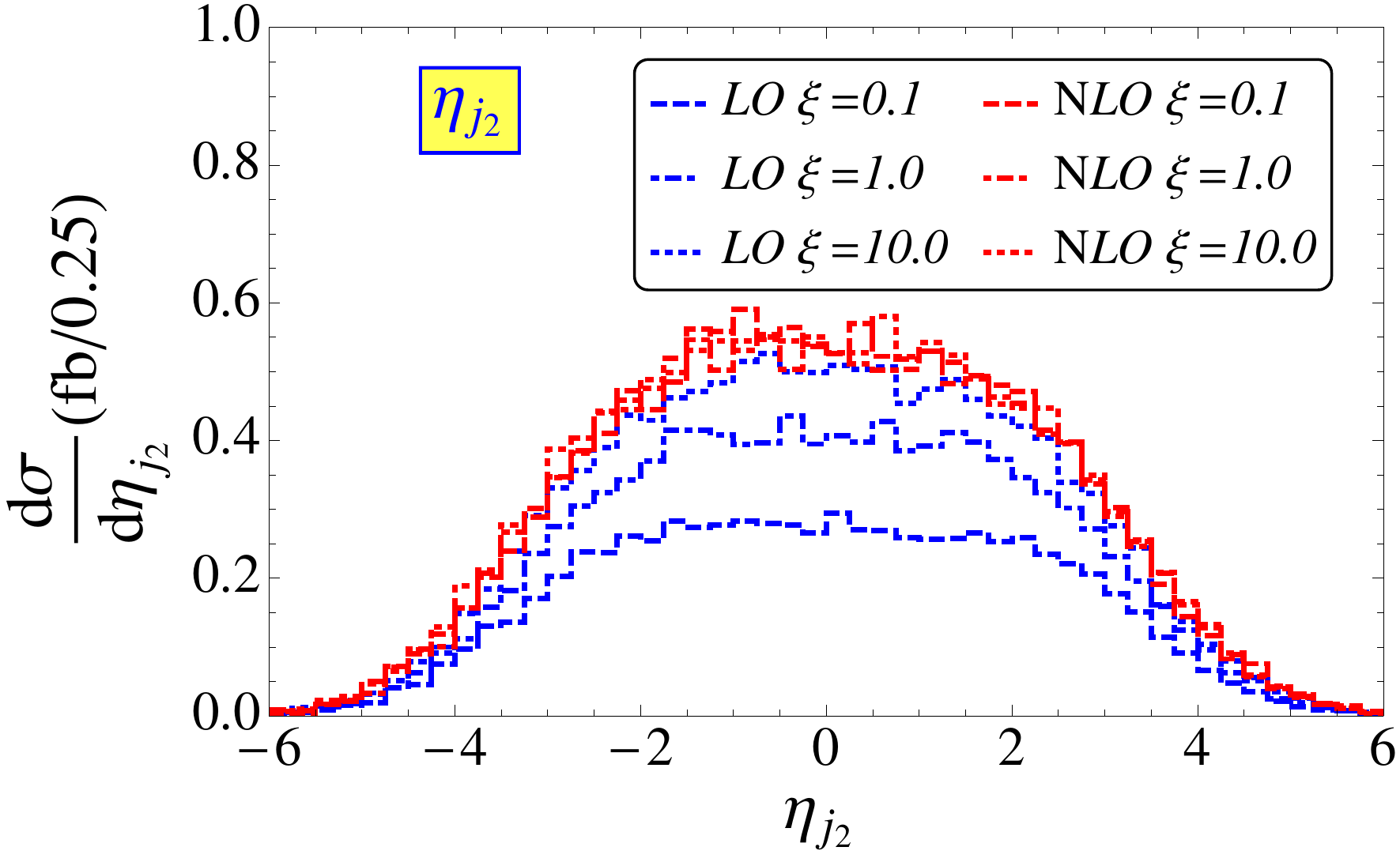}
\end{center}
\caption{Scale dependent LO and NLO-QCD $p_T^{j}$(left column) and $\eta^{j}$ (right column) distributions of the heavy neutrino pair production followed by the decays of the heavy neutrinos into 
$3\ell+\rm{MET}+2j$ channel at the 100 TeV hadron collider for $m_N=300$ GeV.}
\label{HC 300-3l2}
\end{figure} 
\begin{figure}
\begin{center}
\includegraphics[scale=0.265]{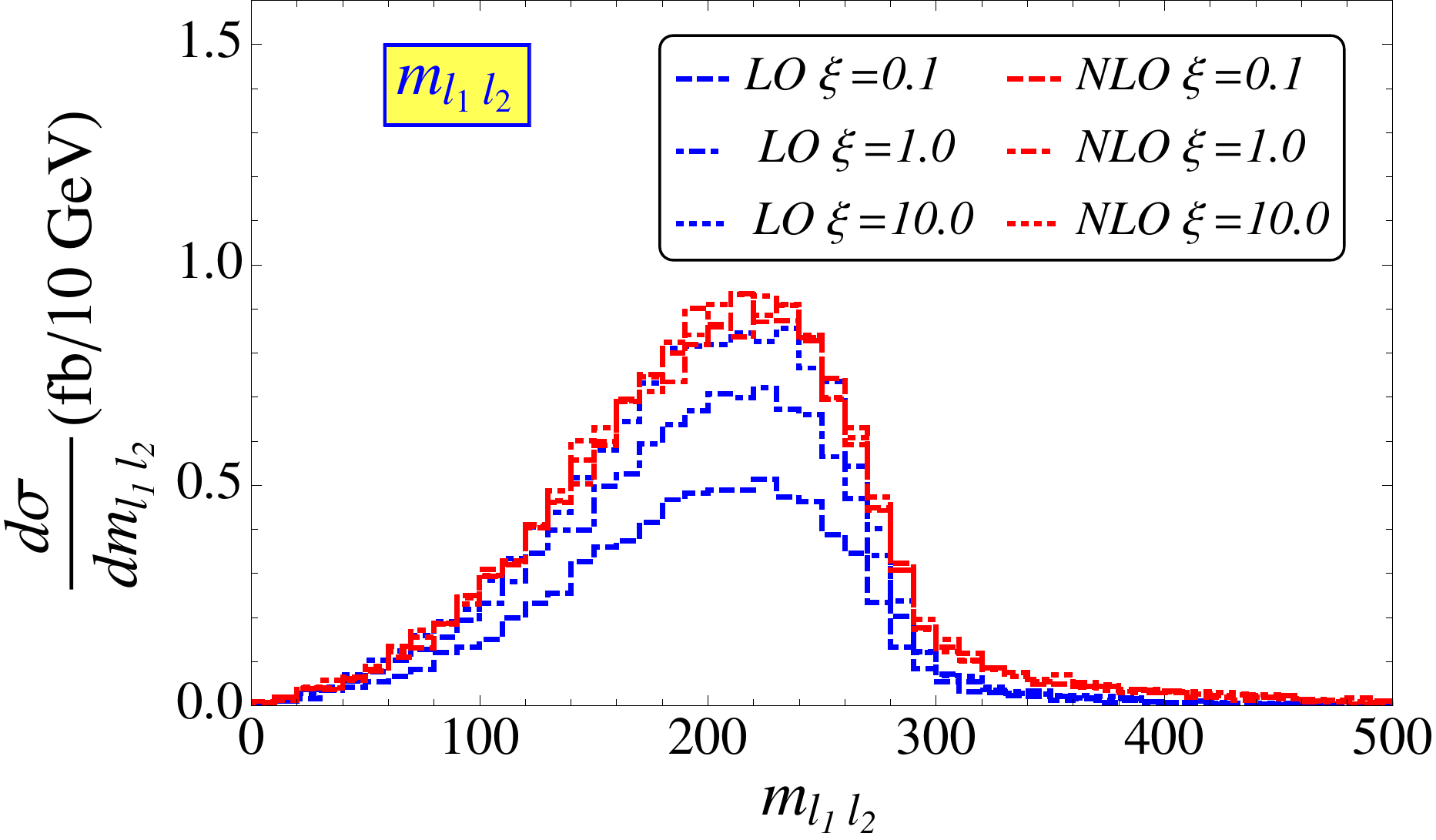}
\includegraphics[scale=0.265]{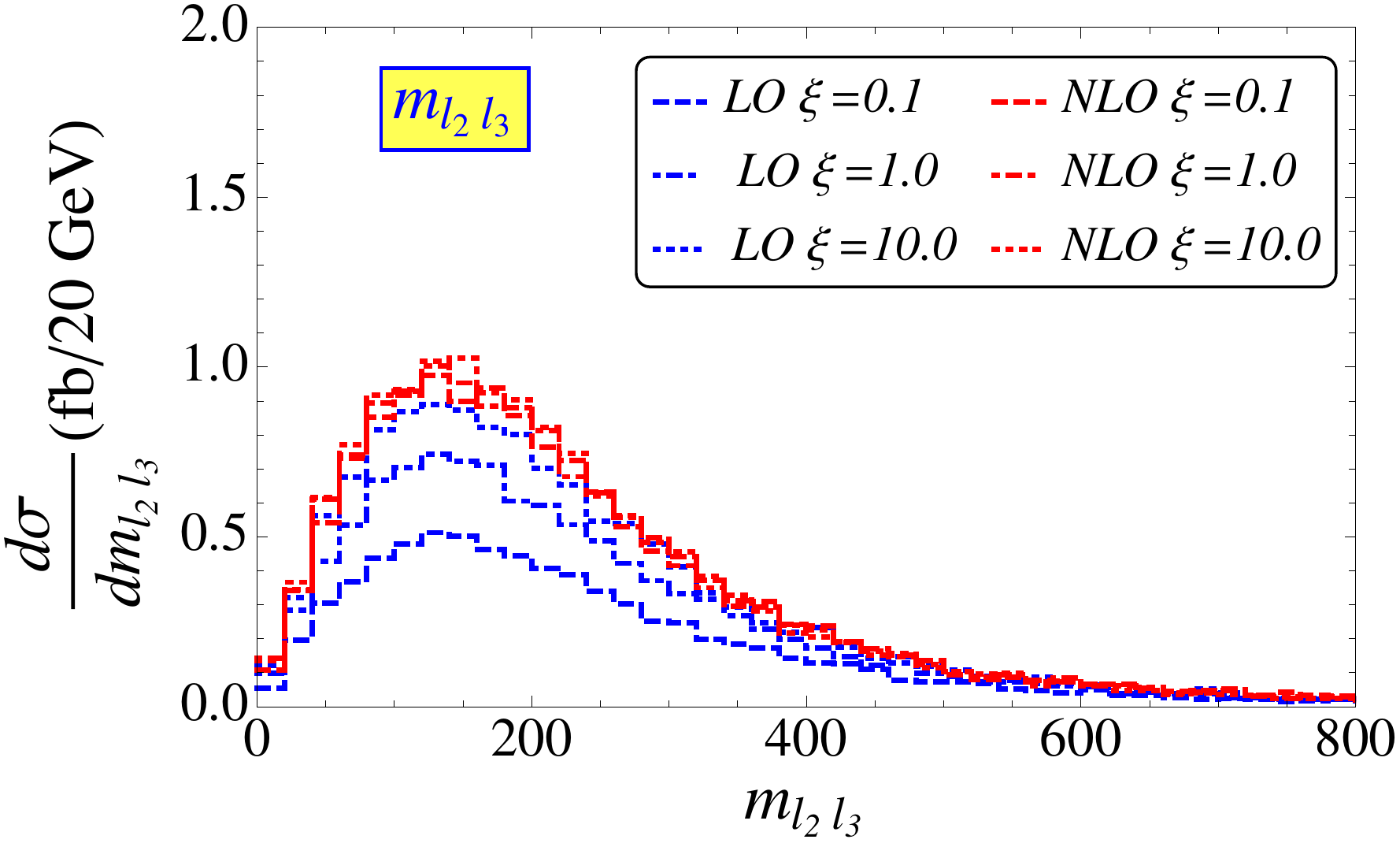}
\includegraphics[scale=0.265]{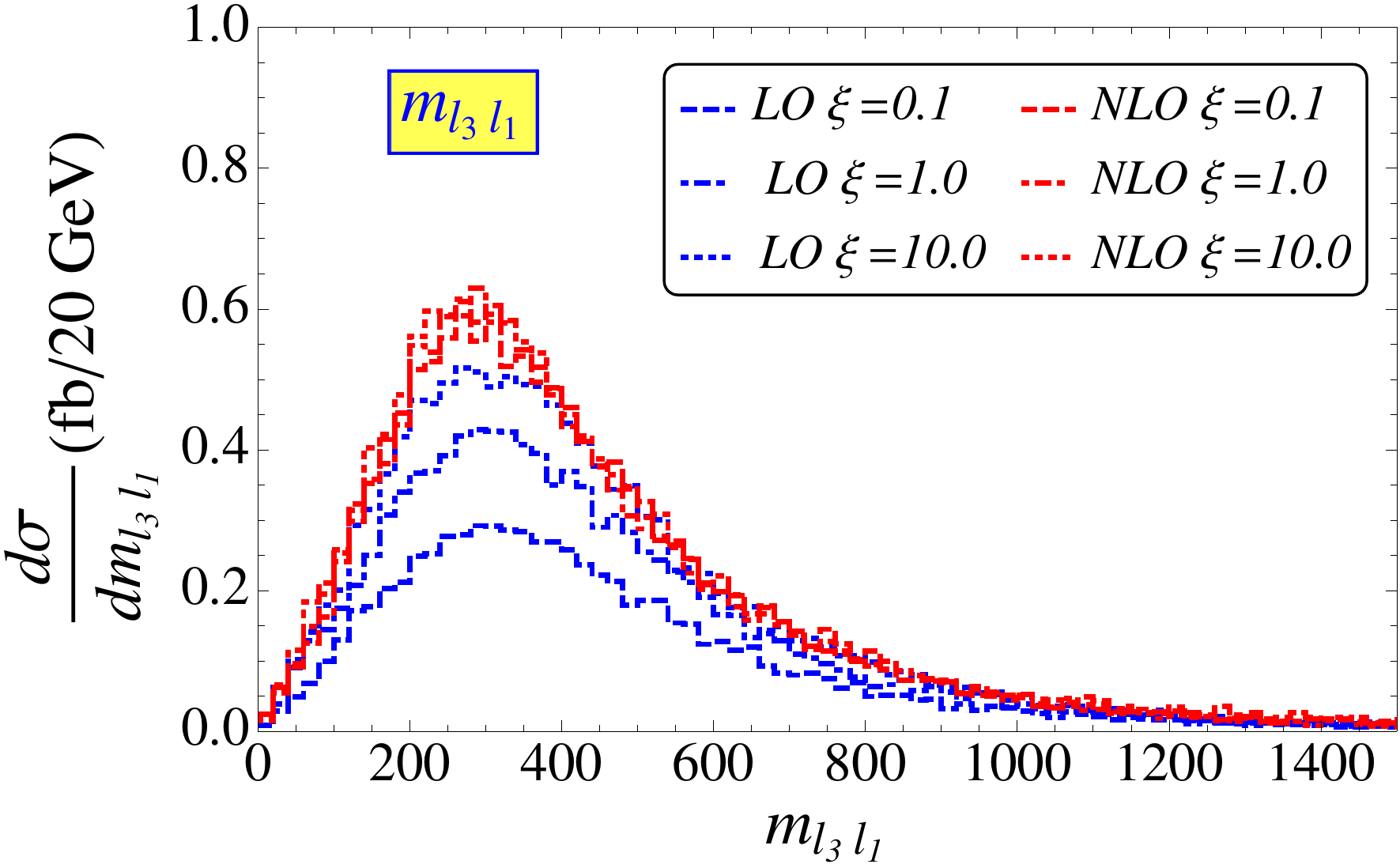}
\end{center}
\caption{Scale dependent LO and NLO-QCD $m_{\ell\ell}$ distributions of the heavy neutrino pair production followed by the decays of the heavy neutrinos into 
$3\ell+\rm{MET}+2j$ channel at the 100 TeV hadron collider for $m_N=300$ GeV.}
\label{HC 300-3l3}
\end{figure}
\begin{figure}
\begin{center}
\includegraphics[scale=0.4]{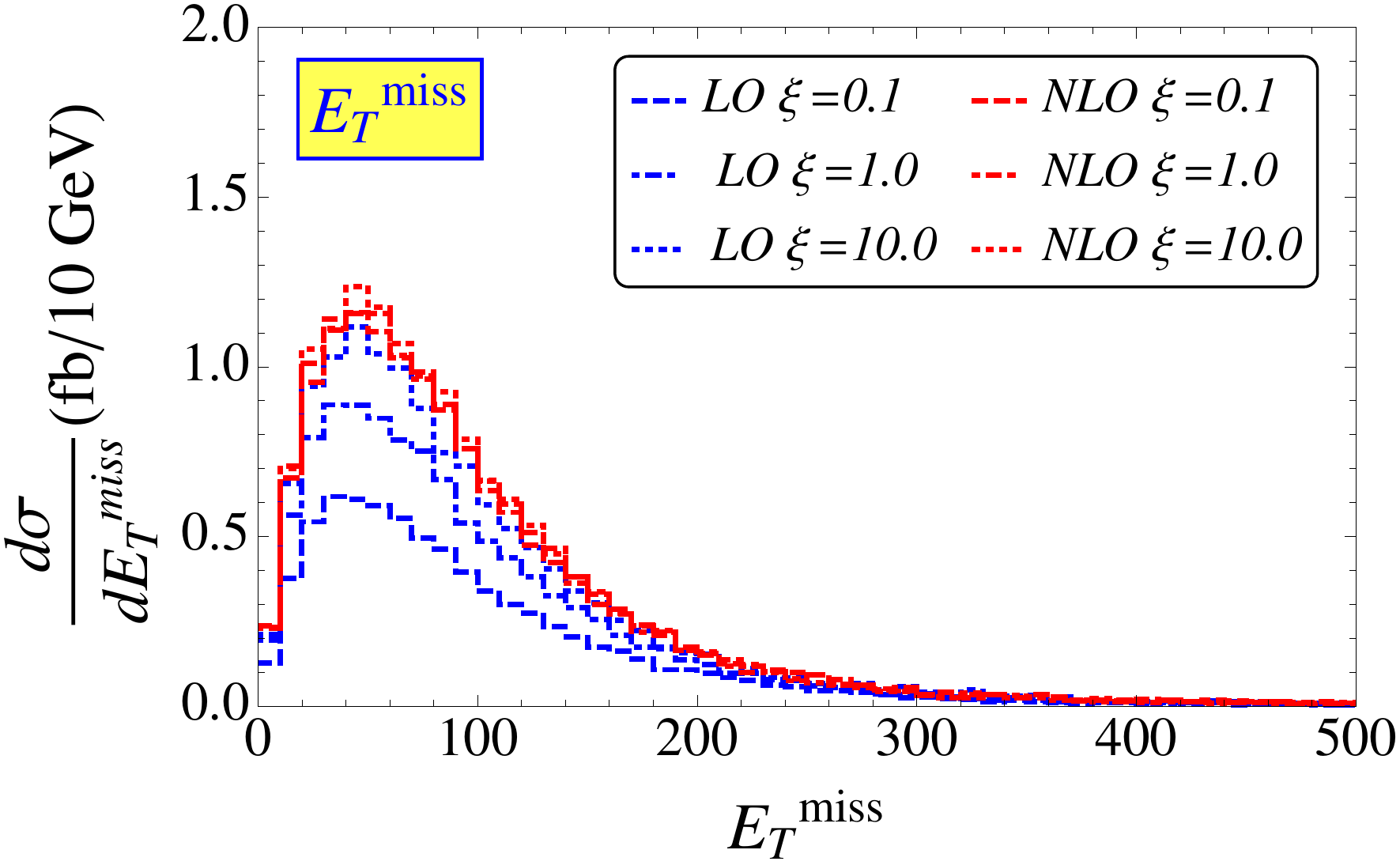}
\end{center}
\caption{Scale dependent LO and NLO-QCD $E_T^{\rm{miss}}$ distributions of the heavy neutrino pair production followed by the decays of the heavy neutrinos into 
$3\ell+\rm{MET}+2j$ channel at the 100 TeV hadron collider for $m_N=300$ GeV.}
\label{HC 300-3l4}
\end{figure} 

A transverse momentum cut $p_{T}^{j} > 30$ GeV and pseudo-rapidity cut $|\eta_{j}| < 2.5$ could be good to consider the jets coming from the $W$ bosons decay in the hadronic mode. The corresponding distributions are plotted in Fig~\ref{HC 300-3l2}. The invariant mass distributions are given in Fig.~\ref{HC 300-3l3} and like the 13 TeV LHC we can construct the OSSF pairs and we can use an invariant mass cut of $m_{\ell \ell} > (m_{Z}+ 15)$ GeV to screen the signal events from the SM backgrounds. 
It will be good to consider the $E_T^{\rm{miss}} > 50$ GeV as plotted in Fig.~\rm{HC 300-3l4}.

\subsection{$4\ell+\rm{MET}$ final state}
\begin{figure}
\begin{center}
\includegraphics[scale=0.30]{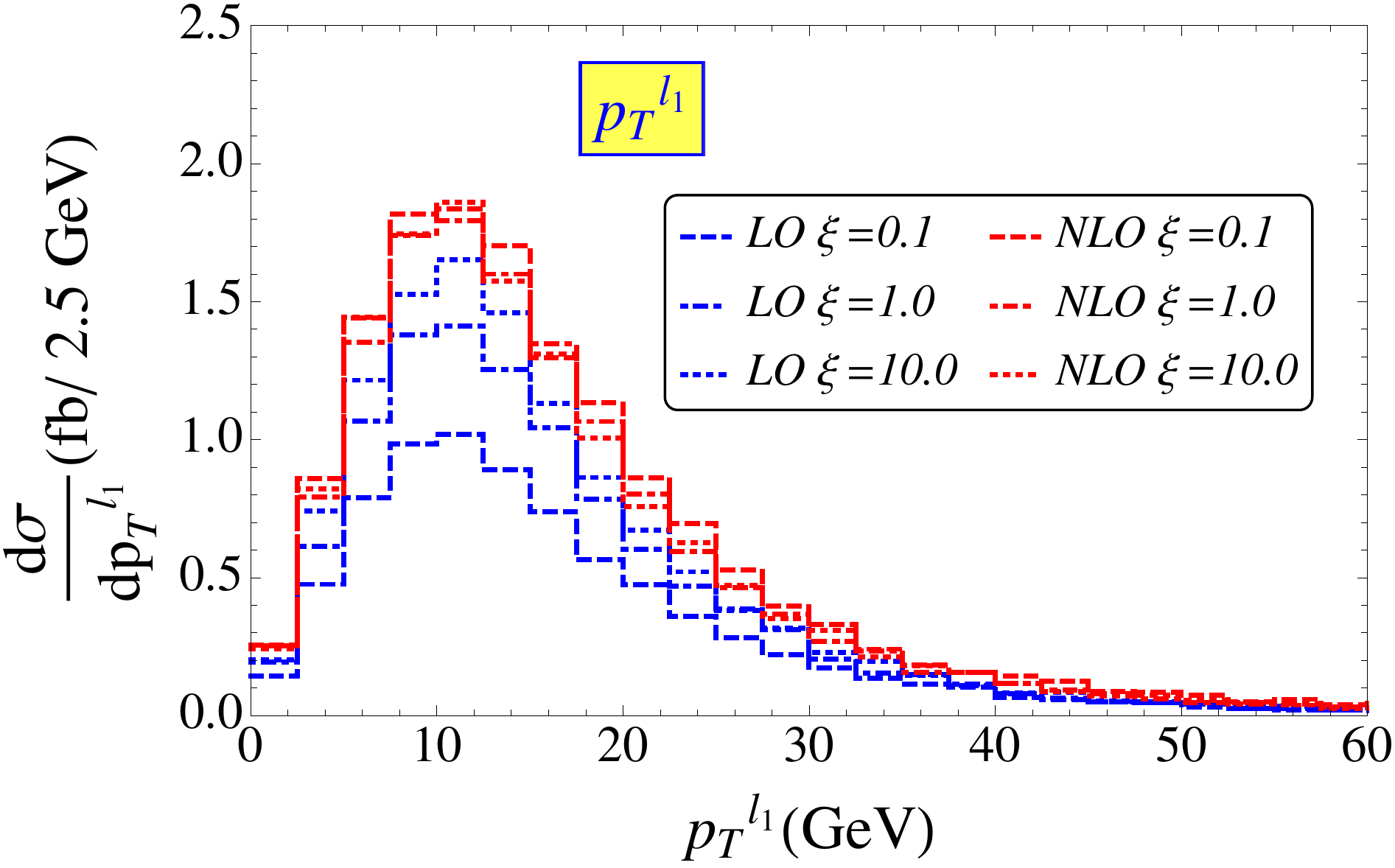}
\includegraphics[scale=0.30]{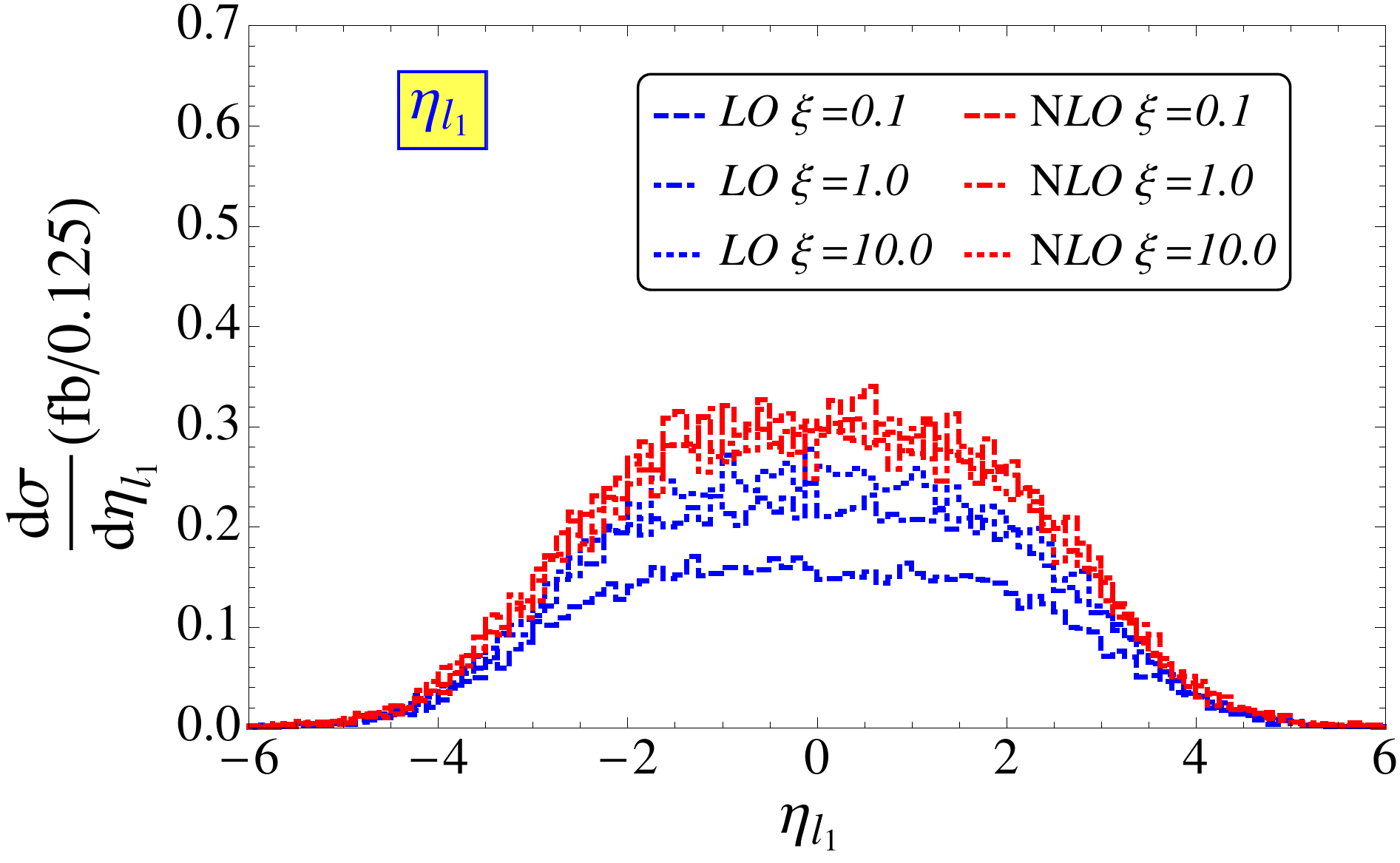}\\
\includegraphics[scale=0.30]{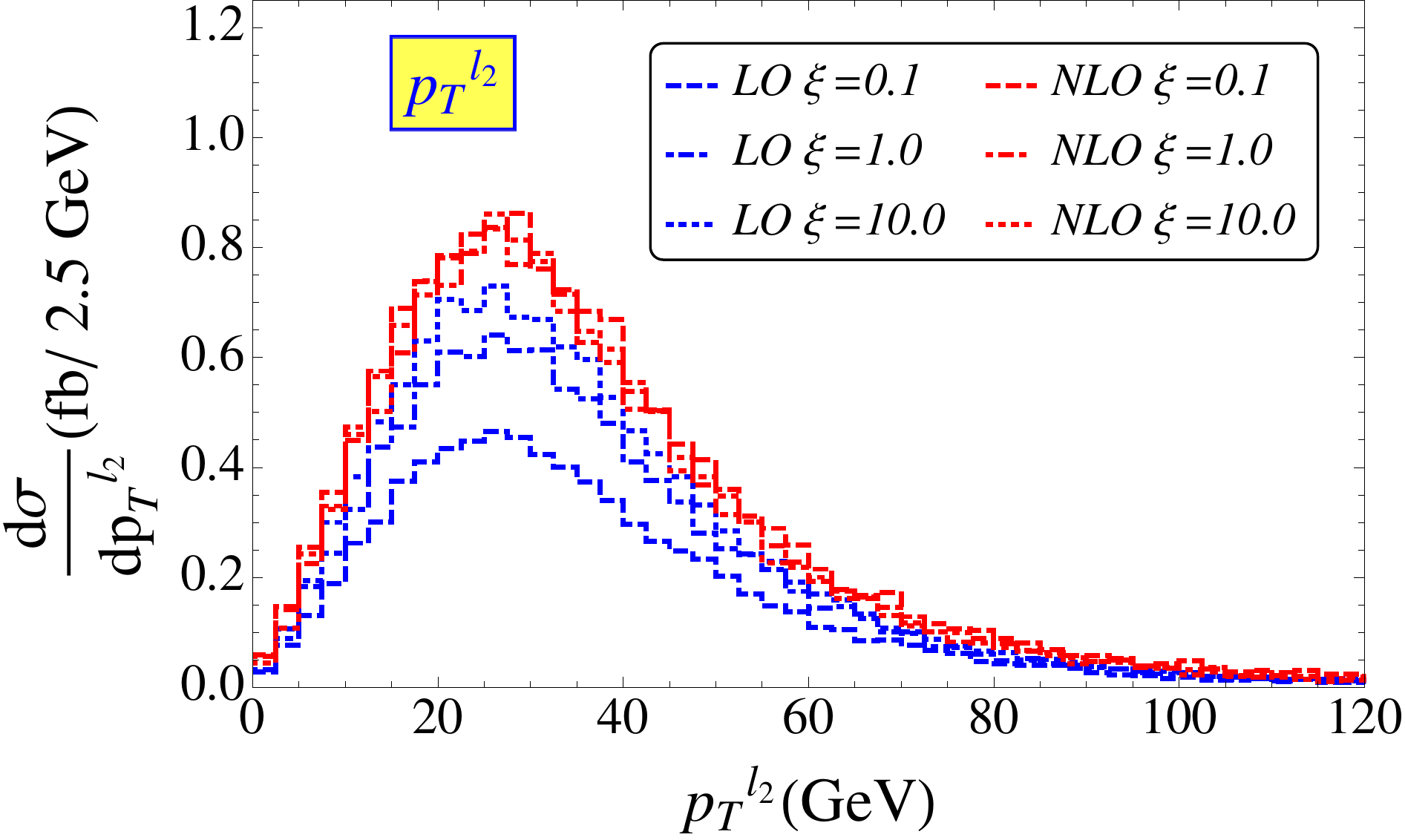}
\includegraphics[scale=0.30]{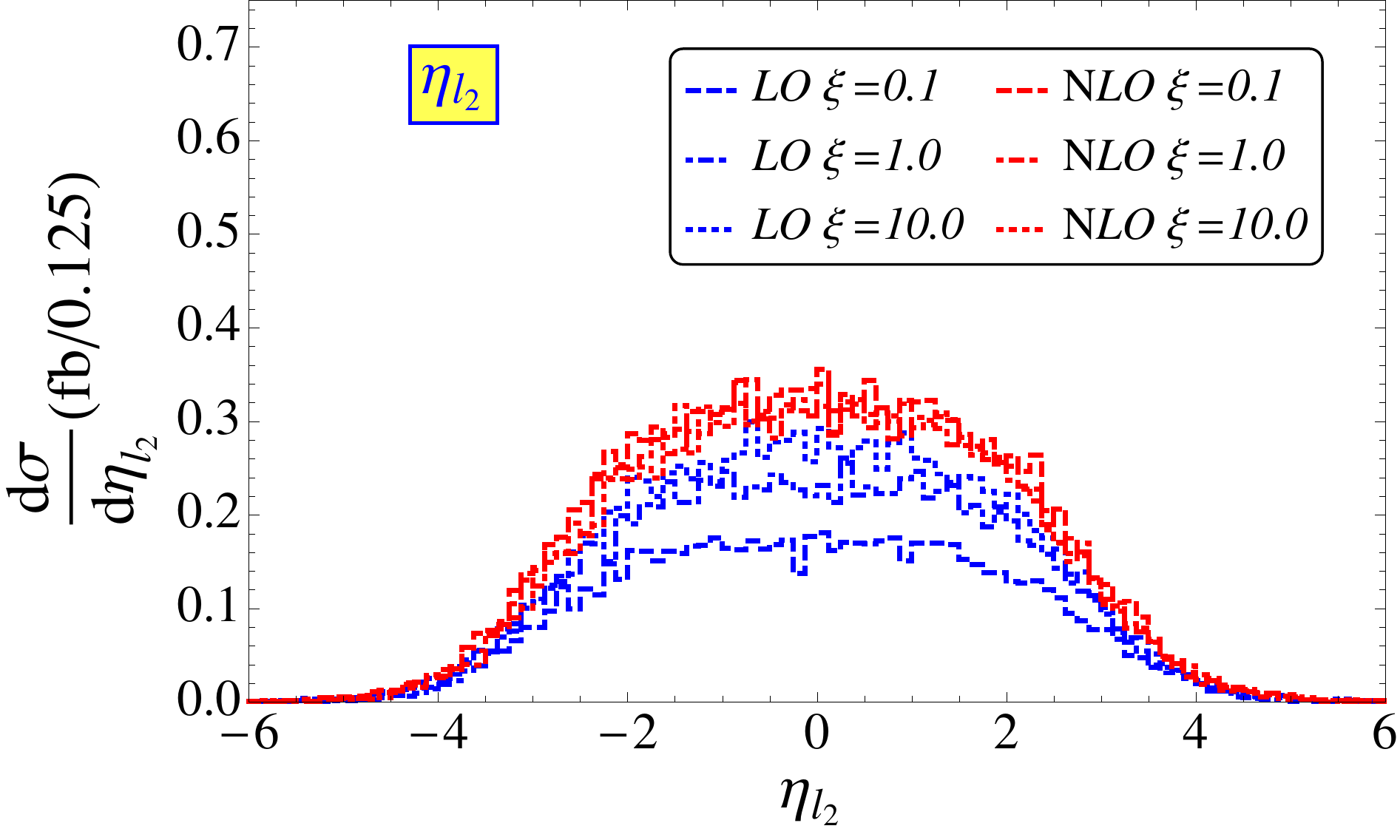}\\
\includegraphics[scale=0.30]{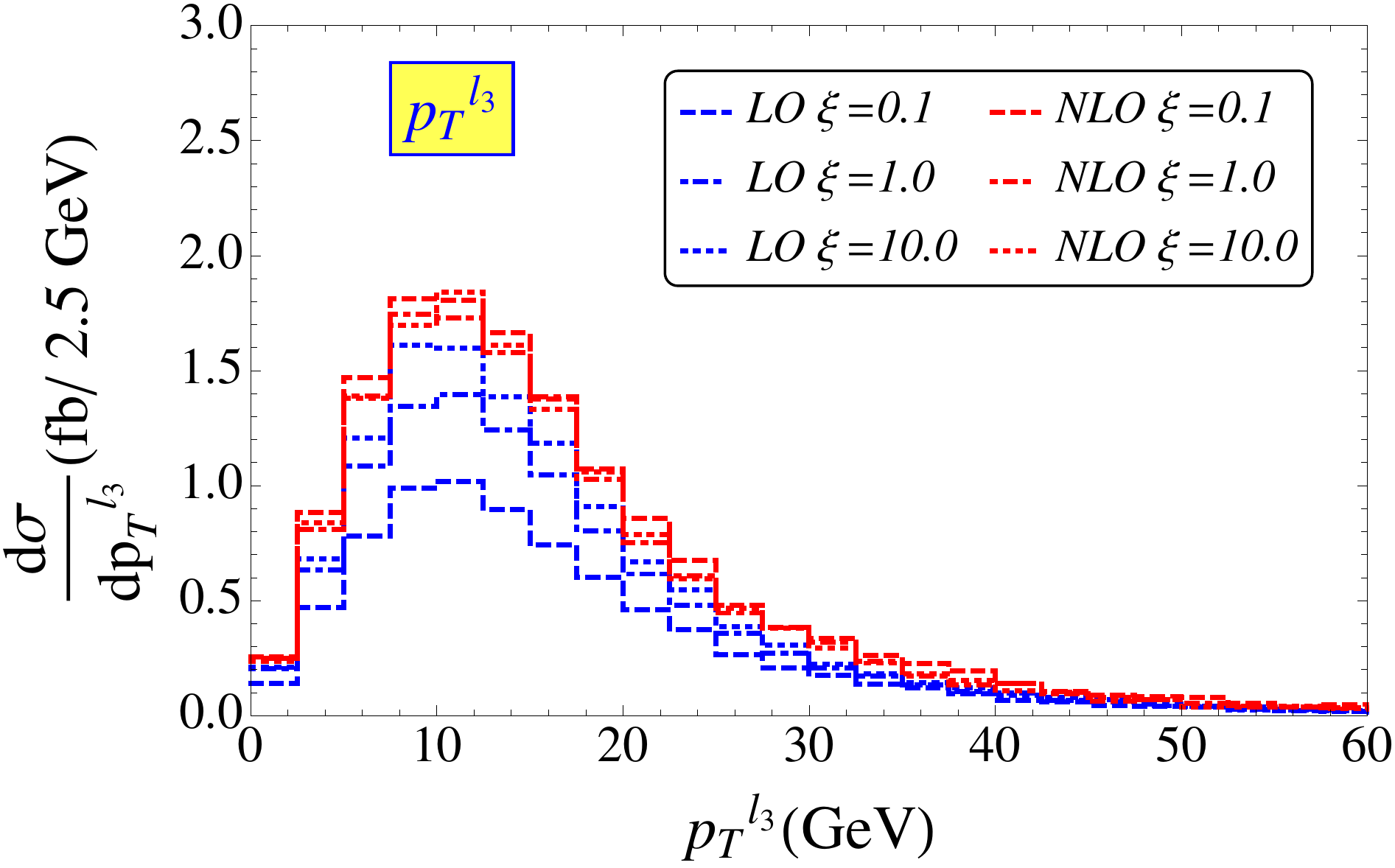}
\includegraphics[scale=0.30]{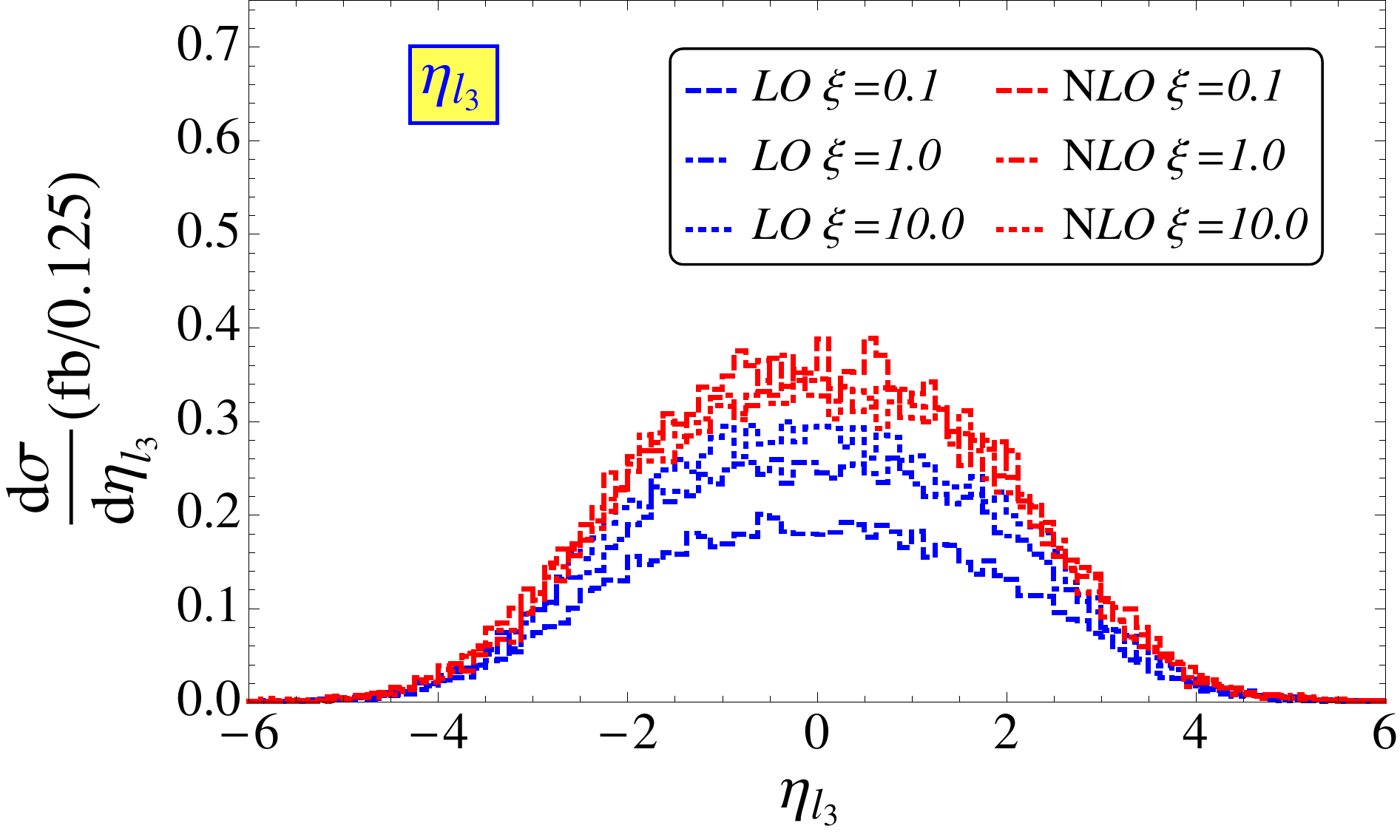}\\
\includegraphics[scale=0.30]{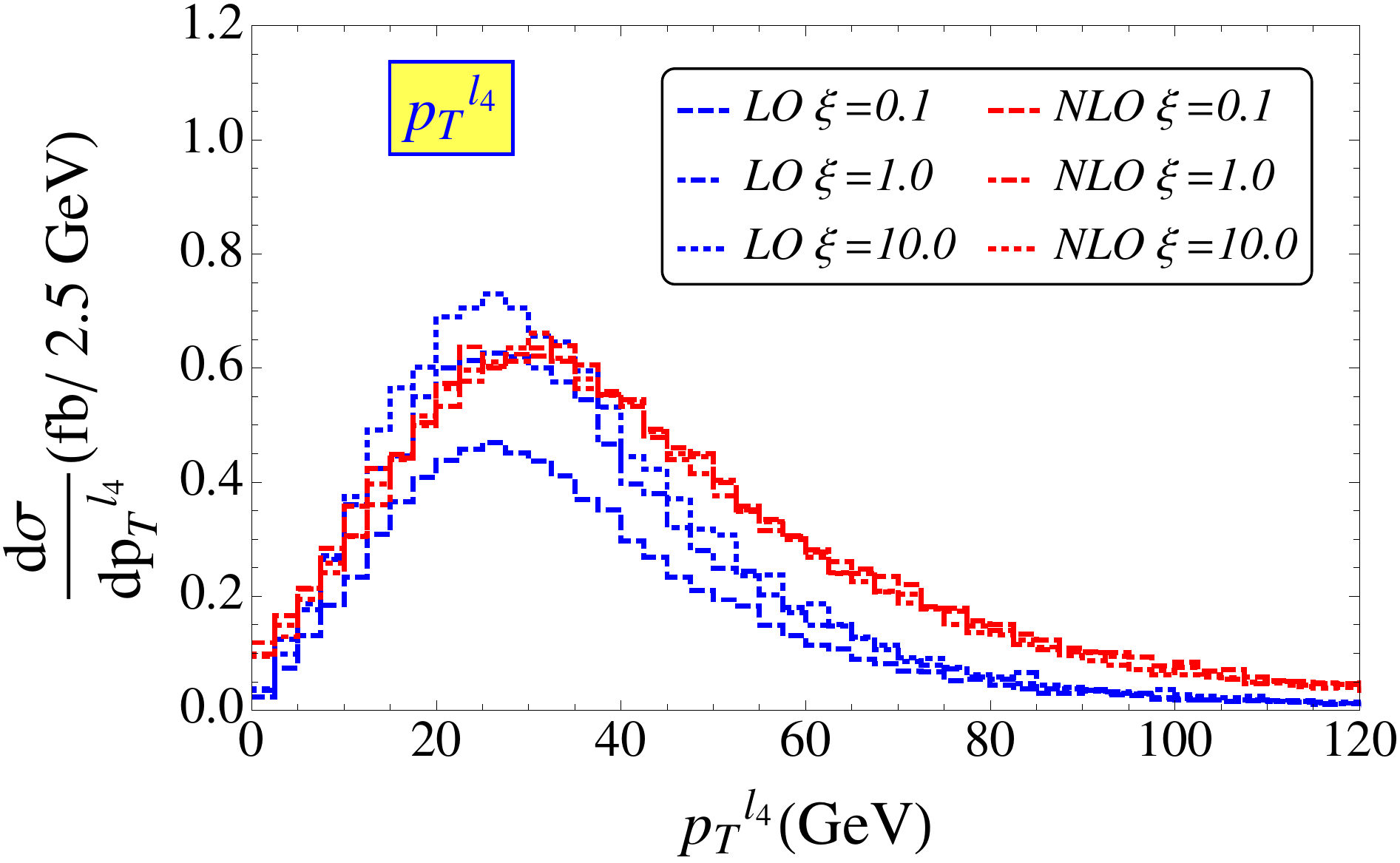}
\includegraphics[scale=0.30]{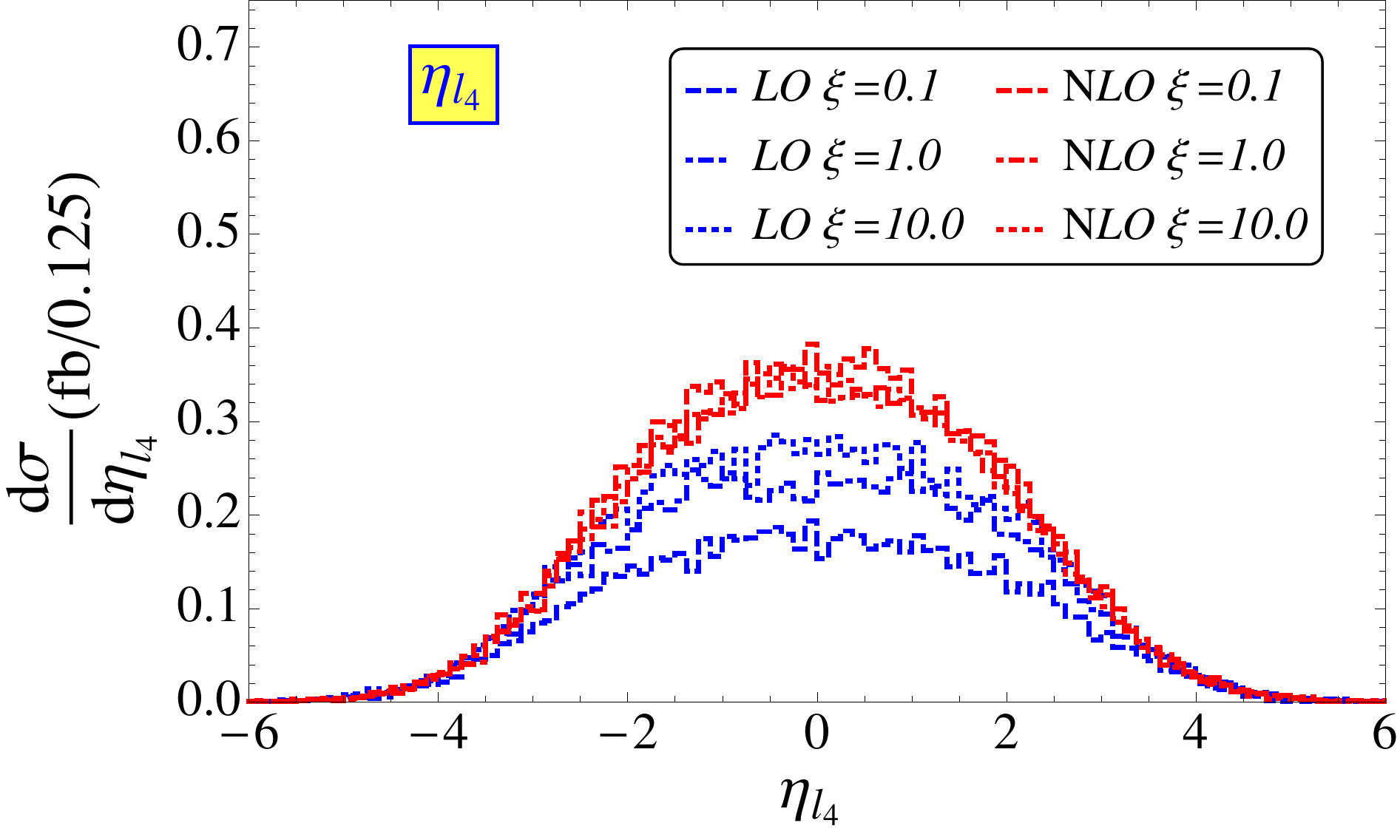}
\end{center}
\caption{Scale dependent LO and NLO-QCD $p_T^{\ell}$ (left column) and $\eta^{\ell}$ (right column) distributions of the heavy neutrino pair production followed by the decays of the heavy neutrinos into 
$4\ell+\rm{MET}$ channel at the 13 TeV LHC for $m_N=95$ GeV.}
\label{HC 95-4l-1}
\end{figure}

In this section we discuss the complete leptonic decay modes of the $W$ bosons produced from the right handed heavy neutrino. As a result the final state will contain $4\ell+\rm{MET}$ signal events. We have chosen $m_{N}=95$ GeV at the $13$ TeV LHC. The $p_T^{\ell}$ and $\eta^{\ell}$ are given in Figs.~\ref{HC 95-4l-1} . The $4\ell$ are coming from the right handed heavy neutrino decay and followed by the leptonic decay of the $W$ boson. In this case there will be several combinations between the leptons as 
\bea
p p &\to& N \overline{N}, \nonumber \\
        \, \, \, \,  \, \, \, \,&&   N \to \ell_{1}^{-} W^{+}, W^{+} \to \ell_{2}^{+}  \nu / \ell^{\prime^{+}}_{2} \nu\nonumber \\
        \, \, \, \,  \, \, \, \,&&   \overline{N} \to \ell_{3}^{+} W^{-}, W^{-} \to \ell_{4}^{-} \overline{\nu} /  \ell^{\prime^{-}}_{4} \overline{\nu}.
        \label{decay100}
\eea
The leading leptons are coming from the decay of the $W$ boson. In such case to survive with  maximum number of the events, a transverse momentum cut on the leading leptons could be $p_{T}^{\ell, \rm{leading}} > 25$ GeV and the trailing leptons from the $N$ decay could have a transverse momentum cut $p_{T}^{\ell, \rm{trailing}} > 15$ GeV.

\begin{figure}
\begin{center}
\includegraphics[scale=0.30]{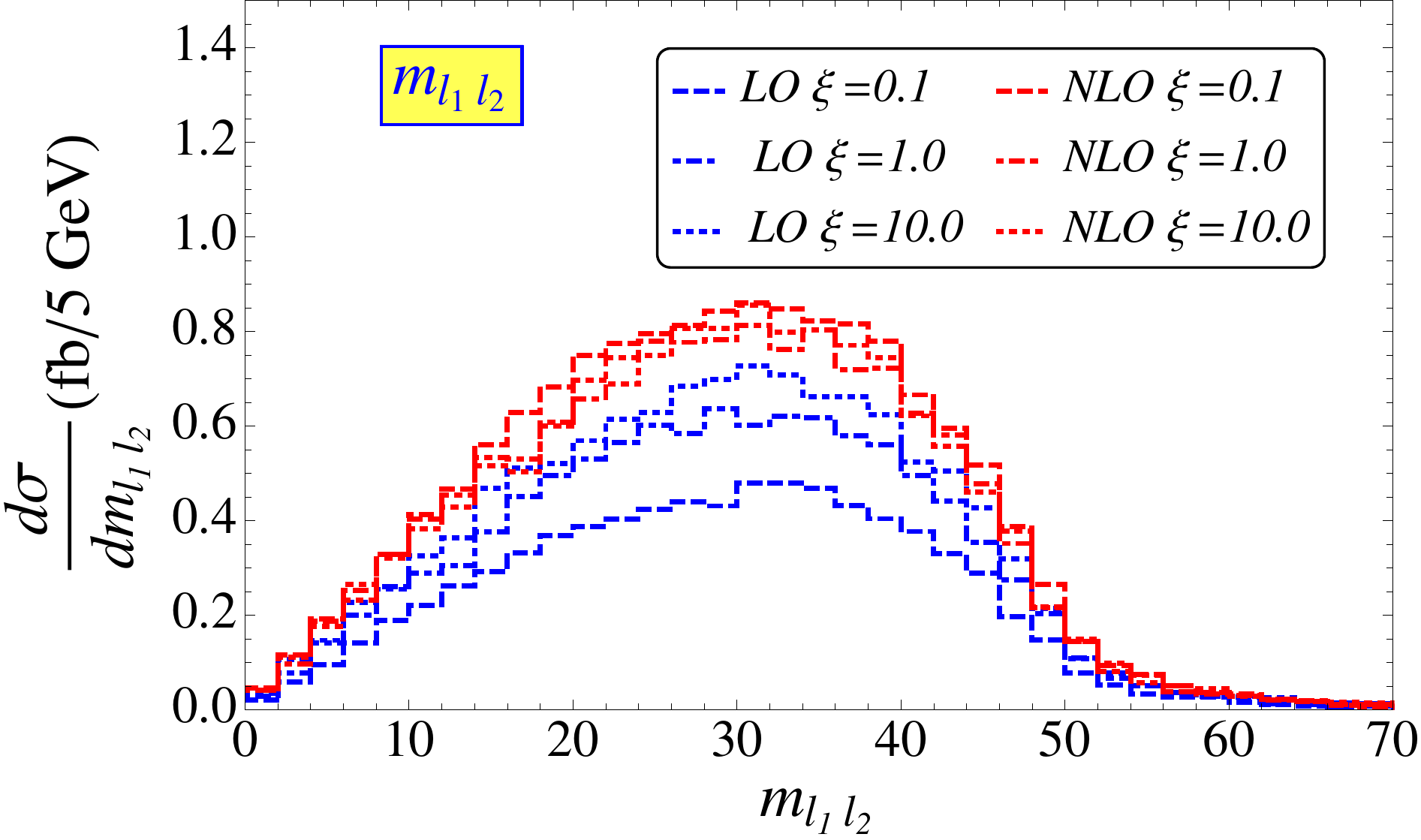}
\includegraphics[scale=0.30]{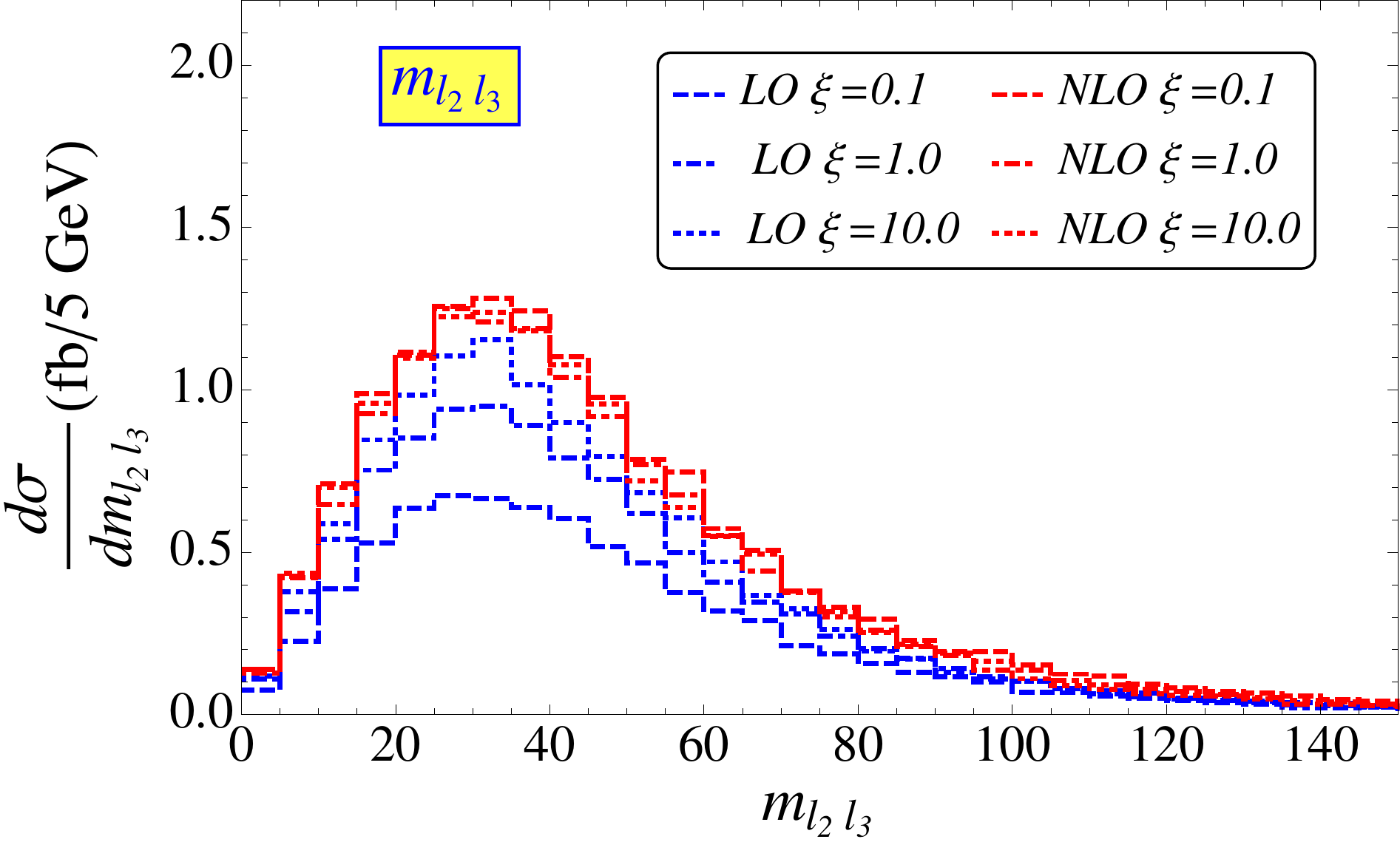}\\
\includegraphics[scale=0.30]{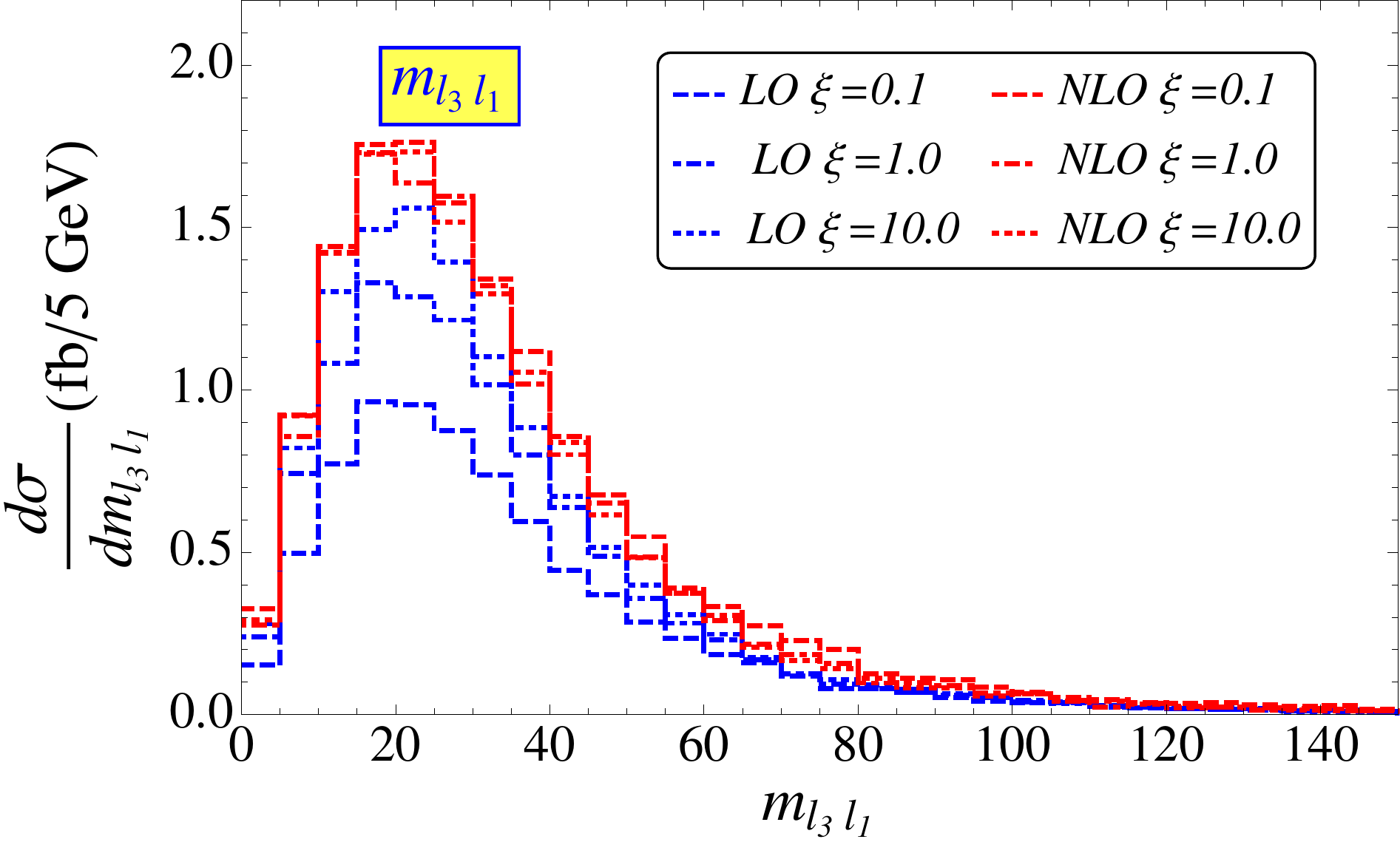}
\includegraphics[scale=0.30]{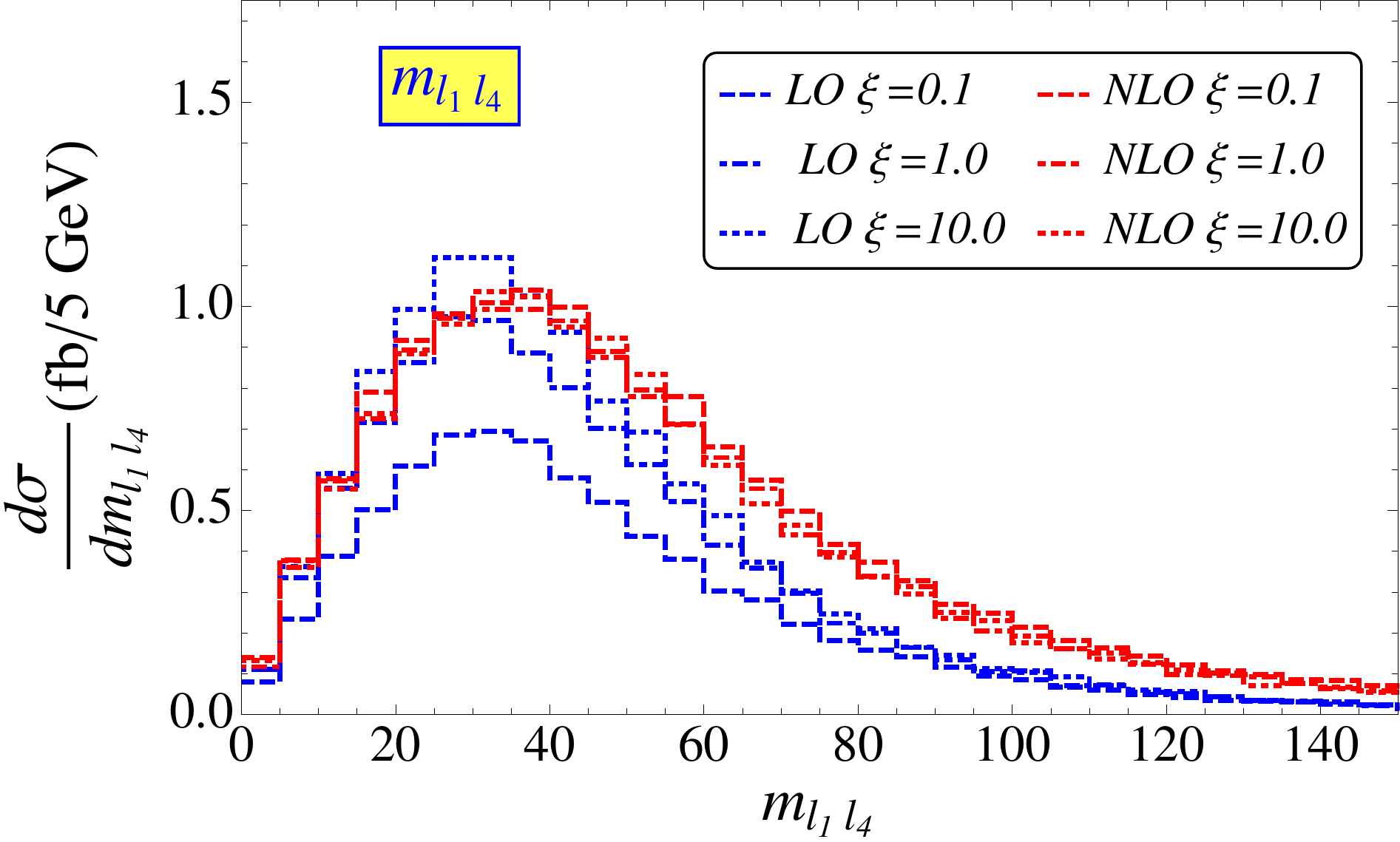}\\
\includegraphics[scale=0.30]{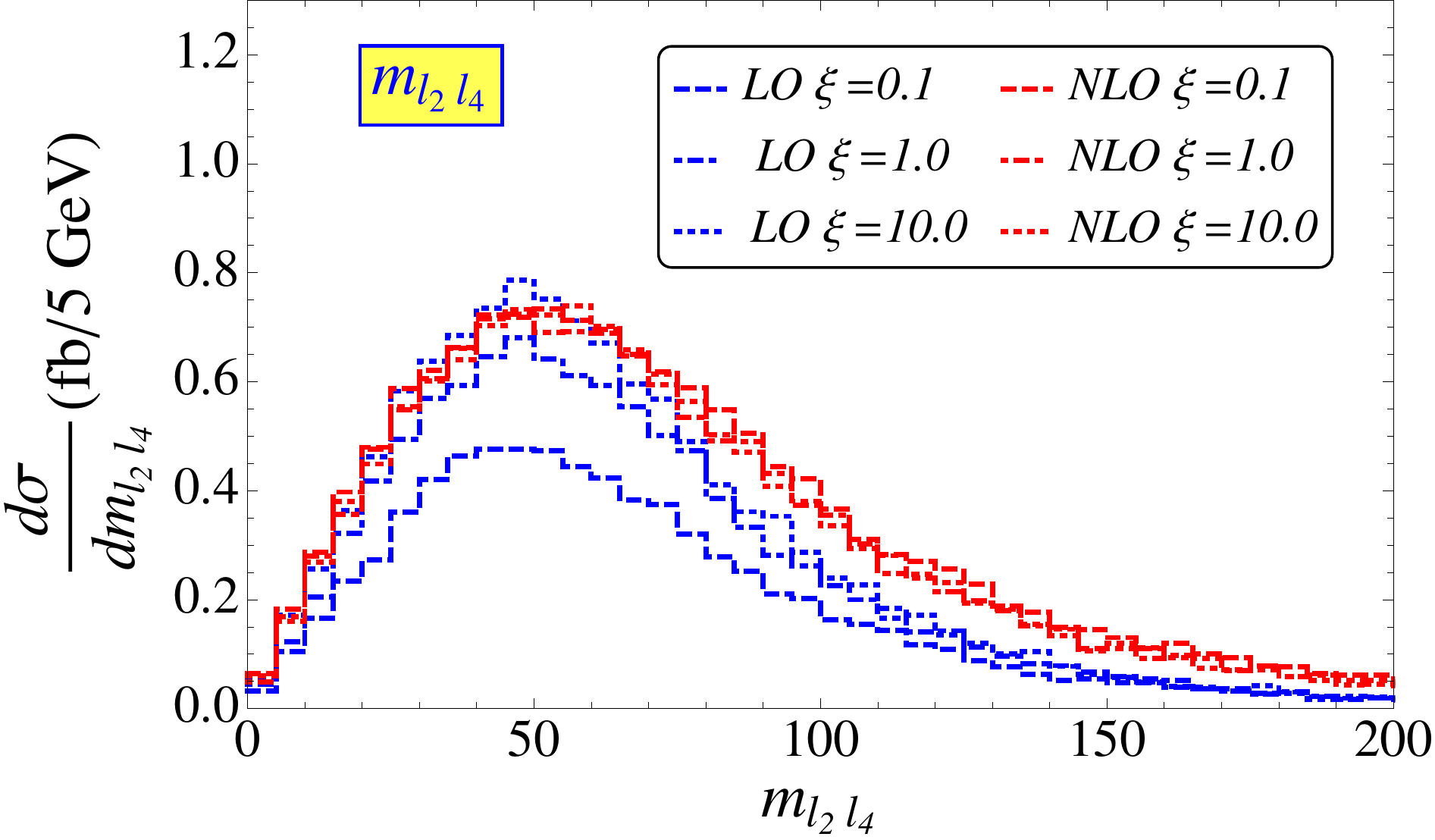}
\includegraphics[scale=0.30]{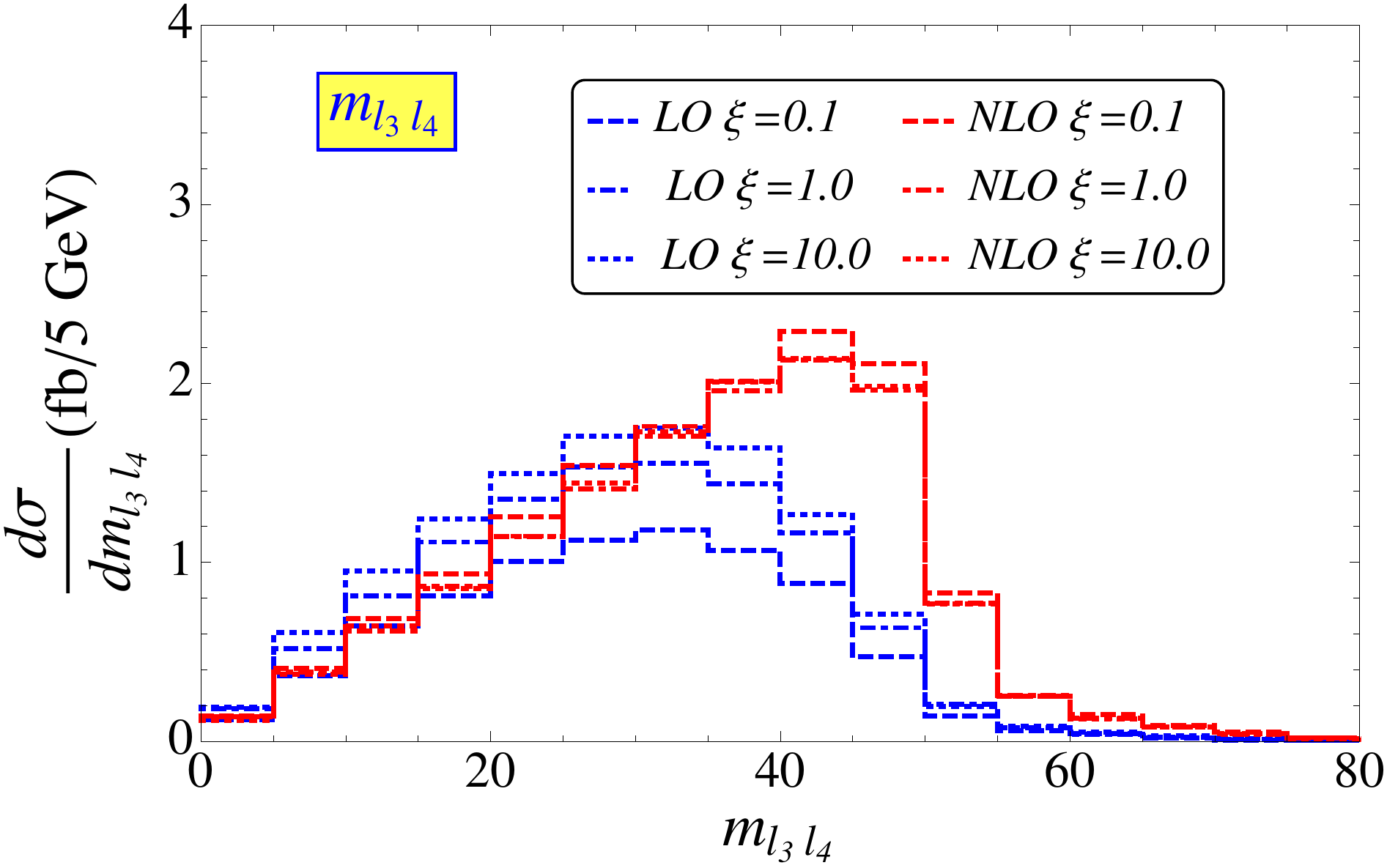}
\end{center}
\caption{Scale dependent LO and NLO-QCD $m_{\ell\ell}$ distributions of the heavy neutrino pair production followed by the decays of the heavy neutrinos into 
$4\ell+\rm{MET}$ channel at the 100 TeV LHC for $m_N=95$ GeV.}
\label{HC 95-4l-2}
\end{figure}

\begin{figure}
\begin{center}
\includegraphics[scale=0.40]{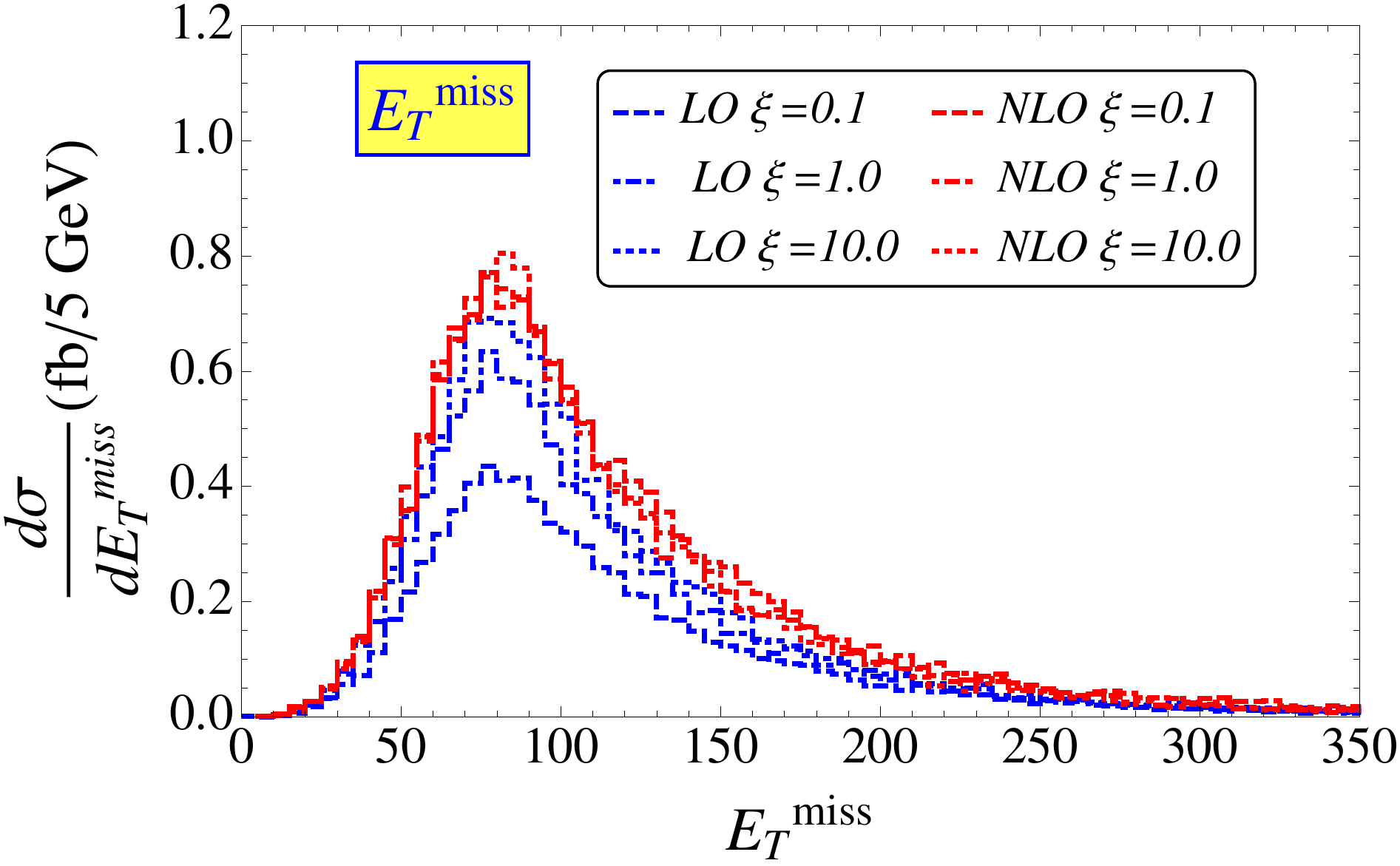}
\end{center}
\caption{Scale dependent LO and NLO-QCD $m_{\ell\ell}$ distribution of the heavy neutrino pair production followed by the decays of the heavy neutrinos into 
$4\ell+\rm{MET}$ channel at the 13 TeV LHC for $m_N=95$ GeV.}
\label{HC 95-4l-4}
\end{figure}

The invariant mass $(m_{\ell\ell})$ distributions of the leptons are given in Fig.~\ref{HC 95-4l-2}. A cut of $m_{\ell\ell} < (m_{Z}-15)$ GeV could be applied to avoid the SM backgrounds coming
from the $Z$ boson decay into OSSF lepton pair. 
\begin{figure}
\begin{center}
\includegraphics[scale=0.30]{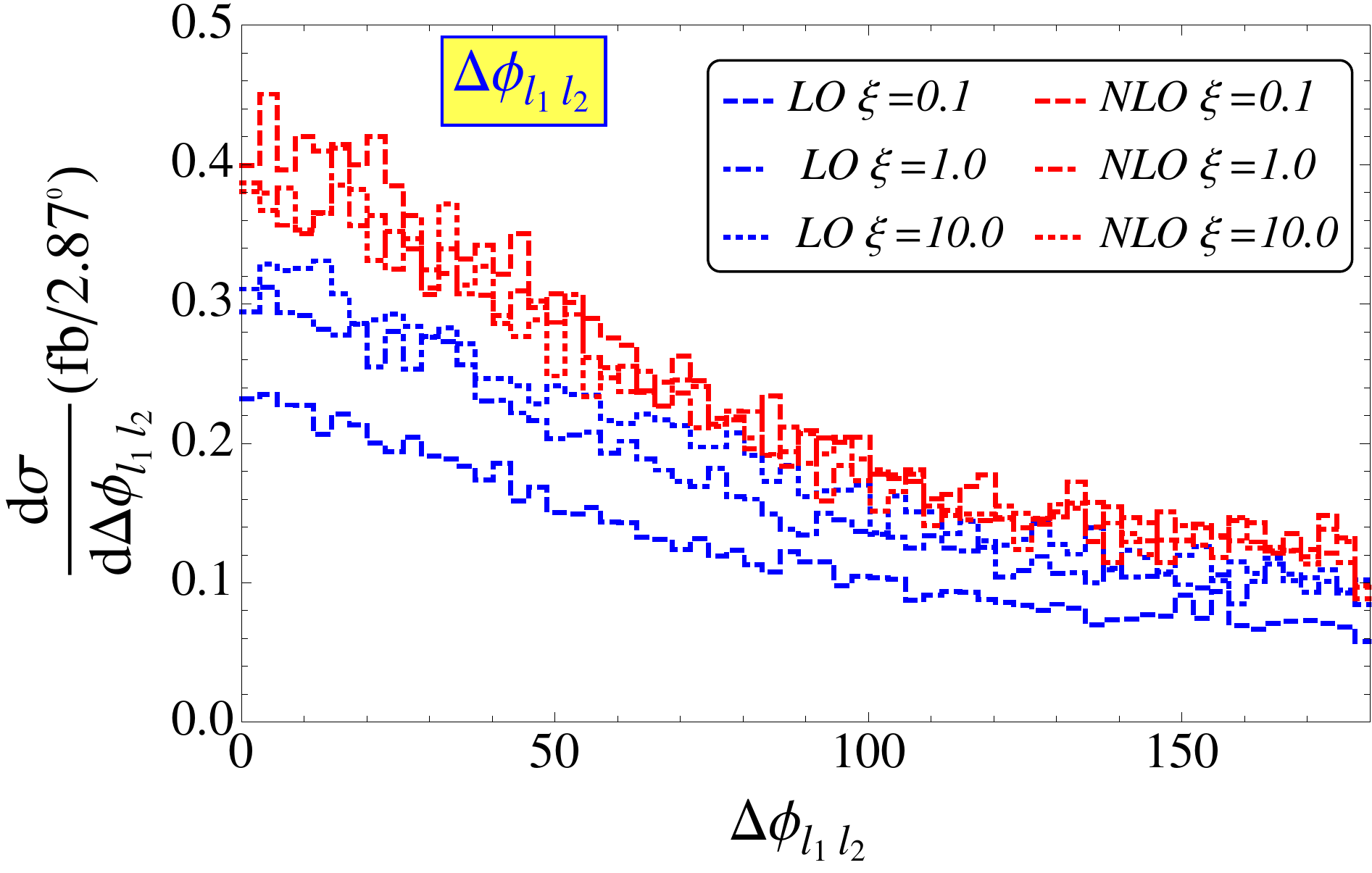}
\includegraphics[scale=0.30]{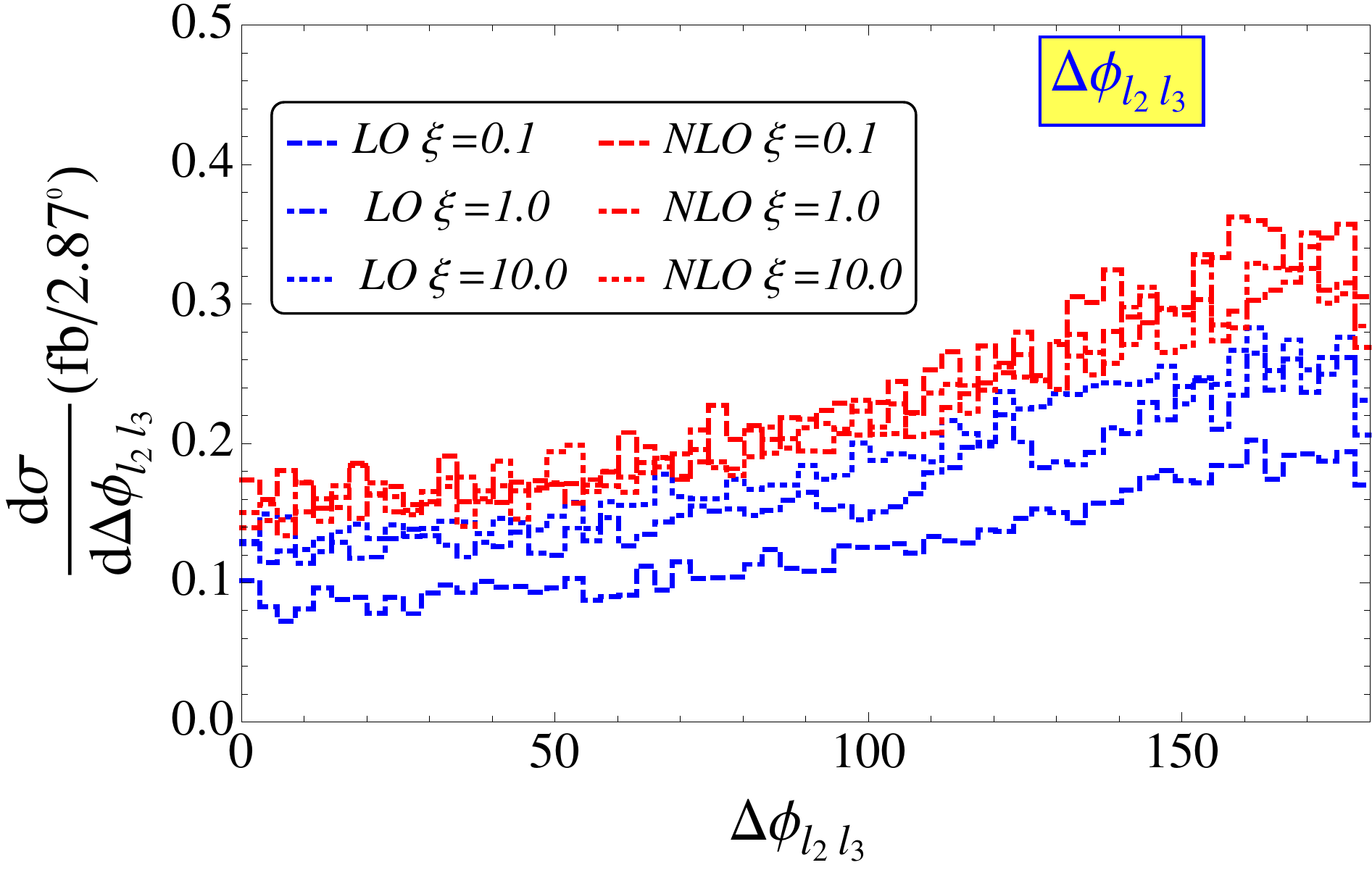}\\
\includegraphics[scale=0.30]{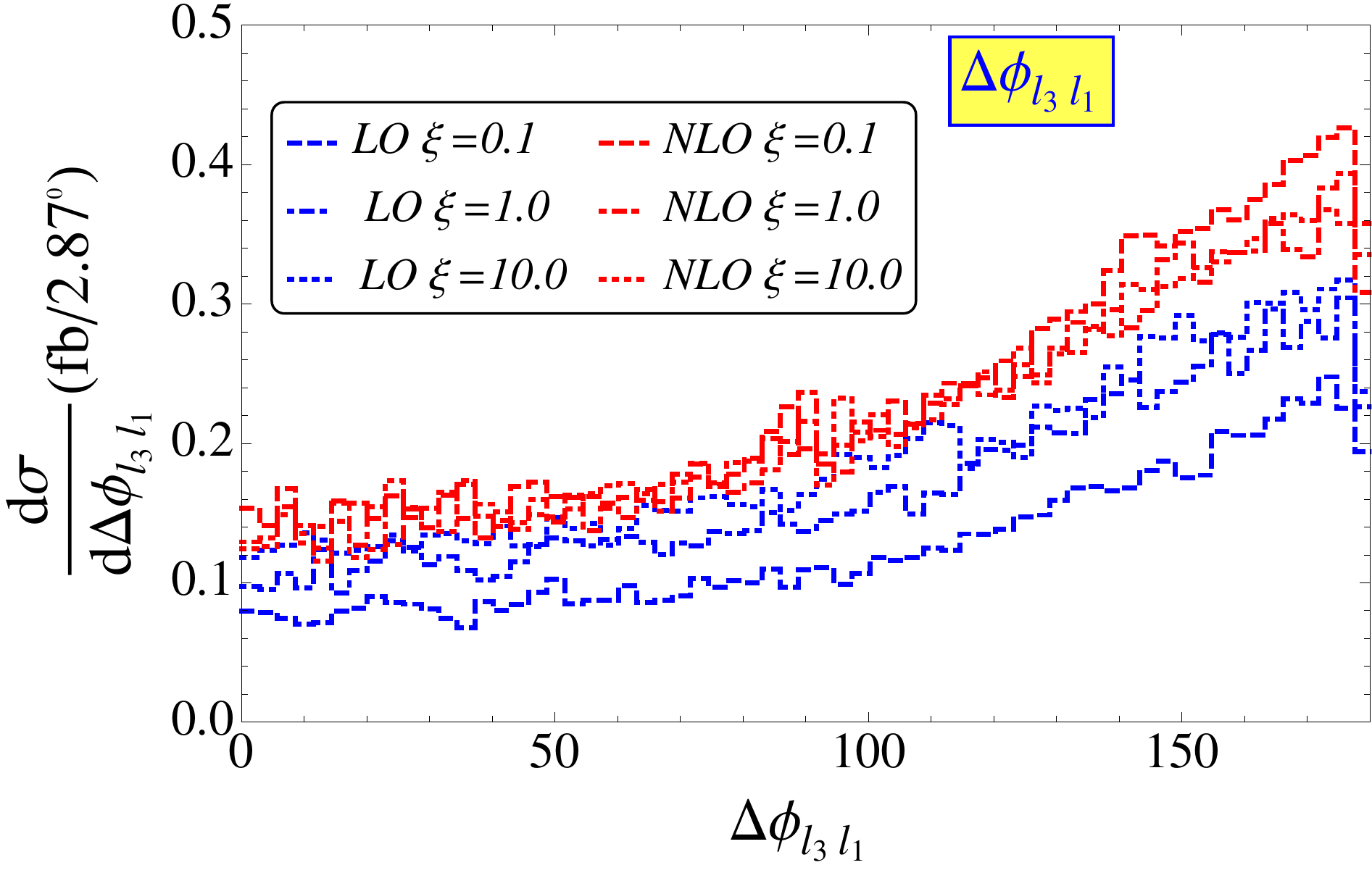}
\includegraphics[scale=0.30]{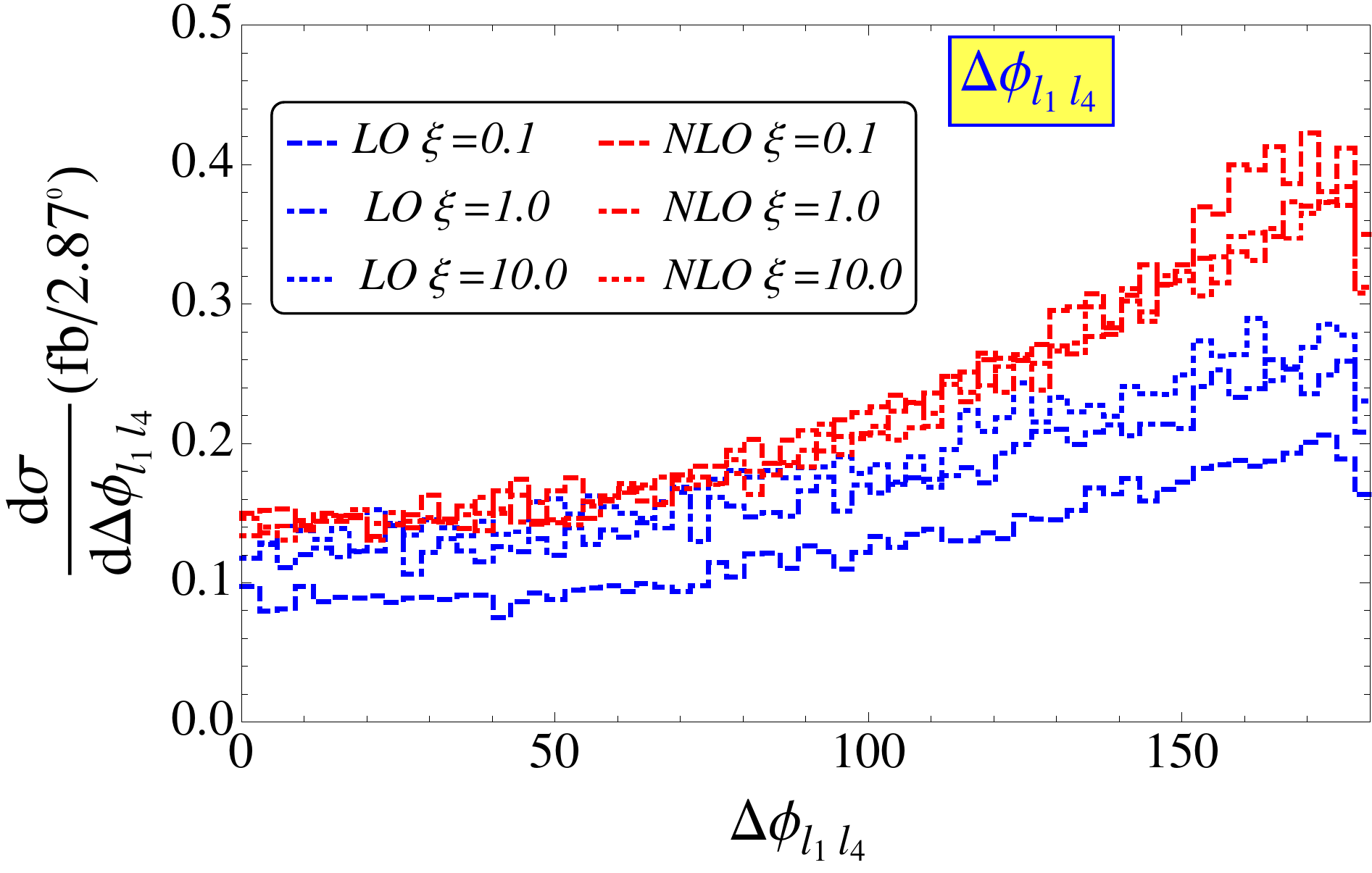}\\
\includegraphics[scale=0.30]{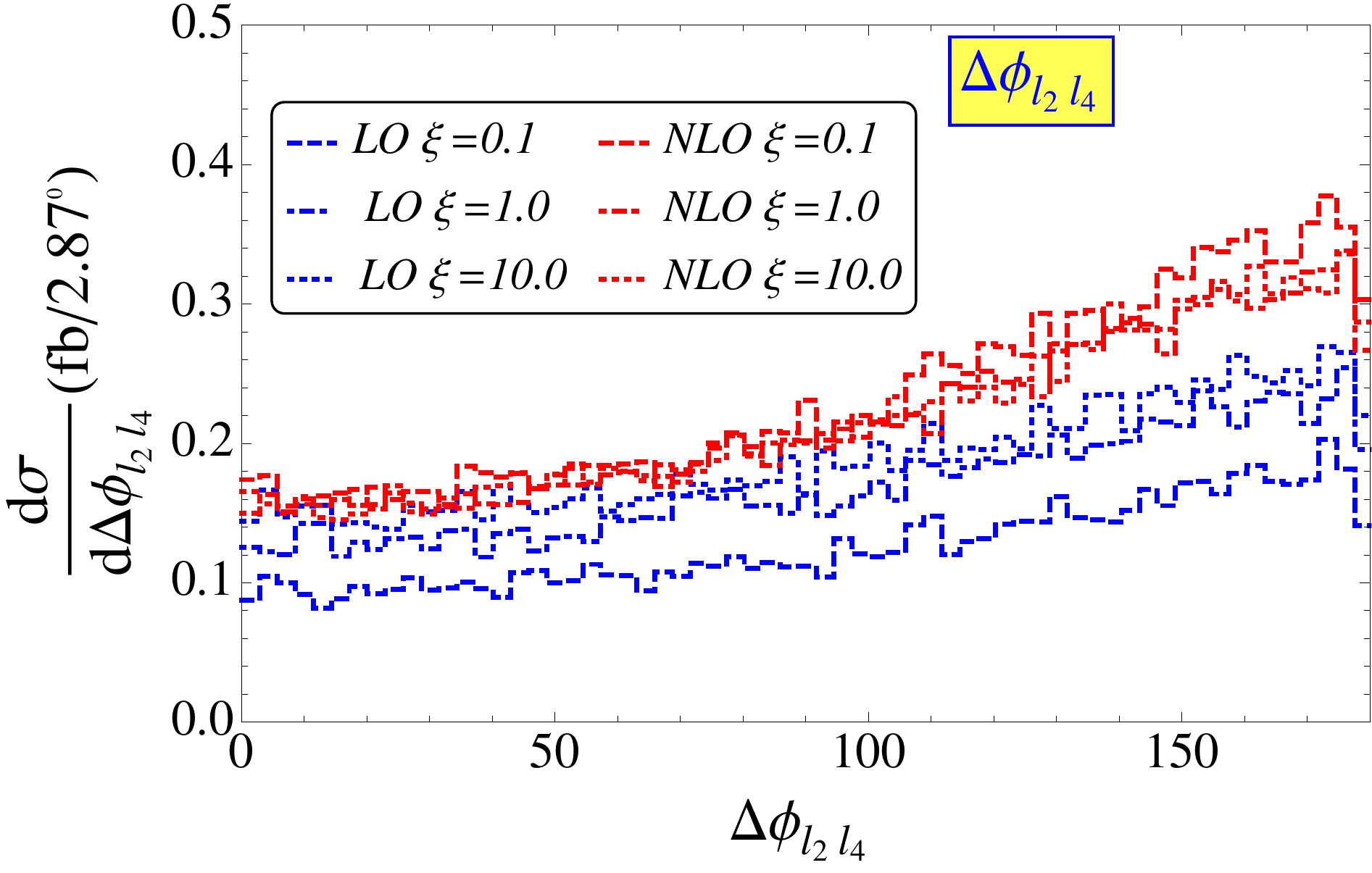}
\includegraphics[scale=0.30]{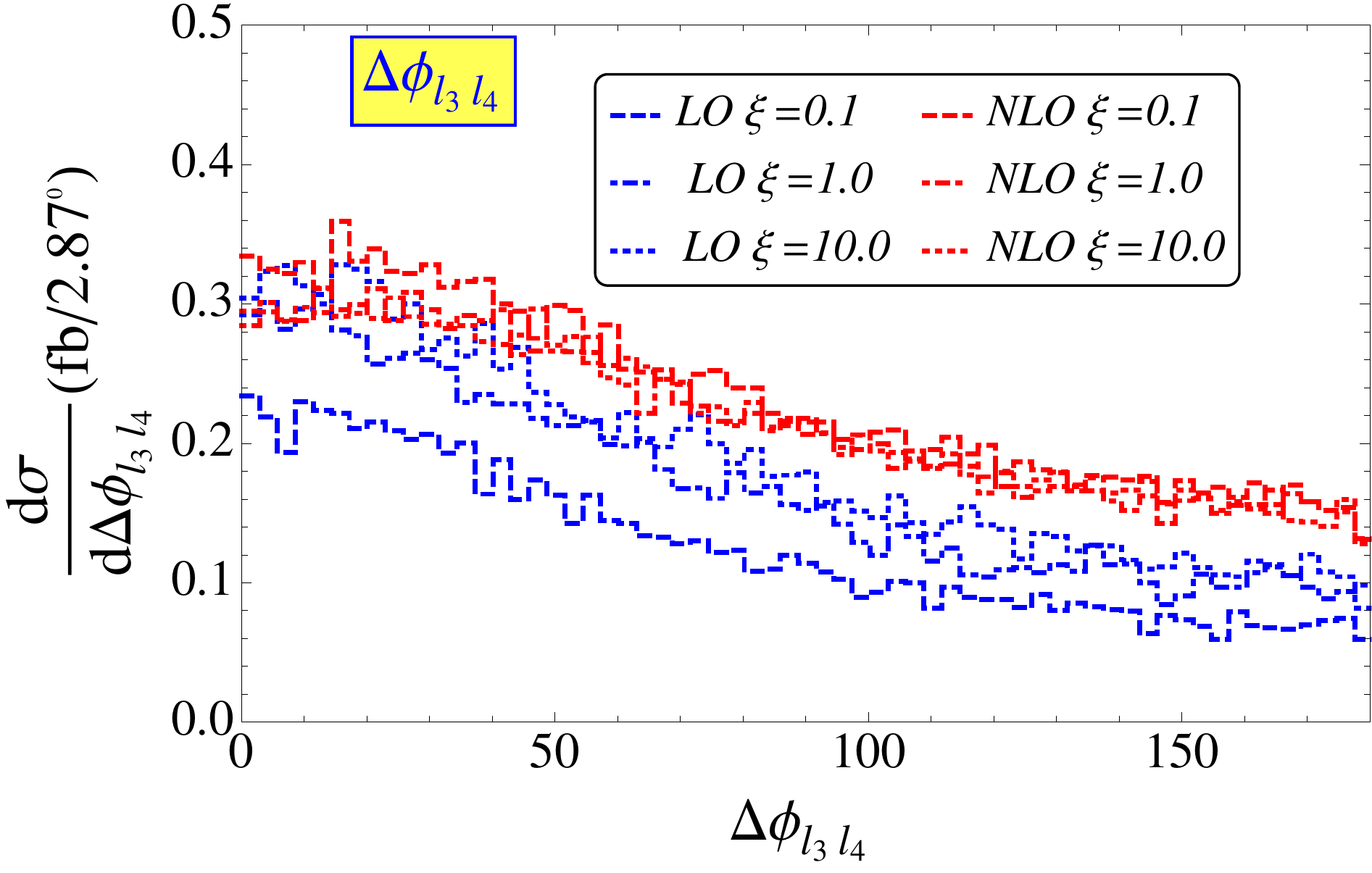}
\end{center}
\caption{Scale dependent LO and NLO-QCD $\Delta\phi_{\ell\ell}$ distribution of the heavy neutrino pair production followed by the decays of the heavy neutrinos into 
$4\ell+\rm{MET}$ channel at the 13 TeV LHC for $m_N=95$ GeV.}
\label{HC 95-4l-3}
\end{figure} 
The LO and NLO-QCD distributions of $\Delta\phi_{\ell\ell}$ between the leptons in the azimuthal plane are plotted in Fig.~\ref{HC 95-4l-3} where the distributions of $\Delta\phi_{\ell_{1}\ell_{2}}$ and $\Delta\phi_{\ell_{1}\ell_{2}}$ 
showed that most of the events are obtained from $\Delta\phi_{\ell\ell}<100^{0}$ where as the for the other combinations, $\Delta\phi_{\ell\ell}> 100^{0}$ could be applicable.
From the MET distribution in Fig.~\ref{HC 95-4l-4}, the selection region is given in Tab.~\ref{R1}.
\begin{table}[ht]
\begin{center}
\begin{tabular}{ccc}
     &&Selection   \\
\hline
Region-I&&$E_{T}^{\rm{miss}} > 50~\rm{GeV}$\\
\hline
\end{tabular}
\end{center}
\caption{Selection regions based on $E_{T}^{\rm{miss}} $}
\label{R1}
\end{table} 
\begin{figure}
\begin{center}
\includegraphics[scale=0.30]{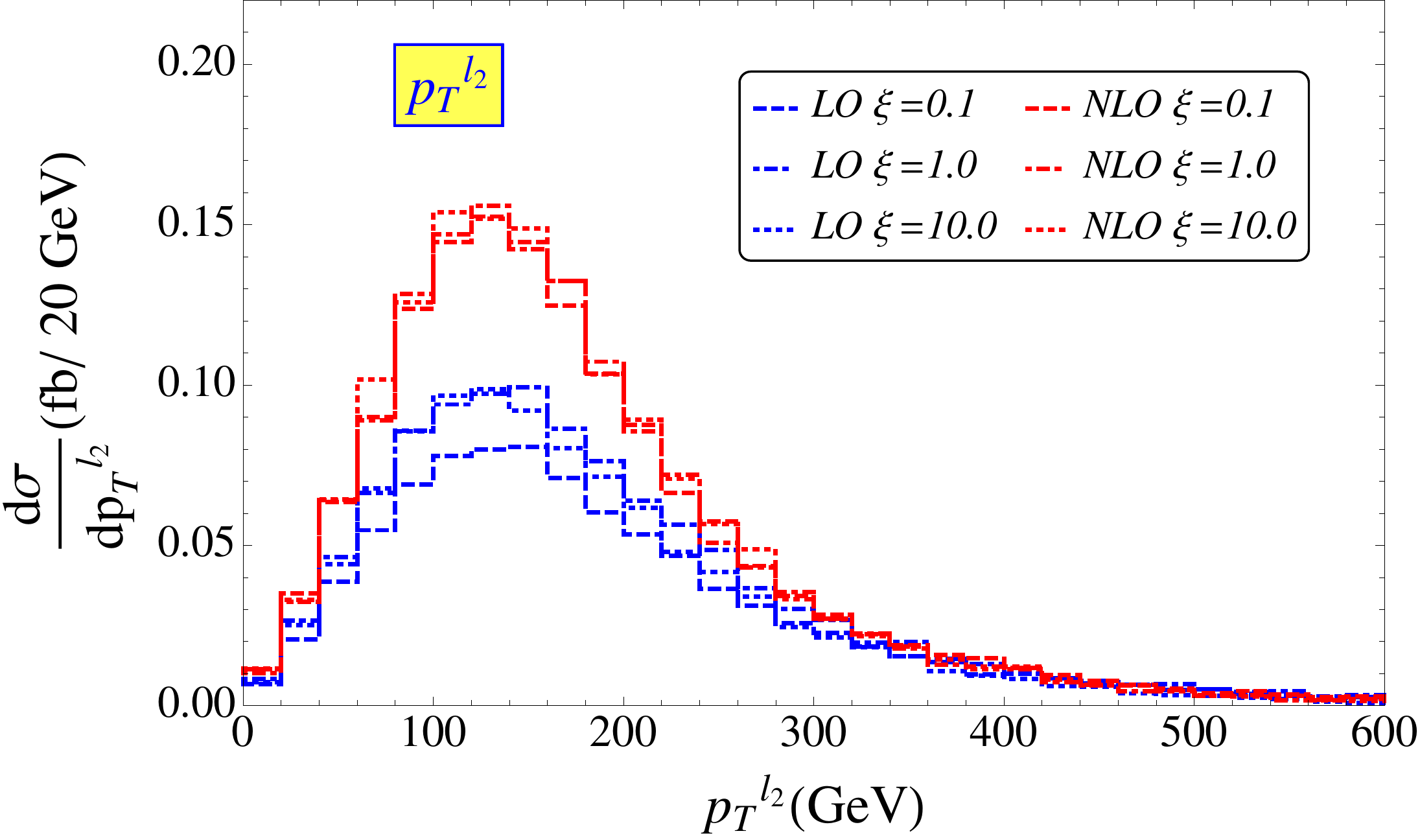}
\includegraphics[scale=0.30]{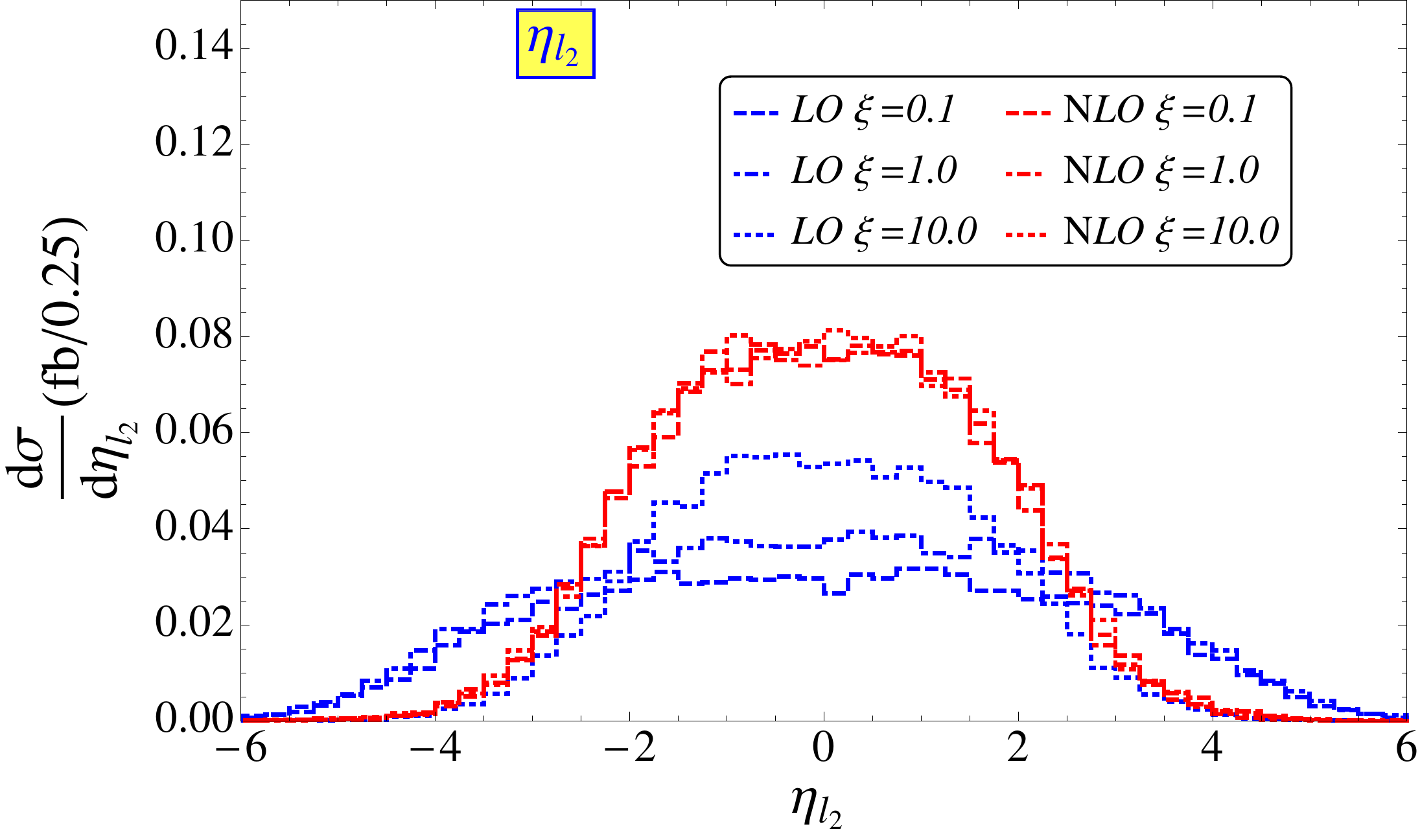}\\
\includegraphics[scale=0.30]{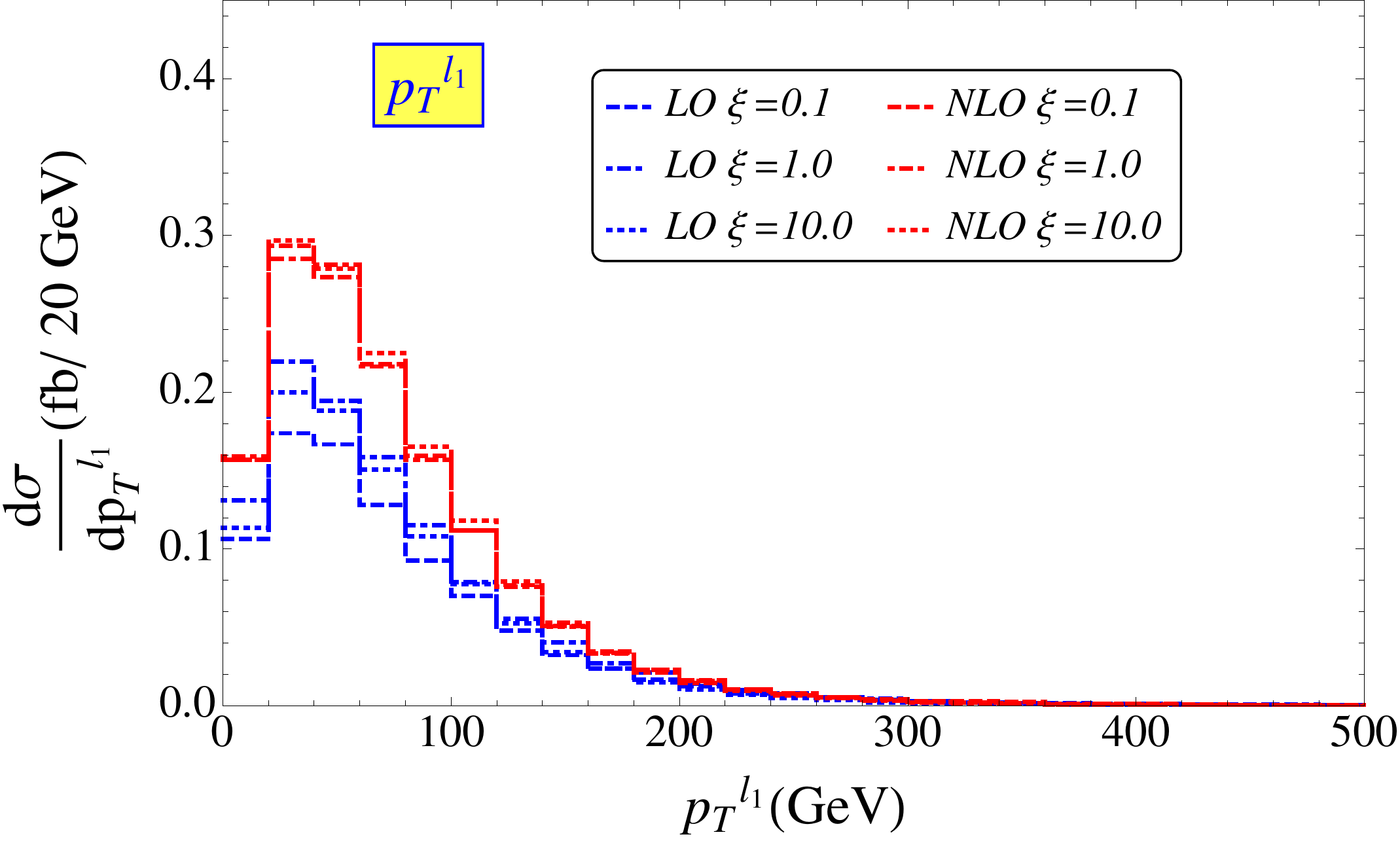}
\includegraphics[scale=0.33]{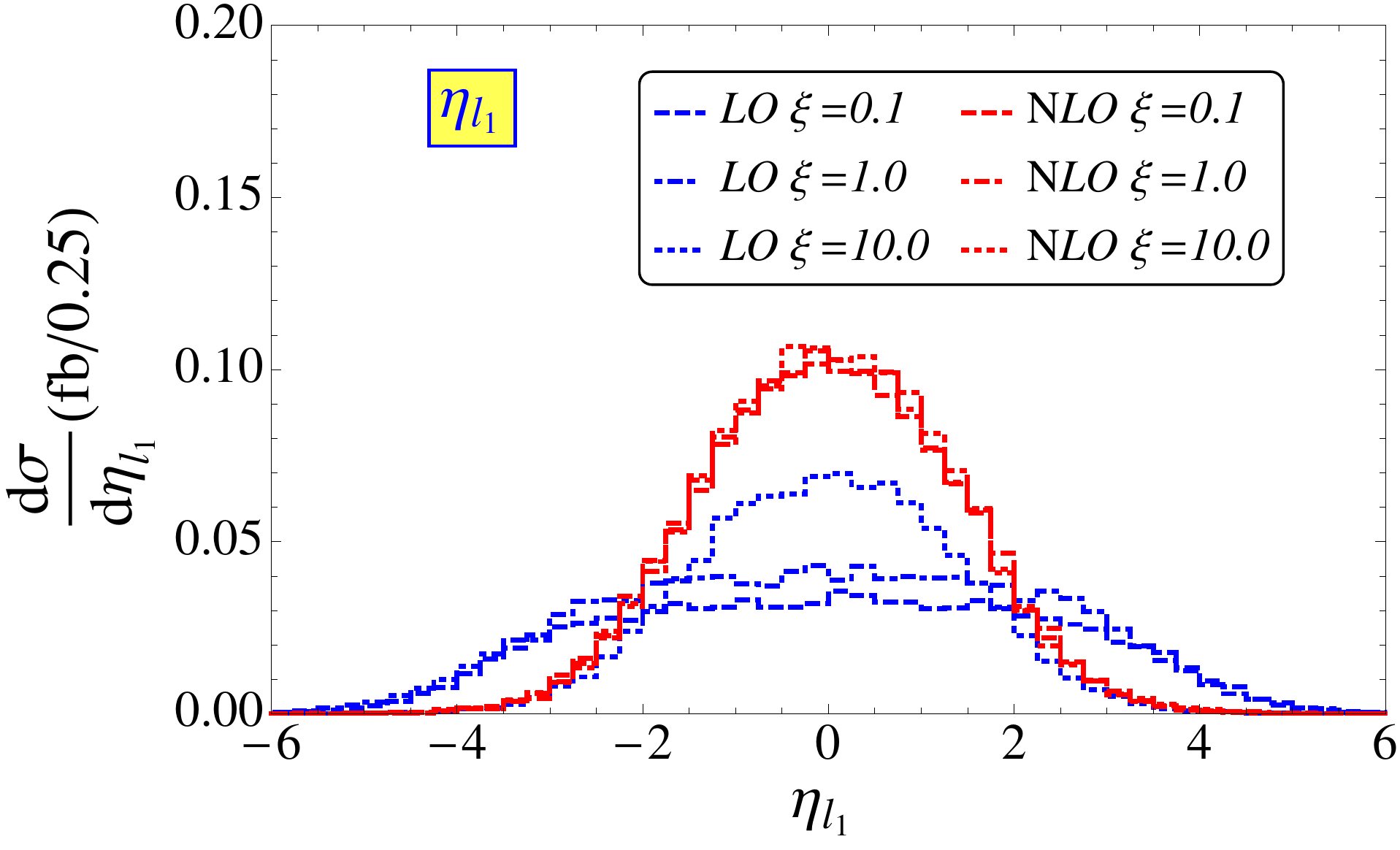}\\
\includegraphics[scale=0.33]{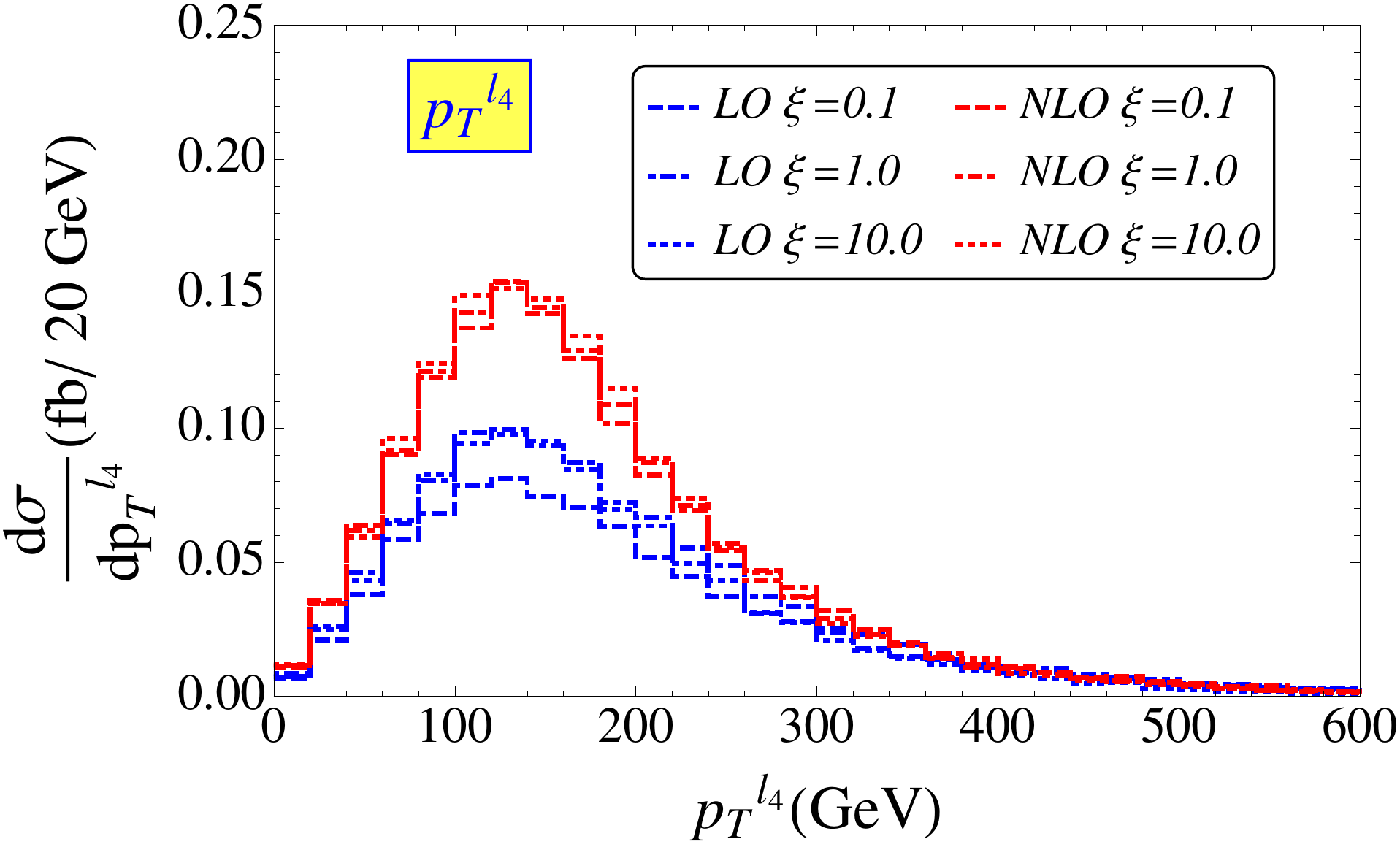}
\includegraphics[scale=0.33]{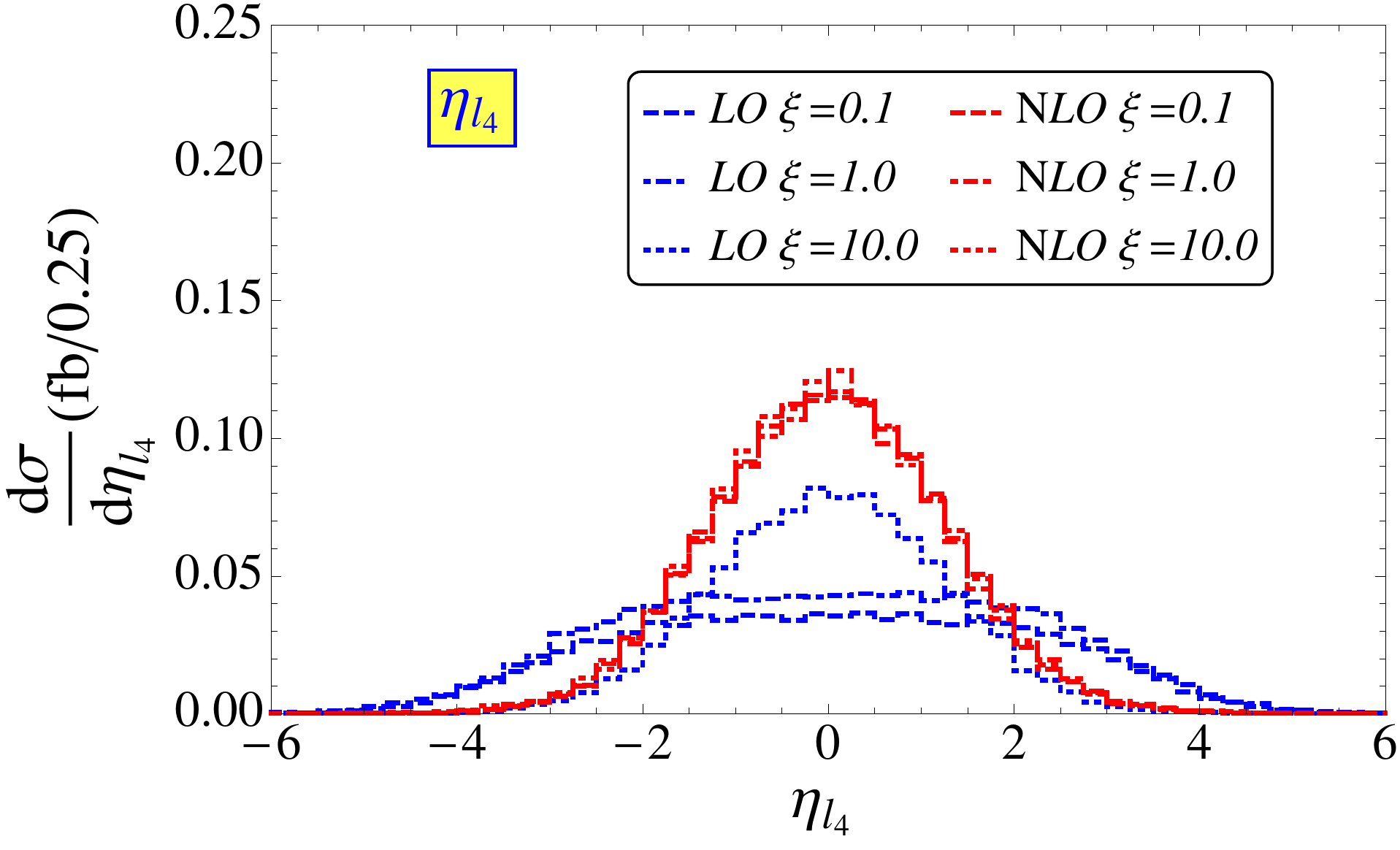}\\
\includegraphics[scale=0.28]{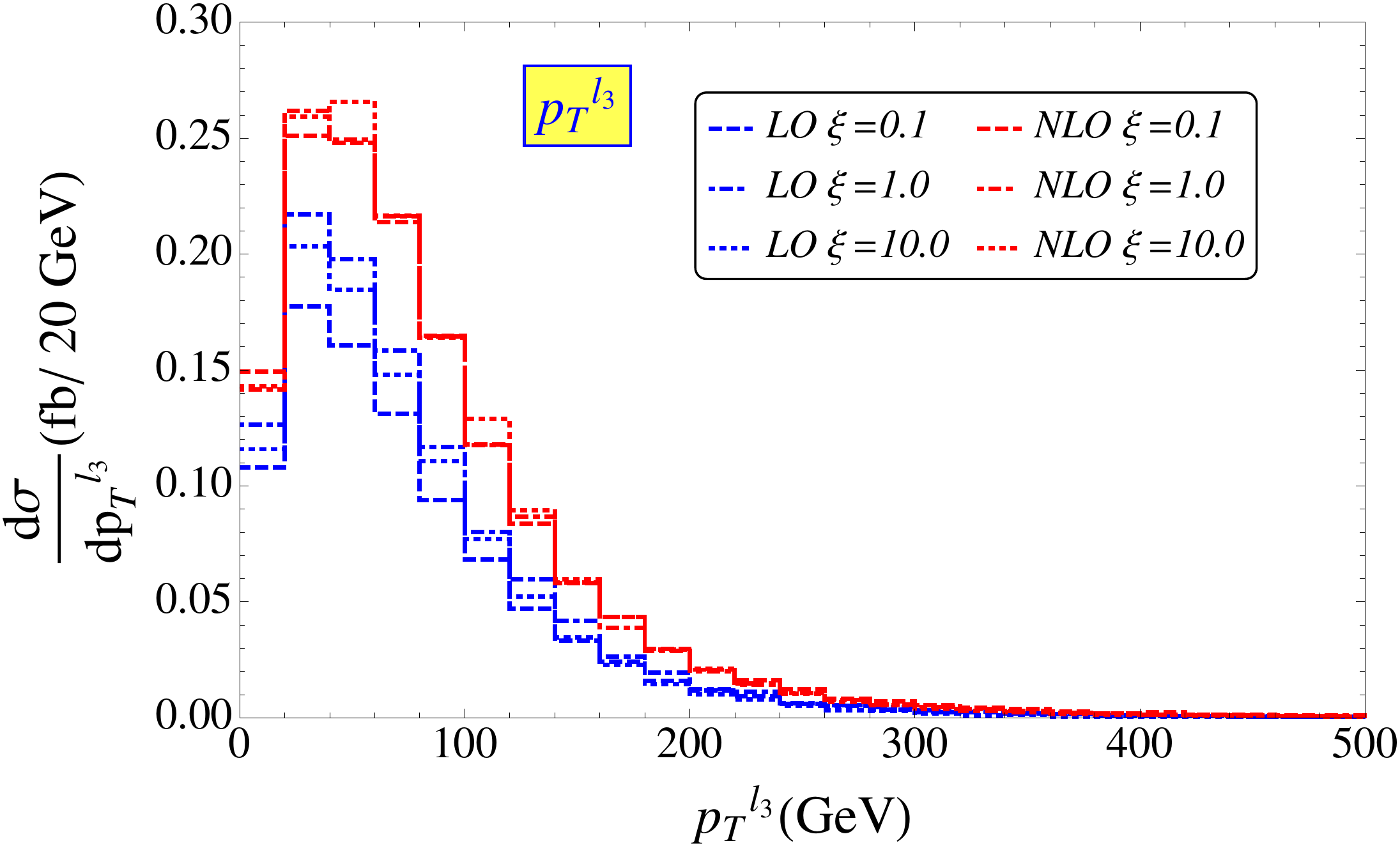}
\includegraphics[scale=0.33]{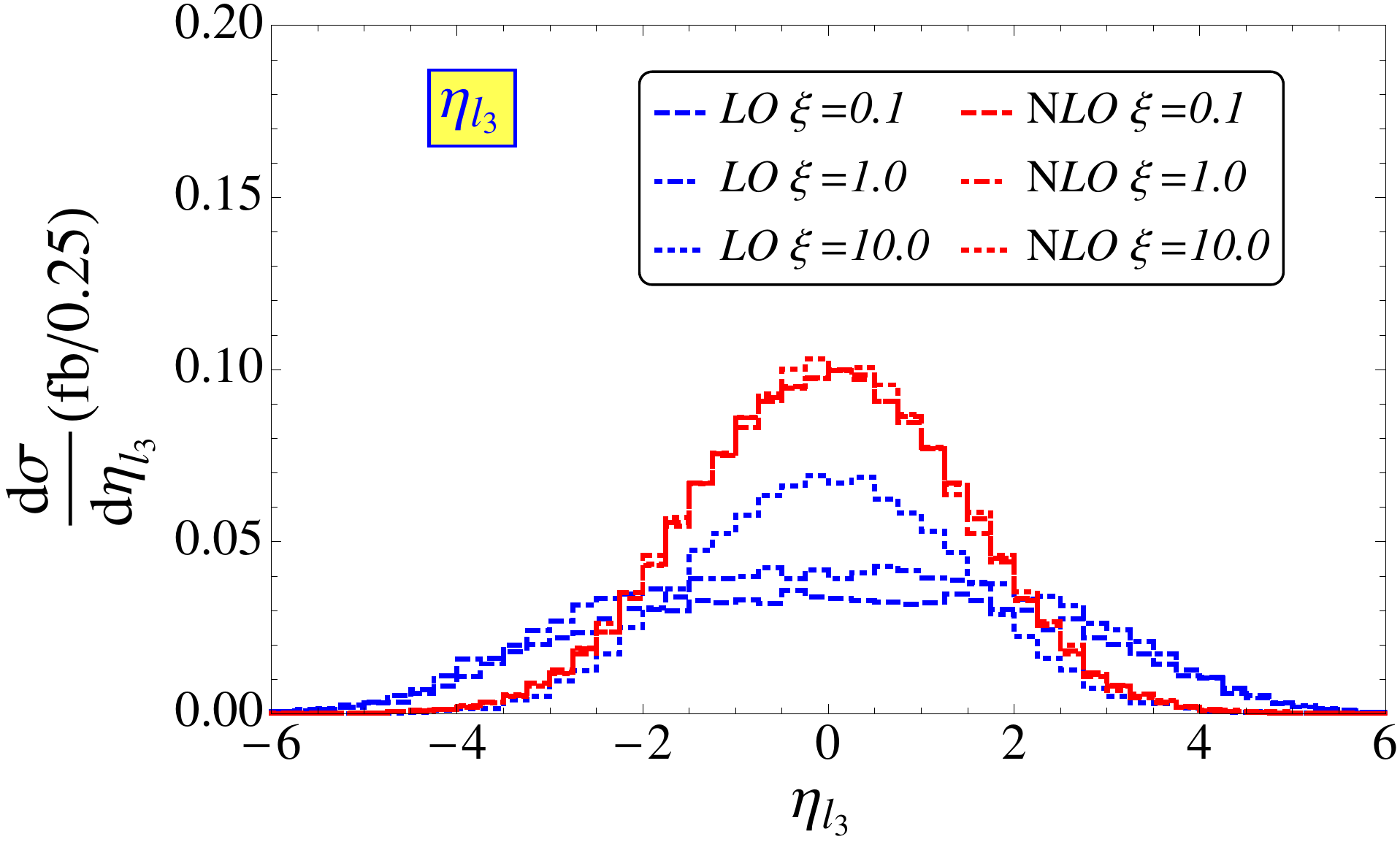}
\end{center}
\caption{Scale dependent LO and NLO-QCD $p_T^{\ell}$ (left column) and $\eta^{\ell}$ (right column) distributions of the heavy neutrino pair production followed by the decays of the heavy neutrinos into 
$4\ell+\rm{MET}$ channel at the 100 TeV hadron collider for $m_N=300$ GeV.}
\label{HC 300-4l-1}
\end{figure} 
\begin{figure}
\begin{center}
\includegraphics[scale=0.3]{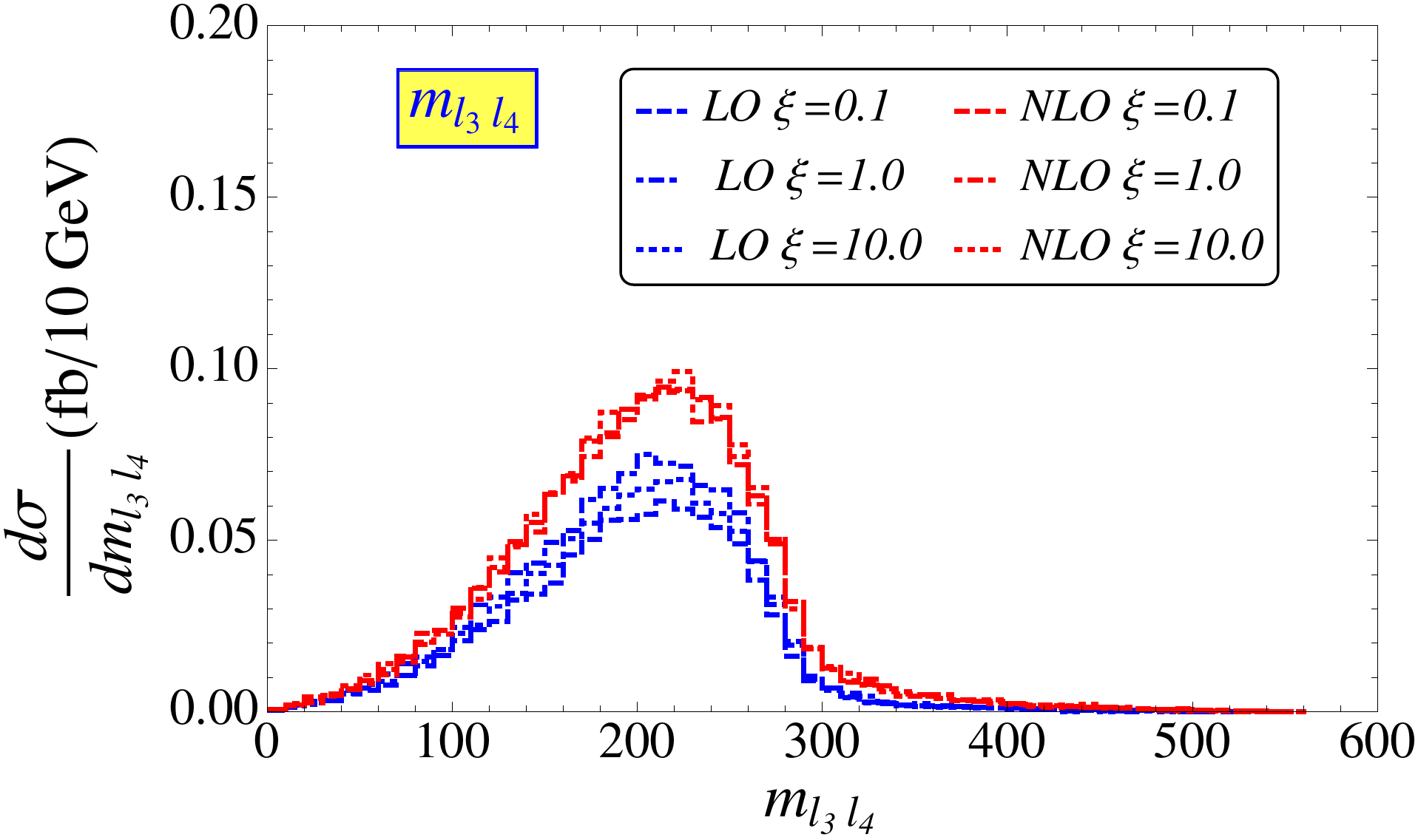}
\includegraphics[scale=0.3]{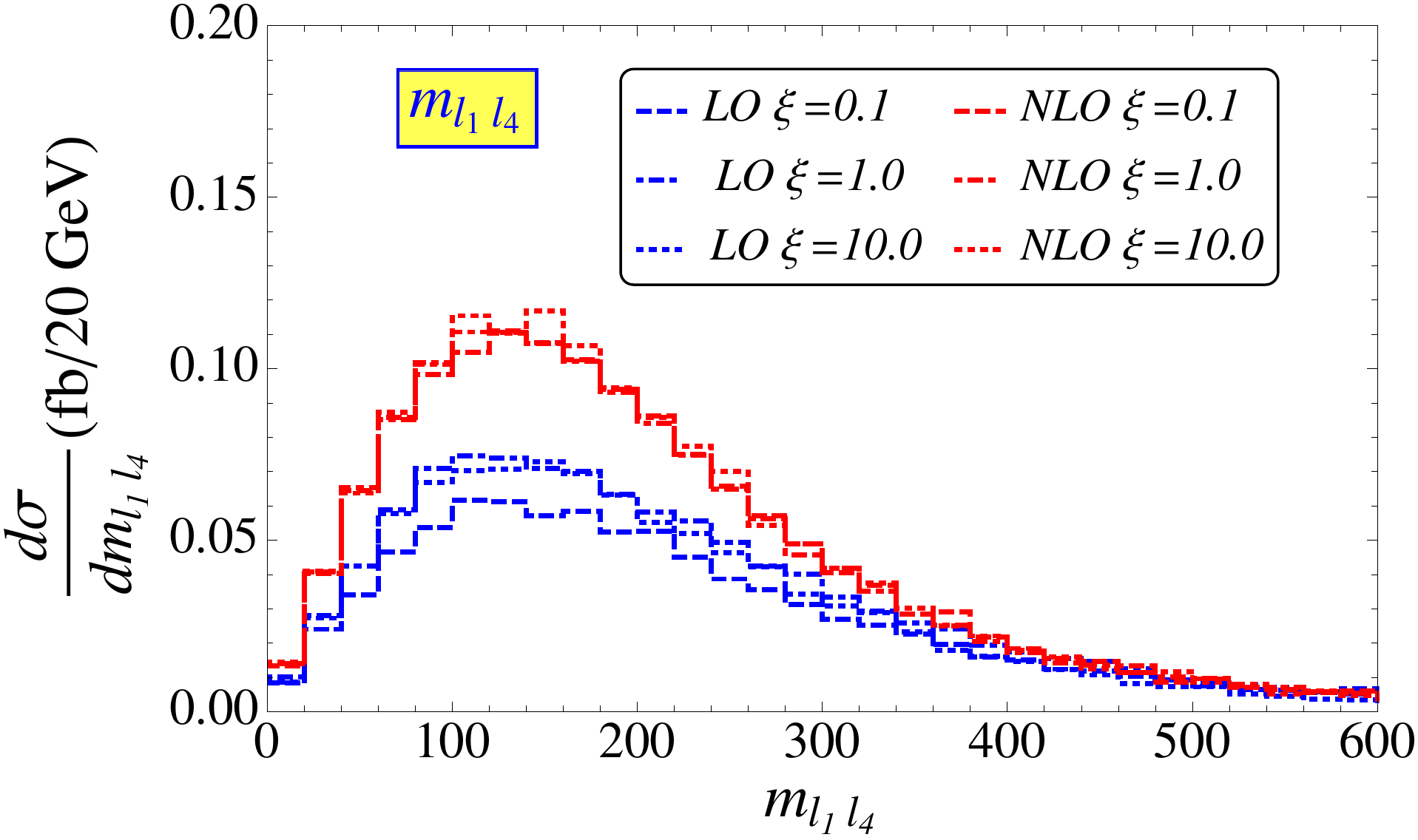}\\
\includegraphics[scale=0.3]{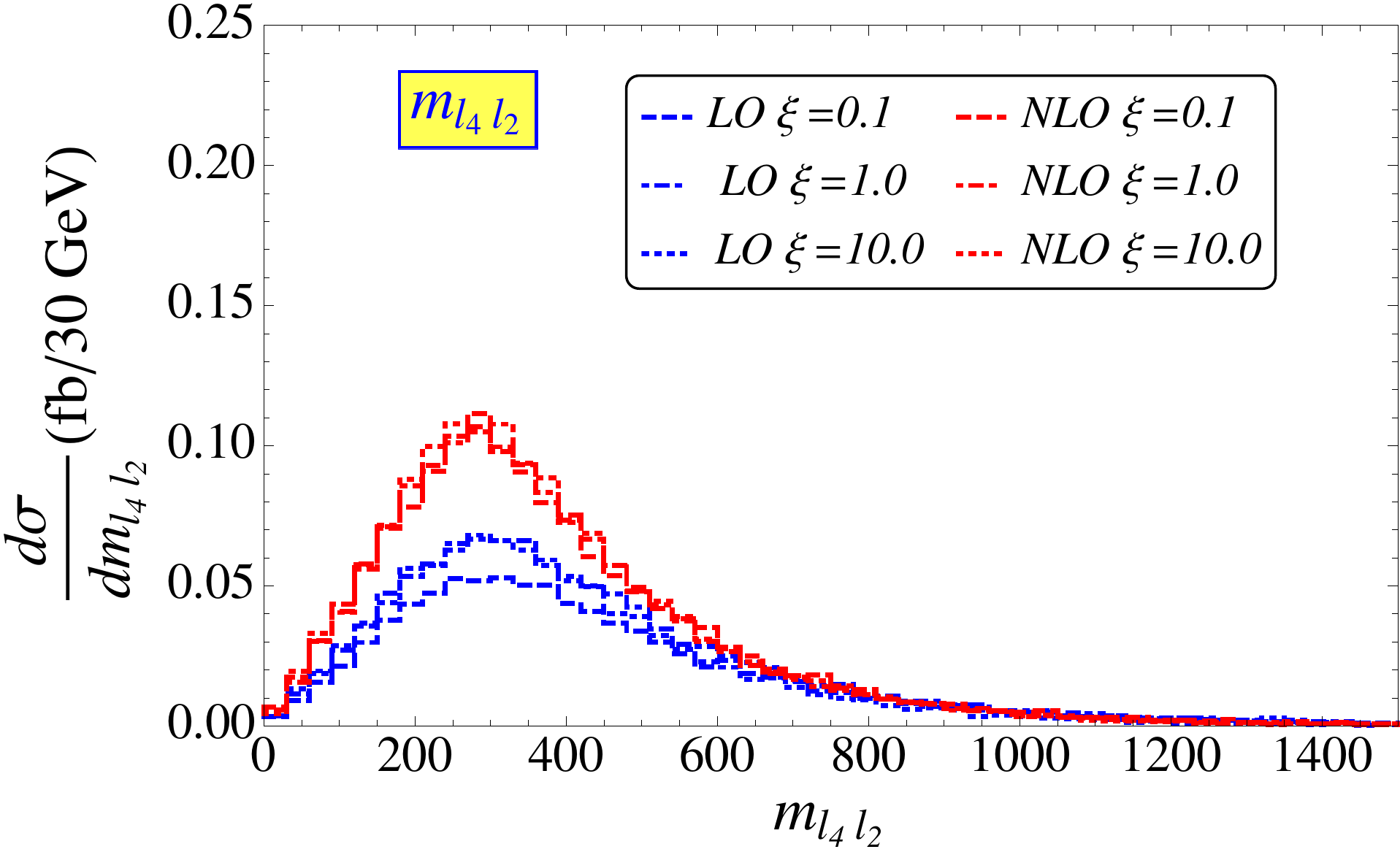}
\includegraphics[scale=0.3]{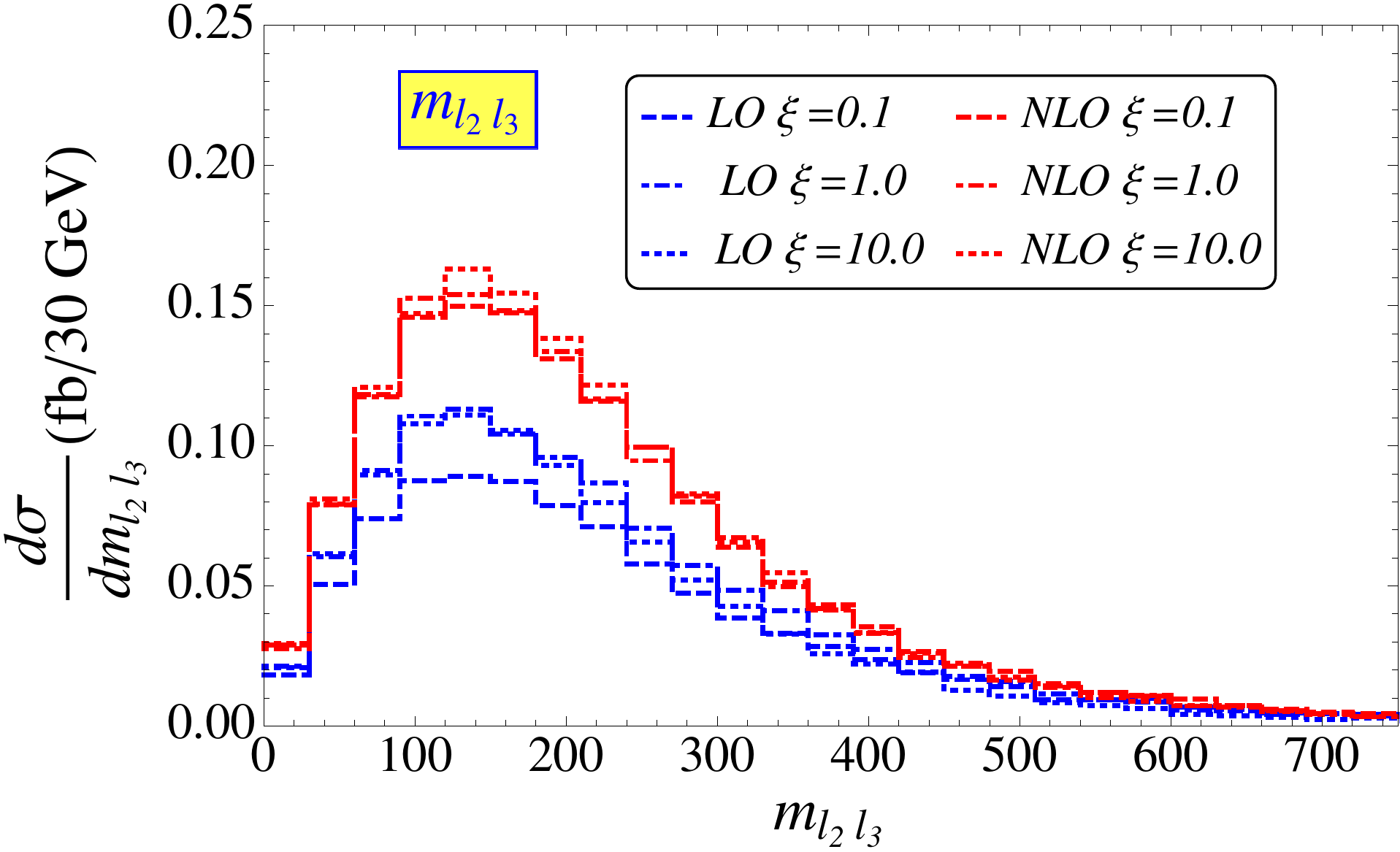}\\
\includegraphics[scale=0.26]{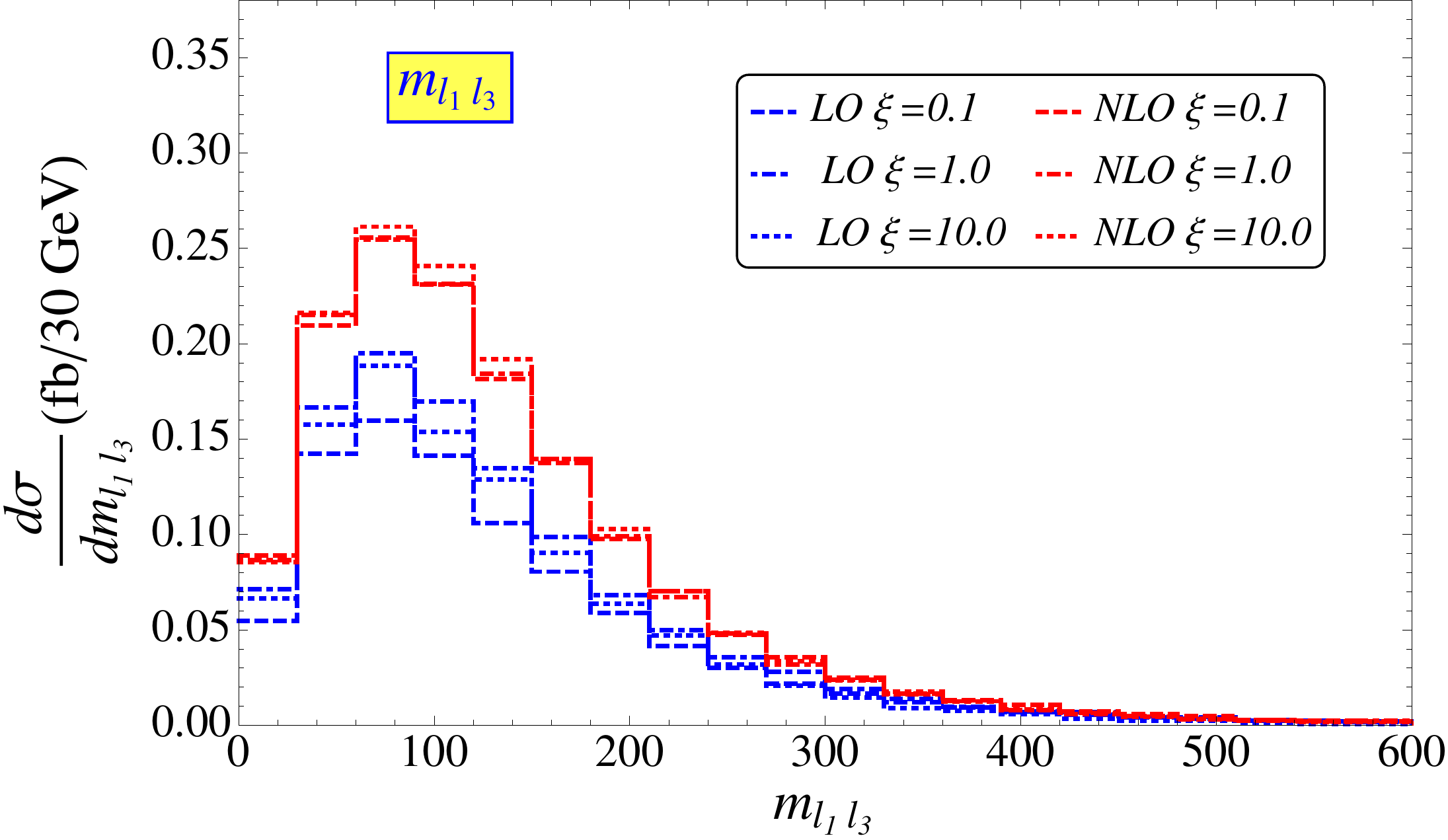}
\includegraphics[scale=0.26]{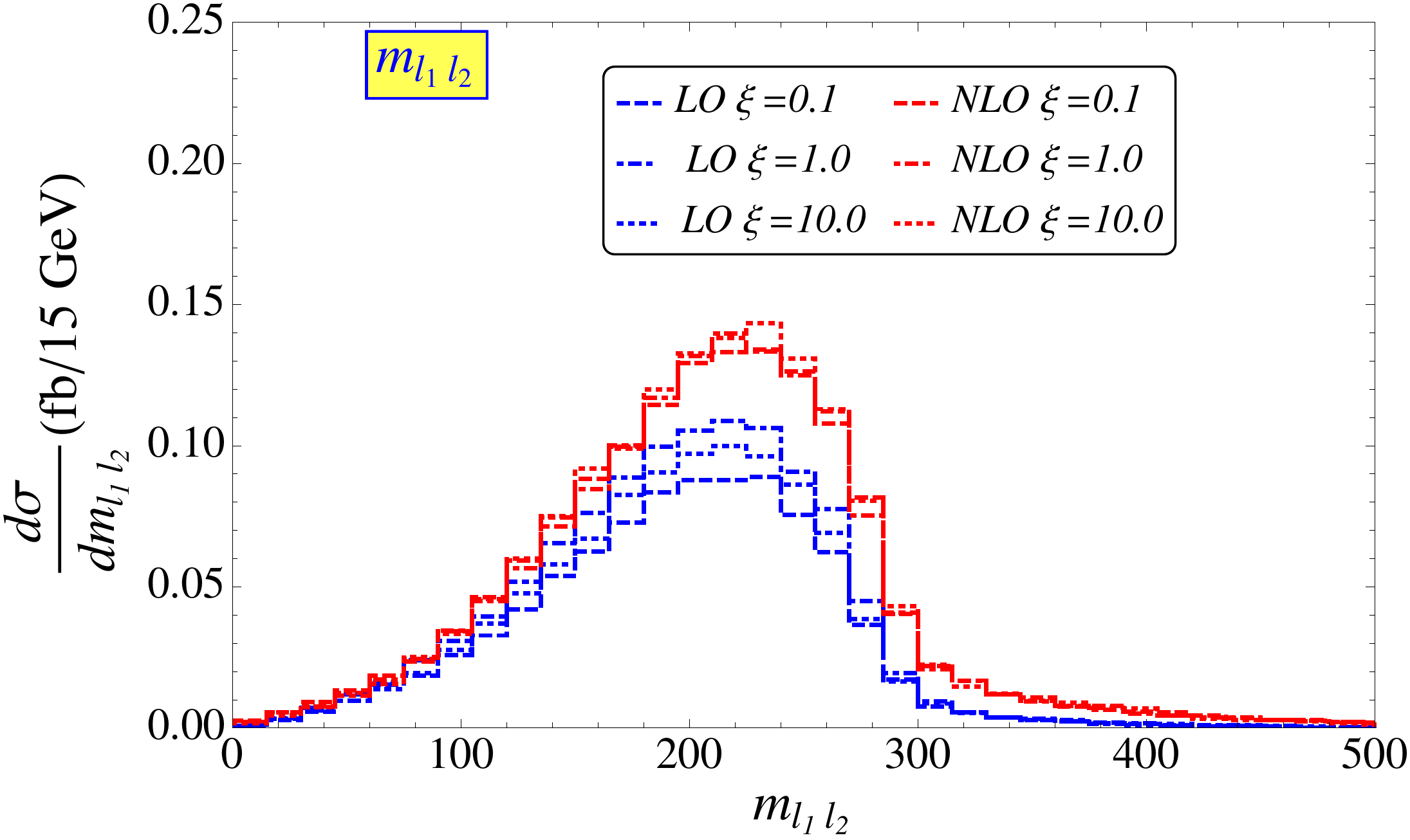}
\end{center}
\caption{Scale dependent LO and NLO-QCD $m_{\ell\ell}$ distribution of the heavy neutrino pair production followed by the decays of the heavy neutrinos into 
$4\ell+\rm{MET}$ channel at the 100 TeV hadron collider for $m_N=300$ GeV.}
\label{HC 300-4l-2}
\end{figure}
\begin{figure}
\begin{center}
\includegraphics[scale=0.50]{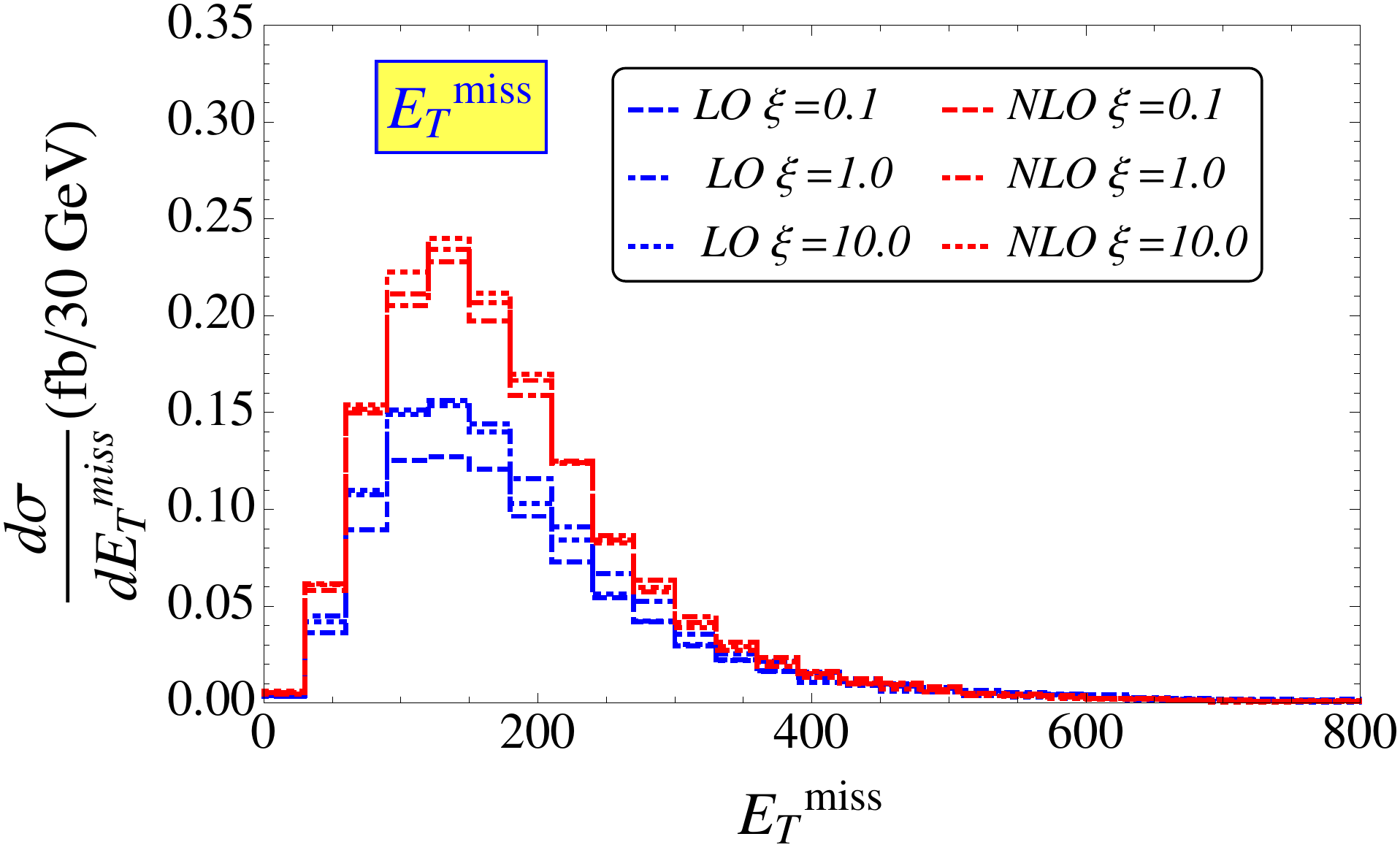}
\end{center}
\caption{Scale dependent LO and NLO-QCD $E_T^{\rm{miss}}$ distribution of the heavy neutrino pair production followed by the decays of the heavy neutrinos into 
$4\ell+\rm{MET}$ channel at the 100 TeV hadron collider for $m_N=300$ GeV.}
\label{HC 300-4l-3}
\end{figure} 

At the 100 TeV hadron collider, we choose $m_{N}=300$ GeV. The scale dependent LO and NLO-QCD distributions for the $p_{T}^{\ell}$ and $\eta^{\ell}$  are shown in Figs.~\ref{HC 300-4l-1}. For the leading lepton we can choose a transverse momentum cut $p_T^{\ell,\rm{leading}} > 90$ GeV and for the transverse momentum of the trailing leptons we use $p_{T}^{\ell, \rm{trailing}} > 40$ GeV.  A pseudo-rapidity cut for the leptons, $|\eta^{\ell} |< 2.5$ can be considered for the signal events.

The invariant mass $(m_{\ell\ell})$ distributions of the leptons are shown in Fig.~\ref{HC 300-4l-2}. A cut of $m_{\ell\ell} > (m_{Z}+15)$ GeV could be applied to accept the leptons above the $Z$ pole and to avoid the SM backgrounds coming from the $Z$ boson decay into the charged lepton pair. 

From the MET distribution in Fig.~\ref{HC 95-4l-3}, the selection region is given in Tab.~\ref{R2}.
\begin{table}[ht]
\begin{center}
\begin{tabular}{ccc}
     &&Selection   \\
\hline
Region-I&&$E_{T}^{\rm{miss}} > 75~\rm{GeV}$\\
\hline
\end{tabular}
\end{center}
\caption{Selection regions based on $E_{T}^{\rm{miss}} $}
\label{R2}
\end{table}

The scale variations according to Eqs.~\ref{muF} and \ref{muR} are used to generate the events for the LO and NLO processes. The NLO processes are softer compared to the LO processes due to the involvement of the strong coupling constant as a function of $\xi$.

For all these processes, the heavy neutrino can be reconstructed from the invariant mass of the $2\ell \nu$ system or $\ell2j$ system using a window cut of 20 GeV- 25 GeV depending upon the detector and its efficiency.
\section{Scale dependent prospective search reaches of the mixing angles from the multilepton final states}
\label{sec:mix}
Only in the inverse seesaw model we have a freedom to consider two typical cases for the flavor structure where the Dirac mass $(m_{D}=\frac{y_{D}v}{\sqrt{2}})$ and the small lepton number violating mass $(\mu)$ terms play important roles as free parameters. In order to keep the discussion simple we consider the degeneracy of the heavy neutrinos in mass such as $m_N= M \bf{1}$ for the following cases: 
\begin{itemize}
\item[(1)] We consider a case where $\mu$ is proportional to the unit matrix, $\mu \propto \mu \bf{1}$. In this case, the flavor structure of the neutrino mass $(m_{\nu}^{\rm{ISS}})$ is
carried out by the Dirac mass term $m_{D}$. We call it Flavor Non-Diagonal (FND) case. An elaborate numerical analysis has been shown for this flavor structure in \cite{Das:2012ze} while satisfying the non-unitarity conditions and as well as Lepton Flavor Violation (LFV) bounds under the general parametrization. However, the Casas- Ibarra \cite{Casas:2001sr} parametrization can be successfully used to study the FND processes to enhance the cross section.

\item[(2)] On the other hand, we have another possibility called Flavor Diagonal (FD) case where $m_{D}$ is proportional to the unit matrix, $m_{D}\propto m_{D} \bf{1}$ and the flavor structure is embedded 
in the small lepton number violating mass term $\mu$. A trilepton analysis from the charged current interaction Eq.~\ref{CC} with associated jets from the quark-gluon fusion $(qg)$ has been studied in \cite{Das:2014jxa}
with a jet transverse momentum cut of $p_{T}^{j} > 30$ GeV.
\end{itemize}

 In this current work we assume FD case\footnote{In case of the inverse seesaw model we can consider the FD case which is not possible in the type-I seesaw framework because in type-I seesaw, the Dirac Yukawa has to carry the flavors. As there is no $\mu$ term in type-I seesaw, the possibility of the FD case will not arise. If we study the type-I case then the signal will be $pp\to NN, N\to\ell^{\pm} W^{\mp}, N\to\ell^{\pm} W^{\mp}$. $W$ can decay leptonically or hadronically depending upon the choice. The same sign leptons coming out of the heavy neutrinos will be the distinguishing mode. A complete study of such model could be interesting in future if a sizable mixing angle is possible. However in Ref.\cite{Rasmussen:2016njh} it has been estimated that the heavy neutrinos in the type-I seesaw model will have the squared of the mixing angle $\mathcal{O}(10^{-12})$ even if Casas- Ibarra parametrization is applied. They have also showed the FCC-ee prospective exclusion limit for the heavy neutrino mass of the $\mathcal{O}(100~\rm{GeV})$. Therefore type-I seesaw case is not discussed here. In addition to that the LFV will be an important issue for the FND cases even in type-I seesaw where $\mu\to e \gamma$ constraints will be strongly applicable even in the general parametrization \cite{DasOkada1}.}$^{,}$ \footnote{In Ref.~\cite{Kang:2015uoc}, the authors have studied the $3\ell$+$\rm{MET}$+$2j$ case and the $N\to h \nu$ followed by $h\to b\overline{b}$ for $m_N=150$ GeV where $\mathcal{BR}(N\to h\nu)\sim 3.86\%$, for $m_N=700$ GeV, $\mathcal{BR}(N\to h \nu)= 23.95\%$ and for $m_N=5$ TeV, $\mathcal{BR}(N\to h \nu)= 25\%$. In our case $3\ell+\rm{MET}+2j$ is coming from the purely inverse seesaw model in FD case. Irrespective of the final states, $\mathcal{BR}(N\to W\ell) \geq 50\%$ in our case always.} with diagonal $m_D$ and $m_N$ matrices to suppress the effects from the LFV processes. We also consider the $3\ell$ and $4\ell$ final states to compare with the anomalous multilepton searches made by the CMS \cite{Chatrchyan:2014aea}. We use the observed number of events and the corresponding SM expectations. Using the events we will estimate a direct upper bound on the elements of the mixing angles for the FD case. In \cite{Das:2012ze}, it is explicitly shown that the light neutrino oscillation data is satisfied for the FD case for a suitable flavor structure of the $\mu$ term. One can even make a simplified benchmark situation considering one of the right handed heavy neutrino at the electroweak scale, when the others are too heavy for the reach of the LHC and even 100 TeV hadron collider. This case can be considered as the Single Flavor (SF) case. As a choice we consider that the heavy neutrino is coupled to the second generation of the lepton $(\mu)$. For the $3\ell$ case the signal events will follow 
\bea         
 p p &\to& N \overline{N}, \nonumber \\
        \, \, \, \,  \, \, \, \,&&   N \to \mu^{-} W^{+}, W^{+} \to \mu^{+} \nu / e^{+} \nu \nonumber \\
        \, \, \, \,  \, \, \, \,&&   \overline{N} \to \mu^{+} W^{-}, W^{-} \to j j .
\label{decay11}        
\eea
and
\bea        
p p &\to& N \overline{N}, \nonumber \\
        \, \, \, \,  \, \, \, \,&&   N \to \mu^{-} W^{+}, W^{+} \to j j \nonumber \\
        \, \, \, \,  \, \, \, \,&&   \overline{N} \to \mu^{+} W^{-}, W^{-} \to  \mu^{-} \overline{\nu} / e^{-} \overline{\nu}.
\label{decay22}
 \eea

 Note that in this case, the lightest heavy-neutrino branching ratio is independent of the square of the mixing parameter $(|V_{\mu N}|^{2})$.  The mixing angle only affects the production cross section through the Eq.~\ref{pair}. As a result the pair production cross section is directly proportional to the fourth power of the mixing angle $(\sigma(m_{N})^{\rm{LO/ NLO}} \propto |V_{\ell N}|^4)$. Using such dependency we can constrain $|V_{\ell N}|^{2}=\sqrt{|V_{\ell N}|^{4}}$ as a function of $m_N$. Similarly for the $4\ell$ case we can do the same while $W \to \mu \nu / e \nu$ will replace $W \to j j$ in Eqs.~\ref{decay11} and 
 \ref{decay22}.

 In the FD case, for simplicity we consider two degenerate heavy neutrinos with the same mass $m_N$. We assume that one of them couples with the electrons, and the other one with the muons, but  both of them have same strength of coupling such that $|V_{eN_1}|^2 = |V_{\mu N_2}|^2 = |V_{\ell N}|^2$. Due to the smallness of lepton number violating mass term $\mu$ in the inverse seesaw, the neutrinoless double beta decay will be suppressed and as well as the LFV constraints. Hence the FD case will be well allowed. In the $3\ell$ and $4\ell$ cases the FD case is twice as large as the SF case. Due to simplicity we will put the upper bounds on the FD case for $\xi=0.1, 1.0, 10.0$ as a function of $m_N$ for the $3\ell$ and $4\ell$ cases.

We use the {\tt CTEQ6l1 PDF (CTEQ6M PDF)} \cite{CTEQ6} for generating the LO (NLO with $\mu_{F}= \mu_{R}$) process to compute the scale dependent $3\ell$ and $4\ell$ 
events with $\xi=0.1$, $1.0$ and $10.0$ at $\sqrt{s}=13$ TeV LHC and 100 TeV hadron collider using {\tt MadGreph-aMC@NLO} bundled with {\tt PYTHIA6Q} using {\tt anti-$k_{T}$} algorithm 
for jet clustering in {\tt FastJet}. We use the hadronized events in {\tt Delphes} \cite{Delphes} to produce events after the detector simulation.
After the detector simulation we have considered the events with $3\ell+ E_{T}^{miss}+ 2-$jets and  $4\ell+ E_{T}^{miss}+n-$jets where $n=0,~1,~2,~3$.

The recent CMS study on the anomalous multilepton search at the $8$ TeV with $19.5$ fb$^{-1}$ luminosity\cite{Chatrchyan:2014aea} has analyzed the $3\ell+ \rm{MET}$ and $4\ell+\rm{MET}$ final states separately.
Adopting the results of their searches for our two different final states we put prospective upper bounds on the mixing angle squared $|V_{\ell N}|^{2}$ at the HL-LHC at the 13 TeV LHC with a luminosity of 3000 fb$^{-1}$ 
and at the 100 TeV hadron collider with luminosities 3000 fb$^{-1}$ and 30000 fb$^{-1}$. The cuts we used from Ref.\cite{Chatrchyan:2014aea} in our analysis are listed as : 
\begin{itemize}
\item [(i)] The transverse momentum of each lepton : $p^\ell_T > 10$ GeV.
\item [(ii)] The transverse momentum of at least one (leading) lepton : $p^{\ell,~{\rm leading}}_{T} > 20$ GeV.
\item [(iii)]  The jet transverse momentum : $p_T^j > 30$ GeV. 
\item [(iv)] The pseudo-rapidity of leptons: $|\eta^\ell| < 2.4$ and of jets : $|\eta^j| < 2.5$.
\item [(v)] The lepton-lepton separation: $\Delta R_{\ell \ell} > 0.1$ and the lepton-jet separation : $\Delta R_{\ell j} > 0.3$. 
\item [(vi)] The invariant mass of each OSSF (opposite-sign same flavor) lepton pair : $m_{\ell^+ \ell^-}< 75$  to avoid the on-$Z$ region which was excluded from the search made the CMS. 
Events with $m_{\ell^+ \ell^-}< 12$ GeV are rejected to eliminate background from low-mass Drell-Yan processes and hadronic decays. 
\item [(vii)]  The scalar sum of the transverse momenta of the jets : $H_{T} <  200$ GeV. 
\item [(viii)] The missing transverse energy : $ E_{T}^{\rm{miss}} < 50$ GeV. 
\end{itemize}

\begin{figure}
\begin{center}
\includegraphics[scale=0.293]{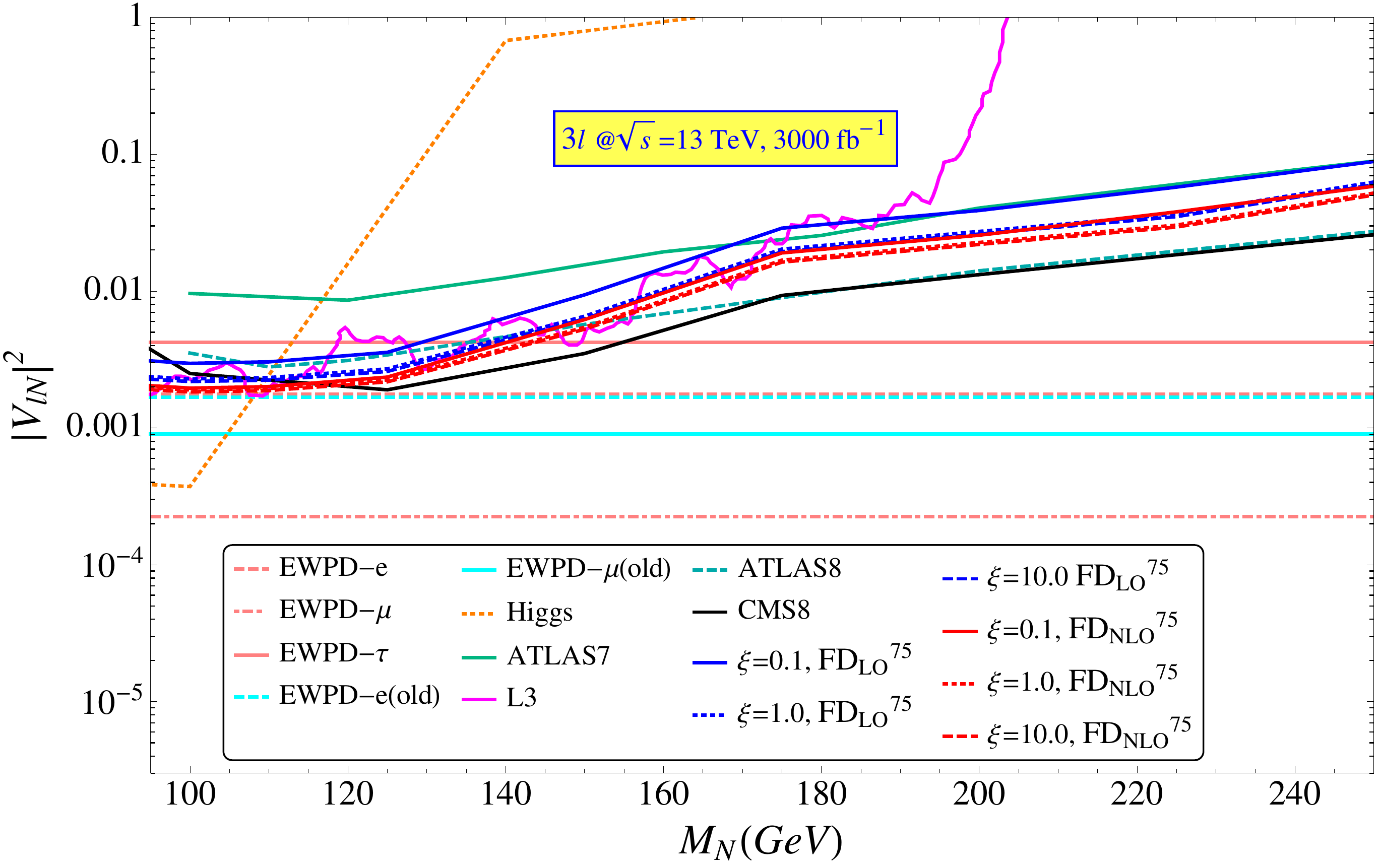}
\includegraphics[scale=0.293]{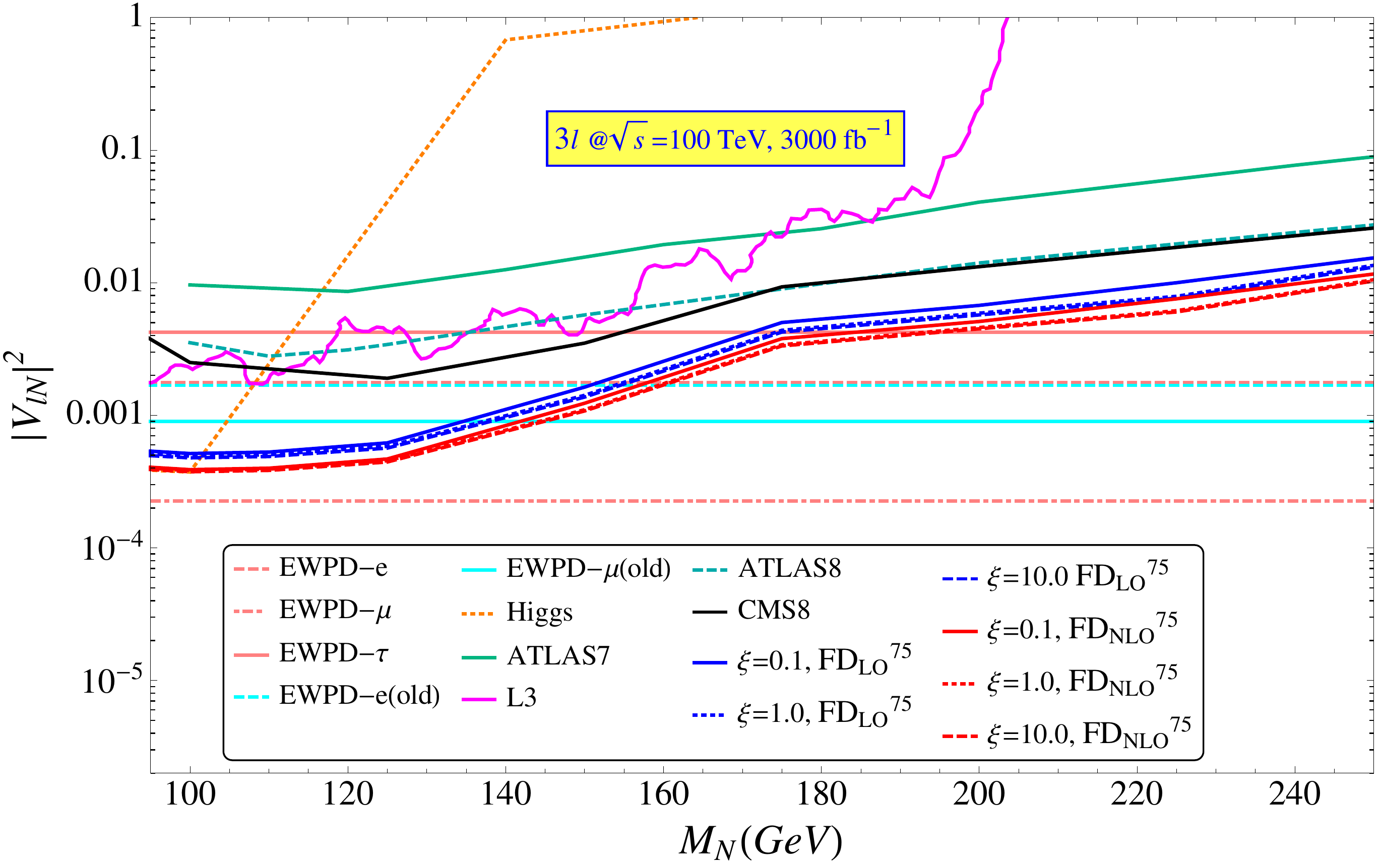}\\
\includegraphics[scale=0.293]{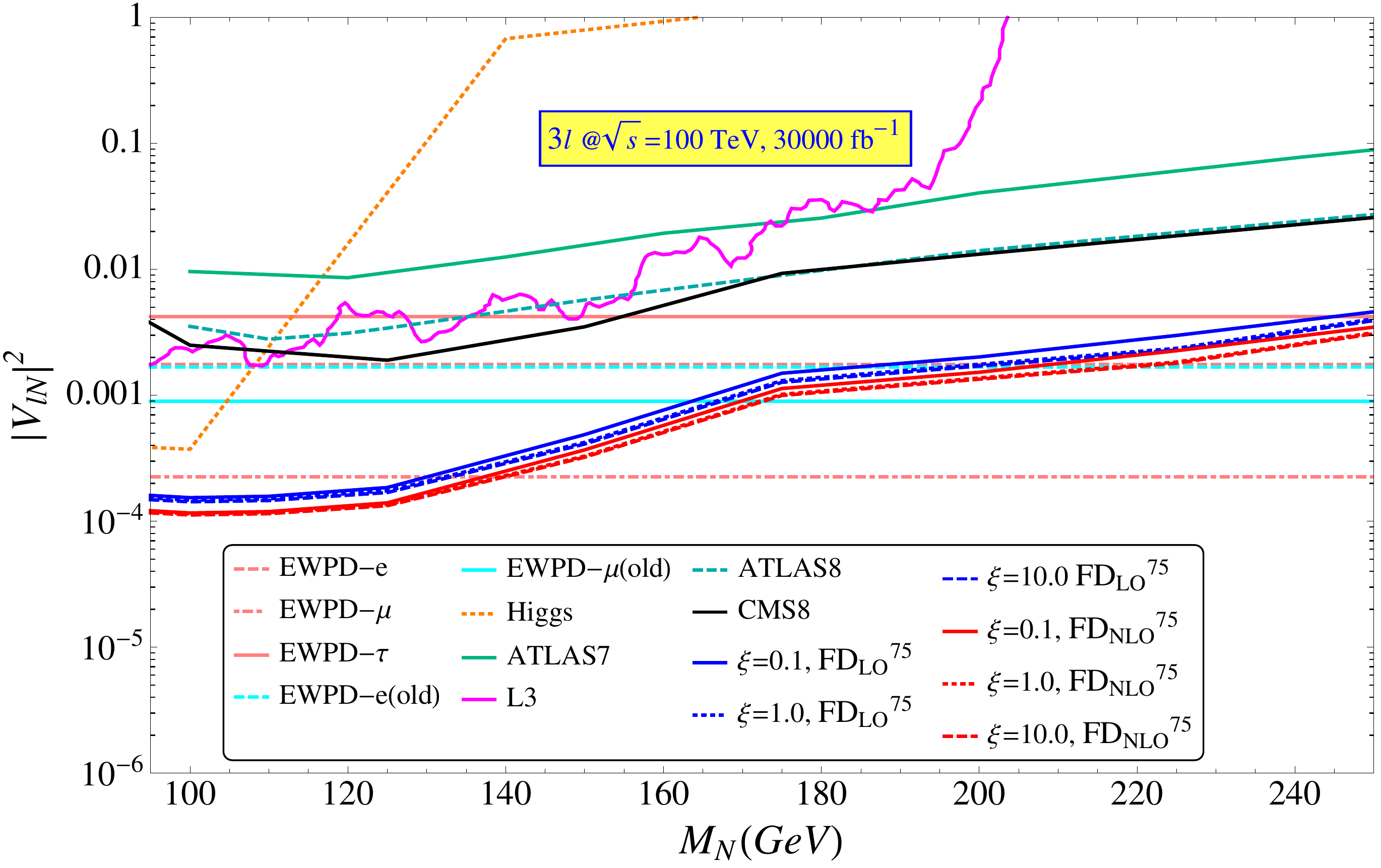}
\end{center}
\caption{The prospective upper bounds on the light-heavy neutrino mixing angles $|V_{\ell N}|^2$ as a function of the right handed heavy pseudo-Dirac neutrino mass $m_{N}$ at with 3000 fb$^{-1}$ luminosity at the 13 TeV HL-LHC (left panel), 100 TeV hadron collider (right panel) and with 30000 fb$^{-1}$ luminosity at the 100 TeV hadron collider (lower panel), derived from the CMS trilepton data at $\sqrt{s}=$8 TeV LHC for 19.5 fb$^{-1}$ luminosity\cite{Chatrchyan:2014aea} at 95 $\%$ CL.
We have considered the scale dependent NLO case$\left(\xi=0.1, 1.0, 10.0\right)$ for the trilepton plus missing energy final state.
 Some relevant existing upper limits (all at 95$\%$ CL) are also shown for comparison: (i) from a $\chi^{2}$-fit to the LHC Higgs data \cite{BhupalDev:2012zg} (Higgs), (ii) from a direct search at LEP \cite{Achard:2001qv}(L3), valid only for the electron flavor, (iii) ATLAS limits from $\sqrt{s}=7$ TeV LHC data \cite{Chatrchyan:2012fla} (ATLAS7) and $\sqrt{s}=$8 TeV LHC data \cite{Aad:2015xaa} (ATLAS8), valid for a heavy Majorana neutrino of the muon flavor, (iv) CMS limits from $\sqrt{s}=$8 TeV LHC data \cite{Khachatryan:2015gha} (CMS8), for a heavy Majorana neutrino of the muon flavor and (v) indirect limits from the global fit to the electroweak precision data (EWPD) from~\cite{deBlas:2013gla, delAguila:2008pw, Akhmedov:2013hec} for electron (cyan, EWPD-e(old)) and muon (cyan, EWPD-$\mu$(old)) flavors(new values can be found from \cite{Antusch:2015gjw}, for tau (dotted, EWPD- $\tau$), electron (solid, EWPD- $e$) and muon (dashed, EWPD- $\mu$) flavors). Here $\rm{FD}^{75}$ is the flavor diagonal cases below the $Z$-pole. }
\label{mix1}
\end{figure}

\begin{figure}
\begin{center}
\includegraphics[scale=0.293]{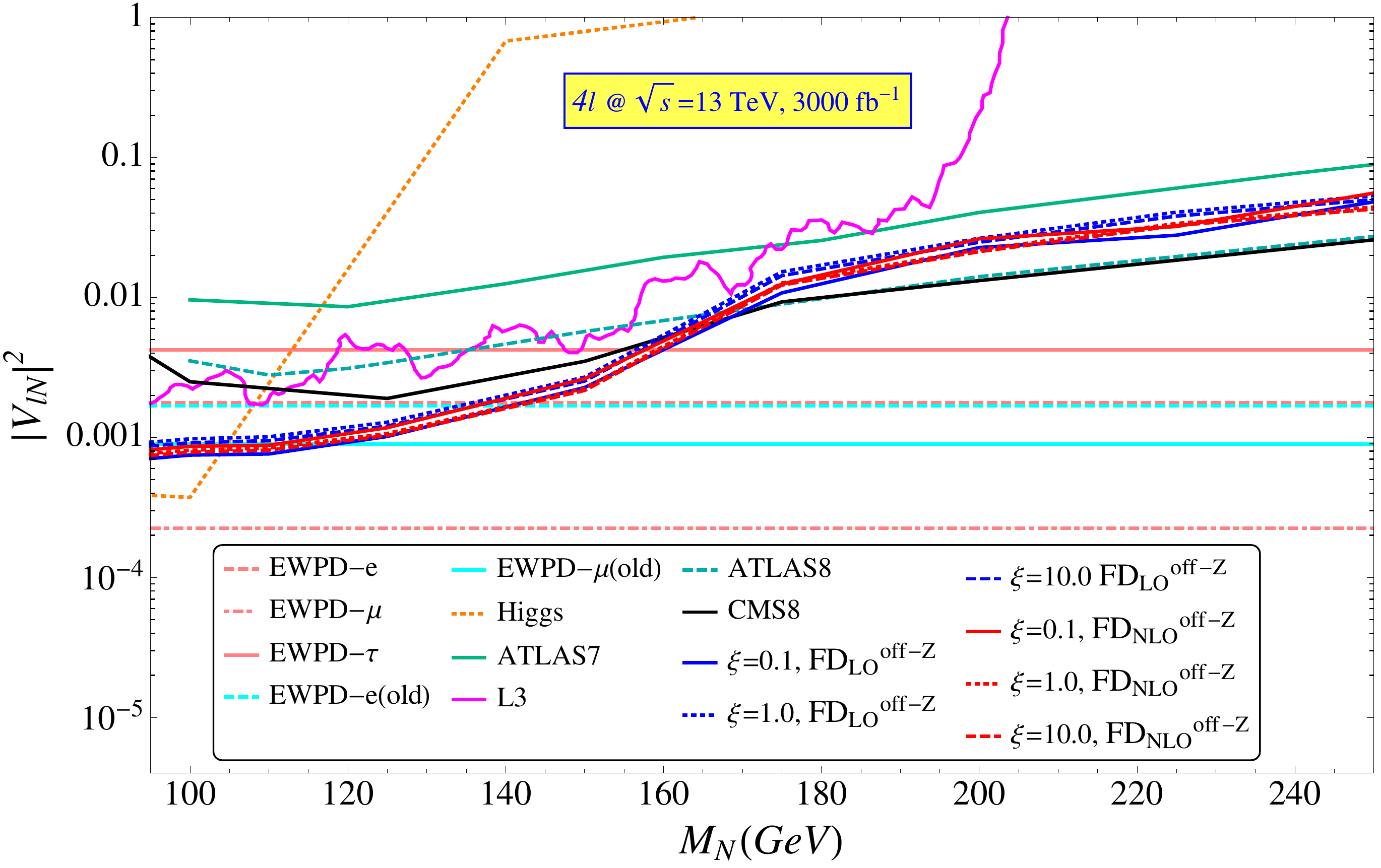}
\includegraphics[scale=0.293]{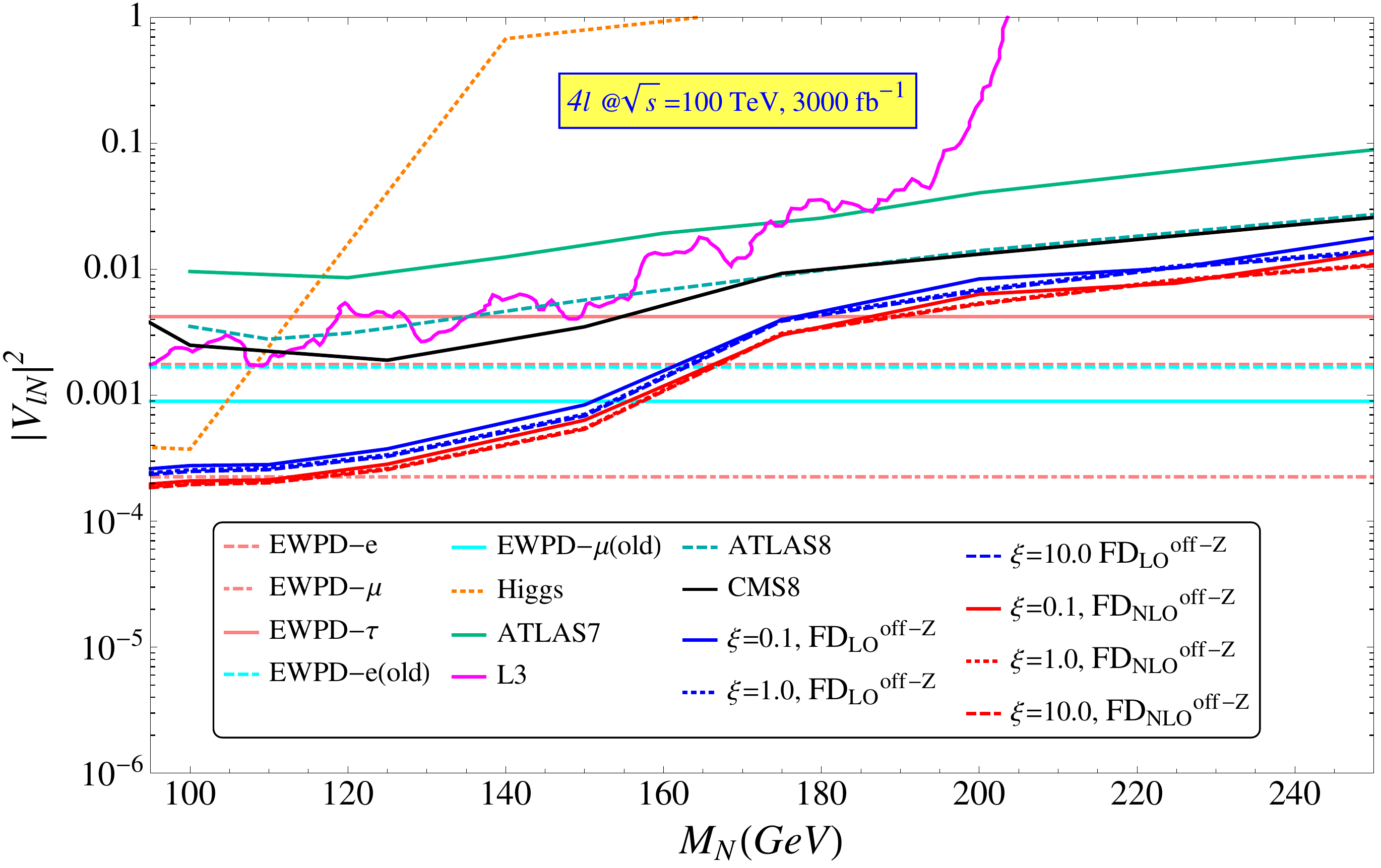}\\
\includegraphics[scale=0.293]{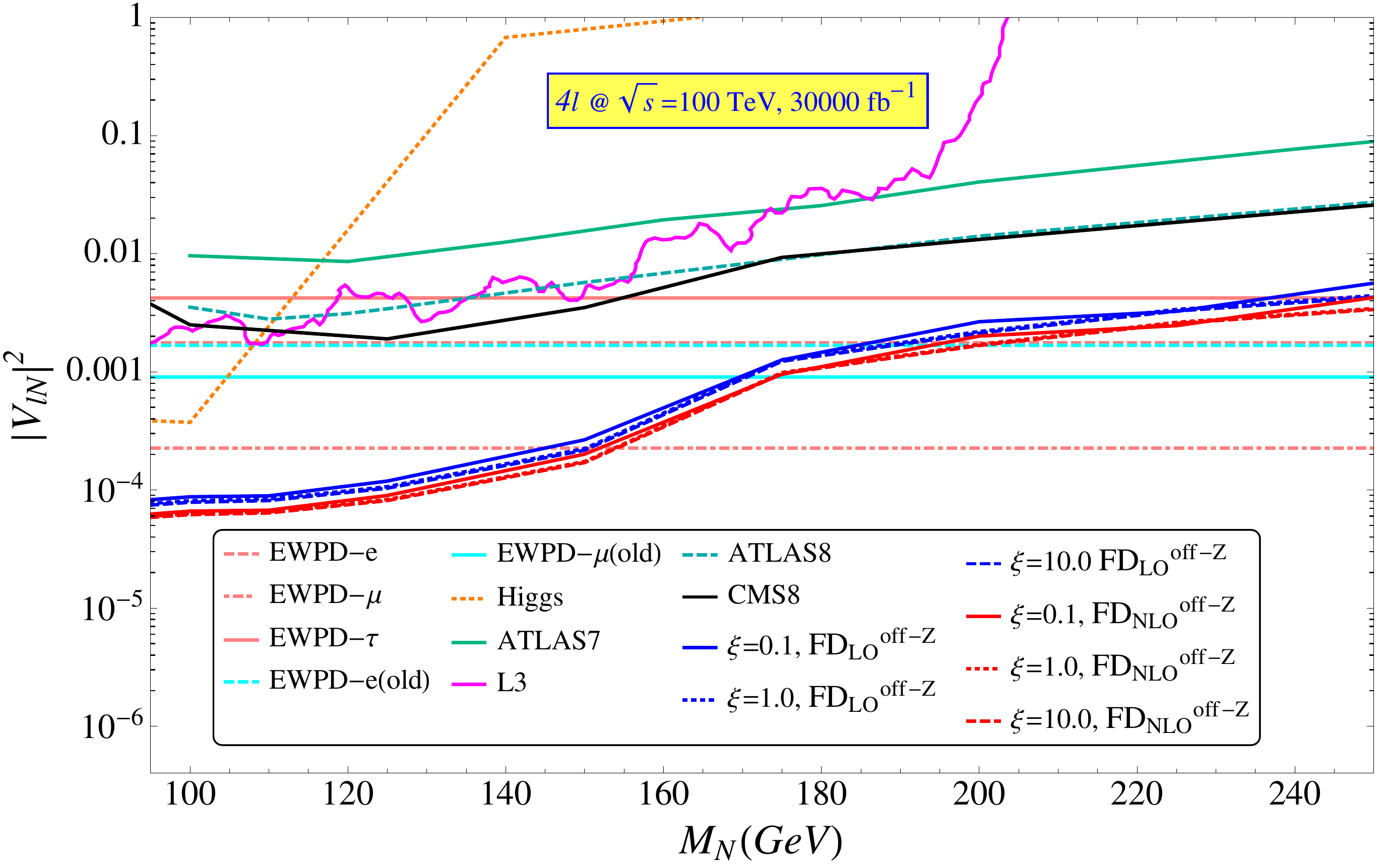}
\end{center}
\caption{The prospective upper bounds on the light-heavy neutrino mixing angles $|V_{\ell N}|^2$ as a function of the right handed heavy pseudo-Dirac neutrino mass $m_{N}$ at with 3000 fb$^{-1}$ luminosity at the 13 TeV HL-LHC (left panel), 100 TeV hadron collider (right panel) and with 30000 fb$^{-1}$ luminosity at the 100 TeV hadron collider (lower panel), derived from the CMS trilepton data at $\sqrt{s}=$8 TeV LHC for 19.5 fb$^{-1}$ luminosity\cite{Chatrchyan:2014aea} at 95 $\%$ CL.
We have considered the scale dependent NLO case$\left(\xi=0.1, 1.0, 10.0\right)$ for the trilepton plus missing energy final state.
 Some relevant existing upper limits (all at 95$\%$ CL) are also shown for comparison: (i) from a $\chi^{2}$-fit to the LHC Higgs data \cite{BhupalDev:2012zg} (Higgs), (ii) from a direct search at LEP \cite{Achard:2001qv}(L3), valid only for the electron flavor, (iii) ATLAS limits from $\sqrt{s}=7$ TeV LHC data \cite{Chatrchyan:2012fla} (ATLAS7) and $\sqrt{s}=$8 TeV LHC data \cite{Aad:2015xaa} (ATLAS8), valid for a heavy Majorana neutrino of the muon flavor, (iv) CMS limits from $\sqrt{s}=$8 TeV LHC data \cite{Khachatryan:2015gha} (CMS8), for a heavy Majorana neutrino of the muon flavor and (v) indirect limits from the global fit to the electroweak precision data (EWPD) from~\cite{deBlas:2013gla, delAguila:2008pw, Akhmedov:2013hec} for electron (cyan, EWPD-e(old)) and muon (cyan, EWPD-$\mu$(old)) flavors(new values can be found from \cite{Antusch:2015gjw}, for tau (dotted, EWPD- $\tau$), electron (solid, EWPD- $e$) and muon (dashed, EWPD- $\mu$) flavors). Here $\rm{FD}^{\rm{off}-Z}$ is the flavor diagonal cases $\rm{off}-Z$-pole. }
\label{mix2}
\end{figure}

To derive the  the limits on $|V_{\ell N}|^{2}$, we calculate the signal cross-section normalized by the fourth power of the mixing angle $(|V_{\ell N}|^4)$ as a function of $m_{N}$ 
in the FD case. Using of the CMS selection criteria listed above for the LO process and the NLO processes using $\xi=0.1,~ 1.0,~10.0$ for the  
detector events from {\tt DELPHES}, we compare them with the observed number of events at the 19.5 fb$^{-1}$ luminosity from Ref.\cite{Chatrchyan:2014aea}.
For the selection criteria listed above, the CMS observed:
\begin{itemize}
\item[(a)] 510 events with the SM background expectation of 560$\pm$87 events for $m_{\ell^{+}\ell^{-}} <$ 75 GeV (Below-$Z$).
\item[(b)] 7 events with the SM background expectation of  8.9$\pm$2.4 events for off-$Z$ (excluding the Z pole by $(m_{Z} \mp 15)~\rm{GeV}$)
\end{itemize}
In case (a) we have an upper limit of 37 signal events and case (b) leads to an upper limit of 0.5 signal events. 
Using these limits, we can set prospective upper bounds on $|V_{\ell N}|^{4}$ as a function of $m_{N}$ for the scale dependent LO and NLO-QCD processes at the HL-LHC and 100 TeV hadron collider for $\xi=0.1,~1.0,~10.0$.
These are displayed in Figs.~\ref{mix1} $(3\ell)$ and \ref{mix2} $(4\ell)$ for simplicity. SF case can easily be estimated diving the FD case by two.
\section{Conclusion}
In this paper we have discussed the possible models for the neutrino mass generation. The type-I and the inverse seesaw models are probably the simplest ideas amongst them to
realize the neutrino mass through the extensions of the SM.

In this paper we have picked up the inverse seesaw model because it could have the order one light-heavy mixing. In the inverse seesaw model
the heavy neutrinos are pseudo-Dirac in nature. Such heavy neutrinos can be produced at the high energy colliders such as 13 TeV HL-LHC and 100 TeV hadron collider with 3000 fb$^{-1}$
and 30000 fb$^{-1}$ luminosities. We have studied specially the neutral current sector where these heavy neutrinos can be produced in a pair through the off-shell $Z$ boson exchange at the hadron colliders.
To produce such processes we have used the LO and NLO-QCD processes where the factorization and renormalization scales were varying from a low value to a high value. Such production cross sections
will be proportional to the fourth power of the mixing angle. We have noticed that the NLO-QCD scale dependence is softer compared to that of the LO scale variation due to the presence of the strong coupling constant such as, $\alpha_s(\mu_{R})$. 
There is one more very clear reason to consider the inverse seesaw model. In this model framework it is only possible to study the FD case. This is not possible in the conventional type-I seesaw model.

 In this paper we have only considered the leading decay mode of the heavy neutrino which can be realized as $N \to W\ell$. Depending upon the hadronic and leptonic
decays of the $W$ boson we get several final states containing two, three and four leptons. We have showed the distributions of different kinematic variables of the leptons, jets,
and the missing transverse energy. We have noticed that the NLO-QCD accuracy is very stable with respect to the large scale variations while the LO suffered from 
substantial changes.

From the $3\ell+2 j+ \rm{MET}$ signal state the jets are coming from the hadronic decay of the $W$, we compare our final state with the anomalous multilepton search made by CMS at the 8 TeV.
Using the selection cuts used by CMS, we have obtained the prospective search reach on $|V_{\ell N}|^{2}$ at the HL-LHC with 13 TeV center of mass energy and 3000 fb$^{-1}$ luminosity. We have also studied the same channel for the proposed 100 TeV hadron collider at the 3000 fb$^{-1}$ and 30000 fb$^{-1}$ luminosities. At the 13 TeV, the result will be comparable to the EWPD for the $\mu$ flavor where as it has a very good possibility to be improved up to an order of magnitude at the 100 TeV so that it can better than the EWPD. The NLO-QCD processes are more stable under the scale variation in comparison to the LO processes.

We have also studied the $4\ell+\rm{MET}$ final state for the HL-LHC at the 13 TeV collider energy at the 3000 fb$^{-1}$ luminosity and  at the 100 TeV hadron collider at the luminosities of 3000 fb$^{-1}$ and 30000 fb$^{-1}$ respectively. 
Comparing the anomalous multilepton search made by CMS at the 8 TeV and the using the selection cuts we obtain the prospective search reaches at the high energy colliders. The 13 TeV result is comparable to the EWPD with the $\mu$ flavor whereas at the 100 TeV collider the search reach can be one order of magnitude improved. The NLO processes are quite stable with respect to the scale variation compared to the LO processes. 

From such channels we can probe the low mass heavy neutrino while the mass of the heavy neutrino ranges between $95$ GeV - $160$ GeV. Currently the LHC is at the phase of Run-II and it has a good prospect at the high luminosity era where a dedicated search for such interactions and the final states can reveal more improved features. Of course, the proposed 100 TeV collider will be a very appropriate machine to probe 
such multilepton channels in future.

In the upcoming paper we will discuss the prospects of the pair-production of the heavy neutrinos followed by its various decay modes at the different high energy colliders from the other extensions of the SM. We will also adress the possible BSM phenomenologies of such models in \cite{Das-Okada}.
\bigskip
\acknowledgments
The work of the author is supported by the Korea Neutrino Research Center which is established by the National Research Foundation of Korea(NRF) grant funded by the Korea government(MSIP) (No. 2009-0083526). The author would like to thank the Theory Division of Saha Institute of Nuclear Physics, Kolkata, India and the organizers of the `III Saha Theory Workshop: Aspects of Early Universe Cosmology' for the hospitality at the final stage of this work.

   \end{document}